\documentclass[a4paper,11pt]{article}
\pdfoutput=1 

\usepackage{jheppub}
\usepackage[T1]{fontenc} 

\usepackage[dvipsnames]{xcolor}
\usepackage{tikz}
\usetikzlibrary{patterns}
\usetikzlibrary{snakes}

\usepackage{enumitem}
\usepackage{mathrsfs}

\newcommand{\bei}{\begin{itemize}}
\newcommand{\eei}{\end{itemize}}
\newcommand{\bee}{\begin{enumerate}}
\newcommand{\eee}{\end{enumerate}}
\newcommand{\beeL}{\begin{enumerate}[label=(\Alph*)]}
\newcommand{\beel}{\begin{enumerate}[label=(\alph*)]}
\newcommand{\beeR}{\begin{enumerate}[label=(\Roman*)]}
\newcommand{\beer}{\begin{enumerate}[label=(\roman*)]}
\newcommand{\beeLd}{\begin{enumerate}[label=\Alph*.]}
\newcommand{\beeld}{\begin{enumerate}[label=\alph*.]}
\newcommand{\beeRd}{\begin{enumerate}[label=\Roman*.]}
\newcommand{\beerd}{\begin{enumerate}[label=\roman*.]}

\newcommand{\bal}{\begin{equation}\begin{aligned}}
\newcommand{\eal}{\end{aligned}\end{equation}}

\newcommand{\ov}{\over}
\newcommand{\g}{\gamma}
\newcommand{\ubr}{\nu}
\newcommand{\lint}{\int\limits}
\newcommand{\xbr}{\xi}

\def\sgn{{\text{sgn}}}

\definecolor{grey}{rgb}{0.4,0.4,0.5}
\definecolor{darkgreen}{rgb}{0,0.5,0}
\definecolor{darkred}{rgb}{0.6,0.0,0}
\definecolor{lightbrown}{rgb}{1,0.9,0.8}
\definecolor{brown}{rgb}{0.6,0.3,0.3}
\definecolor{darkblue}{rgb}{0,0,0.5}
\definecolor{darkmagenta}{rgb}{0.5,0,0.5}

\def\tx{{\tilde x}}
\def\ty{{\tilde y}}
\def\tg{{\tilde\gamma}}

\def\veps{{\varepsilon}}
\def\tPsi{{\tilde\Psi}}
\def\tPhi{{\tilde\Phi}}
\def\tchi{{\tilde\chi}}
\def\barnes{{\text{\tiny odd}}}
\def\besratio{{\text{\tiny even}}}

\newcommand{\adsst}{${\rm  AdS}_3\times {\rm S}^3\times {\rm T}^4\ $}

\def\bR{{\mathbb R}}
\def\bZ{{\mathbb Z}}

\newcommand{\la}{\label}
\def\a {\alpha}
\def\b {\beta}

\def\ors{{\text{\tiny ORS}}}
\def\cdd{{\text{\tiny CDD}}}
\def\bes{{\text{\tiny BES}}}
\def\afs{{\text{\tiny AFS}}}
\def\hl{{\text{\tiny HL}}}
\def\tp{{\widetilde p}}
\def\tE{\widetilde\E}

\newcommand{\E}{\mathcal E}

\renewcommand{\L}{{\scriptscriptstyle\text{L}}}
\newcommand{\R}{{\scriptscriptstyle\text{R}}}

\def\ka {{\kappa}}

\def\s{\sigma}

\def\pa {\partial}

\def\cO{{\cal O}}

\def\cR{{\cal R}}

\def\eps{{\epsilon}}
\newcommand{\de}{\text{d}}

\def\tR{{\tilde R}}

\title{\boldmath Massive dressing factors for mixed-flux AdS$_3$/CFT$_2$}

\author[1,4]{Sergey Frolov,}
\author[2,3,4]{Davide Polvara,}
\author[3,4]{Alessandro Sfondrini}

\affiliation[1]{Hamilton Mathematics Institute and School of Mathematics Trinity College, Dublin 2, Ireland.}
\affiliation[2]{II. Institut f\"ur Theoretische Physik,  Universit\"at Hamburg, Luruper Chaussee 149, 22761 Hamburg, Germany.}
\affiliation[3]{Dipartimento di Fisica e Astronomia, Universit\`a degli Studi di Padova, via Marzolo 8,
35131 Padova, Italy.}
\affiliation[4]{Istituto Nazionale di Fisica Nucleare, Sezione di Padova, via Marzolo 8, 35131 Padova,
Italy.}

\emailAdd{frolovs@maths.tcd.ie}
\emailAdd{davide.polvara@gmail.com}
\emailAdd{alessandro.sfondrini@unipd.it}

\abstract{We follow up on our proposal for dressing factors for the mixed-flux $AdS_3\times S^3\times T^4$ background presented in arXiv:2402.11732. We discuss in detail the analytic properties of the dressing factors in the string and mirror kinematics for fundamental massive particles and bound states. We prove that the dressing factors are unitary and CP-invariant in the string kinematics, parity invariant in the mirror, and solve the crossing equations in both kinematics. In the limit of pure Ramond-Ramond flux they reduce to the known ones.
Finally, we present their expansion at strong tension, as well as in the (small-RR-flux) relativistic limit, finding agreement with the literature.
}

\begin{document} \begin{flushright}\small{ZMP-HH/25-1}\end{flushright}
\maketitle
\section{Introduction}
\label{sec:introduction}
Strings on $AdS_3$ backgrounds are an important example of the holographic correspondence since its discovery~\cite{Maldacena:1997re}.%
\footnote{%
This holographic setup has a long history, which is by now well reviewed. An early review of the holographic setup is~\cite{David:2002wn}. Recent reviews of the integrability approach are~\cite{Demulder:2023bux} for the ``pure Ramond-Ramond'' backgrounds; for what concerns mixed-flux and ``tensionless'' models, another recent review is~\cite{Seibold:2024qkh}.
}
With respect to the best studied holographic setup, the correspondence between type-IIB $AdS_5\times S^5$ strings and $\mathcal{N}=4$ $SU(N_c)$ supersymmetric Yang-Mills theory, they are less supersymmetric (with 16 rather than 32 real supercharges); this makes them both more interesting and harder to study. To begin with, there are several families of maximally supersymmetric $AdS_3$ backgrounds, all taking the form $AdS_3\times S^3\times \mathcal{M}^4$, where we can have $\mathcal{M}^4=T^4$, $\mathcal{M}^4=K3$ or $\mathcal{M}^4=S^3\times S^1$. Among these families, the $T^4$ case is the most studied and best understood.%
\footnote{%
The integrability approach which we employ here works also for the $S^3\times S^1$ background, see~\cite{Sfondrini:2014via} for a review, but it has yet to be fully developed in that case.}
Even when restricting to this case, the physics is quite a bit richer than in the $AdS_5\times S^5$ case. This can be seen from the Green-Schwarz action of the model, which schematically we can write as
\begin{equation}
    S=-\frac{T}{2}\int\de^2\sigma \left[\left( \sqrt{|\gamma|}\gamma^{ab} G_{\mu\nu}(X) +q\varepsilon^{ab} B_{\mu\nu}(X)\right)\partial_aX^\mu\partial_bX^\nu+\text{Fermions}\right],
\end{equation}
where $\gamma^{ab}$ is the inverse-metric on the worldsheet, $\varepsilon^{ab}$ is the Levi-Civita tensor, $G_{\mu\nu}(X)$ is the $AdS_3\times S^3\times T^4$ metric and $B_{\mu\nu}(X)$ is the Kalb-Ramond field.
In perturbative string theory, $T>0$ is the dimensionless tension of the string (expressed in units of the radius of $AdS_3$), which is a continuous parameter. The coefficient of the Kalb-Ramond field must satisfy $-1\leq q\leq 1$, and moreover it is quantised,
\begin{equation}
\label{eq:WZquantised}
    q\,T=\frac{k}{2\pi}\,,\qquad k\in\mathbb{Z}\,.
\end{equation}
To ensure that this background is a solution of the supergravity equations, we have both a $H=\de B$ three-form flux, and a Ramond-Ramond  (RR)  three-form flux. Both are proportional to the sum of the volume forms $AdS_3$ and $S^3$, and the proportionality coefficients go like $q$ and $\sqrt{1-q^2}$, respectively. In other words, we can have a supergravity background with 16 Killing spinors  supported by a mixture of Neveu-Schwarz-Neveu-Schwarz (NSNS) flux and RR flux. This corresponds to the fact that the background can be obtained from the near-horizon limit of a system of D1-D5-F1-NS5 branes.%
\footnote{%
For an overview of the brane construction and its moduli, see~\cite{OhlssonSax:2018hgc} and references therein.}
It turns out that at large tension
\begin{equation}
    T=\sqrt{h^2+\frac{k^2}{4\pi^2}}\,,\qquad k=0,1,2,\dots,\qquad h\geq0\,,
\end{equation}
where the ``coupling'' $k$ is related to the strength of the $B$-field (without loss of generality, we picked a sign for $k$), while the ``coupling'' $h=T\sqrt{1-q^2}+\mathcal{O}(T^{-1})$ is related to the strength of the RR fluxes. Much like it happens for $AdS_5\times S^5$, $h$ is a continuous parameter in perturbative string theory.

There are two special cases in the above setup. If $h=0$, we find a discrete sequence of backgrounds supported by NSNS fluxes only. They can be studied in detail in the RNS formalism~\cite{Maldacena:2000hw}, by means of the current algebras of the worldsheet CFT, which is a supersymmetric WZW model. The relative simplicity of the worldsheet CFTs allowed to conjecture the dual CFTs of the pure-NSNS points~\cite{Eberhardt:2018ouy,Eberhardt:2021vsx}.
If on the other hand $k=0$, we obtain something similar to the $AdS_5\times S^5$ background: the tension is continuous, and it should be identified with some suitable ``'t Hooft coupling'' of the so-far unknown dual CFT.%
\footnote{%
It is a bit improper to talk of a 't~Hooft coupling, firstly because we do not really understand how the large-$N$ expansion should be implemented in the dual CFT, which should be related to a sigma model on an instanton moduli space, see e.g.~\cite{David:2002wn} for a review. Secondly, even once a 't~Hooft-like coupling~$\lambda$ has been identified, it is likely that $h$ should play the role of an interpolating function as it is the case in $AdS_4/CFT_3$, \textit{i.e.}\ $h=h(\lambda)$ is some monotonic function to be determined, with $h(\lambda)\ll 1$ if $\lambda\ll1$ and $h(\lambda)\gg 1$ if $\lambda\gg1$.
} 
In the most general case, $k=1,2,3,\dots$ and $h>0$, we have several one-parameter families of backgrounds --- each of which arising from a pure-NSNS model by turning on a ``'t Hooft coupling''.%
\footnote{%
In the string language, we can turn on the RR fluxes by means of an axion in the brane configuration, see~\cite{OhlssonSax:2018hgc}.
}
In the presence of RR fluxes, it is very difficult to study these models in the RNS formalism. Fortunately and remarkably, the classical GS action is integrable for any value of $k=0,1,2, \dots $ and $h\geq0$~\cite{Cagnazzo:2012se}. This paved the way to solving the free-string spectrum of these models non-perturbatively in $h,k$ in the same spirit as it was done for the spectrum of $AdS_5\times S^5$ strings (\textit{i.e.}, the planar spectrum of local operators in $\mathcal{N}=4$ SYM), see~\cite{Arutyunov:2009ga,Beisert:2010jr} for reviews.

The integrability approach to $AdS_3\times S^3\times T^4$ strings  is by now well reviewed and we will summarise it only briefly, referring the reader to the recent works~\cite{Demulder:2023bux,Seibold:2024qkh} for details. One first considers the GS strings in a suitable (``uniform'') lightcone gauge; this yields a two dimensional non-relativistic QFT in finite volume. The form of the dispersion relation is fixed by (super)symmetry, and in our case it takes the peculiar form~\cite{Hoare:2013lja, Lloyd:2014bsa}
\begin{equation}
\label{eq:dispersion}
    E(p) = \sqrt{\left(M+\frac{k}{2\pi}p\right)^2+4h^2\sin^2\left(\frac{p}{2}\right)}\,.
\end{equation}
This formula represents the energy of a single excitation on the string worldsheet,%
\footnote{%
For comparison,  for superstrings in flat space  this would simply read $E(p)=c\,|p|$ for some constant $c>0$, while for $AdS_5\times S^5$ superstrings it takes the same form as here, with $k=0$ and $M=1,2,3,\dots$.}
where $M\in\mathbb{Z}$ is the sum of the particle's spin in $AdS_3$ and in $S^3$. The problem of finding the finite-volume spectrum is then broken into two steps: determining the S~matrix which scatters two excitations of momenta $(p_1,p_2)$, and writing the ``mirror'' thermodynamic Bethe ansatz (TBA) equations for that S~matrix, which can then be solved numerically.%
\footnote{%
The mirror TBA equations involve an infinite number of unknown functions, the so-called Y-functions. These equations can be sometimes simplified to a set finitely-many relations between a handful of ``Q-functions''. The latter construction has been dubbed ``quantum spectral curve''~\cite{Gromov:2013pga}.
}
The mirror model is related to the original one by a double Wick rotation,~\cite{Arutyunov:2007tc}
\begin{equation}
    E\to i\tilde{p}\,,\qquad p\to i\tilde{E}\,,
\end{equation}
necessary to implement Zamolodchikov's proposal~\cite{Zamolodchikov:1989cf} to relate the finite-volume properties of the model to a finite-temperature computation. Note that such a double-Wick rotation is irrelevant for relativistic models, but it changes drastically the dispersion~\eqref{eq:dispersion}. For consistency of the mirror TBA machinery, the S matrix of the mirror model must be related to the one of the original model (which we call ``string'' model) by a suitable analytic continuation.

For $AdS_3\times S^3\times T^4$ this program has been completed for pure-NSNS backgrounds with $h=0$~\cite{Dei:2018mfl} (where one recovers the worldsheet-CFT results), as well as for pure-RR backgrounds ($k=0$) where one finds a set of mirror TBA~\cite{Frolov:2021bwp} or QSC~\cite{Ekhammar:2021pys,Cavaglia:2021eqr} equations that must then be studied numerically~\cite{Cavaglia:2022xld,Brollo:2023pkl}.
For the mixed-flux case, instead, the program has been delayed by the difficulty in completely determining the worldsheet S~matrix. Most of it is fixed by the symmetries which are linearly realised in the worldsheet lightcone-gauge-fixed theory~\cite{Lloyd:2014bsa}, but some overall prefactors --- the \textit{dressing factors} --- cannot be determined in that way, but have to be fixed by imposing crossing symmetry and unitarity.%
\footnote{This is also the case in relativistic integrable QFTs, but for those models it is relatively simple to fix the dressing factors, up to making appropriate assumptions on the bound-state structure of the model.}
For instance, in the case of $AdS_5\times S^5$ superstrings, whose kinematic is given by~\eqref{eq:dispersion} with $k=0$ and $M=1$, the dressing factor is the celebrated (and highly non-trivial) Beisert-Eden-Staudacher (BES) factor~\cite{Beisert:2006ez}.  
For $AdS_3\times S^3\times T^4$ superstrings at $k=0$, the solution involves the BES factor, supplemented by additional functions with a non-trivial cut structure~\cite{Frolov:2021fmj}. The aim of this paper is to propose the dressing factors \textit{of the mixed-flux model} and perform a series of checks, such as consistency with unitarity, crossing symmetry, and other symmetries of the model as well as compatibility with perturbative computations and special limits.
Moreover, we demand that the ``mirror'' and ``string'' model are related by analytic continuation, whose form we detail, and that \textit{both models} have the necessary self-consistency properties. As a matter of fact, we will construct the dressing factors \textit{starting from the mirror model}, which yields rather convenient integral representation of the dressing factors.

The paper is structured as follows.
In section~\ref{sec:kinematics} we report the kinematical variables used to check the different properties of the string and mirror models. In section~\ref{sec:properties} we list all the properties that the theory must satisfy. In section~\ref{sec:proposal} we present our proposal for the dressing factors of string and mirror particles. 
In section~\ref{sec:verification} we check our proposal against the constraints detailed in section~\ref{sec:properties}. In section~\ref{sec:expansions} we compare our dressing factors with different perturbative computations carried out in the literature. In section~\ref{sec:CDDs} we comment on possible modifications of our proposal through CDD factors. Finally, we conclude in section~\ref{sec:conclusions}, presenting our results and open problems. The paper includes a large number of appendices where we report the technical details of the computations.

\section{Kinematics of the string and mirror model}\la{sec:kinematics}

To formulate and solve the crossing equations for the string and mirror models we use various variables generalising the ones for the pure Ramond-Ramond theory. 
These variables are also needed to prove that the mirror and string models are invariant under parity (P) and a combination of parity and charge conjugation (CP), respectively.

\subsection{Zhukovsky and \texorpdfstring{$u$}{u}-rapidity variables}

Let us recall that the $\ka$-deformed Zhukovsky map is defined through the following equation
\cite{Hoare:2013lja,Frolov:2023lwd} 
\begin{equation}\begin{aligned}
\label{uplane}
u_a(x)&=x+\frac{1}{x}- \frac{\kappa_a}{\pi}\,\ln x \,.
\end{aligned}\end{equation}
Here and in what follows $a=\,$L, R and  if $a=\,$L then $\bar a=\,$R and vice versa, and $\ka_\L=\ka$, $\ka_\R=-\ka$, $\ka\equiv\tfrac{k}{h}>0$.  
The map satisfies the important relation
\begin{equation}
   u_{\bar a}({1/x}) = u_{ a}(x)\,,
\end{equation}
 and therefore, if $x$ solves the equation $u_a(x)=u$ then $1/x$ solves  $u_{\bar a}(x)=u$. We refer to this property of the equation as  \textit{inversion symmetry}. 
As a result, $\ka$-deformed inverse Zhukovsky functions have branch points on a $u$-plane, and we use the two cut structures described in \cite{Frolov:2023lwd}
\begin{equation}\begin{aligned}
\label{uplane2}
u_a(x)&=u  \qquad\Longleftrightarrow \qquad 
x=
\begin{cases}
x_a(u)        \\
\displaystyle
\frac{1}{x_{\bar a}(u)}       
\end{cases}
\qquad \text{or}\qquad x=
\begin{cases}
\tx_a(u)        \\
\displaystyle
\frac{1}{\tx_{\bar a}(u)}       
\end{cases}
\,.
\end{aligned}\end{equation}
Here $x_a(u)$ and $\tx_a(u)$ are the {\it string} and  {\it mirror} inverse 
Zhukovsky functions, respectively. In the limit $\ka\to0$ they become the usual inverse 
Zhukovsky functions with the short cut $(-2,2)$ and  the long cut $(-\infty,-2)\cup (2,\infty)$, respectively.

The potential branch points of these functions correspond to zeroes and poles of $dx/du$  
\bal
{dx_a\ov du} = {x^2\ov (x- \xi_a) (x+{1\ov\xi_a})}\,,\qquad \xi_\L =\xi\,,\quad \xi_\R={1\ov \xi}\,,\qquad  \xi\equiv {\ka\ov2\pi}+ \sqrt{1+{\ka^2\ov4\pi^2}}\,.
\eal 
The point located at
\bal
\label{eq:ubranches}
\ubr = \xi_a+\frac{1}{\xi_a}- {\kappa_a\ov\pi}\,\ln \xi_a &= + 2\sqrt{{\kappa ^2\ov4\pi^2}+1}-{\kappa\ov\pi}  \ln \left({\ka\ov2\pi}+ \sqrt{1+{\ka^2\ov4\pi^2}}\right) 
\eal
 is the branch point  of all the four functions, and it is mapped to the point $\xi_a$ on the $x$-plane: $x_a(\ubr)=\tx_a(\ubr)=\xi_a$. It is of the square-root type, and  going around it  $x_a(u)$ and $\tx_a(u)$ transform  according to the inversion symmetry
\begin{equation}\begin{aligned}
x_a^\circlearrowleft(u)=\frac{1}{x_{\bar a}(u)}\,,
\qquad \tx_a^\circlearrowleft(u)=\frac{1}{\tx_{\bar a}(u)}\,,
\end{aligned}\end{equation}
where $x^\circlearrowleft$ is the result of the analytic continuation along a path $\circlearrowleft$ surrounding  $\nu$. 

\begin{figure}
\begin{center}
\includegraphics[width=5cm]{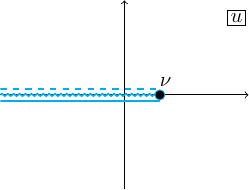}%
\hspace{1cm}%
\includegraphics[width=5cm]{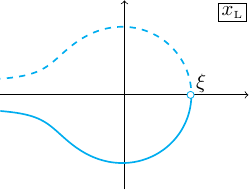}%
\end{center}
\caption{\label{fig:leftplanes}%
Left panel: The branch point (solid dot) and cut (zigzag line) of $x_{\L}(u)$ on the $u$-plane; The cut's edges (solid and dashed lines) are mapped to an unbounded curve in the $x$-plane (right panel).}
\end{figure}

\begin{figure}
\begin{center}
\includegraphics[width=5cm]{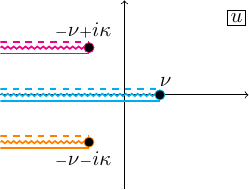}%
\hspace{1cm}%
\includegraphics[width=5cm]{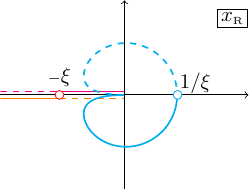}%
\end{center}
\caption{\label{fig:rightplanes}%
Left panel: The three branch points (solid dots) and cuts (zigzag line) of $x_{\R}(u)$ on the $u$-plane. We call the cyan cut the ``main'' one.  Its edges (solid and dashed cyan lines) are mapped to a bounded, open curve in the $x$-plane, while the edges of the other two  cuts are mapped to the half-line in the $x$-plane (right panel). The images of the branch points are denoted by empty dots.}
\end{figure}

\begin{figure}
\centering
\includegraphics[width=5cm]{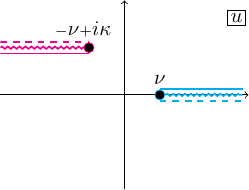}%
\hspace{5mm}%
\includegraphics[width=5cm]{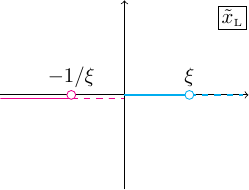}\\%
\includegraphics[width=5cm]{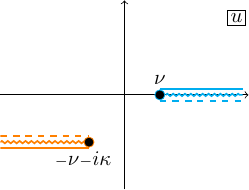}%
\hspace{5mm}%
\includegraphics[width=5cm]{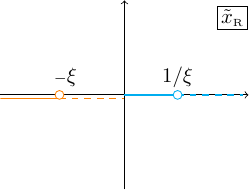}%
\caption{\label{fig:rightmplanes}%
Left: The two branch points (solid dots) and cuts (zigzag line) of $\tx_{\L}(u)$ and $\tx_{\R}(u)$ on the $u$-plane, respectively.
We call the cyan cut the ``main'' one.  Its edges (solid and dashed cyan lines) are mapped to the positive semi-axis in the $x$-plane, while the edges of the second cut are mapped to the lower edge of the $\ln x$ cut (right panels). The images of the branch points are denoted by empty dots.}
\end{figure}

Since $-1/\xi_a$ is negative for $\ka$ real, there may be two branch points located at
\bal
\ubr_{\pm} =-\ubr  \pm i\ka\,.
\eal
As was shown in \cite{Frolov:2023lwd}, $x_\L(u)$ does not have any of those branch points, $x_\R(u)$ has both of them with $x_\R(\ubr_\pm)=-\xi \pm i0$, then $\tx_\L(u)$ has the branch point $\nu_+$ with $\tx_\L(\ubr_+)=-{1/\xi}-i0$, and finally $\tx_\R(u)$ has the branch point $\ubr_-$ with  $\tx_\R(\ubr_-)=-\xi-i0$.
Since the images of 
these two branch points are on the edges of the cut of $\ln x$, moving a point around any of them takes it to a different $x$-plane. These  branch points are also
 of the square-root type, and going around $\ubr_{\pm}$ along a path $\circlearrowleft_{\pm}$ transforms  the $\ka$-deformed  inverse Zhukovsky functions as follows
\bal\label{eq:xatoxakacut}
x_\R^{{\circlearrowleft }_{\pm}}(u)= x_\R(u\mp 2i\ka)\,,\quad \tx_\L^{{\circlearrowleft }_{+}}(u)= {1\ov \tx_\R(u- 2i\ka)}\,,\quad  \tx_\R^{{\circlearrowleft }_{-}}(u)={1\ov \tx_\L(u+ 2i\ka)}\,.
\eal
Finally, there is a branch point at $u=\infty$ corresponding to $x=0$ and $x=\infty$ which is of the logarithmic type. The result of the analytic continuation along a path   surrounding $u=\infty$ depends on the cut structure of a $u$-plane, the orientation of the path and the initial point of the path. The $x$ and $u$ planes for the string and mirror Zhukovsky variables ($x_\L$, $x_\R$, $\tx_{\L}$ and $\tx_\R$) are shown in Figures~\ref{fig:leftplanes}, \ref{fig:rightplanes} and \ref{fig:rightmplanes}.
 

\smallskip

 Let us now summarise the defining properties of the $\ka$-deformed  inverse Zhukovsky functions. We always choose all cuts on a $u$-plane to be horizontal. Then,
the string functions $x_a(u)$ satisfy the condition  $x_a(+\infty)=+\infty$,  and they map their $u$-planes to the \textit{string physical regions} of  left and right particles.\footnote{A very different cut structure of $x_a(u)$ was used in \cite{OhlssonSax:2023qrk}. Their identification of the physical regions also dramatically differs from ours.}. The function $x_\L(u)$ has one cut on its $u$-plane which runs from $-\infty$ to $\ubr$ (Figure~\ref{fig:leftplanes}) while $x_\R(u)$ has three cuts on its $u$-plane which run from $-\infty$ to $\ubr$,  from $-\infty$ to $-\ubr +i\ka$ and  from $-\infty$ to $-\ubr -i\ka$  (Figure~\ref{fig:rightplanes}). 
We refer to the common cut from $-\infty$ to $\ubr$ as the main string cut, and the cuts  from $-\infty\pm i\ka$ to $\ubr_\pm$ as the $\pm\ka$-cuts. The cuts of the string $u$-planes are mapped to the boundaries $\pa\cR_\L$ and  $\pa\cR_\R$ of the physical regions $\cR_\L$ and $\cR_\R$ of left and right particles, respectively. 
Moving through the main cut one gets to an antistring $u$-plane which is mapped to an anti-string region $\cR_a^{\rm as}$. Since the regions and the boundaries are mapped to each other under the inversion map $x\to 1/x$, we can use $x_\L(u)$ and $x_\R(u)$ to cover the whole $x$-plane with the $\ln x$ cut from $-\infty$ to 0. The functions $x_a(u)$ satisfy the complex conjugation condition
\bal
x_a(u)^* = x_a(u^*)\,,
\eal
which is the same as in the pure-RR case. 
Then, the function $x_\L(u)$ has a very simple asymptotic behaviour at large $u$
\bal\la{eq:xL_large_u}
x_\L(u) = u+ {\ka\ov\pi}\ln u+\frac{\kappa ^2 \log u}{\pi ^2u}+ \cdots \,,\quad |u|\gg 1\,,
\eal
independent of the direction. 
\begin{figure}
\centering
\includegraphics[width=5cm]{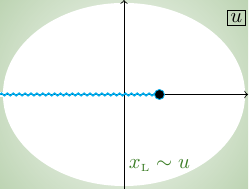}%
\hspace{5mm}%
\includegraphics[width=5cm]{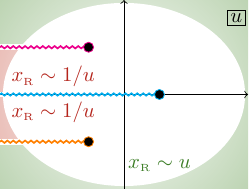}\\%
\includegraphics[width=5cm]{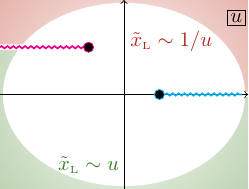}%
\hspace{5mm}%
\includegraphics[width=5cm]{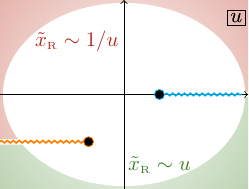}%
\caption{\label{fig:x-asymptotics}%
The asymptotic behaviour of $x_a(u)$ and $\tx_a(u)$ at large~$|u|$ depends on the cut structure of the functions, as evidenced by the shading in different colours.
The top-left and top-right panels depict the string variables and correspond to eq.~\eqref{eq:xL_large_u} and eq.~\eqref{eq:xR_large_u}, respectively. The bottom-left and bottom-right panels depict the mirror variables and correspond to eq.~\eqref{eq:txL_large_u} and~\eqref{eq:txR_large_u}, respectively.}
\end{figure}
On the other hand the asymptotic behaviour of $x_\R(u)$ depends on the direction, and for $|u| \gg 1$ we find
\bal\la{eq:xR_large_u}
x_\R(u) = 
\begin{cases}
{1\ov u} - {\ka\ov\pi u^2}\ln u+{2i\ka\ov u^2}\sgn(\Im(u))+ \cdots  &\text{if}  \   |\Im(u)|<\ka\ \text{and}\   \Re(u)\ll -1\,,  \\
u - {\ka\ov\pi}\ln u+\frac{\kappa ^2 \log u}{\pi ^2u}+ \cdots  &\ \ \text{otherwise}\,,
\end{cases}
\eal
see Figure~\ref{fig:x-asymptotics}.
The different asymptotic behaviour is clearly related to the different number of cuts of the functions $x_a(u)$.

The mirror functions $\tx_a(u)$ satisfy the condition  $\Im\big(\tx_a(u)\big)<0$,  and they map their $u$-planes to the mirror physical region of  left and right mirror particles that coincides with the lower half-plane. Each of the mirror functions has two cuts. The function $\tx_\L(u)$ has the main mirror cut that runs from $\ubr$ to $+\infty$ and the $+\ka$-cut 
while $\tx_\R(u)$ has the main mirror cut and the $-\ka$-cut (Figure~\ref{fig:rightmplanes}). The cuts of the mirror $u$-planes are mapped to the common boundary $\pa\cR_-$ of the physical mirror region $\cR_-$ of left and right mirror particles that coincides with the lower edge of the $\ln x$ cut and the semi-line $x>0$. The anti-mirror region $\cR_+$ is the upper half-plane, and its boundary  $\pa\cR_+$ is  the upper edge of the $\ln x$ cut and the semi-line $x >0$. 
Since the mirror and anti-mirror regions and their boundaries are mapped to each other under the inversion map $x\to 1/x$, we can use $\tx_\L(u)$ and $\tx_\R(u)$ to cover the whole $x$-plane with the $\ln x$ cut from $-\infty$ to 0. The functions $\tx_a(u)$ satisfy the complex conjugation condition
\bal
\tx_a(u)^* = {1\ov \tx_{\bar a}(u^*)}\,,
\eal
which relates the left and right $\tx_a$ functions, and  reduces to the pure RR one if $\ka=0$.
The functions $\tx_a(u)$ have similar asymptotic behaviour at large $u$, $ |u|\gg 1$, as depicted in Figure~\ref{fig:x-asymptotics}. We have
\bal\la{eq:txL_large_u}
\tx_\L(u) = 
\begin{cases}
  u + {\ka\ov\pi}\ln u+ \cdots & \text{if}\   \Im(u)<0 \ \text{or}\  \Im(u)<\ka\,,\, \Re(u)<-\nu\,,  \\
 {1\ov u} + {\ka\ov\pi u^2}\ln u+ \cdots & \text{if}\    \Im(u)>\ka \ \text{or}\  \Im(u)>0\,,\, \Re(u)>-\nu\,,
\end{cases}
\eal
and
\bal\la{eq:txR_large_u}
\tx_\R(u) = 
\begin{cases}
  u -{\ka\ov\pi}\ln u+ \cdots & \text{if}\  \Im(u)<-\ka \ \text{or}\  \Im(u)<0\,,\, \Re(u)>-\nu\,,  \\
 {1\ov u} - {\ka\ov\pi u^2}\ln u+ \cdots & \text{if}\    \Im(u)>0\ \text{or}\  \Im(u)>-\ka\,,\, \Re(u)<-\nu\,.
\end{cases}
\eal
The similar asymptotic behaviour is clearly related to the equal number of cuts of the functions $\tx_a(u)$.

String and mirror  $\ka$-deformed  inverse Zhukovsky functions can be expressed through each other as follows
\bal\label{eq:txvsx}
\tx_a(u) = 
\begin{cases}
 x_a(u) & \text{if} \quad\Im(u)<0,  \\
 \frac{1}{x_{\bar a}(u)} & \text{if} \quad\Im(u)>0, 
\end{cases}
\eal
and it is easy to check that they are compatible with the properties of the functions described above.

Let us recall that the $\ka$-deformed string Zhukovsky variables of a string particle
  of charge $M$  for the mixed-flux  \adsst background satisfy \cite{Hoare:2013lja,Lloyd:2014bsa}
\begin{equation}
\label{xpmshortening}
\begin{aligned}
x^{+m}_{a}+\frac{1}{x^{+m}_{a}}-x^{-m}_{a}-\frac{1}{x^{-m}_{a}}&=\frac{2i}{h}m + i\frac{\ka_a}{\pi}p\,,\quad e^{ip}={x^{+m}_{a}\ov x^{-m}_{a}}\,,\end{aligned}
\end{equation}
where  $m=|M|$, $p$ plays the role of the momentum of a string particle.
 The Zhukovsky variables  can be parametrised as follows
\begin{equation}
\label{eq:xpmofp_main}
\begin{aligned}
    x^{\pm m}_{\L} (p)&=& \frac{e^{\pm i p /2}}{2h\,\sin \tfrac{p}{2}} \Big(m+\tfrac{k}{2\pi}p+\sqrt{\big(m+\tfrac{k}{2\pi}p\big)^2+4h^2\sin^2\tfrac{p}{2}}\Bigg)\,,\\
    x^{\pm m}_{\R}(p) &=& \frac{e^{\pm i p /2}}{2h\,\sin \tfrac{p}{2}} \Big(m-\tfrac{k}{2\pi}p+\sqrt{\big(m-\tfrac{k}{2\pi}p\big)^2+4h^2\sin^2\tfrac{p}{2}}\Bigg)\,.
\end{aligned}
\end{equation}
For any $p$ they satisfy~\eqref{xpmshortening}, and are related to the energy of a state as
\begin{equation}\label{energyE}
 E_*
 = \sqrt{\big(m\pm\tfrac{k}{2\pi}p\big)^2+4h^2\sin^2\tfrac{p}{2}} \,,
\end{equation}
where we indicated with ``$*$'' the symbols L, R,  and the sign $+$ under the square  root  corresponds to L.  
They also satisfy the following relations
\bal\label{twopirel}
 x^{\pm m}_{\L} (p+2\pi)&=  x^{\pm (m+k)}_{\L} (p)\,,\quad   &x^{\pm (m+k)}_{\R} (p+2\pi)&=x^{\pm m}_{\R} (p)\,,
 \\
x^{\pm m}_\L(-p)&= -x^{\mp m}_\R(p)\,,\quad &x^{\pm m}_\R(-p)&=  -x^{\mp m}_\L(p)\,, 
\\
x^{\pm m}_\L(2\pi-p) &= -x^{\mp (m+k)}_\R(p)\,,\qquad &x^{\pm (m+k)}_\R(2\pi-p) &= -x^{\mp m}_\L(p)\,,
\\
x^{\pm m}_\R(2\pi-p)&=-{1\ov x^{\pm (k-m)}_\R(p)}\,,\quad &m=0,1,\ldots,k&
 \,,
\eal
which are useful to understand the CP invariance of the string model.

String and mirror functions $x_a(u)$, $\tx_a(u)$ can  be used to 
 provide a $u$-plane parametrisation of the  Zhukovsky variables for string and mirror fundamental particles and their bound states. 
 We set for a string particle\footnote{In the case $m=1$ we just write $x_{\L}^{\pm}(u)$, $x_{\R}^{\pm}(u)$.} 
  of charge $M$
\begin{equation}\begin{aligned}
x_{a}^{\pm m}(u)=x_{a}(u\pm {i\ov h}m)\,,
\end{aligned}\end{equation} 
and  refer to  the variables 
\begin{equation}\begin{aligned}
p_a = i\,(\ln x_a^{-m} - \ln x_a^{+m})
\end{aligned}\end{equation}
as momenta of the particles.  The momenta are real for real $u$, and their range is from 0 to $2\pi$. To get particles with arbitrary momenta one should analytically continue $x_a^{\pm m}$ variables to other $x$-planes through the $\ln x$ cut or, equivalently, through the $\pm\ka$-cuts. We discuss how to describe particles with momenta in the range $(-2\pi,0)$ in section~\ref{sec:cont-p1}. 

In terms of $x^{\pm m}_a$ the energy of a string particle is
\begin{equation}
    E_a=\frac{h}{2i}\left(x^{+m}_a-\frac{1}{x^{+m}_a}-x^{-m}_a+\frac{1}{x^{-m}_a}\right)\,,
\end{equation}
and it is positive for  real $u$.

It is clear from these formulae that $m=0$ and $m=k$ are special cases. If $m=0$ then the momentum and the energy do not vanish only if $u$ is on a string cut. The positivity of the energy then requires $u$ to be on the upper edge of the main cut. On the other hand if $m=k$ then for a right particle of charge $M=-k$ the full range of $p$ is covered for $u\ge -\nu$. Since both $x^{\pm k}_\R(u)$ have a cut for $u\le -\nu$ one has to move $u$ either to the upper or lower edge of the cut. As a result, both momenta and energy become complex. Thus, for a right particle of charge $M=-k$ the physical range of $u$ is $u\ge -\nu$. Similar considerations hold when $m$ is a multiple of~$k$ if we continue the momentum beyond $0<p<2\pi$ (we will discuss this continuation in Section~\ref{sec:cont-p1}).
Hence, in this paper we restrict ourselves to the values of $M$ different from $0$~mod$\,k$.

The  $\ka$-deformed Zhukovsky variables for a mirror particle of charge $M$ is defined in the same way
\begin{equation}\begin{aligned}
\label{eq:boundstatemirror}
\tx_{a}^{\pm m}(u)=\tx_{a}(u\pm {im\ov h})\,,
\end{aligned}\end{equation} 
and the analytic continuation of $p_a$ and $E_a$ gives the mirror energy and momentum
\bal\label{eq:def_mirror_en_mom}
{1\ov i}p_a\ \mapsto\  \tE_a = \ln \tx_a^{-m} - \ln \tx_a^{+m} \,,\qquad
{1\ov i}E_a \ \mapsto\  \tp_a &= {h\ov2}\left( \tx_a^{-m} -{1\ov \tx_a^{-m}} - \tx_a^{+m} +{1\ov \tx_a^{+m}} \right)
\,.
\eal
They satisfy the complex conjugation conditions
\bal
\left(\tE_a(u) \right)^* = \tE_{\bar a}(u^*)\,, \quad \left(\tp_a(u) \right)^* = \tp_{\bar a}(u^*)\,.
\eal
Thus, for real $u$ neither the mirror energy nor momentum are real. 

\subsection{String and mirror \texorpdfstring{$\gamma$}{gamma}-rapidity variables}

To solve the crossing equations in the RR case one uses $\g$-rapidity variables~\cite{Beisert:2006ib,Fontanella:2019baq,Frolov:2021fmj}. 

\paragraph{String gamma's.} In the mixed-flux case the string $\g$-rapidities are defined as follows
\bal
x_a(\g)&={\xi_a+e^{\g} \ov 1 - \xi_a e^{\g} }\,,\qquad  
\g_a(x) = \ln{x - \xi_a\ov x\,\xi_a +1}\,,
 \eal
 and the cut  of $\g_a(x)$ is the interval $(-1/\xi_a,\xi_a)$. Clearly, $\g_a(x)$ satisfies the following complex conjugation condition
 \bal
 \big(\g_a(x)\big)^* = \g_a(x^*)\,,
 \eal
 and therefore
 \bal
 \big(\g_a(x_a^\pm)\big)^* = \g_a(x_a^\mp)\,,\quad u\in\bR\,.
 \eal
As will be discussed in detail in the next section, under the crossing transformation we move $x^{\pm m}_a$ to $1/x^{\pm m}_{\bar a}$. One then uses the following relations. If $\Im(x)>0$ then we do not cross the cut, and get
\bal
\Im(x)>0\,: \quad\g_a(x) \xrightarrow{\text{crossing}} \g_a(1/x) &= \ln\left({1 - x\,\xi_a\ov x+\xi_a} \right) = \g_{\bar a}(x) -i\pi\,.
\eal
If $\Im(x)<0$ then we cross the cut, and get 
\bal
\Im(x)<0\,: \quad\g_a(x) \xrightarrow{\text{crossing}} \g_a(1/x) -2i\pi &= \ln\left({1 - x\,\xi_a\ov x+\xi_a} \right)-2i\pi = \g_{\bar a}(x) -i\pi\,.
\eal
Thus, under the crossing $\g$'s transform as $\g_a\to \g_{\bar a} - i\pi$.
 
If we consider $x$ as a function of $u$ then we get the functions $\g_a(u)$ defined by
\bal
\g_a(u) =  \ln{x_a(u) - \xi_a\ov x_a(u)\, \xi_a +1}\,,
\eal
and the crossing transformation  corresponds to crossing the string main cut from below.

\paragraph{Mirror gamma's.} The mirror $\g$-rapidities are defined similarly
\bal\label{mirrorgamma1b}
\tx_a(\tg)&={\xi_a-i\,e^{\tg} \ov 1 + i\,\xi_a e^{\tg} }\,,
\qquad
e^{\tg_a(\tx)} &= {1\ov i}\,{\xi_a -\tx \ov \tx\,\xi_a +1}\,,
 \eal
The cuts  for $\tg_a(\tx)$ are chosen to be   $(-\infty,-1/\xi_a)$  and $(\xi_a,+\infty)$,  and since for $\Im(\tx)<0$ we have
\bal
 \ln\left({1\ov i}\,{\xi_a -\tx \ov \tx\,\xi_a +1} \right)=  \ln\left({\xi_a -\tx \ov \tx\,\xi_a +1} \right)-{i\pi\ov2}\,,\quad \Im(\tx)<0\,,
\eal
because
\bal
\text{sign}\left(\Im\left({\xi_a -\tx \ov \tx\,\xi_a +1} \right)\right) >0\,,\quad \text{if}\quad \Im(\tx)<0\,,
\eal
we can define the principal branch of $\tg_a(\tx)$ as
\bal
\tg_a(\tx) &= \ln\left({\xi_a -\tx \ov \tx\,\xi_a +1} \right)-{i\pi\ov2}\,.
\eal
Under the crossing transformation we move $\tx_a^{\pm m}$ to $1/\tx_{\bar a}^{\pm m}$ between $(-1/\xi_a,\xi_a)$. Since there is no cut we get
\bal
\tg_a(\tx) \xrightarrow{\text{crossing}} \tg_a(1/\tx) &= \ln\left({\tx\,\xi_a -1 \ov \xi_a +\tx} \right)-{i\pi\ov2} = \tg_{\bar a}(\tx) -i\pi\,.
\eal
Thus, under the crossing $\tg$'s transform as $\tg_a\to \tg_{\bar a} - i\pi$, which is the same transformation as for the string $\g$'s.

On the other hand under the analytic continuation to the string region we move $\tx_a^{+ m}$ to $1/\tx_{\bar a}^{+ m}=x_a^{+ m}$  through the cut $(\xi_a,+\infty)$ (for $\tg_a$ as a function of $\tx$ it is in fact not important through which cut we move). 
Since $\text{sign}\left(\Im\left({\xi_a -\tx \ov \tx\,\xi_a +1} \right)\right) >0$ if $ \Im(\tx)<0$ moving through the cut adds additional $2\pi i$, and we get
\bal
\tg_a(\tx) \xrightarrow{\text{to string}} \tg_a(1/\tx)+2i\pi &= \ln\left({\tx\,\xi_a -1 \ov \xi_a +\tx} \right)-{i\pi\ov2} +2i\pi= \tg_{\bar a}(\tx) +i\pi\,.
\eal
If we consider $\tx$ as a function of $u$ then we get the functions 
\bal\label{mirrorgamma3b}
\tg_a(u) &=  \ln\left({\xi_a -\tx_a(u) \ov \tx_a(u)\,\xi_a +1} \right)-{i\pi\ov2}\,,
\eal
and the crossing transformation  corresponds to crossing the mirror main cut from above while
 the analytic continuation to the string region corresponds to crossing the mirror main cut from below.
 
  \paragraph{String vs mirror gamma's.}
It is also easy to relate $\tg_a(u)$ with $\g_a(u)$. We find
\bal
\tg_a(u) =
\left\{
\begin{array}{ccc}
\g_a(u) +{i\pi\ov2} & \text{if}  &  \Im(u)<0 \\
\g_{\bar a}(u) -{i\pi\ov2} & \text{if}  &  \Im(u)>0 
\end{array}
\right.\,.
\eal
By using these formulae, we get ($u\in\bR$)
\bal
\label{eq:stringmirrorgamma}
\tg_a(u+{i\ov h}m) \xrightarrow{\text{to string}}  \tg_{\bar a}(u+{i\ov h}m) +i\pi =\g_a(u+{i\ov h}m) +{i\pi\ov2} \,,
\\
\tg_a(u-{i\ov h}m) \xrightarrow{\text{to string}}  \tg_{ a}(u-{i\ov h}m)  =\g_a(u-{i\ov h}m) +{i\pi\ov2}\,.
\eal
Thus, the difference of mirror gamma's becomes the difference of string gamma's.

In what follows we will be using the following notations
\bal
\g_a^{\pm m} \equiv \g_a(x_a^{\pm m})\,,\qquad \tg_a^{\pm m} \equiv \tg_a(\tx_a^{\pm m})\,,
\eal
where depending on the context $\g_a^{\pm m} $ are considered either as functions of $x_a^{\pm m}$ or as functions of $u$, and similarly for $\tg_a^{\pm m} $.

\begin{figure}
\begin{center}
\includegraphics[width=5cm]{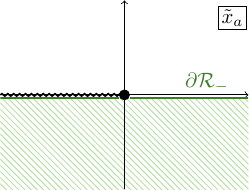}%
\hspace{1cm}%
\includegraphics[width=5cm]{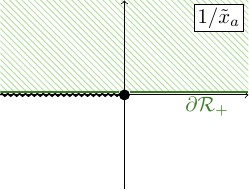}
\end{center}
\caption{Left: the mirror region (shaded in green) with its boundary $\partial\mathcal{R}_-$. Right: the anti-mirror region (shaded in green) with its boundary $\partial\mathcal{R}_+$. Note that each boundary approaches $\tilde{x}_a=0$, where there is the branch point of the logarithm. 
}
\label{fig:DRpm_bdy}
\end{figure}

\subsection{Discrete transformations}
\label{sec:discrete-transforms}

The boundaries $\pa\cR_\pm$ of the mirror and anti-mirror regions depicted in Figure~\ref{fig:DRpm_bdy} are mapped to either themselves or each other under three discrete transformations of the $x$-plane which are important for proving crossing symmetry and CP invariance of the string model. 
Here we summarise the transformation properties of various rapidity variables under the transformations.

\paragraph{Inversion: $x\mapsto 1/x$.} Under the inversion transformation $x\mapsto 1/x$ the mirror and string regions and their boundaries are mapped to each other. Let us recall that
\bal\label{eq:x-inversion}
u_a({1/x}) = u_{\bar a}(x)\,,
\eal
and the inversion symmetry follows from this equation. 
Then,
\bal\label{eq:g-inversion}
\g_a(1/x) &= \g_{\bar a}(x) -i\pi\,\sgn\left(\Im(x)\right)\,,\quad \tg_a(1/x) = \tg_{\bar a}(x) +i\pi\,\sgn\left(\Im(x)\right)\,.
\eal

\paragraph{Reflection: $x\mapsto -x$.} Under the reflection transformation $x\mapsto -x$ only the mirror regions and their boundaries are mapped to each other.  We  find
\bal\label{eq:x-reflection}
u_a(-x)=-u_{\bar a}(x) +i\ka_a\,\sgn(\Im(x))\,.
\eal
Choosing $x= -\tx_a(u)$, we get
\bal\label{eq:u-reflection}
\tx_a(-u+i\ka_a)=-{1\ov \tx_a(u)}\quad \Longleftrightarrow\quad \tx_a(-u)=-{1\ov \tx_a(u+i\ka_a)}\,.
\eal
Thus, the reflection in the $x$-plane is related to the reflection in the mirror $u$-plane. Then,
\bal\label{eq:g-reflection}
\g_a(-x) &= -\g_{\bar a}(x)\,,\quad \tg_a(-x) = -\tg_{\bar a}(x)-i\pi \,.
\eal

\paragraph{Composition of inversion and reflection: $x\mapsto -1/x$.} Under the composition of inversion and reflection transformations $x\mapsto -1/x$ the mirror regions and their boundaries are mapped to themselves.  We  find
\bal\label{eq:x-inversion-reflection}
u_a(-1/x)=-u_{a}(x) -i\ka_a\,\sgn(\Im(x))\,.
\eal
Choosing $x= -1/\tx_a(u)$, we get again eq.\eqref{eq:u-reflection}, and therefore, the composition of inversion and reflection in the $x$-plane is also related to the reflection in the mirror $u$-plane. Then,
\bal\label{eq:g-inversion-reflection}
\g_a\big(-{1\ov x}\big) = - \g_a\big(x\big) + i\pi \,\sgn\big(\Im(x)\big)\,,\quad \tg_a\big(-{1\ov x}\big) = - \tg_a\big(x\big) -i\pi- i\pi \,\sgn\big(\Im(x)\big)\,.
\eal
By using \eqref{eq:txvsx}, one can derive  transformation rules for string $x_a(u)$ under  the reflection in the string $u$-plane. They depend on the location of $u$, and in practice it is easier to use \eqref{eq:u-reflection} for the cases which we will consider later.

The reflection transformations in the mirror and string $u$-plane are used to prove the parity and the  CP invariance of the mirror and string models, respectively. 
The parity transformation in the mirror theory 
on a certain particle is given by
\bal
\tp_a\to -\tp_a\,, \qquad  \tE_a \to \tE_a \,,
\eal
and it is easy to implement it both on the $x$ and $u$-planes.
It is clear from the definitions of the mirror energy and momentum given in~\eqref{eq:def_mirror_en_mom} that parity is implemented by sending
\bal
\tx^{\pm}_a \to -\frac{1}{\tx^{\mp}_a}\,,
\eal
which is just the composition of inversion and reflection transformations. If we now consider the mirror energy and momentum as functions of the mirror $u$-plane rapidity then by using  \eqref{eq:u-reflection}, we find that the parity transformation in the mirror theory  is implemented by the reflection in the $u$-plane and a shift
\bal
&u\mapsto -u +i\ka_a\,,\quad 
&&\tx_a(-u +i\ka_a) = -{1\ov\tx_a(u)}\,,\\
&\tp_a(-u +i\ka_a) = -\tp_a(u)\,,\qquad &&\tE_a(-u +i\ka_a) = +\tE_a(u)\,. 
\eal

The parity transformation in the string theory is more difficult to implement, and it will be discussed in section \ref{sec:cont-p3}.

\section{Properties of S matrix}
\label{sec:properties}

Because of the simpler analytic structure of the mirror $u$-plane (compared to the string $u$-plane), we first construct the dressing factors, and hence the S matrix, in the mirror model. Once such an S-matrix is known, we can obtain the S-matrix of the worldsheet string model by analytically continuing our solution to the string region. 
In this section, we present a natural normalisation of the S-matrix in the mirror region and the constraints this S-matrix must satisfy.

\subsection{S-matrix normalisation}

Particles in mirror theory are in correspondence with four-dimensional short representations of $psu(1 | 1)_\L \oplus psu(1 | 1)_\R$. 
Following the notation of~\cite{Frolov:2023lwd} we split the particles into left and right, having quantum numbers $m=+1$ and $m=-1$ respectively. The bosonic content of the theory is given by
the left particles $Y\,,\, Z$, and the right particles $\bar Y\,,\, \bar Z$.

The S-matrix for the scattering of fundamental massive particles is fully determined by symmetries up to four scalar functions, the so-called dressing factors. Up to these four functions, we fix the normalisation so that the poles corresponding to bound states would be manifest.
In particular, we require to have simple poles in the S-matrix elements associated with the scattering of particles of type $Z$ and $\bar{Z}$.
We define 
\begin{equation}
\label{eq:massivenormmir}
    \begin{aligned}
    \mathbf{S}\,\big|Y_{1}Y_{2}\big\rangle&=&
    {\tx_{\L1}^+\ov \tx_{\L1}^-}   {\tx_{\L2}^-\ov \tx_{\L2}^+}
  \left( \frac{\tx_{\L1}^- - \tx_{\L2}^+}{\tx_{\L1}^+ - \tx_{\L2}^- }\right)^2   \frac{u_1-u_2 + {2i\ov h}}{u_1-u_2 - {2i\ov h}}
  \big(\Sigma^{11}_{\L\L} \big)^{-2}\,
    \big|Y_{1}Y_{2}\big\rangle,\\
    \mathbf{S}\,\big|Y_{1}\bar{Z}_{2}\big\rangle&=&
     {\tx_{\R2}^-\ov \tx_{\R2}^+}
    \frac{1-\frac{1}{\tx_{\L1}^-\tx_{\R2}^-}}{1-\frac{1}{\tx_{\L1}^+x_{\R2}^+}}
    \frac{1-\frac{1}{\tx_{\L1}^+\tx_{\R2}^-}}{1-\frac{1}{\tx_{\L1}^-\tx_{\R2}^+}}\big(\Sigma^{11}_{\L\R}\big)^{-2}\,
    \big|Y_{1}\bar{Z}_{2}\big\rangle,\\
    \mathbf{S}\,\big|\bar{Z}_{1}Y_{2}\big\rangle&=&
 {\tx_{\R1}^+\ov \tx_{\R1}^-} 
   \frac{1-\frac{1}{\tx_{\R1}^+\tx_{\L2}^+}}{1-\frac{1}{\tx_{\R1}^-\tx_{\L2}^-}}
    \frac{1-\frac{1}{\tx_{\R1}^+\tx_{\L2}^-}}{1-\frac{1}{\tx_{\R1}^-\tx_{\L2}^+}}\big(\Sigma^{11}_{\R\L}\big)^{-2}\,
    \big|\bar{Z}_{1}Y_{2}\big\rangle,\\
    \mathbf{S}\,\big|\bar{Z}_{1}\bar{Z}_{2}\big\rangle&=&
    \frac{u_1-u_2 + {2i\ov h}}{u_1-u_2 - {2i\ov h}}\big(\Sigma^{11}_{\R\R}\big)^{-2}\,
    \big|\bar{Z}_{1}\bar{Z}_{2}\big\rangle,
    \end{aligned}
\end{equation}
For $k=0$ the normalisation above reduces to the one of the mirror-theory for pure Ramond-Ramond, as written in appendix B of~\cite{Frolov:2021bwp} and the four dressing factors $\Sigma^{11}_{\L\L}$, $\Sigma^{11}_{\L\R}$, $\Sigma^{11}_{\R\L}$ and $\Sigma^{11}_{\R\R}$ reduce to the dressing phases of the Ramond-Ramond model in the mirror region. The dressing factors depend on the kinematics of the scattering particles through the $u$ variables; while we sometimes omit this dependence, $\Sigma^{11}_{ab}$ is a function of the external kinematics
\bal
\Sigma^{11}_{ab}=\Sigma^{11}_{ab}(u_1, u_2)\,.
\eal
Depending on whether the external particles are evaluated at mirror or string kinematics we also use the notation
\bal
\Sigma^{11}_{ab}(\tx^\pm_{a1}, \tx^\pm_{b2}) \qquad  \text{or} \quad\Sigma^{11}_{ab}(x^\pm_{a1}, x^\pm_{b2})\,,
\eal
where $\tx^\pm_{aj}=\tx_a(u_j\pm \frac{i}{h})$ and $x^\pm_{aj}=x_a(u_j\pm \frac{i}{h})$, with $j=1,2$.
The dressing factors can be fixed through symmetry considerations, as discussed in the remaining part of this section.

\subsection{Discrete symmetries}

We require the S-matrix to satisfy several symmetries, which translate into constraints on the dressing factors.

\paragraph{Unitarity in the string model.}
In the mirror region the S~matrix is not expected to satisfy unitarity. Indeed the Lagrangian of the mirror theory (obtained from the worldsheet Lagrangian by performing a double Wick rotation of space and time) is not real. We must anyway recover this property after continuing the S-matrix to the string region, where the Lagrangian is real.  Then, for $x^\pm_{a1}$ and $x^\pm_{b2}$ in the string region and $u_1$, $u_2 \in \mathbb{R}$ (where the energies and momenta of the particles are real) it must hold that
\bal
|\Sigma^{11}_{ab}(x^\pm_{a1},x^\pm_{b2})|^2=1 \,, \qquad a, b=\text{L}, \, \text{R}\,. 
\eal

\paragraph{Braiding unitarity.}
Braiding unitarity is a consistency condition of the Zamolodchikov-Faddeev algebra and differently from standard unitarity is expected to hold both in the string and mirror region. This property requires that
\bal
\Sigma^{11}_{ab}(u_1,u_2) \Sigma^{11}_{ba}(u_2,u_1)=1 \quad \forall \ u_1,\, u_2 \in \mathbb{C} \quad \text{and} \quad a, b=\text{L}, \, \text{R}\,. 
\eal
Note that in this case $u_1$ and $u_2$ can be any points in the complex plane and starting from the string region we can continue this equality to the mirror region, where it also must be satisfied.

\paragraph{P invariance in the mirror model.}
The mirror Lagrangian is invariant under exchanging the sign of the space coordinate. Then we expect the S-matrix to be parity (P) invariant in the mirror region, meaning that \footnote{Parity is separately satisfied by the normalisation factors in all S-matrix elements and the dressing factors must therefore satisfy this symmetry.}
\bal
\label{eq:parity_momentum_representation_mir}
\Sigma^{11}_{ab}(\tilde{p}_1, \tilde{p}_2) = \Sigma^{11}_{ba}(-\tilde{p}_2, -\tilde{p}_1) \,, \qquad a, b=\text{L}, \, \text{R}\,.
\eal
Using the observations presented in section~\ref{sec:discrete-transforms}, equation~\eqref{eq:parity_momentum_representation_mir} can then be equivalently written in the $x$ and $u$ plane as
\bal
\label{eq:parity_constraint_Sigma11}
\Sigma^{11}_{ab}(\tx^\pm_{a1}, \tx^\pm_{b2}) = \Sigma^{11}_{ba}(-\frac{1}{\tx^\mp_{b2}}, -\frac{1}{\tx^\mp_{a1}}) \,,
\eal
and
\bal
\Sigma^{11}_{ab}(u_1, u_2) = \Sigma^{11}_{ba}(-u_2 + i \ka_b,-u_1 + i \ka_a) \,,
\eal
respectively.

\paragraph{CP invariance in the string model.}
In the presence of a $B$-field the string theory is not parity invariant; instead, it is invariant under a combination of parity and charge conjugation (CP), which changes the sign of momenta, and exchanges left and right particles. Consider  S-matrix elements $S_{AB\to CD}(p_1,p_2)$ which describe the scattering of a $m_1$-particle bound state with momentum $p_1$ and a $m_2$-particle bound state with momentum $p_2$ where the pairs of indices $A,B$ and $C,D$ are any states from either  short left  or right massive multiplets $Y,Z,\eta^1,\eta^2$ and $\bar Y,\bar Z, \bar\eta^1,\bar\eta^2$.
The CP invariance imposes the following condition on the S-matrix elements
\bal
\label{eq:CP_invariance_general}
S_{AB\to CD}(-p_1,-p_2)=S_{\bar B \bar A\to \bar D\bar C}(p_2,p_1)\,,
\eal
where the bar over the indices means the charge conjugation, i.e. if $A\in\{ Y,Z,\eta^1,\eta^2\}$ then $\bar A\in\{ \bar Y,\bar Z, \bar\eta^1,\bar\eta^2\}$ and vice versa, and $p_1,p_2$ are arbitrary (positive or negative) momenta. 
While this constraint is particularly simple written in momentum representation, it involves a completely non-trivial analytic continuation in the $u$ plane. This is clear by the fact that our fundamental string region was defined so that if $u\in \mathbb{R}$ then $p \in (0, 2\pi)$. Of course if $p_1 \in (0, 2\pi)$ on the RHS of~\eqref{eq:CP_invariance_general} then $-p_1 \in (-2\pi,0)$ on the LHS of the equation. Reaching the region of negative momenta will require to continue the dressing factors outside the fundamental string region and will provide a completely non-obvious consistency check of our solutions. We will return to this point in sections~\ref{sec:cont-p1} and~\ref{sec:cont-p3}.

\subsection{Crossing invariance}

Among all symmetries, a special role is played by the \textit{crossing invariance}, stating that S-matrix elements are left unchanged by switching incoming particles with outgoing anti-particles with opposite energies and momenta.  The explicit form of the crossing equations depends on the matrix structure of the S~matrix under consideration, and as such it provides a crucial input for determining the dressing factors of the model. However, the solution of crossing in not unique, as we will discuss in section~\ref{sec:verification}.
In our normalisation crossing symmetry yields the following crossing equations
\bal
\label{eq:mirror_crossing_eq}
 \big(\Sigma^{11}_{aa}(u_1,u_2)\big)^{2} \big(\Sigma^{11}_{\bar{a} a}(\bar u_1,u_2)\big)^{2}&=
\frac{(\tx_{a1}^- - \tx_{a2}^+)(\tx_{a1}^+ - \tx_{a2}^-)}{(\tx_{a1}^- - \tx_{a2}^-)(\tx_{a1}^+ - \tx_{a2}^+)}
  \frac{u_1-u_2 + {2i\ov h}}{u_1-u_2 - {2i\ov h}}
\,,\\
 \big(\Sigma^{11}_{aa}(\bar u_1,u_2)\big)^{2} \big(\Sigma^{11}_{\bar{a} a}( u_1,u_2)\big)^{2}&=
\frac{\big(1-\frac{1}{\tx^+_{\bar{a}1}\tx^+_{a2}}\big)\big(1-\frac{1}{\tx^-_{\bar{a}1}\tx^-_{a2}}\big)}{\big(1-\frac{1}{\tx^+_{\bar{a}1}\tx^-_{a2}}\big)\big(1-\frac{1}{\tx^-_{\bar{a}1}\tx^+_{a2}}\big)}
 \frac{u_1-u_2 + {2i\ov h}}{u_1-u_2 - {2i\ov h}}
 \,,
\eal
which can be read from the ones in~\cite{Lloyd:2014bsa} after changing the normalisation.
In the formulas above $\bar{u}_1$ is the same as $u_1$ but it is reached by crossing the mirror theory cut with both $\tx^+_1$ and $\tx^-_1$, so that the mirror energy and momentum have an opposite sign compared to the one in the mirror region. We will describe in detail the continuation path in the next sections. In~\eqref{eq:mirror_crossing_eq} we wrote the crossing equations for mirror kinematics; these equations can be trivially continued to the string region just by replacing $\tx$ with $x$.
For convenience, we split dressing factors solving these equations into  \textit{even} and \textit{odd} parts  \cite{Beisert:2006ib}, which satisfy separately
\bal
\label{eq:even_crossing_equation}
\left( \Sigma^\besratio_{\bar{a} a}(\bar{u}_1, u_2) \right)^2 \left( \Sigma^\besratio_{aa}(u_1, u_2) \right)^2&= \left( \Sigma^\besratio_{a a}(\bar{u}_1, u_2) \right)^2 \left( \Sigma^\besratio_{\bar{a} a}(u_1, u_2) \right)^2=\frac{u_1- u_2+\frac{2i}{h}}{u_1- u_2-\frac{2i}{h}}\,,
\eal
and
\bal
\left( \Sigma^\barnes_{\bar{a} a}(\bar{u}_1, u_2) \right)^2 \left( \Sigma^\barnes_{aa}(u_1, u_2) \right)^2&=  \frac{\left(\tx^-_{a 1} -\tx^+_{a 2} \right) \left(\tx^+_{a 1} -\tx^-_{a 2} \right) }{\left(\tx^-_{a 1} -\tx^-_{a 2} \right) \left(\tx^+_{a 1} -\tx^+_{a 2} \right) }\,,\\
\left( \Sigma^\barnes_{aa}(\bar{u}_1, u_2) \right)^2 \left( \Sigma^\barnes_{\bar{a} a} (u_1, u_2)  \right)^2&= \frac{\left(1- \frac{1}{\tx^+_{\bar{a} 1} \tx^+_{a 2} }\right) \left(1- \frac{1}{\tx^-_{\bar{a} 1} \tx^-_{a 2} }\right) }{\left(1- \frac{1}{\tx^+_{\bar{a} 1} \tx^-_{a 2} }\right) \left(1- \frac{1}{\tx^-_{\bar{a} 1} \tx^+_{a 2} }\right) }\,.
\eal
In this manner the combination
\bal
\Sigma^{11}_{ab}(u_1,u_2) = \Sigma^{\besratio}_{ab}(u_1,u_2) \ \Sigma^{\barnes}_{ab}(u_1,u_2)
\eal
provides a solution to the crossing equations.

Expressed in terms of the mirror $\tg$ variables introduced in the previous section the odd part of the crossing equations takes the following simple form
\bal
\label{eq:odd_mir_crossing1}
\left(\Sigma^\barnes_{\bar{a} a}(\bar{u}_1, u_2) \right)^2 \left( \Sigma^\barnes_{aa}(u_1, u_2) \right)^2=  \frac{\sinh \left( \frac{\tg^{+-}_{aa}}{2} \right) \sinh \left(\frac{\tg^{-+}_{aa}}{2} \right) }{\sinh \left(\frac{\tg^{--}_{aa}}{2} \right) \sinh \left(\frac{\tg^{++}_{aa}}{2} \right) }\,,
\eal
\bal
\label{eq:odd_mir_crossing2}
\left( \Sigma^\barnes_{aa}(\bar{u}_1, u_2) \right)^2 \left( \Sigma^\barnes_{\bar{a} a}(u_1, u_2) \right)^2=\frac{\cosh \left( \frac{\tg^{--}_{\bar aa}}{2} \right) \cosh \left(\frac{\tg^{++}_{\bar aa}}{2} \right) }{\cosh \left(\frac{\tg^{-+}_{\bar aa}}{2} \right) \cosh \left(\frac{\tg^{+-}_{\bar aa}}{2} \right) }\,.
\eal
As we will see, these equations can be solved in terms of a combination of Barnes G-functions depending on differences of $\gamma$-rapidities (formally, taking the same form as the solution of the odd part of crossing  proposed in \cite{Frolov:2021fmj} for the pure RR case). The odd part of the crossing equation for mixed-flux was already solved in~\cite{OhlssonSax:2023qrk} where it was expressed in terms of an integral representation, equivalent to the Barnes G-function representation. The solution of the even part is substantially more involved as it requires introducing a suitable deformation of the BES phase, as we shall see in the next section, where we provide a minimal solution for the (even and odd) crossing equation.

\section{Proposal for the massive dressing factors}\la{sec:proposal}

In analogy with the pure-RR case~\cite{Frolov:2021fmj}, we will break down the proposal for the dressing factors into three pieces: the BES part and the HL part, which together provide a minimal solution for the even part  of the dressing factors
\bal
\label{eq:even_Sigma_mirror}
\Sigma^{\besratio}_{ab}(\tx^\pm_{a1}, \tx^\pm_{b2})= \frac{\Sigma^{\bes}_{ab}(\tx^\pm_{a1}, \tx^\pm_{b2})}{\Sigma^{\hl}_{ab}(\tx^\pm_{a1}, \tx^\pm_{b2})}\,,
\eal
and the remaining odd part (which can be written in terms of a difference-form expression). Each of these pieces will be defined in terms of a function $\tPhi^{\text{any}}_{ab}(\tx_{a1},\tx_{b2})$, related to the dressing phase, valid when both excitations take values in the mirror region. (``Any'' is a stand-in for ``even'', ``odd'', ``BES'', ``HL'', and so on.) The indices $a,b$ could take the values L or R. The full dressing phase is then given by the decomposition~\cite{Arutyunov:2006iu}
\begin{equation}
\label{eq:fourPhidecomp}
\tilde{\theta}_{ab}^{\text{any}}(\tx^\pm_{a1},\tx^\pm_{b2})= 
\tPhi^{\text{any}}_{ab}(\tx_{a1}^+,\tx_{b2}^+)-\tPhi^{\text{any}}_{ab}(\tx_{a1}^+,\tx_{b2}^-)-\tPhi^{\text{any}}_{ab}(\tx_{a1}^-,\tx_{b2}^+)+\tPhi^{\text{any}}_{ab}(\tx_{a1}^-,\tx_{b2}^-)\,,
\end{equation}
where ``any'' denotes any of the phases. This decomposition is very convenient when dealing with bound states, as we shall see. From the phases we obtain the dressing factors as
\bal
\label{eq:dressingphases}
\Sigma^{\text{any}}_{ab}(\tx^\pm_{a1}, \tx^\pm_{b2})=\exp\left[{i\, \tilde{\theta}_{ab}^{\text{any}}(\tx^\pm_{a1}, \tx^\pm_{b2})}\right]\,.
\eal

\subsection{\texorpdfstring{$\ka$}{kappa}-deformed improved BES factor}
\label{mirrorPhi}

The Beisert-Eden-Staudacher (BES) phase~\cite{Beisert:2006ez} plays a central role in solving the crossing equations of different theories. In this section, we introduce a $\ka$-deformed version of BES for the mirror theory with mixed flux. To do that we use an integral representation of BES, as already proposed in~\cite{Dorey:2007xn}, and write the phase as an integral over the boundary between the mirror and antimirror region of our theory, see Figure~\ref{fig:DRpm_bdy}. 

To construct the phase, we start introducing the following building blocks 
\bal\label{Phiabmunu}
\tPhi_{ab}^{\alpha \beta}(x_{1},x_{2}) 
=& -  \lint_{{ \pa\cR_\alpha}} \frac{{\rm d} w_1}{2\pi i} \lint_{{ \pa\cR_\beta}} \frac{{\rm d} w_2}{2\pi i} {1\ov w_1-x_{1}}{1\ov w_2-x_{2}} K^\bes(u_a(w_1)-u_b(w_2))\,,
\eal
where $x_i$ can be anywhere on the $x$-plane, and
$$
a,b=\text{L},\, \text{R} \,,\quad \alpha,\beta=\pm
$$
and
\bal
\label{eq:Kbes_v_mirror}
K^\bes(v)= i \log \frac{\Gamma \left(1+\frac{ih}{2}v \right)}{\Gamma \left(1-\frac{ih}{2}v \right)}\,.
\eal
The integration paths $\pa\cR_-$ and $\pa\cR_+$ are the boundaries of the mirror and anti-mirror regions, as shown in figure~\ref{fig:DRpm_bdy}, and are defined as follows\footnote{More precisely they are defined through the principal value prescription
\bal
\nonumber
\pa\cR_\pm:& \quad (-R\pm i 0, -\frac{1}{R} \pm i 0) \ \cup \ (+\frac{1}{R}, +R)\,,
\eal
where we take the limit $R \to \infty$ after integrating.}
\bal
\pa\cR_-:& \quad (-\infty-i 0, +\infty-i 0)\,, \qquad \pa\cR_+:& \quad (-\infty+i 0, +\infty+i 0)\,.
\eal
For $\ka=0$ the contours $\pa\cR_-$ and $\pa\cR_+$ are equivalent and no matter the choice of $a$, $b$ and contours, the $\tPhi$ function~\eqref{Phiabmunu} becomes a building block for the mirror BES phase defined in appendix~\ref{app:stringvsmirror}. However, as soon as $ \ka \ne 0$, the integration over the two contours differs since $u_a(w_1)$ and $u_b(w_2)$ are functions with cuts on the negative real axis. 
The $\kappa$-deformed BES phases will be written in terms of the semi-sums
\bal\label{Phiabara}
\tPhi_{aa}(x_{1},x_{2}) &=  \frac{1}{2} \left(\tPhi_{aa}^{--}(x_{1},x_{2})+\tPhi_{aa}^{++}(x_{1},x_{2}) \right)\,,
\\
\tPhi_{\bar a a}(x_{1},x_{2}) &= \frac{1}{2} \left( \tPhi_{\bar a a}^{-+}(x_{1},x_{2})+\tPhi_{\bar a a}^{+-}(x_{1},x_{2})  \right)
\,.
\eal
Similarly to the RR case, the  functions $\tPhi_{ab}^{\alpha\beta}(x_{1},x_{2})$ are not analytic in the whole $x$-plane (for either variable). They are however analytic 
if  both points $x_1$ and $x_2$ are in the mirror region, i.e.\ $\Im(x_{1})<0$ and $\Im(x_{2})<0$. Starting from this region we can extend these functions --- and clearly $\tPhi_{ab}(x_1,x_2)$ of eq.~\eqref{Phiabara} --- to the cover of the $x$ plane; to make things clearer we denote such an extension by
\bal
\tchi_{ab}^{\alpha \beta}(x_{1},x_{2}) \quad\text{and}\quad \tchi_{ab}(x_{1},x_{2}) \,,
\qquad a,b=\text{L}, \, \text{R} \,, \quad \alpha, \beta=\pm\,.
\eal
We define the complete BES mirror phase for mirror fundamental particles with  Zhukovsky rapidities $\tx^\pm_{a1},\tx^\pm_{b2}$ as $\tilde{\theta}_{ab}^{\bes}(\tx^\pm_{a1},\tx^\pm_{b2})$ using eq.~\eqref{eq:fourPhidecomp}. Similarly we may also define~$\tilde{\theta}_{ab}^{\alpha \beta}(\tx^\pm_{a1},\tx^\pm_{a2})$ so that
\bal
\la{eq:ttheta_definition_mir_fund}
&\tilde{\theta}_{aa}^{\bes}(\tx^\pm_{a1},\tx^\pm_{a2})= \frac{1}{2} \left(\tilde{\theta}_{aa}^{--}(\tx^\pm_{a1},\tx^\pm_{a2})+\tilde{\theta}_{aa}^{++}(\tx^\pm_{a1},\tx^\pm_{a2}) \right) \,,\\
&\tilde{\theta}_{\bar a a}^{\bes}(\tx^\pm_{\bar a1},\tx^\pm_{a2})=  \frac{1}{2} \left(\tilde{\theta}_{\bar aa}^{-+}(\tx^\pm_{\bar a1},\tx^\pm_{a2})+\tilde{\theta}_{\bar aa}^{+ -}(\tx^\pm_{\bar a1},\tx^\pm_{a2}) \right) \,.
\eal
The mirror BES dressing factors are then defined by~\eqref{eq:dressingphases}.

\paragraph{Dressing factors on the $u$ plane.}
For the purpose of continuing the functions $\tPhi^{\a\b}_{ab}$ outside the mirror region it is often convenient to write them  as integrals in the $u$-plane 
\bal\label{eq:tPhi_integrated_in_u_planes}
\tPhi_{aa}^{--}(x_{1},x_{2}) 
=& -  \lint_{\widetilde{ \rm cuts}} \frac{{\rm d} v_1}{2\pi i} \lint_{\widetilde{ \rm cuts}} \frac{{\rm d} v_2}{2\pi i} { \tx'_a(v_1) \tx'_a(v_2)\ov (\tx_a(v_1)-x_1) \ (\tx_a(v_2)-x_2) } \ K^\bes(v_1-v_2) \,,\\
\tPhi_{aa}^{++}(x_{1},x_{2}) 
=& -  \lint_{\widetilde{ \rm cuts}} \frac{{\rm d} v_1}{2\pi i} \lint_{\widetilde{ \rm cuts}} \frac{{\rm d} v_2}{2\pi i} { \left(\frac{1}{\tx_{\bar{a}}(v_1)} \right)' \  \left(\frac{1}{\tx_{\bar{a}}(v_2)} \right)' \ov (x_{1}-\frac{1}{\tx_{\bar{a}}(v_1) }) \ (x_{2}- \frac{1}{\tx_{\bar{a}}(v_2)}) } \ K^\bes(v_1-v_2) \,,\\
\tPhi_{\bar{a}a}^{-+}(x_{1},x_{2}) 
=& -  \lint_{\widetilde{ \rm cuts}} \frac{{\rm d} v_1}{2\pi i} \lint_{\widetilde{ \rm cuts}} \frac{{\rm d} v_2}{2\pi i} { \tx'_{\bar{a}}(v_1) \  \left(\frac{1}{\tx_{\bar{a}}(v_2)} \right)' \ov (\tx_{\bar{a}}(v_1)-x_1) \ (x_{2}- \frac{1}{\tx_{\bar{a}}(v_2)}) } \ K^\bes(v_1-v_2) \,,\\
\tPhi_{\bar{a}a}^{+-}(x_{1},x_{2}) 
=& -  \lint_{\widetilde{ \rm cuts}} \frac{{\rm d} v_1}{2\pi i} \lint_{\widetilde{ \rm cuts}} \frac{{\rm d} v_2}{2\pi i} { \left(\frac{1}{\tx_{a}(v_1)} \right)' \  \tx'_{a}(v_2) \ov (x_{1}-\frac{1}{\tx_{a}(v_1) }) \ ( \tx_{a}(v_2)-x_2) } \ K^\bes(v_1-v_2) \,.
\eal
In the expressions above $\tx_a$ and $\tx_{\bar{a}}$ correspond to the inverse of the mirror Zhukovsky maps and take values in the mirror region. By `$\widetilde{ \rm cuts}$' we mean that the integrals are performed over the cuts of these functions so that on the lower edges of the cuts we move to the right. For example in the third row of the expression above we integrate $v_1$ around the cuts of $\tx_{\bar{a}}(v_1)$ and $v_2$ around the cuts of $\tx_{\bar{a}}(v_2)$.

\begin{figure}
\begin{center}
\includegraphics[width=5cm]{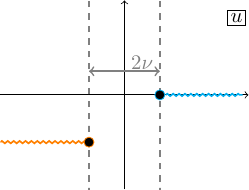}%
\hspace{1cm}%
\includegraphics[width=5cm]{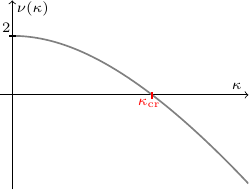}
\end{center}
\caption{Left: The branch cuts of the $\log\Gamma$ function in $K^\bes$ do not intersect the integration contour $\partial\mathcal{R}_-$  as long as the projections of the mirror branch cuts are separated by a positive distance; the distance is $2\nu$. (The same goes for $\partial\mathcal{R}_+$ and the anti-mirror  branch cuts, not depicted here.) Right: we plot $\nu$ as a function of~$\kappa$, see~\eqref{eq:ubranches}. The axes are not to scale. It is a monotonically decreasing function, which starts out as $\nu=2$ for $\kappa=0$ (that is, in the pure-RR case), vanishes at $\kappa_{\text{cr}}\approx9.48$, and goes to $\nu=-\infty$ as $\kappa\to+\infty$. 
}
\label{fig:kappacritical}
\end{figure}

\paragraph{Branch-cut structure and critical~$\kappa$.}

Let us discuss the branch cut structure of the BES kernel~\eqref{eq:Kbes_v_mirror}. It is easiest to consider the branch cuts of the kernel when the integrals are defined in terms of the $v_1$ and $v_2$~rapidities. 
The BES kernel has branch points in the $v_1$ plane located at
\bal
\la{eq:K_br_points}
v^{(+)}_{1 j} = v_2 + \frac{2i}{h} j \,, \qquad
v^{(-)}_{1j} = v_2 - \frac{2i}{h} j \,, \qquad j=1, 2 , \dots \,.
\eal
Following the standard choice of branch cut for the logarithm,  the branch cuts run vertically to $+i \infty$ and $-i \infty$ respectively. For each pair of branch points, identified by a value of $j=1, 2, \dots$, the associated branch cuts are then given by
\bal
v_2 + \frac{2i}{h} j + i t \,, \qquad 
v_2 - \frac{2i}{h} j - it \,, \qquad t \ge0\,.
\eal
Let us integrate $v_1$ for some fixed value of $v_2$. Note that a necessary condition for being on a branch cut is that the difference $v_{12}= v_1 - v_2$ has a vanishing real part.
In the $v_1$ plane the integration contours run around the cuts of the mirror region (for $\mathcal{R}_-$) or anti-mirror region (for $\mathcal{R}_+$). 
In the cases that we depicted so far, see Figure~\ref{fig:rightmplanes}, the projection of these cuts onto the real line gives two disjoint intervals. In other words, in this scenario if $v_{1}$ and $v_{2}$ are on the cuts, the real part of $v_{12}$ can vanish only if they are both on the same cut, in which case the imaginary part vanishes too and $v_1$ cannot be on a branch cut. The double integral for the improved BES phase is then well defined.
This is one of the reasons why it is easier to work with mirror contours rather than with string ones. However, note that this scenario hinges on the position of the branch points, and specifically on the sign of~$\ubr$. By using the explicit expression~\eqref{eq:ubranches}  we find that
\begin{equation}
    \ubr>0\qquad\Leftrightarrow\qquad
    0\leq \ka<\ka_{\text{cr}}\approx 9.48\,,
\end{equation}
where the critical value $\ka_{\text{cr}}$ is the solution of a transcendental equation.  

As long as $\ka< \ka_{\text{cr}}$ the improved BES phase is well defined and can be treated as a continuous function of $\ka$. Because the parameter~$\ka$ can be increased continuously by decreasing~$h$ at $k$~fixed, one should require the dressing factors to change continuously as $\ka$ becomes larger than~$\ka_{\text{cr}}$. If the branch \textit{points} do not hit the integration contour, it is sufficient to continuously deform the branch \textit{cuts} of $K^\bes$. Note however that for even values of~$k$ the branch points at $-\nu\pm i k/h$ will hit those in eq.~\eqref{eq:K_br_points} as $\ka$ approaches $\ka_{\text{cr}}$.%
\footnote{In fact, the same problem appears for $k$ odd too when considering the continuation to other regions, in particular to the string region. The culprit in that case is the integrand of the $\tPsi$ functions which we introduce in section~\ref{sec:cont_to_string}.}
Let $k=2r$ be even. In this case if $v_2$ is on the main mirror cut $(\ubr, +\infty)$ then the branch point $v^{(+)}_{1 j}$ with $j=r$ in~\eqref{eq:K_br_points} is on the $+\ka$ cut in the left $v$ plane and the branch point $v^{(-)}_{1 j}$ with $j=r$ is on the $-\ka$ cut in the right $v$ plane. We regularise this by the principal value prescription, i.e.\ by replacing the integration kernel in \eqref{eq:tPhi_integrated_in_u_planes} as
\bal
\label{eq:Kbes_pv_prescription_kgkc}
K^\bes(v_{12}) \to \frac{1}{2}\left( K^\bes(v_{12}+ i \eps) +K^\bes(v_{12}- i \eps)\right),
\eal
where $\eps$ is a small parameter that should be sent to zero after integration. In this way, all branch points are always away from the integration contours and the cut of $\log \Gamma$ can be deformed to avoid any overlap with the contours.
The result of this integral has the same form as a function of~$k$ as we would find in the odd-$k$ case, at least in the cases where we could explicitly perform the integration.%
\footnote{The same prescription applies also to $k$ odd where this problem appears in other integrals, the $\tPsi$ functions of section~\ref{sec:cont_to_string}.}
This prescription is important for the computation of the relativistic limit of the dressing factors, see appendix~\ref{app:rel_limit}.

\subsection{Modified Hern\'andez-L\'opez factor in the mirror model}

Starting from the mirror BES phase defined in the previous section we can obtain a $\ka$-deformed Hern\'andez-L\'opez factor (HL) by considering the limit $h \gg1$, $k \gg 1$ with $\ka=\frac{k}{h}$ fixed.
The HL phase corresponds to the order $h^0$ of this limit; in particular, the large $h$ expansion leads to 
\bal
\label{eq:BES-expansion}
\tPhi_{ab}(\tx_1, \tx_2)=h\,\tPhi^{\afs}_{ab}(\tx_1, \tx_2)+\tPhi^{\hl}_{ab}(\tx_1, \tx_2)+ \mathcal{O}(\frac{1}{h})\,.
\eal
In appendix~\ref{app:HL} we provide a detailed derivation of HL, while here we just report the result of the computation. Taking the limit just described we obtain (up to terms that cancel in the full phase when we consider both points in the mirror region)
\bal
\tPhi_{aa}^{\hl}(\tx_{1},\tx_{2}) &=- \frac{1}{4 \pi}   \lint_{\widetilde{ \rm cuts}} {\rm d} v \frac{\tx'_a(v)}{\tx_a(v) - \tx_{1}} \left( \log\left(\tx_a(v - i \eps)-\tx_{2} \right) - \log\left(\tx_a(v + i \eps)-\tx_{2} \right) \right)\,,\\
\tPhi_{\bar a a}^{\hl}(\tx_{1}, \tx_{2})&=+\frac{1}{4 \pi}   \lint_{\widetilde{ \rm cuts}} {\rm d} v \frac{\tx'_{\bar{a}}(v)}{\tx_{\bar{a}}(v) - \tx_{1}} \left( \log\left(\frac{1}{\tx_{\bar{a}}(v-i \eps)}-\tx_{2} \right) - \log\left(\frac{1}{\tx_{\bar{a}}(v+ i \eps)}-\tx_{2} \right) \right) \, .
\eal
As before the integrals are performed around the cuts of $\tx_a$ and $\tx_{\bar{a}}$, see Figure~\ref{fig:rightmplanes} so that on the lower edges of the cuts we move to the right. In the expressions above $\eps$ is a small parameter which we send to zero after integrating. This means that if $v$ is a point on the lower edge of a cut then $v+ i \eps$ is on the upper edge of the cut.
The functions $\tPhi_{aa}^{\hl}(x, y)$ and $\tPhi_{\bar{a}a}^{\hl}(x, y)$ are originally defined for points $x$ and $y$ in the mirror region. As for BES, we define $\tilde{\chi}^\hl_{aa}(x, y)$ and  $\tilde{\chi}^\hl_{\bar{a} a}(x, y)$
to be the continuations of the functions $\tPhi$'s to the entire infinite cover of the $u$ plane.
The improved HL phases and factors are then given by the decomposition~\eqref{eq:fourPhidecomp} and by~eq.\eqref{eq:dressingphases}, respectively.

\subsection{Odd dressing factor}

We introduce finally the simplest ingredient to solve crossing, which is provided by the following combination of Barnes Gamma function $G$:
\bal
R (\g)\equiv {G(1- \frac{\g}{2\pi i})\ov G(1+ \frac{\g}{2\pi i}) } =   \left({e\ov 2\pi}\right)^{+\frac{\gamma}{2\pi i}}\prod_{\ell=1}^\infty \frac{\Gamma(\ell+\frac{\gamma}{2\pi i})}{\Gamma(\ell-\frac{\gamma}{2\pi i})}\,e^{-\frac{\gamma}{\pi i}\,\psi(\ell) }\,.
\eal
In the expression above $\psi(\ell) = \frac{d}{d \ell} \log \Gamma(\ell)$.
The function $R$ satisfies a simple monodromy relation, namely
\bal
\label{eq:monodromy_R}
R (\g-2\pi i) =i\,  {\pi \ov \sinh{\g\ov2}}R(\g)\,,\ \ R (\g+2\pi i) = i\, { \sinh{\g\ov2}\ov \pi }R(\g)\,,\ \ R (\g+\pi i) =  { \cosh{\g\ov2}\ov \pi }R(\g-\pi i)\,.
\eal
It also satisfies the following relations
\bal
\label{eq:brunit_unit_R}
R(\g)R(-\g)=1\,, \qquad R(\g)^* R(\g^*)=1\,,
\eal
necessary to check unitarity (in the string model) and braiding unitarity.
This function can be used to construct dressing factors which solve the odd part of the crossing equation, in the sense of~\cite{Beisert:2006ib}. We set 
\bal
\tPhi_{aa}^\barnes(\tx_{a1},\tx_{a2})=& +i \ln R(\tg_{a1}-\tg_{a2})\,,\\
\tPhi_{\bar{a}a}^\barnes(\tx_{\bar{a}1},\tx_{a2})=& -\frac{i}{2} \ln R(\tg_{\bar{a}1}-\tg_{a2}+i\pi)R(\tg_{\bar{a}1}-\tg_{a2}-i\pi)\,,
\eal
where the $\tg$ variables were introduced in section~\ref{sec:kinematics}. The phases and dressing factors follow from~\eqref{eq:fourPhidecomp} and~\eqref{eq:dressingphases}. We can also write the final result compactly by introducing the short-hand
\bal
\tg^{\alpha \beta}_{ab} \equiv \tg^{\alpha}_{a1}-\tg^{\beta}_{b2}\,, \quad \alpha, \beta=\pm, \quad a, b= \text{L}, \text{R},
\eal
and writing
\bal
\label{eq:odd_Sigma_mirror}
&\left(\Sigma^{\barnes}_{a a} (\tx^\pm_{a1}, \tx^\pm_{a2}) \right)^{-2}=\frac{R^2(\tg^{--}_{aa}) R^2(\tg^{++}_{aa})}{R^2(\tg^{-+}_{aa}) R^2(\tg^{+-}_{aa}) }\,,\\
&\left( \Sigma^{\barnes}_{\bar{a} a} (\tx^\pm_{\bar{a}1}, \tx^\pm_{a2}) \right)^{-2}=\frac{R(\tg^{-+}_{\bar a a}+ i \pi) R(\tg^{-+}_{\bar a a}- i \pi) R(\tg^{+-}_{\bar a a}+ i \pi)R(\tg^{+-}_{\bar a a}- i \pi)}{R(\tg^{--}_{\bar a a}+ i \pi) R(\tg^{--}_{\bar a a}- i \pi) R(\tg^{++}_{\bar a a}+ i \pi) R(\tg^{++}_{\bar a a}- i \pi)} \,,
\eal
One can check numerically that this proposal matches the integral representation for the odd part of the dressing factors early advanced in~\cite{OhlssonSax:2023qrk}.

\subsection{Full dressing factor for mirror excitations}
\label{sec:mirror-dressing-bound}
The full dressing factor for fundamental particles which appear in~\eqref{eq:massivenormmir} can be obtained from the BES, HL and odd phases introduced above by setting
\bal
\label{eq:complete_dressing_ph_mir_reg}
&\left(\Sigma^{11}_{a b} (\tx^\pm_{a1}, \tx^\pm_{b2})\right)^{-2}=
\left(\Sigma^{\barnes}_{a b} (\tx^\pm_{a1}, \tx^\pm_{b2}) \right)^{-2} \ \left( \frac{\Sigma^{\bes}_{a b} (\tx^\pm_{a1}, \tx^\pm_{b2})}{\Sigma^{\hl}_{a b} (\tx^\pm_{a1}, \tx^\pm_{b2}) }\right)^{-2}\,,
\eal
where the ratio in the last bracket solves the ``even'' part of the crossing equation. It is straightforward to extend this to bound states. A mirror bound state composed of $m$ constituents with rapidities $u_1, \, \dots, \, u_m$  satisfies the conditions
\begin{equation}
\label{eq:mirror-bs-fusion}
    \tx_a^-(u_{j})=\tx_a^+(u_{j+1}),\qquad j=1,\dots, m-1\,.
\end{equation}
The dressing factor for the scattering of a $m_1$-particle bound state with a $m_2$-particle bound state is found by taking the product of $m_1\cdot m_2$ dressing factors for fundamental particle, each evaluated at the kinematics of the associated constituent. However, owing to~\eqref{eq:fourPhidecomp} it is easy to see that the double product telescopes, leaving only the ``outermost'' Zhukovsky variables
\begin{equation}
    \tx^{+ m}_a\equiv \tx_{a}^+(u_1)\,,\qquad
    \tx^{- m}_a\equiv \tx_{a}^-(u_m)\,,
\end{equation}
which obey the same kinematics as~\eqref{eq:boundstatemirror},
so that we get 
\begin{equation}
\label{eq:mirror-factor-bound-state}
\left(\Sigma^{m_1m_2}_{a b} (\tx^{\pm m_1}_{a1}, \tx^{\pm m_2}_{b2})\right)^{-2}=
\left(\Sigma^{\barnes}_{a b} (\tx^{\pm m_1}_{a1}, \tx^{\pm m_2}_{b2})\right)^{-2} \ \left( \frac{\Sigma^{\bes}_{a b} (\tx^{\pm m_1}_{a1}, \tx^{\pm m_2}_{b2})}{\Sigma^{\hl}_{a b} (\tx^{\pm m_1}_{a1}, \tx^{\pm m_2}_{b2}) }\right)^{-2}\,,
\end{equation}
where the functions previously introduced are just evaluated at points satisfying the (mirror) bound state kinematics~\eqref{eq:boundstatemirror}. 

\subsection{Analytic continuation to the string region}
\la{sec:cont_to_string}

\begin{figure}
\begin{center}
\includegraphics[width=5cm]{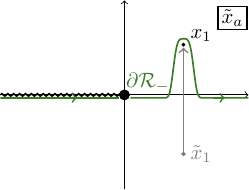}%
\hspace{1cm}%
\includegraphics[width=5cm]{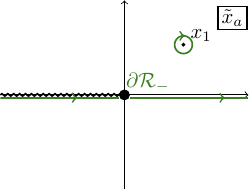}%
\end{center}
\caption{\label{fig:continuation}%
When analytically continuing $\tilde{x}$ to the antimirror region (for instance to reach a point $x_{ja}^+$) we need to be mindful of poles crossing the integration contours of the dressing phases. For instance, the integrand $\tPhi^{- \beta}_{ab}(\tilde{x}_a,\tilde{x}_b)$ develops a pole when $\tilde{x}_a$ approaches the integration contour. To analytically continue $\tPhi^{-\beta}_{ab}$ we can deform the integration contour (left), or equivalently write the continued function in terms of the same integral on the original contour, plus a residue (right). In the latter way, we generate $\tilde{\Psi}$ functions as residues with respect to the integration in either the first or the second variable, and the kernel $K^\bes$ as a residue with respect to both integrations.}
\end{figure}

Once we have solutions for the crossing equations in the mirror region we may continue them  to the string region as we show here. Let us consider an excitation of the mirror model so that both $\tilde{x}_a^+$ and $\tilde{x}_a^-$ lie in the lower-half plane. We want to continue these points to some string excitation parametrised by ${x}_a^\pm$. Now ${x}_a^\pm$ lie in the string physical region, but they do not necessarily lie in the mirror physical region too. In fact, for $u \in \mathbb{R}$ (where the energy and momentum of the string particle are both real) this is impossible since $x^+_a=1/{\tx^+_{\bar{a}}}$; in this case, $x^-_a$ lies in the intersection of the string and mirror region, while $x_a^+$ lies in the intersection of the string and anti-mirror region. To reach such a point, we need to move $\tx^+_{a}$ through one of the boundaries of the mirror region. The continuation path can be defined in terms of the $u$-rapidity, by parameterising $\tx^+_{a}$ (or any other function of interest such as the dressing factors) as functions of~$u$, i.e.~$\tx^+_{a}=\tx_a(u+\frac{i}{h})$; alternatively, it can be described in terms of~$\tilde{x}_a$. In terms of~$u$, we begin from the mirror $u$-plane, and we cross the main mirror cut $(\ubr, +\infty)$ from below, see figure~\ref{fig:rightmplanes}. This is a cut for $\tilde{x}_a(u)$; note also that~$\tilde{\Phi}^{\alpha\beta}_{ab}(\tx_1,\tx_2)$ is discontinuous there.  Our prescription is to analytically continue all such functions along a path going through the main cut from below on the mirror $u$-plane. On the $\tilde{x}$ plane, this results in a path crossing the real line to the right of $\xbr_a$. Of course there is no cut for $\tilde{x}$ itself, but we may encounter cuts of other functions. In particular, it is clear that \textit{e.g.}\ $\tilde{\Phi}^{\alpha\beta}_{ab}(\tx_1,\tx_2)$ needs to be continued when $\tx_1$ or $\tx_2$ crosses the integration contour, due to a pole in the integrand, see figure~\ref{fig:continuation}. 
Using this reasoning, we obtain the formulae for the continuation of the dressing factors to the string region for fundamental particles of real momentum and energy, when $\tilde{x}_{a1}^+,\tilde{x}_{a2}^+$ are taken to ${x}_{a1}^+,{x}_{a2}^+$ across their cuts as described.

For the sake of clarity let us describe in more detail how the continuation is done on the function $\tchi^{\alpha \beta}_{ab}(\tx^+_{a1}, \tx^-_{b2})$. Originally, both entries of this function are in the mirror region and we have
\bal
\tchi^{\alpha \beta}_{ab}(\tx^+_{a1}, \tx^-_{b2})=\tPhi^{\alpha \beta}_{ab}(\tx^+_{a1}, \tx^-_{b2})\,.
\eal
For $u_2 \in \mathbb{R}$ then $\tx^-_{b2}=x^-_{b2}$ and the second entry is already in the string region. On the contrary, we need to cross the real-$x$ line to continue $\tx^+_{a1}$ to the string region. 
In performing this continuation we cross the integration contour of $\tPhi$ and we need to pick up a residue coming from the continuation (see figure~\ref{fig:continuation}). Then after continuing $\tx^+_{a1}$ to the string region we obtain
\bal
\tchi^{\alpha \beta}_{ab}(x^+_{a1}, x^-_{b2})=\tPhi^{\alpha \beta}_{ab}(x^+_{a1}, x^-_{b2})- \tPsi^{\beta}_{b}(x^+_{a1}, x^-_{b2})\,,
\eal
where $\tPsi$ is the  analytic continuation of the discontinuity of the $\tPhi$ function,
\begin{equation}
\label{PsiBESkamirror}
\begin{aligned}
&\tPsi_{b}^{\beta}(x_{1},x_{2}) \equiv -\lint_{\pa\cR_\beta}\frac{{\rm
d}w_2}{2\pi i}\,\frac{1}{w_2-x_{2}} K^\bes(u_a(x_{1})-u_b(w_2))\,.
\end{aligned}
\end{equation}
Note that writing $x_{1}$ as $x_{1}=\tx_a(u_1)$ or $x_{1}=x_a(u_1)$ one gets $u_a(x_{1})=u_1$, and therefore $\tPsi_{b}^{\beta}$ depends in fact on $u_1$. Because of that we use interchangeably the notations $\tPsi_{b}^{\beta}(x_{1},\tx_{b2})$ and $\tPsi_{b}^{\beta}(u_{1},\tx_{b2})$ for one and the same $\tPsi_{b}^{\beta}$ function.

The continuation for the remaining pieces of BES can be found similarly and it yields
\bal
\tchi^{\alpha \beta}_{ab}(x^+_{a1}, x^-_{b2})&= \tPhi^{\alpha \beta}_{ab}(x^+_{a1}, x^-_{b2})- \tPsi^{\beta}_{b}(x^+_{a1}, x^-_{b2})\,,\\
\tchi^{\alpha \beta}_{ab}(x^-_{a1}, x^+_{b2})&= \tPhi^{\alpha \beta}_{ab}(x^-_{a1}, x^+_{b2})+ \tPsi^{\alpha}_{a}(x^+_{b2},x^-_{a1})\,,\\
\tchi^{\alpha \beta}_{ab}(x^+_{a1}, x^+_{b2})&= \tPhi^{\alpha \beta}_{ab}(x^+_{a1}, x^+_{b2})+ \tPsi^{\alpha}_{a}(x^+_{b2},x^+_{a1})- \tPsi^{ \beta}_{b}(x^+_{a1}, x^+_{b2})- K^\bes(u_a(x^+_{a1}) - u_b(x^+_{b2})) \,.
\eal
Introducing the function
\bal
\label{eq:tPsi_as_half_sum}
\tPsi_{b}(\tx_{a1},\tx_{b2}) \equiv \frac{1}{2} \left(\tPsi_{b}^{-}(\tx_{a1},\tx_{b2})+\tPsi_{b}^{+}(\tx_{a1},\tx_{b2}) \right)
\eal
and using the definitions in~\eqref{Phiabara}, the $\ka$-deformed improved BES phases are then analytically continued to the string region as
\bal
\label{eq:impr_bes_in_str_region}
\tilde{\theta}_{a b}^{\bes}(x^\pm_{a1}, x^\pm_{b2})&=\tPhi_{ab}(x^+_{a1},x^+_{b2})+\tPhi_{ab}(x^-_{a1},x^-_{b2})-\tPhi_{ab}(x^+_{a1},x^-_{b2})-\tPhi_{ab}(x^-_{a1},x^+_{b2})\\
&+\tPsi_{a}(x^+_{b2}, x^+_{a1})-\tPsi_{b}(x^+_{a1}, x^+_{b2})+\tPsi_{b}(x^+_{a1}, x^-_{b2})-\tPsi_{a}(x^+_{b2}, x^-_{a1})\\
&- K^\bes(u_a(x^+_{a1}) - u_b(x^+_{b2}))\,.
\eal

Similarly, we can find the continuation of the HL phase. In this case we obtain
\bal
\label{eq:newHL_string_aa}
\tilde{\theta}^{\hl}_{aa}(x^{\pm}_{ a1},x^{\pm}_{ a2})=&+\tPhi_{aa}^{\hl}(x^+_{a1},x^+_{a2})+\tPhi_{aa}^{\hl}(x^-_{a1},x^-_{a2})-\tPhi_{aa}^{\hl}(x^+_{a1},x^-_{a2})-\tPhi_{aa}^{\hl}(x^-_{a1},x^+_{a2})\\
&+ \frac{1}{2i} \log  \frac{x^+_{a1}}{x^-_{a1}} \ 
\frac{x^-_{a2}}{x^+_{a2}}\frac{ \left( x^+_{a1} - x^-_{a2} \right) \left( \frac{1}{x^+_{\bar{a}1}}-x^+_{a2} \right)  \left( x^-_{a1} - \frac{1}{x^+_{\bar{a}2}} \right)}
{ \left( x^+_{a2} -x^-_{a1}\right) \left(x^+_{a1}- \frac{1}{x^+_{\bar{a}2}} \right)  \left( \frac{1}{x^+_{\bar{a}1}}-x^-_{a2} \right)}  \,,
\eal
and
\bal
\label{eq:newHL_string_baa}
\tilde{\theta}^{\hl}_{\bar{a} a}(x^{\pm}_{ \bar{a} 1},x^{\pm}_{ a2})=&+\tPhi_{\bar{a} a}^{\hl}(x^+_{\bar{a} 1},x^+_{a2})+\tPhi_{\bar{a} a}^{\hl}(x^-_{\bar{a}1},x^-_{a2})-\tPhi_{\bar{a} a}^{\hl}(x^+_{\bar{a} 1},x^-_{a2})-\tPhi_{\bar{a} a}^{\hl}(x^-_{\bar{a} 1},x^+_{a2})\\
& + \frac{1}{2i} \log \frac{x^-_{a2}}{x^-_{\bar{a}1}}  
\frac{  \left( x^+_{a1} - x^-_{a2} \right) \left( x^+_{\bar{a}1} - x^+_{\bar{a}2} \right) \left( x^-_{\bar{a}1} - \frac{1}{x^+_{a2}} \right)}
{ \left( x^+_{\bar{a}2} -x^-_{\bar{a}1} \right) \left( x^+_{a2} - x^+_{a1} \right) \left(\frac{1}{x^+_{\bar{a}1}}-x^-_{a2}  \right)} \,.
\eal
The full expression for the HL phase (including the terms which would cancel in the mirror-mirror region) is given in~\eqref{eq:PhiHL_as_functions_of_IHL1} and~\eqref{eq:PhiHL_as_functions_of_IHL2};  its analytic continuation is discussed in detail in appendix~\ref{app:discontinuities_HL}. 
Notice how e.g.\ the function in~\eqref{eq:newHL_string_aa}, which is evaluated at $x_{aj}^\pm$, also depends explicitly on~$1/x_{\bar{a}j}$. Here $1/x_{\bar{a}j}$ is a solution of
\begin{equation}
    u_{\bar{a}}(x_{\bar{a}j}^\pm) = u_{a}(x_{aj}^\pm)\,,
\end{equation}
so that in the limit $\kappa\to0$ we would have $x_{aj}^\pm\to x_j^\pm$ and $1/x_{\bar{a}j}^\pm\to 1/x_j^\pm$. At $\kappa>0$ the function $x_{\bar{a}j} (x_{aj})$ (i.e. $x_{\bar{a}j}$ as a function of $x_{aj}$) cannot be expressed in terms of elementary functions, and it is a remarkable self-consistency check of our construction that it does not contribute to crossing nor to the perturbative expansion of the dressing factors, once all terms are accounted for. We will explicitly show this later.

Finally, as for the odd dressing factors, recall that by virtue of eq.~\eqref{eq:stringmirrorgamma} the analytic continuation from the mirror to the string region corresponds to an overall shift of the $\tg$ functions by $+\frac{i \pi}{2}$. Due to this fact the odd part of the phases, which only depends on the difference of the rapidities, is unchanged  under this analytic continuation.

\paragraph{String bound-state dressing factors.}
Starting from the dressing factors of fundamental particles in the string region, it is possible to define those of bound states. This can be done by fusion, in a way similar to what we described for the mirror model in section~\ref{sec:mirror-dressing-bound} above. Note however that the string bound-state fusion condition differs from~\eqref{eq:mirror-bs-fusion} and it is instead
\begin{equation}
    x_a^+(u_{j})=x_a^-(u_{j+1}),\qquad j=1,\dots, m-1\,.
\end{equation}
The string bound-state dressing factors are presented in appendix~\ref{app:crossing-string-bs}, see also appendix~\ref{app:string-bound-states} for the S-matrix normalisation.

\section{Verifying the proposal}\label{sec:verification}
In this section we will check that our proposal is consistent with the various symmetries which we listed in section~\ref{sec:properties}.

\subsection{Crossing in the mirror region}

The crossing transformation in the mirror theory is defined by moving
first $\tx^-_{a1}$ and then  $\tx^+_{a1}$ to the upper-half plane through the interval $(0,\xbr_a)$. On the $u$-plane this corresponds to crossing first the main cut of $\tx^-_{a}(v)$ (i.e. the line $(\frac{i}{h}+\ubr,\frac{i}{h}+\infty)$) from above, and then the main cut of $\tx^+_{a}(v)$ (i.e. the line $(-\frac{i}{h}+\ubr,-\frac{i}{h}+\infty)$) also from above. We then take the points $\tilde{x}^\pm_{a}$ to the values~$1/\tilde{x}^\pm_{\bar{a}}$, as needed under crossing.

\paragraph{Mirror crossing for improved BES.}
After continuing 
\bal
\tx^-_{a1} \to \frac{1}{\tx^-_{\bar{a}1}}
\eal
we obtain
\bal
\label{eq:tchi_crossing_step1}
&\tchi^{\a \b}_{ab}(\frac{1}{{\tx}^-_{\bar{a}1}},\tx_{b2})=\tPhi^{\a \b}_{ab}(\frac{1}{{\tx}^-_{\bar{a}1}},\tx_{b2})-\tPsi^{\b}_{b}(\frac{1}{{\tx}^-_{\bar{a}1}},\tx_{b2})\,,\\
&\tchi^{\a \b}_{ab}(\tx^+_{a1},\tx_{b2})=\tPhi^{\a \b}_{ab}(\tx^+_{a1},\tx_{b2})\,.
\eal
Now we must repeat the same procedure for $\tx^+_{a1}$.
Analogously to what happens for $k=0$ \cite{Arutyunov:2009kf}, in doing this second continuation we are forced to cross a cut of the function 
$$
\tPsi^{\b}_{b}(\frac{1}{{\tx}^-_{\bar{a} 1}},\tx_{b2})
$$
and an additional contribution is generated. We refer the reader to appendix~\ref{app:tPsi_disc} for a more detailed explanation of how additional contributions from the cuts of $\tPsi$ are generated.
After continuing in order $\tx^-_{a1}$ and $\tx^+_{a1}$ to the anti-mirror region we end up with
\bal
&\tchi^{\a \b}_{ab}(\frac{1}{\tx^-_{\bar a1}},\tx_{b2})=\tPhi^{\a \b}_{ab}(\frac{1}{\tx^-_{\bar a1}},\tx_{b2})-\tPsi^{\b}_{b}(\frac{1}{\tx^-_{\bar a1}},\tx_{b2})-\frac{1}{i} \log\frac{\frac{1}{\tx^+_{\bar{b}1}}-\tx_{b2}}{\tx^+_{b1}-\tx_{b2}}\,,\\
&\tchi^{\a \b}_{ab}(\frac{1}{\tx^+_{\bar a1}},\tx_{b2})=\tPhi^{\a \b}_{ab}(\frac{1}{\tx^+_{\bar a1}},\tx_{b2})-\tPsi^{\b}_{b}(\frac{1}{\tx^+_{\bar a1}},\tx_{b2})\,,
\eal
where the additional $\log$ in the first line of the expression above comes from the continuation of $\tPsi_{b}^{\b}(\frac{1}{\tx^-_{\bar{a}1}},\tx_{b2})$ when moving $\tx^+_{a1}$ to the anti-mirror region, as described in appendix~\ref{app:tPsi_disc}.

Using that $\forall \ x$ and $y \in \mathbb{C}$
\bal
\tPhi^{\a \b}_{ab}(\frac{1}{x}, y)+\tPhi^{-\a \, \b}_{\bar a b}(x, y)=\tPhi^{-\a \, \b}_{\bar a b}(0, y)\,,
\eal 
and the analytic continuation just derived, then the crossing equations for the improved BES phases take the form
 \bal\label{creqthmir}
 \tilde{\theta}_{ab}^{\a\b}({1\ov \tx_{\bar a 1}^\pm}, \tx_{b2}^\pm) + \tilde{\theta}_{\bar a b}^{-\a \, \b}( \tx_{\bar a 1}^\pm, \tx_{b2}^\pm)
=&-\Delta^{\b}_{b} (u_1 \pm \frac{i}{h}, \tx_{b2}^\pm)-{ 1\ov i}\log \frac{{1\ov \tx_{\bar b 1}^+}-\tx_{b2}^-}{\tx_{b 1}^+-\tx_{b2}^-}+{ 1\ov i}\log \frac{{1\ov \tx_{\bar b 1}^+}-\tx_{b2}^+}{\tx_{b 1}^+-\tx_{b2}^+}\,,
 \eal
 where we used the definition~\eqref{eq:Dab_definition}. 
 For $\b=-$ we use indentity~\eqref{eq:Deltaepsm_2mm} and find
\bal
 \tilde{\theta}_{a b}^{ \a \, -}({1\ov \tx_{\bar{a} 1}^\pm}, \tx_{b 2}^\pm) &+ \tilde{\theta}_{\bar{a} b}^{-\a, \, -}( \tx_{ 1}^\pm, \tx_{b2}^\pm)={ 1\ov i}\log 
 \frac{\tx_{b 1}^--\tx_{b 2}^+}{{ \tx_{b1}^-}-\tx_{b2}^-} \ \frac{1-{1\ov \tx_{\bar{b} 1}^+\tx_{b2}^+}}{1-{1\ov \tx_{\bar{b} 1}^+\tx_{b2}^-}} \ {u_{12}+{2i\ov h}\ov u_{12}-{2i\ov h}}  \,.
\eal
For $\nu=+$ we use~\eqref{eq:Deltaepsp_2mm} and find
 \bal
 \tilde{\theta}_{a b}^{ \a \, +}({1\ov \tx_{\bar{a} 1}^\pm}, \tx_{b 2}^\pm) &+ \tilde{\theta}_{\bar{a} b}^{-\a, \, +}( \tx_{ 1}^\pm, \tx_{b2}^\pm)={ 1\ov i}\log \frac{\tx_{b 1}^+ -\tx_{b 2}^-}{{ \tx_{b1}^+}-\tx_{b2}^+} \ \frac{1-{1\ov \tx_{\bar{b} 1}^- \tx_{b2}^-}}{1-{1\ov \tx_{\bar{b} 1}^- \tx_{b2}^+}}   \,.
\eal
Then, using the definitions in~\eqref{eq:ttheta_definition_mir_fund}, we get the following crossing equation for the $\ka$-deformed improved BES phases
\bal
\label{eq:crossng_impr_bes_mirror_reg}
&2\tilde \theta^\bes_{a a}( \tx_{a 1}^\pm, \tx_{a 2}^\pm)+2\tilde \theta^\bes_{\bar a a}({1\ov \tx_{a 1}^\pm}, \tx_{a 2}^\pm) =2\tilde \theta^\bes_{a a}({1\ov \tx_{\bar a 1}^\pm}, \tx_{a 2}^\pm) + 2\tilde \theta^\bes_{\bar a a}( \tx_{\bar a 1}^\pm, \tx_{a 2}^\pm)\\
&={ 1\ov i}\log  \frac{\tx_{a 1}^+-\tx_{a 2}^-}{{ \tx_{a1}^+}-\tx_{a2}^+}\ \frac{\tx_{a 1}^--\tx_{a2}^+}{{ \tx_{a1}^-}-\tx_{a2}^-}\ \frac{1-{1\ov \tx_{\bar a 1}^-\tx_{a2}^-}} {1-{1\ov \tx_{\bar a 1}^-\tx_{a2}^+}}\ \frac{1-{1\ov \tx_{\bar a 1}^+\tx_{a2}^+}}{1-{1\ov \tx_{\bar a 1}^+\tx_{a2}^-}}\ {u_{12}+{2i\ov h}\ov u_{12}-{2i\ov h}}  \,.
\eal

\paragraph{Mirror crossing for improved HL.}
Let us show how we can continue the first variable in the improved HL phase to the anti-mirror region. Using~\eqref{eq:PhiHL_as_functions_of_IHL1}, for $\tx_{a1}$ and $\tx_{a2}$ in the mirror region it holds that
\bal
&\tPhi_{aa}^{\hl}(\tx_{a1},\tx_{a2})=I^\hl_{aa}(\tx_{a1},\tx_{a2})+{i\ov 4}\ln(-\tx_{a2}) +{\pi\ov 8}\,.
\eal
Using~\eqref{eq:IHL_continuation_properties_main_above} we can continue $\tx_{a1}\to \frac{1}{\tx_{\bar{a}1}}$ as follows 
\bal
&\tPhi_{aa}^{\hl}(\tx_{a1},\tx_{a2})\to I^\hl_{aa}(\frac{1}{\tx_{\bar{a}1}},\tx_{a2})+\frac{1}{2i} \log  \frac{\frac{1}{\tx_{\bar a 1}} - \tx_{a 2}}{\tx_{a 1} - \tx_{a 2}} +{i\ov 4}\ln(-\tx_{a2}) +{\pi\ov 8}\,.
\eal
Now using~\eqref{eq:PhiHL_as_functions_of_IHL1} we know that
 \bal
I^\hl_{aa}(\frac{1}{\tx_{\bar{a}1}},\tx_{a2})=\tPhi_{aa}^{\hl}(\frac{1}{\tx_{\bar{a}1}},\tx_{a2})+{i\ov 4}\ln(-\tx_{a2}) +{\pi\ov 8}
\eal
and therefore the continuation is given by
\bal
\tPhi_{aa}^{\hl}(\tx_{a1},\tx_{a2}) \to \tPhi_{aa}^{\hl}(\frac{1}{\tx_{\bar{a}1}},\tx_{a2})+\frac{1}{2i} \log  \frac{\frac{1}{\tx_{\bar a 1}} - \tx_{a 2}}{\tx_{a 1} - \tx_{a 2}}  + \frac{i}{2}\ln(-\tx_{a2}) + \frac{\pi}{4} \,.
\eal
Similarly we obtain
\bal
\tPhi_{\bar a a}^{\hl}(\tx_{\bar{a}1}, \tx_{a2}) \to \tPhi_{\bar a a}^{\hl}(\frac{1}{\tx_{a1}}, \tx_{a2}) +\frac{1}{2i} \log  \frac{\frac{1}{\tx_{\bar a 1}} - \tx_{a 2}}{\tx_{a 1} - \tx_{a 2}} +\frac{i}{2}\ln(-\tx_{a2}) + \frac{\pi}{4} \,.
\eal
Plugginig these continuations into the full phase we obtain
\bal
\label{eq:crossng_impr_hl_mirror_reg}
&2\tilde{\theta}^{\hl}_{aa}(\frac{1}{\tx^\pm_{\bar a1}},\tx^\pm_{a2}) + 2\tilde{\theta}^{\hl}_{\bar a a}(\tx^\pm_{\bar a1},\tx^\pm_{a2})=2\tilde{\theta}^{\hl}_{aa}(\tx^{\pm}_{ a1},\tx^{\pm}_{a2}) + 2\tilde{\theta}^{\hl}_{\bar a a}(\frac{1}{\tx^{\pm}_{a1}},\tx^{\pm}_{a2})\\
&=-{1\ov  i}  \log  \frac{\left(\tx^+_{ a 1} - \tx^+_{a 2} \right)  \left(\tx^-_{ a 1} - \tx^-_{a 2} \right) \left(1- \frac{1}{\tx^+_{\bar{a}1} \tx^-_{a2}} \right) \left(1- \frac{1}{\tx^-_{\bar{a}1} \tx^+_{a2}} \right)}{\left(\tx^+_{ a 1} - \tx^-_{a 2} \right) \left(\tx^-_{ a 1} - \tx^+_{a 2} \right)  \left(1- \frac{1}{\tx^+_{\bar{a}1} \tx^+_{a2}} \right)  \left(1- \frac{1}{\tx^-_{\bar{a}1} \tx^-_{a2}} \right)}    \,.
\eal
Combining~\eqref{eq:crossng_impr_bes_mirror_reg} and~\eqref{eq:crossng_impr_hl_mirror_reg} we see that the crossing equation~\eqref{eq:even_crossing_equation} for the ratio between BES and HL is satisfied.

\paragraph{Mirror crossing for the odd part.}
Finally, the odd parts of the phases defined in~\eqref{eq:odd_Sigma_mirror} satisfy the equations~(\eqref{eq:odd_mir_crossing1}, \eqref{eq:odd_mir_crossing2}). This is immediately verified by using the monodromy properties of the $R$ functions (see~\eqref{eq:monodromy_R}) combined with the fact that under crossing 
\bal
\tg^\pm_a \to \tg^\pm_{\bar a} -i\pi
\eal
as explained in section~\ref{sec:kinematics}.
This concludes our check of the crossing equations for the phases~\eqref{eq:complete_dressing_ph_mir_reg} for mirror kinematics.

\subsection{Crossing in the string region}

Starting from string kinematics, the analytic continuation to the anti-string region is performed as follows. We start with
\bal
x^{+}_{a1}=x_{a}(u_1+\frac{i}{h}), \qquad x^{-}_{a1}=x_{a}(u_1-\frac{i}{h})\,, \quad u_1 \in \mathbb{R} \,,
\eal
which have long string theory cuts $(-\infty-\frac{i}{h}, \ubr-\frac{i}{h})$ and $(-\infty+\frac{i}{h}, \ubr+\frac{i}{h})$, respectively. Originally we have $u_1 \in \mathbb{R}$; starting from this point, we follow the clockwise direction and go across the following cuts. In order we cross the mirror theory cut $(\ubr-\frac{i}{h},+\infty-\frac{i}{h})$ from above, the string theory cut $(-\infty-\frac{i}{h}, \ubr-\frac{i}{h})$ from below, the string theory cut $(-\infty+\frac{i}{h}, \ubr+\frac{i}{h})$ from below and finally the mirror theory cut $(\ubr+\frac{i}{h},+\infty+\frac{i}{h})$ from above.

\begin{figure}
\begin{center}
\includegraphics{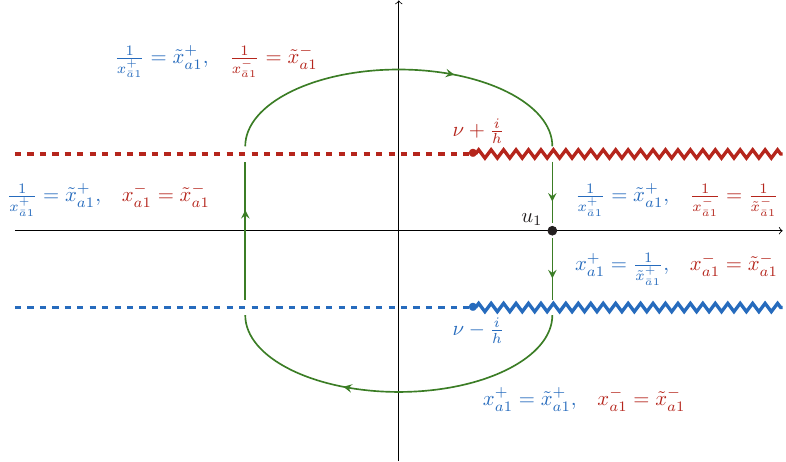}
\caption{Crossing path in the string model. The blue and red zigzag lines correspond to the mirror theory cuts of $\tilde{x}^+_{a}$ and $\tilde{x}^-_{a}$ respectively. The dashed blue and red lines correspond to the string theory cuts of $x^+_{a}$ and $x^-_{a}$. The continuation to the anti-string region can be defined by using either the mirror parameterisation $\tilde{x}^\pm_{a}$ or the string one $x^\pm_{a}$. The figure shows the crossing path for both parameterisations.}
\label{fig:continuation_crossing_string}
\end{center}
\end{figure}
Performing this path in the $u$ plane we obtain that first
\bal
x^{+}_{a1} \to \frac{1}{x^{+}_{\bar{a}1}}
\eal
and then
\bal
x^{-}_{a1} \to \frac{1}{x^{-}_{\bar{a}1}}\,.
\eal
This continuation can be performed using either the string parameterisation $x^\pm_{a1}$ or the mirror parameterisation $\tx^\pm_{a1}$. In the latter case the continuation reads
\bal
\label{eq:str_to_antistr_using_mir}
\frac{1}{\tx^{+}_{\bar{a}1}} \to \tx^{+}_{a1}, \qquad \tx^{-}_{a1} \to \frac{1}{\tx^{-}_{\bar{a}1}}\,.
\eal
Even if we are in the string region, it may be helpful to write the continuation using the mirror parameterisation since the BES and HL phases are written as integrals around the mirror theory cuts. The continuation path in the $u$ plane is shown in figure~\ref{fig:continuation_crossing_string}. In the $x$ plane, the path corresponds to moving first $x^+_{a1}$ to the lower half of the complex plane through the half line $x>\xbr_a$ and entering the anti-string region with $x^+_{a1}$ from below. Then we move $x^-_{a1}$ to the anti-string region from below and cross the interval $(0, \xbr_a)$.

\paragraph{String crossing for improved BES.}
If we consider fundamental particles, 
the continuation of $x_{a1}^{+}$ to the anti-string region is the opposite of what we did to go from the mirror to the string region in section~\ref{sec:cont_to_string} (as shown in the expression on the l.h.s. of~\eqref{eq:str_to_antistr_using_mir}); as a consequence we just need to remove from~\eqref{eq:impr_bes_in_str_region}  the terms arising from the continuation of the first variable to the mirror region and obtain
\bal
2\tilde{\theta}_{a b}^{\bes}(\frac{1}{x^{+}_{\bar{a} 1}},x^{-}_{a1}; x^{\pm}_{b2})&=\tPhi_{ab}(\frac{1}{x^{+}_{\bar{a} 1}},x^{+}_{b2})+\tPhi_{ab}(x^{-}_{a1},x^{-}_{b2})-\tPhi_{ab}(\frac{1}{x^{+}_{\bar{a}1}},x^{-}_{b2})-\tPhi_{ab}(x^{-}_{a1},x^{+}_{b2})\\
&+\tPsi_{a}(x^{+}_{b2}, \frac{1}{x^{+}_{\bar{a}1}})-\tPsi_{a}(x^{+}_{b2}, x^{-}_{a1}) \,.
\eal
Now we continue $x^{-}_{a1}$ and we enter first its string theory cut from below (this is not a real cut of $x^{-}_{a1}$) and then its mirror theory cut from above. The continuation of the $\tPhi$ and $\tPsi$ functions is given by
\bal
\tPhi_{ab}(x^{-}_{a1},\tx_{2}) &\to \tPhi_{ab}(\frac{1}{x^{-}_{\bar{a} 1}},\tx_{2}) - \tPsi_{b}(\frac{1}{x^{-}_{\bar{a} 1}},\tx_{2})\,,\\
\tPsi_{a}(x^{+}_{b2}, x^{-}_{a1}) &\to \tPsi_{a}(x^{+}_{b2}, \frac{1}{x^{-}_{\bar{a}1}}) +  K^\bes(u_b(x^{+}_{b2})-u_{\bar{a}}(x^{-}_{\bar{a}1}))\,.
\eal
Then the continuation of the improved BES phase under a crossing transformation of the string model is given by
\bal
\label{eq:continuation_BES_antistring}
\tilde{\theta}_{a b}^{\bes}(\frac{1}{x^{\pm}_{\bar{a} 1}}, x^{\pm}_{b2})&=\tPhi_{ab}(\frac{1}{x^{+}_{\bar{a} 1}},x^{+}_{b2})+\tPhi_{ab}(\frac{1}{\tx^{-}_{\bar{a} 1}},x^{-}_{b2})-\tPhi_{ab}(\frac{1}{x^{+}_{\bar{a}1}},x^{-}_{b2})-\tPhi_{ab}(\frac{1}{x^{-}_{\bar{a} 1}},x^{+}_{b2})\\
&+\tPsi_{a}(x^{+}_{b2}, \frac{1}{x^{+}_{\bar{a}1}})-\tPsi_{a}(x^{+}_{b2}, \frac{1}{x^{-}_{\bar{a} 1}})- \tPsi_{b}(\frac{1}{x^{-}_{\bar{a} 1}},x^{-}_{b2})+ \tPsi_{b}(\frac{1}{x^{-}_{\bar{a} 1}},x^{+}_{b2})\\
& +K^\bes\big( u_{12}-\frac{2i}{h} \big)\,.
\eal
Combining~\eqref{eq:impr_bes_in_str_region} and~\eqref{eq:continuation_BES_antistring} we obtain the following crossing equations for the BES phases
\bal
&\tilde{\theta}_{a b}^{\bes}(\frac{1}{x^{\pm}_{\bar{a} 1}}, x^{\pm}_{b2})+\tilde{\theta}_{\bar{a} b}^{\bes}(x^{\pm}_{\bar{a}1}, x^{\pm}_{b2})=\\
&+\Psi_{ b}(x^{+}_{\bar{a} 1}, x^{-}_{b2})- \Psi_{b}(\frac{1}{x^{-}_{\bar{a} 1}},x^{-}_{b2}) +\Psi_{ b}(\frac{1}{x^{-}_{\bar{a} 1}},x^{+}_{b2}) -\Psi_{ b}(x^{+}_{\bar{a} 1}, x^{+}_{b2}) \\
&+\Psi_{a}(x^{+}_{b2}, \frac{1}{x^{+}_{\bar{a}1}}) +\Psi_{\bar{a}}(x^{+}_{b2}, x^{+}_{\bar{a} 1})-\Psi_{a}(x^{+}_{b2}, \frac{1}{x^{-}_{\bar{a} 1}})-\Psi_{ \bar{a}}(x^{+}_{b2}, x^{-}_{\bar{a} 1})\\
&- \frac{1}{i} \log \left( \frac{h}{2} \right)^{2} -  \frac{1}{i} \log \left[ \left( u_{12}-\frac{2i}{h}  \right)   u_{12}   \right]\,,
\eal
where to generate the last line we used the factorial properties of the $\Gamma$ functions.
Noting that
\bal
\Psi_{b}^{\beta} \left(x,\frac{1}{y} \right)+\Psi_{\bar{b}}^{-\beta} \left(x,y \right)=\Psi_{\bar{b}}^{-\beta} \left(x,0\right)\,,
\eal
the third line of the expression above vanishes. Moreover from appendix~\ref{app:psi-identities} we recognise
\bal
&\Psi_{ b}(x^{+}_{\bar{a} 1}, x^{-}_{b2})- \Psi_{b}(\frac{1}{x^{-}_{\bar{a} 1}},x^{-}_{b2}) +\Psi_{b}(\frac{1}{x^{-}_{\bar{a} 1}},x^{+}_{b2}) -\Psi_{b}(x^{+}_{\bar{a} 1}, x^{+}_{b2})=\\
&-\frac{1}{2} \left(\Delta_{b}^{-}(u_1\pm {i\ov h},x_{b 2}^{\pm })+\Delta_{b}^{+}(u_1\pm {i\ov h},x_{b 2}^{\pm}) \right)=\frac{1}{2i}\log \left( u_{12}+\frac{2i}{h} \right) \left( u_{12}-\frac{2i}{h} \right) u^2_{12}\\
&\qquad\qquad+\frac{2}{i} \log  \frac{h}{2} -\frac{1}{2i}\log  \frac{\left(x^{+}_{b1} -x^{+}_{b2} \right) \left(x^{-}_{b1} -x^{-}_{b2} \right) \left(1-\frac{1}{x^{+}_{\bar{b} 1} x^{-}_{b2}} \right) \left(1-\frac{1}{x^{-}_{\bar{b} 1} x^{+}_{b2}} \right)}{\left(x^{+}_{b1} -x^{-}_{b2} \right) \left(x^{-}_{b1} -x^{+}_{b2} \right) \left(1-\frac{1}{x^{+}_{\bar{b} 1} x^{+}_{b2}} \right) \left(1-\frac{1}{x^{-}_{\bar{b} 1} x^{-}_{b2}} \right)} \,,
\eal
and we end up with
\bal
\label{eq:crossing_string_BES}
&2\tilde{\theta}_{a b}^{\bes}(\frac{1}{x^{\pm}_{\bar{a} 1}}, x^{\pm}_{b2})+2\tilde{\theta}_{\bar{a} b}^{\bes}(x^{\pm}_{\bar{a}1}, x^{\pm}_{b2})=\\
&+ \frac{1}{i}\log \frac{ u_{12}+\frac{2i}{h}   }{ u_{12}-\frac{2i}{h}  }-\frac{1}{i}\log \frac{\left(x^{+}_{b1} -x^{+}_{b2} \right) \left(x^{-}_{b1} -x^{-}_{b2} \right) \left(1-\frac{1}{x^{+}_{\bar{b} 1} x^{-}_{b2}} \right) \left(1-\frac{1}{x^{-}_{\bar{b} 1} x^{+}_{b2}} \right)}{\left(x^{+}_{b1} -x^{-}_{b2} \right) \left(x^{-}_{b1} -x^{+}_{b2} \right) \left(1-\frac{1}{x^{+}_{\bar{b} 1} x^{+}_{b2}} \right) \left(1-\frac{1}{x^{-}_{\bar{b} 1} x^{-}_{b2}} \right)} \,.
\eal

\paragraph{String crossing for improved HL.}

Let us now focus on the crossing equations for the improved HL phases.
In this case, using the results of appendix~\ref{app:discontinuities_HL} we find that
\bal
&\tPhi_{aa}^{\hl}(x^-_{a1},x^\pm_{a2}) \to \tPhi_{aa}^{\hl}(\frac{1}{x^-_{\bar{a}1}},x^\pm_{a2})+ \frac{1}{i} \log \frac{\frac{1}{x^-_{\bar{a}1}}- x^\pm_{a2}}{x^-_{a1} - x^\pm_{a2}} + i \log(-x^\pm_{a2}) \mp \frac{\pi}{2}\,,\\
&\tPhi_{aa}^{\hl}(x^+_{a1},x^\pm_{a2}) \to \tPhi_{aa}^{\hl}(\frac{1}{x^+_{\bar{a}1}},x^\pm_{a2})+ \frac{1}{i} \log  \frac{\frac{1}{x^+_{\bar{a}1}}- x^\pm_{a2}}{x^+_{a1} - x^\pm_{a2}}  -i \log(-x^\pm_{a2})\pm \frac{\pi}{2}\,,
\eal
and
\bal
&\tPhi_{\bar{a}a}^{\hl}(x^-_{\bar{a}1},x^\pm_{a2}) \to \tPhi_{\bar{a}a}^{\hl}(\frac{1}{x^-_{a1}},x^\pm_{a2})+ \frac{1}{i} \log  \frac{\frac{1}{x^-_{\bar{a}1}} -x^\pm_{a2}}{x^-_{a1} - x^\pm_{a2}}  + i \log(-x^\pm_{a2}) \mp \frac{\pi}{2} \,,\\
&\tPhi_{\bar{a}a}^{\hl}(x^+_{\bar{a}1},x^\pm_{a2}) \to \tPhi_{\bar{a}a}^{\hl}(\frac{1}{x^+_{a1}},x^\pm_{a2})+ \frac{1}{i} \log  \frac{\frac{1}{x^+_{\bar{a}1}} -x^\pm_{a2}}{x^+_{a1} - x^\pm_{a2}}   -i \log(-x^\pm_{a2})\pm \frac{\pi}{2}\,.
\eal
Recalling that the HL phases in the string region are given by~\eqref{eq:newHL_string_aa} and~\eqref{eq:newHL_string_baa} we obtain the following crossing equations for fundamental string particles
\bal
\label{eq:crossing_string_HL}
&2 \tilde{\theta}_{aa}^\hl (\frac{1}{x^\pm_{\bar{a}1}}, x^\pm_{a2}) + 2 \tilde{\theta}_{\bar{a} a}^\hl (x^\pm_{\bar{a}1}, x^\pm_{a2})=2 \tilde{\theta}_{\bar{a} a}^\hl (\frac{1}{x^\pm_{a1}}, x^\pm_{a2}) + 2 \tilde{\theta}_{a a}^\hl (x^\pm_{a1}, x^\pm_{a2})=\\
&- \frac{1}{i} \log \frac{ \left( x^+_{a1} - x^+_{a2} \right) \ \left( x^-_{a1} - x^-_{a2} \right) \ \left( 1 - \frac{1}{x^+_{\bar{a}1} x^-_{a2}} \right) \ \left( 1 - \frac{1}{x^-_{\bar{a}1} x^+_{a2}} \right)}{\left( x^+_{a1} - x^-_{a2} \right) \left( x^-_{a1} - x^+_{a2} \right)\ \left( 1 - \frac{1}{x^+_{\bar{a}1} x^+_{a2}} \right) \ \left( 1 - \frac{1}{x^-_{\bar{a}1} x^-_{a2}} \right)} \,.
\eal

As for the crossing equations in the mirror region even in this case the difference between the crossing equations of BES and HL, i.e. the difference between~\eqref{eq:crossing_string_BES} and~\eqref{eq:crossing_string_HL}, satisfies
\bal
&2\tilde{\theta}_{a b}^{\bes}(\frac{1}{x^{\pm}_{\bar{a} 1}}, x^{\pm}_{b2})-2\tilde{\theta}_{a b}^{\hl}(\frac{1}{x^{\pm}_{\bar{a} 1}}, x^{\pm}_{b2})+2\tilde{\theta}_{\bar{a} b}^{\bes}(x^{\pm}_{\bar{a}1}, x^{\pm}_{b2})-2\tilde{\theta}_{\bar{a} b}^{\hl}(x^{\pm}_{\bar{a}1}, x^{\pm}_{b2})= \frac{1}{i}\log  \frac{ u_{12}+\frac{2i}{h} }{ u_{12}-\frac{2i}{h}}\,,
\eal
and the dressing factors~\eqref{eq:even_Sigma_mirror} satisfy the even part of the crossing equations also in the string region.

Finally, the odd dressing factors in~\eqref{eq:odd_Sigma_mirror} satisfy the odd crossing equation no matter whether the crossing transformation is performed in the mirror or string model. Indeed in both cases, crossing corresponds to shifting the $\g$ rapidities by $-i \pi$; this can be
\bal
\tg^\pm_{a1} \to \tg^\pm_{\bar{a}1} - i\pi\,,
\eal
if we consider crossing in the mirror model, or
\bal
\g^\pm_{a1} \to \g^\pm_{\bar{a}1} - i\pi\,,
\eal
if we consider crossing in the string model. In both cases, the crossing equations are satisfied thanks to the monodromy relations of the $R$ functions (see~\eqref{eq:monodromy_R}). We conclude that the dressing factors~\eqref{eq:complete_dressing_ph_mir_reg} satisfy the crossing equations both in the string and mirror region, as expected.

\subsection{Unitarity in the string model}

Braiding unitarity is satisfied by construction both in the string and mirror model due to the antisymmetry properties of the different pieces of the dressing factors.
The check of unitarity is nontrivial and we present it here.
Recall that in the string model for $u \in \mathbb{R}$ it holds that $(x^\pm_{a})^*=x^\mp_{a}$.
Then starting from~\eqref{eq:impr_bes_in_str_region} and using the conjugacy conditions~\eqref{eq:app_tPhi_compl_conj} and \eqref{eq:app_tPsi_compl_conj} together with the definition~\eqref{eq:app_delta_string} we obtain
\bal
\left( \tilde{\theta}_{a b}^{\bes}(x^\pm_{a1}, x^\pm_{b2}) \right)^*= \tilde{\theta}_{a b}^{\bes}(x^\pm_{a1}, x^\pm_{b2})&+\frac{1}{2} \left( \Delta_b^+(u_1 \pm \frac{i}{h}, x^\pm_{b2}) +\Delta_b^-(u_1 \pm \frac{i}{h}, x^\pm_{b2}) \right)\\
&-\frac{1}{2} \left( \Delta_a^+(u_2 \pm \frac{i}{h}, x^\pm_{a1}) +\Delta_a^-(u_2 \pm \frac{i}{h}, x^\pm_{a1}) \right) \,.
\eal
Substituting the relations~\eqref{eq:Deltaepsm_2ms} and~\eqref{eq:Deltaepsp_2ms} into the expressions above we get
\bal
\left( \tilde{\theta}_{a a}^{\bes}(x^\pm_{a1}, x^\pm_{a2}) \right)^*&= \tilde{\theta}_{a a}^{\bes}(x^\pm_{a1}, x^\pm_{a2})\\
&+\frac{1}{2i}\log  \frac{ \left(1-\frac{1}{x^{+}_{\bar{a} 1} x^{-}_{a2}} \right) \left(1-\frac{1}{x^{-}_{\bar{a} 1} x^{+}_{a2}} \right) \left(1-\frac{1}{x^{+}_{a1} x^{+}_{\bar{a} 2}} \right) \left(1-\frac{1}{x^{-}_{a1} x^{-}_{\bar{a} 2}} \right)}{ \left(1-\frac{1}{x^{+}_{\bar{a} 1} x^{+}_{a2}} \right) \left(1-\frac{1}{x^{-}_{\bar{a} 1} x^{-}_{a2}} \right) \left(1-\frac{1}{x^{-}_{a 1} x^{+}_{\bar{a} 2}} \right) \left(1-\frac{1}{x^{+}_{a1} x^{-}_{\bar{a} 2}} \right)}\,,\\
\left( \tilde{\theta}_{\bar{a} a}^{\bes}(x^\pm_{\bar{a}1}, x^\pm_{a2}) \right)^*&= \tilde{\theta}_{\bar{a} a}^{\bes}(x^\pm_{\bar{a}1}, x^\pm_{a2})\\
&+\frac{1}{2i}\log  \frac{\left(x^{+}_{a1} -x^{+}_{a2} \right) \left(x^{-}_{a1} -x^{-}_{a2} \right) \left(x^{-}_{\bar{a}1} -x^{+}_{\bar{a}2} \right) \left(x^{+}_{\bar{a}1} -x^{-}_{\bar{a}2} \right) }{\left(x^{+}_{a1} -x^{-}_{a2} \right) \left(x^{-}_{a1} -x^{+}_{a2} \right) \left(x^{-}_{\bar{a} 1} -x^{-}_{\bar{a} 2} \right) \left(x^{+}_{\bar{a} 1} -x^{+}_{\bar{a} 2} \right) }\,.
\eal
Due to the logarithms in the expressions above $|\Sigma^\bes_{ab}(u_1, u_2)|^2 \ne 1$ for $u_1, u_2 \in \mathbb{R}$ in the string region. This is a peculiarity of the mixed flux; indeed the log contributions in the expressions above cancel in the Ramond-Ramond case, where $x^\pm_{aj}=x^\pm_{\bar{a}j}$ ($j=1,2$). It is easy to show that, starting from~\eqref{eq:newHL_string_aa}, \eqref{eq:newHL_string_baa}, the same logarithms are also generated in the complex conjugation of the HL phase and therefore they cancel in the ratio between BES and HL. Due to this fact $|\Sigma^\besratio_{ab}(u_1, u_2)|^2=1$. Using the second relation in~\eqref{eq:brunit_unit_R}, together with $(\g^\pm_a )^* =\g^\mp_a$, it is easy to show that the odd part of the dressing factor, evaluated in the string model, also satisfies unitarity. Then the string theory S-matrix is  unitary as it must be.
While in the proof we restricted to fundamental particles, it is possible to show that the S-matrix of string bound states is also unitary. This can be easily shown from equations~\eqref{eq:app_SYY_indep} and~\eqref{eq:app_SYbZ_indep}.

\subsection{P invariance in the mirror model}

We show that the S~matrix is parity invariant in the mirror theory, which corresponds to showing that the constraint~\eqref{eq:parity_constraint_Sigma11} is satisfied.
The check of parity on the odd part of the dressing factors is immediate. Indeed from~\eqref{eq:g-inversion-reflection} we see that under a parity transformation for mirror kinematics we have 
\bal
\tg_a (-\frac{1}{\tx})=-\tg_a (\tx) + i \pi\,.
\eal
Since the odd part of the phases is of difference form in the $\tg$ rapidities, the constraint~\eqref{eq:parity_constraint_Sigma11} is satisfied for the odd dressing factors.

To check parity on the even part of the phases we start considering two points $\tx$ and $\ty$ in the mirror region. Then we want to evaluate
\bal
&\tPhi^{\a \a}_{aa} (- \frac{1}{\tx},- \frac{1}{\ty})= -  \lint_{{ \pa\cR_\a}} \frac{{\rm d} w_1}{2\pi i} \lint_{{ \pa\cR_\a}} \frac{{\rm d} w_2}{2\pi i} {1\ov -\frac{1}{\tx}-w_1}{1\ov -\frac{1}{\ty}- w_2} K^\bes(u_a(w_1)-u_a(w_2)) \,,
\eal
for $\a = \pm$.
Changing integration variable $w_1 \to -\frac{1}{w_1}$ and $w_2 \to -\frac{1}{w_2}$ both the contours $\pa\cR_-$ and the contours $\pa\cR_+$ are mapped to themselves.
Due to property~\eqref{eq:x-inversion-reflection} and the fact that $K^\bes(v)$ is odd in $v$ we obtain
\bal
\tPhi^{\a  \a}_{aa} (- \frac{1}{\tx},- \frac{1}{\ty})=- \tPhi^{\a \a}_{aa} (\tx,\ty)+\tPhi^{\a \a}_{aa} (0,\ty)+\tPhi^{\a \a}_{aa} (\tx,0)-\tPhi^{\a \a}_{aa} (0,0) \,.
\eal
Note that for this property to be satisfied we need to require $w_1$ and $w_2$ to be both in $\cR_-$ or both in $\cR_+$. If they were on different sides of the cut $(-\infty, 0)$ in the $x$-plane, $u_a(w_1)-u_a(w_2)$ would transform badly. Similarly we get
\bal
\tPhi^{\a , -\a}_{\bar aa} (- \frac{1}{\tx},- \frac{1}{\ty})=- \tPhi^{\a, -\a}_{\bar aa} (\tx,\ty)+\tPhi^{\a,  -\a}_{\bar aa} (0,\ty)+\tPhi^{\a,  -\a}_{\bar aa} (\tx,0)-\tPhi^{\a,  -\a}_{\bar aa} (0,0) \,.
\eal
It is then clear that the full improved BES phases constructed from the $\tPhi$ functions satisfy
\bal
\tilde{\theta}_{b a}^{\bes}(-\frac{1}{\tx^\mp_{b2}},-\frac{1}{\tx^\mp_{a1}})=\tilde{\theta}_{ab}^{\bes}(\tx^\pm_{a1},\tx^\pm_{b2})
\eal
and are therefore invariant under parity. The same is applied to the HL phases, which are obtained from the subleading order of BES at large $h$ and $k$ (with $\ka=\frac{k}{h}$ fixed). The constraint~\eqref{eq:parity_constraint_Sigma11} is then satisfied for the full dressing factors (comprising both the even and odd parts).

\subsection{Continuation to an arbitrary momentum region}\label{sec:cont-p1}

Let us recall that the momentum and energy are defined in the string region as follows
\bal\label{paeastr}
p_a( x_a^{\pm m})=i(\ln x_a^{-m} -\ln x_a^{+m})\,, \quad
E_a( x_a^{\pm m})={ih\ov2}\left( x_a^{-m} -{1\ov x_a^{-m}} - x_a^{+m} +{1\ov x_a^{+m}} \right)
\,,
\eal
and the range of $p_a$ is $(0,2\pi)$. In this section we always use the convention that any $p_a$ is given by \eqref{paeastr}, and use the following notation for momenta in other momentum regions
\bal
p_a^{(n)}\equiv p_a( x_a^{\pm m}) +2\pi n\,,\quad n\in \bZ\,.
\eal
To reach a different momentum region one  has to cross the $\ln x$ cut, and since we have two variables $x_a^{-m}\,,\, x_a^{+m}$ there are infinitely many ways to do so. For example the simplest two ways to get to the region of $p_a^{(-1)}$ are either \textit{i)} $x^+_a$ crosses  the $\ln x$ cut from below, or \textit{ii)} $x^-_a$ crosses  the $\ln x$ cut from above. Then, the dressing factors have additional cuts due to the $\tPsi$ functions, and we may cross them too. Here we discuss an analytic continuation path which is consistent with the CP invariance of the string model. We do it in the following two steps by using the mirror $u$-plane variable, so that $x_a^{+ m}=x_a(u+{i\ov h}m)=1/\tx_{\bar a}(u+{i\ov h}m)$,  $x_a^{- m}=x_a(u-{i\ov h}m)=\tx_a(u-{i\ov h}m)$, $u\in\bR$. The steps are:
\bee
\item  Move $u$ through the main mirror cut to the mirror $u$-plane. Then, $x_{a}^{\pm m}$ become $\tx_{a}^{\pm m}$.

Do not cross any cut of $\tPsi$'s. 
\item  Move  
  the resulting $\tx_{a}^{+m}$  through the semi-line $x<0$ in the $x$-plane or  through the $\ka_a$-cut in the $u$-plane to the anti-mirror $u$-plane, and shift $u$ appropriately. This continuation shifts momentum by $-2\pi$: $p_a\to p_a^{(-1)}=p_a-2\pi$. 
  
 Do not cross any cut of $\tPsi$'s. 
\eee

Let us now discuss the two steps in more detail.

\paragraph{Step 1.}
The analytic continuation of $x_a^{+m}=1/\tx_{\bar a}^{+m}$ through the semi-line $x>\xi_a$  in the $x$-plane or  through the lower edge of the main mirror cut  in the $u$-plane just replaces  $x_a^{\pm m}$ with  $\tx_a^{\pm m}$
 because
\bal
x_a^{+ m}&={1\ov \tx_{\bar a}^{+ m}}
\ \xrightarrow{\text{mirror}, \, x>\xi_a}\ 
\tx_a^{+ m}\,,
\quad 
x_a^{- m}={ \tx_{a}^{- m}}
\ \xrightarrow{\text{mirror}, \, x>\xi_a} 
\ \tx_a^{- m}\,.
\eal
Then, the string $\g_a^{\pm m}=\g_a(x_a^{\pm m})$ has the cut $(-1/\xi_a\,,\, \xi_a)$,
and therefore 
\bal
F( x_a^{\pm m}) \xrightarrow{\text{mirror}, \, x>\xi_a} F( \tx_a^{\pm m})
\,,
\eal
where $F= \{ p_a\,,\, E_a\,,\, \g_a \}$

The next  step of the analytic continuation leads to different results for right and left particles.

\paragraph{Step 2 --- Right particles.}
The analytic continuation of $\tx_\R^{+m}$ through the semi-line $x<-\xbr$ in the $x$-plane or  through the lower edge of the $-\ka$-cut  in the $u$-plane shifts the string momentum by $-2\pi$. Note that for the analytic continuation of $\g_\R(\tx_\R^{+m})$ it is important to know whether it is done through $(-\infty,-\xi)$ or through $(-\xi,0)$, and our choice guarantees that we do not cross the cut of $\g_\R$. 

So, we move $\tx_\R^{+m}$ through the $-\ka$-cut, and using \eqref{eq:xatoxakacut}, we get
\bal 
\tx_\R^{+m}&
\  \xrightarrow{-\ka\text{-cut}}
\ {1\ov \tx_\L(u+ {i\ov h}(m+k) +{i\ov h}k)}\,,
\quad
\tx_\R^{-m}
\ \xrightarrow{-\ka\text{-cut}} 
\ { \tx_\R(u- {i\ov h}(m+k) +{i\ov h}k)}\,.
\eal
Then, we shift $u$ by $-ik/h$, and obtain
\bal 
\tx_\R^{+m}&
\  \xrightarrow{-\ka\text{-cut and shift of}\ u}
\ x_\R^{+(m+k)} \,,
\quad
\tx_\R^{-m}
\ \xrightarrow{-\ka\text{-cut and shift of}\ u}
\ x_\R^{-(m+k)}\,.
\eal
Thus,
\bal
p_\R( \tx_\R^{\pm m})  \xrightarrow{-\ka\text{-cut and shift of}\ u}  p_\R( x_\R^{\pm (m+k)}) -2\pi \,, \quad
F( \tx_\R^{\pm m}) \xrightarrow{-\ka\text{-cut and shift of}\ u}  F( x_\R^{\pm (m+k)}) 
\,,
\eal
where $F= \{ E_\R\,,\, \g_\R \}$.

Since the range of the momentum of the right $m+k$-particle bound state is $(0,2\pi)$, 
the momentum of the analytically continued right $m$-particle bound state takes values in the interval $(-2\pi,0)$.

\paragraph{Step 2 --- Left particles.} 
The analytic continuation of $\tx_\L^{+m}$ through the semi-line $x<-1/\xbr$ in the $x$-plane or through the lower edge of the $+\ka$-cut  in the mirror $u$-plane shifts the string momentum by $-2\pi$, and does not cross the cut of $\g_\L$.

So, we move $\tx_\L^{+m}$ through the $+\ka$-cut, and using again \eqref{eq:xatoxakacut}, we get
\bal 
\tx_\L^{+m}&
\ \xrightarrow{+\ka\text{-cut}} 
\ {1\ov \tx_\R(u+ {i\ov h}(m-k) -{i\ov h}k)}\,,
\quad
\tx_\L^{-m}
\ \xrightarrow{+\ka\text{-cut}} 
\ { \tx_\L(u- {i\ov h}(m-k) -{i\ov h}k)}\,.
\eal
Then, we shift $u$ by $+ik/h$, and get
\bal 
\tx_\L^{+m}&
 \xrightarrow{+\ka\text{-cut and shift of}\ u} {1\ov \tx_\R(u+ {i\ov h}(m-k))}\,,
\ \
\tx_\L^{-m}
 \xrightarrow{+\ka\text{-cut and shift of}\ u} { \tx_\L(u- {i\ov h}(m-k))} \,.
\eal

Next, if $m>k$ then
\bal 
\tx_\L^{+m}&
\ \xrightarrow{+\ka\text{-cut and shift of}\ u} 
\ x_\L^{+(m-k)}\,,
\quad
\tx_\L^{-m}
\ \xrightarrow{+\ka\text{-cut and shift of}\ u} 
\ x_\L^{-(m-k)}\,,
\eal
and ($m=k+1,k+2,\ldots$)
\bal
p_\L( \tx_\L^{\pm m})  \xrightarrow{+\ka\text{-cut and shift of}\ u}  p_\L( x_\L^{\pm (m-k)}) -2\pi \,, \quad
F( \tx_\L^{\pm m}) \xrightarrow{+\ka\text{-cut and shift of}\ u}  F( x_\L^{\pm (m-k)}) 
\,,
\eal
where $F= \{ E_\L\,,\, \g_\L \}$.

Since the range of the momentum of the left $m-k$-particle bound state is $(0,2\pi)$, 
the momentum of the analytically continued left $m$-particle bound state takes values in the interval $(-2\pi,0)$.

On the other hand if $m<k$ then
\bal 
\tx_\L^{+m}&
\ \xrightarrow{+\ka\text{-cut, and shift of}\, u} 
\ {1\ov x_\R^{-(k-m)}}\,,
\quad
\tx_\L^{-m}
\  \xrightarrow{+\ka\text{-cut, and shift of}\, u} 
\ {1\ov x_\R^{+(k-m)}}\,,
\eal
and ($m=1,2,\ldots , k-1$)
\bal
p_\L( \tx_\L^{\pm m}) & \xrightarrow{+\ka\text{-cut, and shift of}\ u}  p_\L\big( {1\ov x_\R^{\mp(k-m)}}\big) -2\pi= p_\R( x_\R^{\pm (k-m)}) -2\pi \,, \quad
\\
E_\L( \tx_\L^{\pm m}) &\xrightarrow{+\ka\text{-cut, and shift of}\ u} E_\L\big( {1\ov x_\R^{\mp(k-m)}}\big) = E_\R( x_\R^{\pm (k-m)}) 
\,, \quad
\\
\g_\L( \tx_\L^{\pm m}) &\xrightarrow{+\ka\text{-cut, and shift of}\ u} \g_\L\big( {1\ov x_\R^{\mp(k-m)}}\big) = \g_\R( x_\R^{\mp (k-m)})\pm i\pi
\,.
\eal
Note that in terms of the momentum the transformation of $\g_{a}$ can be written in the form
\bal
\g^{\pm m}_\R (p -2 \pi) = \g^{\pm (k+m)}_\R (p)\,,\quad \g^{\pm m}_\L (p -2 \pi) = \g^{\mp (k-m)}_\R (p) \pm i \pi\,.
\eal
We see that the analytic continuation for left particles, when $m<k$, produces a right $k-m$-particle bound state with momentum in $(0,2\pi)$. It is interesting that the analytically continued $\tx_\L^{\pm m}$ are in the anti-string right region but the energy is positive because $\tx_\L^{\pm m}\to {1\ov x_\R^{\mp(k-m)}}$.

This completes the description of the analytic continuation to the region of $p_a^{(-1)}$. Obviously, if we want to get to the region of $p_a^{(-n)}$, $n\ge1$ we repeat the procedure $n$ times, and to get to the region of $p_a^{(+n)}$, $n\ge1$ we do the two steps in the opposite order.

The analytic continuation to the region of $p_a^{(-1)}$ of the diagonal S-matrix elements $S_{YY}^{m_1m_2}$, $S_{\bar Y\bar Y}^{m_1m_2}$, $S_{\bar Y Z}^{m_1m_2}$, $S_{Y\bar Z}^{m_1m_2}$ is performed in appendix \ref{app:pto2pimp}. We find that, as a result of the continuation to the region of string negative momentum, for $0<p_1<2 \pi$ the right particles obey the following ``periodicity''
\bal
\label{eq:2pishift-nomonodromy}
&S^{m_1 \, m_2}_{\bar{Y} \bar{Y}} (p_1-2 \pi, p_2) = S^{m_1+k \, m_2}_{\bar{Y} \bar{Y}} (p_1, p_2)\,,\\
&S^{m_1 \, m_2}_{\bar{Y} Z} (p_1-2 \pi, p_2) = S^{m_1+k \, m_2}_{\bar{Y} Z} (p_1, p_2) \,.
\eal
The left particles satisfy similar equations for $m_1=k+1,k+2,\ldots$ 
\bal
\label{eq:2pishift-nomonodromy2}
&S^{m_1 \, m_2}_{{Y} {Y}} (p_1-2 \pi, p_2) = S^{m_1-k \, m_2}_{{Y} {Y}} (p_1, p_2)\,,\\
&S^{m_1 \, m_2}_{{Y} \bar Z} (p_1-2 \pi, p_2) = S^{m_1-k \, m_2}_{{Y} \bar Z} (p_1, p_2) \,.
\eal
On the other hand the left particles satisfy instead for $m_1=1,2,\ldots, k-1$
\bal
\label{eq:2pishift-yesmonodromy}
&S^{m_1 \, m_2}_{Y Y} (p_1-2 \pi, p_2) =
\frac{\alpha_{\L}(x_{\L 2}^{-m_2})}{\alpha_{\L}(x_{\L 2}^{+m_2})}\ S^{k-m_1 \, m_2}_{\bar{Z} Y} (p_1, p_2)\,,\\
&S^{m_1 \, m_2}_{Y \bar{Z}} (p_1-2 \pi, p_2) = 
\frac{\alpha_{\R}(x_{\R 2}^{+m_2})}{\alpha_{\R}(x_{\R 2}^{-m_2})}\  S^{k-m_1 \, m_2}_{\bar{Z} \bar{Z}} (p_1, p_2) \,,
\eal
where
\begin{equation}
\label{eq:defalpha}
    \alpha_a(x)=\left(1-\frac{\xi_a}{x}\right)\left(x+\frac{1}{\xi_a}\right)\,.
\end{equation}

\subsection{CP invariance of the string model}\label{sec:cont-p3}

It is easy to check that for all S-matrices the CP invariance condition~\eqref{eq:CP_invariance_general} is satisfied up to an overall factor. Thus, it is sufficient to prove the condition only for the diagonal S-matrix elements $S_{YY}^{m_1m_2}$, $S_{\bar Y\bar Y}^{m_1m_2}$, $S_{\bar Y Z}^{m_1m_2}$, $S_{Y\bar Z}^{m_1m_2}$.
Since we only know the S-matrix elements with $p_i$ in the range $(0,2\pi)$ we need to use the analytic continuation to the region of $p_a^{(-1)}$ described in the previous subsection. As a result of the continuation a momentum $p$ is mapped to $p-2\pi$. We need therefore in addition to the two steps described in the previous subsection to perform an additional transformation which maps $p$ to $2\pi-p$, and therefore the original $p$ to $-p$.
This additional transformation is the reflection in the $u$ plane: $u\mapsto -u$.

\paragraph{ Right particles.} By using the formula \eqref{eq:u-reflection} we find that under the reflection  $u\to -u$ the right  $m+k$-particle bound state Zhukovsky variables transform as 
\bal
x_{\R}^{\pm (k+m)} 
\ \xrightarrow{u\to-u} 
\ -x_{\L}^{\mp m}\,.
\eal
Thus,
\bal
\label{eq:secCP_pm2p_to_mp}
&p_\R( x_\R^{\pm (m+k)}) -2\pi\xrightarrow{u\to-u} - p_\L( x_\L^{\pm m}) \,, \quad
E_\R( x_\R^{\pm (m+k)}) \xrightarrow{u\to - u}  E_\L( x_\L^{\pm m}) 
\\
&\g_\R( x_\R^{\pm (m+k)}) \xrightarrow{u\to - u}  -\g_\L( x_\L^{\mp m}) 
\,.
\eal
Combining all the steps, we find that the analytic continuation from $p$ to $-p$ transforms the right variables into the left ones\footnote{We should stress that the continuation is $p \to -p'$, with $p\ne p'$, as evident from~\eqref{xrmtoxlm}.}
\bal\label{xrmtoxlm}
x_{\R}^{\pm m} \to -x_{\L}^{\mp m}\,,&\quad \g_\R( x_\R^{\pm m}) \to -\g_\L( x_\L^{\mp m}) \,,
\\
p_\R( x_\R^{\pm m}) \to - p_\L( x_\L^{\pm m}) \,, &\quad
E_\R( x_\R^{\pm m}) \to   E_\L( x_\L^{\pm m}) 
\,,
\eal
as is necessary to satisfy the CP invariance.

\medskip

\paragraph{Left particles.} Similarly, under the reflection  $u\to -u$ the left  $m-k$-particle bound state Zhukovsky variables with $m>k$ transform as 
\bal 
x_\L^{\pm(m-k)}&
\ \xrightarrow{u\to-u} 
\ -x_\R^{\mp m}\,.
\eal
Thus, 
\bal
& p_\L( x_\L^{\pm (m-k)}) -2\pi  \xrightarrow{u\to-u} -p_\R( x_\R^{\pm m})  \,, \quad
E_\L( x_\L^{\pm (m-k)})  \xrightarrow{u\to-u} E_\R( x_\R^{\pm m}) 
\\
&\g_\L( x_\R^{\pm (m-k)}) \xrightarrow{u\to - u}  -\g_\R( x_\L^{\mp m}) 
\,.
\eal
Combining the three steps, we find (for $m=k+1, \, k+2, \, \ldots$)
\bal\label{xlmtoxrm}
x_{\L}^{\pm m} \to -x_{\R}^{\mp m}\,,&\quad \g_\L( x_\L^{\pm m}) \to -\g_\R( x_\R^{\mp m}) \,,
\\
 p_\L( x_\L^{\pm m})  \to - p_\R( x_\R^{\pm m})\,, &\quad
E_\L( x_\L^{\pm m})  \to E_\R( x_\R^{\pm m})   
\,.
\eal

\medskip

If $m<k$  then under the reflection $u\to -u$ we get that the right  $k-m$-particle bound state Zhukovsky variables, momentum, energy and $\g_\R$'s transform as 
\bal
x_{\R}^{\pm (k-m)} 
\  \xrightarrow{u\to-u}\ 
 -{1\ov x_{\R}^{\pm m}}\,,
\eal 
\bal
&p_\R( x_\R^{\pm (k-m)}) -2\pi  \xrightarrow{u\to-u} -p_\R( x_\R^{\pm m})  \,, \quad
E_\R( x_\R^{\pm (k-m)})  \xrightarrow{u\to-u} E_\R( x_\R^{\pm m}) \,,
\\
&\g_\R( x_\R^{+(k-m)}) \xrightarrow{u\to-u} \g_\R\big( -{1\ov x_\R^{+m}}\big) = -\g_\R( x_\R^{+m})+i\pi
\,, \quad
\\
&\g_\R( x_\R^{- (k-m)}) \xrightarrow{u\to-u} \g_\R\big( {1\ov x_\R^{-m}}\big) = -\g_\R( x_\R^{-m})-i\pi
\,.
\eal
Combining the three steps, we again find \eqref{xlmtoxrm}, as is needed for the CP invariance.

Since $S_{AB}(p_{a1},p_{b2})$ and  $S_{BA}(p_{b2},p_{a1})$ have the same analytic structure,  the braiding unitarity holds under the analytic continuation. Then, it is sufficient to check the following two relations 
\bal
S_{\bar Y\bar Y}^{m_1m_2}(-p_{\L1},p_{\R2})&=S_{ Y Y}^{m_2m_1}(-p_{\R2},p_{\L1})\,,
\quad
S_{\bar Y Z}^{m_1m_2}(-p_{\L1},p_{\L2})=S_{ \bar Z Y}^{m_2m_1}(-p_{\L2},p_{\L1})\,,
\eal
where $p_{aj}=p_a(x_{aj}^{\pm m_j})=p_a(u_j)$ are positive and take values between $0$ and $2\pi$. All the other relations follow from those two by the analytic continuation of either $p_{a1}$ or $p_{a2}$ to the region of $p_a^{(-1)}$.
The proof of these two relations can be found in appendix \ref{app:CP}.

\section{Perturbative expansions}
\label{sec:expansions}
Here we will compare our proposal with existing perturbative computations, as well as with the relativistic limit of the model considered in~\cite{Frolov:2023lwd}\footnote{A similar version of the limit was considered in~\cite{Fontanella:2019ury}, but the bound states and dressing factors were not worked out.}.

\subsection{Large-tension expansion}
The perturbative computations have been done in the weakly-coupled string non-linear sigma model, that is at large tension. We set
\begin{equation}
\label{eq:largetension}
k = 2\pi q\,T\,,\qquad h=\sqrt{1-q^2}\,T+\mathcal{O}(T^{-1})\,,
\end{equation}
where $T\gg1$ is the string tension. We then expand the dressing factors at large~$T$ while keeping $q$ fixed, with $0<q<1$ for mixed-flux models. The subleading terms~$\mathcal{O}(T^{-1})$ are not relevant at the order of our interest, which is up to one~loop at large tension.

\paragraph{Semiclassical expansion.}
In refs.~\cite{Babichenko:2014yaa,Stepanchuk:2014kza} the dressing factor was computed in a semiclassical expansion up to one loop. To match our proposal with the kinematics employed in those papers, we expand
\begin{equation}
    u_a(x^+_a)-u_a(x^-_a)=\frac{2i}{h}\,,
\end{equation}
order by order in the tension, which fixes 
\begin{equation}
\label{eq:xpm-semiclassical}
    x^\pm_{a}=x_a \pm\frac{i}{T}\frac{1}{\sqrt{1-q^2}\,u_a'(x_a)}+\mathcal{O}(T^{-2})\,,
\end{equation}
up to terms which we will not need. In other words, we want to expand the dressing factors when $x_{a}^\pm$ are close to some point $x_{a1}$ in the string physical region (and similarly for $x_{b2}^\pm$).

\paragraph{Near-BMN expansion, small and positive momentum.}
Other perturbative results~\cite{Hoare:2013pma,Engelund:2013fja,Hoare:2013lja,Roiban:2014cia,Bianchi:2014rfa,Sundin:2014ema,Baglioni:2023zsf} have been computed at a large-tension expansion around the pp-wave geometry. This allows to pertubratively compute the S~matrix up to one loop in the string model~\cite{Hoare:2013pma,Engelund:2013fja,Hoare:2013lja,Roiban:2014cia,Bianchi:2014rfa,Sundin:2014ema}, and at tree-level in the mirror model~\cite{Baglioni:2023zsf}. To match with that computation we use~\eqref{eq:largetension} and assume the momentum of the excitations to be small, of order $1/T$. For $p$ small \textit{and positive} we get that the Zhukovsky variables of fundamental particles are
\begin{equation}
    x_{a}^{\pm}\left(\frac{p}{T}\right)=x_a(p)\left(1\pm \frac{ip}{2T}\right)+O(T^{-2}),
\end{equation}
with
\begin{equation}
\label{eq:xpm-BMN-piszero}
    x_{\L}(p)=\frac{1+q\,p+\omega_{\L}(p)}{\sqrt{1-q^2}\,p}\,,\qquad
    x_{\R}(p)=\frac{1-q\,p+\omega_{\R}(p)}{\sqrt{1-q^2}\,p}\,,
\end{equation}
where
\begin{equation}
\label{eq:perturbativedispersion}
    \omega_{\L}(p)=\sqrt{1+2q\,p+p^2}\,,\qquad
    \omega_{\R}(p)=\sqrt{1-2q\,p+p^2}\,,    
\end{equation}
are the leading order expressions for the dispersion relations.
Once again this means that the phases will have to be expanded when $x^\pm_{a}$ are close to each other. The same argument applies for $\tilde{x}_a^\pm$ in the mirror model, and in fact the mirror formulae can be obtained by the substitution
\begin{equation}
\label{eq:mirrormaptree}
    p\to i\tilde{\omega}_a\,,\qquad
    \omega_a\to i\tilde{p}\,,
\end{equation}
where $\tilde{\omega}_a(p)$ is the mirror dispersion~\cite{Baglioni:2023zsf}
\begin{equation}
    \tilde{\omega}_{\L}(\tilde{p})=\sqrt{\tilde{p}^2+1 -q^2}+i q\,,\qquad
    \tilde{\omega}_{\R}(\tilde{p})=\sqrt{\tilde{p}^2+1 -q^2}-i q\,.
\end{equation}

\paragraph{Near-BMN expansion, small and negative momentum.}
The near-BMN perturbative S-matrix is constructed starting from an interacting Hamiltonian, polynomial in the fields and their derivatives. At leading order, the dispersion is indeed~\eqref{eq:perturbativedispersion}. Hence, perturbatively there is absolutely nothing special about taking $p<0$. On the other hand, in our finite-coupling description we have a fundamental region with $0<p<2\pi$; considering negative momentum requires a non-trivial analytic continuation described in section~\ref{sec:cont-p1}. Therefore, matching our negative-momentum results with the near-BMN expansion is an important check of our construction. By using the results~\eqref{eq:2pishift-nomonodromy} and~\eqref{eq:2pishift-yesmonodromy} we can relate the expansion at negative momentum $-p/T$ with that at positive momentum $2\pi-p/T$, with $p>0$ fixed and $T\gg1$, if we also suitably shift $m$ by $k$. In particular, we will need the following kinematics:
\begin{equation}
\label{eq:xpm-BMN-pis2pi}
\begin{aligned}
x^{\pm(k+1)}_{\R}\left(2\pi-\frac{p}{T}\right)&=&x_{\L}(p)\left(-1\pm \frac{ip}{2T}\right)+O(T^{-2}),\\
x^{\pm(k-1)}_{\R}\left(2\pi-\frac{p}{T}\right)&=&\frac{1}{x_{\R}(p)}\left(-1\pm \frac{ip}{2T}\right)+O(T^{-2}),
\end{aligned}
\end{equation}
where the expansion is around a negative real number.
Hence, the two points~$x^{\pm(k+1)}$ (or $x^{\pm(k-1)}$) are on opposite sides of the cut of $\ln x$, very close to the cut. This will require special care when expanding the dressing factors.

\begin{figure}[t]
\begin{center}
\includegraphics[width=5cm]{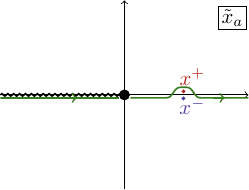}%
\hspace{1cm}%
\includegraphics[width=5cm]{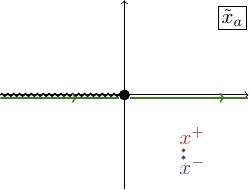}
\end{center}
\caption{\label{fig:xpm-BMN}%
In the near-BMN and semiclassical limits, the Zhukovsky variables take values at $x^\pm \approx x\pm i\delta x/T$, with $x>0$ and $\delta x>0$ (left panel). For $x^-$, this lies in the intersection of the string and mirror region, and the dressing factors can be evaluated straightforwardly. In the case of $x^+$,  the dressing factors are defined by analytic continuation (see also Figure~\ref{fig:continuation}) which can be implemented by deforming the integration contour (green line) like in the figure. This is related by analytic continuation to taking both $x^\pm$ in the intersection of the string and mirror region (right panel).}
\end{figure}

\paragraph{Expansion of the dressing factors away from the $\ln x$ cut.}
Let us consider the expansion for the cases of~\eqref{eq:xpm-semiclassical} and~\eqref{eq:xpm-BMN-piszero}. For fundamental particles of real energy and momentum we are interested in taking $x^\pm$ just above / below a positive and real $x$.
The setup is illustrated in Figure~\ref{fig:xpm-BMN} and, as explained in the caption, the computation of the phases can be related by analytic continuation to taking both $x^\pm$ in the intersection of the string and mirror region (and at distance $\mathcal{O}(T^{-1})$ to each other),
\begin{equation}
\label{eq:schematic-expansion}
    x^\pm_{aj}=\tx_j \pm \frac{i}{T}\delta x_j+\mathcal{O}(T^{-2})\,,
\end{equation}
with $\tx_j$ in the mirror region and $\delta x_j>0$.
Hence, we can use the mirror expressions and compute, schematically,
\begin{equation}
\label{eq:double-der-expansion}
\begin{aligned}
    \tilde{\theta}(x_1^\pm, x_2^\pm)&=\tPhi(x_1^+,x_2^+)-\tPhi(x_1^+,x_2^-)-\tPhi(x_1^-,x_2^+)+\tPhi(x_1^-,x_2^-)\\
    &=\frac{\partial^2\tPhi(\tx_1,\tx_2)}{\partial \tx_1\partial \tx_2}\left(-4\frac{\delta x\,\delta y}{T^2}+\mathcal{O}(T^{-4})\right)\,,
\end{aligned}
\end{equation}
where the phase is the BES, HL, or the odd phase. Then, we can straightforwardly take $\tx_j\to x_j$, where $x_j$ is given by \eqref{eq:xpm-semiclassical} or~\eqref{eq:xpm-BMN-piszero}. To simplify the computation of the expansion~\eqref{eq:double-der-expansion} of the full phase it is worth recalling the asymptotic expansion of the BES phase,
\begin{equation}
    \tPhi_{ab}^\bes(\tx_1,\tx_2)=\sqrt{1-q^2}\,T\, \tPhi_{ab}^\afs(\tx_1,\tx_2)+\tPhi_{ab}^\hl(\tx_1,\tx_2)+\mathcal{O}(T^{-1})\,,
\end{equation}
so that in the full phase the $\tPhi_{ab}^\hl(\tx_1,\tx_2)$ contribution cancels due to the HL phase which appears with the opposite sign with respect to BES. It is also easy to check that the odd phase $\tPhi^\barnes_{ab}(\tx_1,\tx_2)$ starts out at order $\mathcal{O}(T^0)$. All in all, the expansion of the full phase is
\begin{equation}
\begin{aligned}
    -i\ln\Sigma^{11}_{ab}(x_1^\pm,x_2^\pm)=&-\frac{4\sqrt{1-q^2}}{T}\, \frac{\partial^2\tPhi^\afs_{ab}(\tx_1,\tx_2)}{\partial \tx_1\partial \tx_2}\,\delta x_1\,\delta x_2\\
    &-\frac{4}{T^2}\,
    \frac{\partial^2\tPhi^\barnes_{ab}(\tx_1,\tx_2)}{\partial \tx_1\partial \tx_2}\,\delta x_1\,\delta x_2+\mathcal{O}(T^{-3})\,.
\end{aligned}
\end{equation}
In other words, the AFS phase contributes to the S~matrix from tree level, and the odd phase contributes from one loop. From two loops, we would have to account from contributions due to the subleading terms in the expansion~\eqref{eq:schematic-expansion} as well as for the subleading terms in the BES phase. The computation of the second derivative of the AFS phase is performed in Appendix \ref{app:mirPhiafs}, and when both variables are in the mirror region it gives
\begin{equation}
\begin{aligned}
\frac{\partial^2\tPhi^\afs_{aa}(\tx_1,\tx_2)}{\partial \tx_1\partial \tx_2}&=
\frac{1}{4}\frac{u_1+u_2}{\tx_1-\tx_2}-\frac{1}{2}\frac{u_1'\,u_2'}{u_1-u_2}+\frac{\tx_1-\tx_2}{4(\tx_1)^2(\tx_2)^2}\,,\\
\frac{\partial^2\tPhi^\afs_{a\bar{a}}(\tx_1,\tx_2)}{\partial \tx_1\partial \tx_2}&=
\frac{1}{2}\frac{\tx_1-\tx_2}{\tx_1\tx_2(1-\tx_1\tx_2)}+\frac{\kappa_a}{4\pi}\frac{1+\tx_1\tx_2}{\tx_1\tx_2(1-\tx_1\tx_2)}\,.
\end{aligned}
\end{equation}
The expansion of the odd phase is rather straightforward if one recalls that the ratio of Barnes functions appearing in $R(\gamma)$ satisfies
\begin{equation}
    -i\frac{\de^2}{\de \gamma^2}\ln R^2(\gamma)=\frac{\gamma-\sinh\gamma}{4\pi \sinh^2(\gamma/2)}\,.
\end{equation}
Using this, and taking $\tx_j\to x_j$ at the end of the computation, we can compute the one-loop expansion of our dressing factors. To avoid confusion with the choice of normalisation, it is convenient to give the expression for whole S-matrix elements. We have
\begin{equation}
\label{eq:oneloopresult}
\begin{aligned}
&\frac{1}{i}\ln S^{11}_{\bar{Y}\bar{Y}}\left(\frac{p_1}{T},\frac{p_2}{T}\right)=\frac{p_1^2\omega_2+p_2^2\omega_1}{(p_1-p_2)\,T}-\frac{p_1^2p_2^2(\omega_1+\omega_2)}{2\pi T^2\,(p_1-p_2)^2}
\\
&\qquad-\frac{p_1^2p_2^2[1-q^2+(p_1-q)(p_1-q)+\omega_1\omega_2]}{2\pi T^2\,(p_1-p_2)^2}\log\frac{p_1-q+\omega_1}{p_2-q+\omega_2}
+\mathcal{O}(T^{-3})\,,\\
&\frac{1}{i} \ln S_{Z \bar{Y}}^{11}\left(\frac{p_1}{T}, \frac{p_2}{T}\right)=-\frac{p_1\omega_2}{T}
-\frac{p_1^2p_2^2(\omega_1-\omega_2)}{2\pi T^2\,(p_1+p_2)^2}\\
&\qquad+\frac{p_1^2p_2^2(q^2-1+(p_1+q)(p_2-q)+\omega_1\omega_2)}{2\pi T^2\,(p_1+p_2)^2}\log\frac{p_1+q+\omega_1}{p_2-q+\omega_2}+\mathcal{O}(T^{-3})\,.
\end{aligned}
\end{equation}
These results can be compared with the perturbative computations of~\cite{Hoare:2013pma,Engelund:2013fja,Hoare:2013lja,Roiban:2014cia,Bianchi:2014rfa,Sundin:2014ema}. The results match at tree level, but there is a discrepancy at one loop, which is the same as in the pure-RR case~\cite{Frolov:2021bwp},
\begin{equation}
\label{eq:mismatch}
    \frac{p_1p_2(p_1\omega_2-p_2\omega_1)}{4\pi T^2}+\mathcal{O}(T^{-3})\,.
\end{equation}
This term can be reabsorbed by a local counterterm.
It is also possible to compare with the perturbative computations done in the mirror model, which are known at tree level~\cite{Baglioni:2023zsf}. To obtain the result in the mirror kinematics, it is sufficient \textit{not} to continue $\tx_j\to x_j$ but instead use the mirror versions of $\tx_j$ and~$\delta x_j$. By construction, this is tantamount to taking~\eqref{eq:oneloopresult} and using the mirror map~\eqref{eq:mirrormaptree}. This matches with~\cite{Baglioni:2023zsf}.
To compare with the semiclassical results of~\cite{Babichenko:2014yaa,Stepanchuk:2014kza}, it can be useful to use the primitives
\begin{equation}
\begin{aligned}
\tPhi_{aa}^{\afs}(\tx_1,\tx_2)
&=-\frac{\kappa_a}{4 \pi }\text{Li}_2\frac{\tx_1-\tx_2}{\tx_1} +\frac{\ln \tx_2}{2\tx_1}+\frac{u_{1}}{2}  \ln \frac{u_{1}-u_{2}}{\tx_1-\tx_2}   -   (1\leftrightarrow 2)
 \,,\\
\tPhi_{a\bar a}^{\afs}(\tx_1,\tx_2)
&= \frac{\kappa_a}{4\pi}\left[\ln \tx_1\ln \tx_2+2\text{Li}_2\frac{\tx_1\tx_2-1}{\tx_1\tx_2}\right]\\
&\qquad+\frac{1}{2}
   \left[\frac{\ln \tx_2}{\tx_1}+\frac{\tx_1{}^2+1}{\tx_1}\ln\frac{\tx_1\tx_2-1}{\tx_1\tx_2}-(1\leftrightarrow2)\right].
\end{aligned}
\end{equation}
It is important to note that~\cite{Babichenko:2014yaa,Stepanchuk:2014kza} use a different normalisation than ours.

\begin{figure}[t]
\begin{center}
\includegraphics[width=4.9cm]{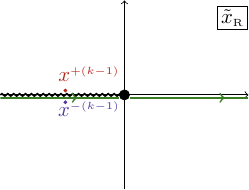}%
~\includegraphics[width=4.9cm]{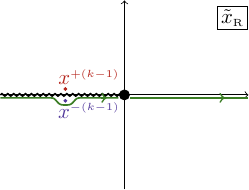}%
~\includegraphics[width=4.9cm]{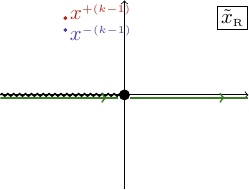}
\end{center}
\caption{\label{fig:xpm-BMN-negative}%
We sketch the strategy to evaluate the phases when in the kinematics~\eqref{eq:xpm-BMN-pis2pi}. Originally (left) we have the bound-state $x^+$ evaluated in the string region, close to the cut of $\ln x$;  for definiteness, we depict the case of $\tilde{x}_{\R}$ and of the $\partial \mathcal{R}_-$ contour. 
Up to some residues we can relate this picture to the middle panel, where both $x^+$ and $x^-$ are on the same side of the contour. They are on opposite sides of the $\ln x$ cut, but \textit{as long as the dressing factor which we are evaluating has no branch cut there}, we can evaluate by an expansion of the type~\eqref{eq:schematic-expansion}, where $\tilde{x}_j$ is now taken to be in the \textit{antimirror} region, like in the rightmost panel.}
\end{figure}

\paragraph{Expansion around the cut of $\ln x$.}
The strategy which helped us was to evaluate the phases at points $x^\pm$ which are nearby and in the same physical region. In this case the computation is more involved. We are dealing with bound states, which means that the full phases must be obtained by fusing circa~$k$ constituents. Moreover, the fused variables $x^{\pm(k-1)}$ or $x^{\pm(k+1)}$ are evaluated close to the cut of $\ln x$. The setup, and the strategy which we follow to simplify the computation, are sketched in Figure~\ref{fig:xpm-BMN-negative}. Following the steps outlined in the figure's caption, we can arrive at a configuration where $x^{\pm(k-1)}$ (or $x^{\pm(k+1)}$) are on the same side of the integration contour --- both above. Remarkably, when properly taking care of the various residues generated by the analytic continuation, we find that the fused and continued dressing factor depend only on $x^{\pm(k-1)}$ (or $x^{\pm(k+1)}$), \textit{not on the bound state constituents}. Moreover, the dependence on the Zhukovsky variables can be expressed entirely in terms of the integral of $\tPhi$ and of rational functions the Zhukovsky variables. In other words, the points $x^{\pm(k-1)}$ (or $x^{\pm(k+1)}$) can be continued to the anti-mirror region, and we can write them as in~\eqref{eq:schematic-expansion}, where now the expansion is around a point in the antimirror region. The whole procedure is worked out in detail in Appendix~\ref{app:bmn_neg_p_new} for the processes
\begin{equation}
    S^{k+1,1}_{\bar{Y}\bar{Y}}\left(2\pi-\frac{p_1}{T},\frac{p_2}{T}\right)\,,\qquad
     S^{k-1,1}_{\bar{Y}\bar{Y}}\left(2\pi-\frac{p_1}{T},\frac{p_2}{T}\right)\,.
\end{equation}
These are related by the identities~\eqref{eq:2pishift-nomonodromy} and~\eqref{eq:2pishift-yesmonodromy} to the processes which we computed above when the first momentum is negative, i.e.~to
\begin{equation}
    S^{11}_{\bar{Y}\bar{Y}}\left(-\frac{p_1}{T},\frac{p_2}{T}\right)\,,\qquad
     S^{11}_{Z\bar{Y}}\left(-\frac{p_1}{T},\frac{p_2}{T}\right)\,,
\end{equation}
respectively. There are a few points on which it is worth remarking, referring the reader to the appendix for details. The computation of the BES and HL phase again just boils down to computing the double derivative of the AFS part. However, we now need this function in a different domain, namely when $\Im[\check{x}_1]>0$ and $\Im[\tx_2]<0$, which gives
\begin{equation}
\label{eq:doublederivativeAFSmixed}
\frac{\partial^2\tPhi^\afs_{aa}(\check{x}_1,\tx_2)}{\partial \check{x}_1\partial \tx_2}=
\frac{1}{4}\frac{u_1+u_2}{\check{x}_1-\tx_2}+\frac{\check{x}_1-\tx_2}{4(\check{x}_1)^2(\tx_2)^2}\,,
\end{equation}
see~\eqref{eq:ddtPhiafsaa_arb_points}.

We also stress that, after defining (see~\eqref{eq:xpm-BMN-piszero})
\bal
\gamma_a(p) \equiv \ln \frac{x_{a}(p) - \xbr_a}{x_{a}(p) \xbr_a +1}\,, 
\eal
the near-BMN expansion of the $\gamma$-rapidities is
\begin{equation}
    \gamma_a^{\pm}\left(\frac{p}{T}\right)=\gamma_a(p)\pm \frac{i p^2}{2T}+\mathcal{O}(T^{-2})\,,\qquad
    \gamma_a^{\pm(k+1)}\left(2\pi-\frac{p}{T}\right)=\gamma_a(-p)\pm \frac{i p^2}{2T}+\mathcal{O}(T^{-2})\,.
\end{equation}
For the $(k-1)$-particle bound state we find
\begin{equation}
    \gamma_a^{\pm(k-1)}\left(2\pi-\frac{p}{T}\right)=\pm i\pi+\gamma_{\bar{a}}(-p)\mp \frac{i p^2}{2T}+\mathcal{O}(T^{-2})\,.
\end{equation}
The additional shift of $\pm i \pi$ means that, for this process, the odd dressing factor actually contributes at order $\mathcal{O}(T^{-1})$, that is at tree level (instead of one loop).
Keeping all this into account, as well as accounting for the monodromy factor in~\eqref{eq:2pishift-yesmonodromy}, we find that indeed the very same result as in~\eqref{eq:oneloopresult} with $p_1\to-p_1$. This is a very non-trivial check of our prescription for analytic continuation to the negative-momentum region.

\subsection{Relativistic limit}
\label{sec:rel_limit}
The relativistic limit of this model was discussed in~\cite{Frolov:2023lwd}. Note that if we expand the string dispersion relation at small~$h$ around%
\footnote{The relativistic limit can actually be performed for any~$M$, but here we focus on $M\neq0$ mod $k$, which is what is relevant for our comparison.}
\begin{equation}
\label{eq:relativisticpexpansion}
    p=-\frac{2\pi M}{k}+\frac{4\pi h}{k}\sin\left|\frac{\pi M}{k}\right|\,\sinh\theta+\mathcal{O}(h^2)\,,
\end{equation}
we obtain a relativistic model. Hence, the S~matrix is of difference form, 
\begin{equation}
S^{M_1,M_2}(\theta_1,\theta_2)=S^{M_1,M_2}(\theta_1-\theta_2,0)\equiv S^{M_1,M_2}(\theta_{12})\,,
\end{equation}
and the crossing transformation is realised on the relativistic rapidity in the usual way, with the whole S~matrix transforming as
\begin{equation}
\label{eq:relativisticcrossing}
    S^{M_1,M_2}(\theta_{12})\quad\to \quad S^{k-M_1,M_2}(\theta_{12}+i\pi)\,,
\end{equation}
up to a suitable charge conjugation of the internal indices.
The dressing factor can then be fixed by the usual relativistic bootstrap arguments, as carried out in~\cite{Frolov:2023lwd}.
Let us consider the case of a right particle with $M=-1$. This is convenient as the momentum~\eqref{eq:relativisticpexpansion} falls in our fundamental region $0<p<2\pi$. Remark that the relativistic crossing~\eqref{eq:relativisticcrossing} is different from, and possibly nonequivalent to the one which we considered in the full model. Another difference is that, in the relativistic limit, particles can be identified modulo~$k$: for instance a right particle of mass $m=1$ is perfectly equivalent to a left particle of mass~$m=k-1$ \cite{Frolov:2023lwd}. We have seen that this is not so in the full model, due to monodromies such as the ones appearing in~\eqref{eq:2pishift-yesmonodromy}. 
However, working out the relativistic expansion of the kinematic variables, which at leading order gives
\begin{equation}
\label{eq:xmpm_in_rel_lim}
x^{\pm m}_{\R} = e^{-\theta \pm \frac{i \pi}{k} m}+\mathcal{O}(h^{1}),\qquad
\xi=\frac{k}{\pi h}+\mathcal{O}(h^0),
\end{equation}
we see that the monodromies~\eqref{eq:2pishift-yesmonodromy} disappear in this limit.
It is then intriguing to compare an S-matrix element with its relativistic limit. For instance, in the relativistic limit we would expect~\cite{Frolov:2023lwd}
\begin{equation}
    S_{\bar{Z}\bar{Z}}^{11}(x_{1}^\pm,x_{2}^\pm)\to
    S_{YY}^{k-1,k-1}(\theta_{12})=
    \frac{\sinh\left(\tfrac{1}{2}\theta_{12}-\tfrac{i\pi}{k}\right)}{\sinh\left(\tfrac{1}{2}\theta_{12}+\tfrac{i\pi}{k}\right)}
    \frac{R^2\left(\theta_{12}+ \frac{2i \pi}{k} \right)\,R^2\left(\theta_{12}- \frac{2i \pi}{k}\right)}{R^2\left(\theta_{12}\right)}.
\end{equation}
Owing to a number of remarkable cancellations the two expressions indeed match. The detailed discussion of the limit is presented in appendix~\ref{app:rel_limit}. Here we will briefly remark on some aspects that underlie its non-trivial nature. To begin with, note that the~$\bar{Z}\bar{Z}$ is normalised as in~\eqref{eq:massivenormmir}, which yields  the rational prefactor
\begin{equation}
    \frac{u_1-u_2+\tfrac{2i}{h}}{u_1-u_2+\tfrac{2i}{h}}\to\frac{\theta_{12}+\tfrac{2\pi i}{k}}{\theta_{12}-\tfrac{2\pi i}{k}}\,,
\end{equation}
that needs to be canceled in the final result. Indeed, a careful computation shows that this term is canceled by the limit of the terms arising from the analytic continuation of $\tPhi^\bes(\tilde{x}_1,\tilde{x}_2)$ to the string region. Interestingly, a large part of the contribution of $\tPhi^\bes(\tilde{x}_1,\tilde{x}_2)$ cancels against $\tPhi^\hl(\tilde{x}_1,\tilde{x}_2)$, similarly to what happens in the $h\gg1$ limit discussed in the previous subsection --- even though we are in a rather opposite kinematics.
We regard the matching with this relativistic limit as further evidence for our proposal.

\section{On possible CDD factors}
\label{sec:CDDs}
The crossing equations allow for infinitely many solutions. Given a ``minimal'' solution, more can be obtained by multiplying it by solutions of the homogeneous crossing equation --- also called CDD factors~\cite{Castillejo:1955ed}. In two-dimensional relativistic integrable QFTs, CDDs can be constrained quite straightforwardly in terms of the poles of the S~matrix and of its fall-off conditions at large rapidity. In our case, due to the substantially more intricate analytic structure,  a larger number of possibilities are allowed. It is therefore natural to wonder if we missed anything.

The rather non-trivial checks passed by our proposal reduce substantially the number of sensible CDD factors. One suggestive idea however is that we may be able to set the monodromy of~\eqref{eq:2pishift-yesmonodromy} to~$1$ by a judicious choice of the CDD factors, without spoiling any other property of the model.
A CDD factor was recently conjectured by Ohlsson-Sax, Riabchenko and Stefa{\'n}ski (ORS)~\cite{OhlssonSax:2023qrk} with the purpose of obtaining a nice behavior under fusion of $(k-1)$ excitations. It takes the form
\begin{equation}\label{eq:CDD-ORS}
    \left(\Sigma^{\ors}_{ab}(x_1^\pm,x_2^\pm)\right)^{-2}=\left(\frac{\alpha_a(x_1^-)\,\alpha_b(x_2^+)}{\alpha_a(x_1^+)\,\alpha_b(x_2^-)}\right)^{1/k},\qquad
    \alpha_a(x)=\left(1-\frac{\xi_a}{x}\right)\left(x+\frac{1}{\xi_a}\right),
\end{equation}
and it has to be multiplied or divided out from the S~matrix (depending on whether we consider the $aa$ or $a\bar{a}$ sector). Note that $\alpha_a(x)$ already appeared in eq.~\eqref{eq:2pishift-yesmonodromy}.
One immediate issue with this expression is its large-tension behaviour, e.g.
\begin{equation}
-i\ln\Sigma^{\ors}_{\L\L}(x_1^\pm,x_2^\pm)=-\frac{p_1\omega_2-p_2\omega_1}{2\pi\,q\,T^2}+\mathcal{O}(T^{-3})\,.
\end{equation}
While this term can be removed by a local counterterm, notice that it is singular as $q\to0$. This seems problematic, as the original near-BMN Lagrangian, as well as anything thus far perturbatively computed from it, are regular in this limit.
Another concern is that, while a modification of this sort may affect~\eqref{eq:2pishift-yesmonodromy} in a desirable way, it would also spoil~\eqref{eq:2pishift-nomonodromy}. 

A perhaps more intriguing expression can be constructed out of
\begin{equation}
    \beta_{a}(\tx)\equiv\ln \alpha_{a}(\tx)\,.
\end{equation}
Starting from the mirror region, consider the combinations
\begin{equation}
\begin{aligned}
    &\Phi_{aa}^\cdd(\tx_1,\tx_2)=+\frac{1}{2\pi}\left(\tg_a(\tx_1)\beta_a(x_2)-\tg_a(\tx_2)\beta_a(x_1)\right),\\
    &\Phi_{a\bar{a}}^\cdd(\tx_1,\tx_2)=-\frac{1}{2\pi}\left(\tg_a(\tx_1)\beta_{\bar{a}}(x_2)-\tg_{\bar{a}}(\tx_2)\beta_a(x_1)\right),
\end{aligned}
\end{equation}
which can be used to construct
\begin{equation}
\label{eq:cddwithoutgamma}
    \theta^\cdd_{ab}(\tx_1^\pm,\tx_2^\pm)=\Phi^\cdd_{ab}(x_1^+,x_2^+)+\Phi^\cdd_{ab}(x_1^-,x_2^-)-\Phi^\cdd_{ab}(x_1^+,x_2^-)-\Phi^\cdd_{ab}(x_1^-,x_2^+).
\end{equation}
This CDD factor has a familiar expansion, e.g.
\begin{equation}
    \theta^\cdd_{\L\L}(\tx_1^\pm,\tx_2^\pm)=-\frac{p_1p_2(p_1\omega_2-p_2\omega_1)}{2\pi\,T^2} +\mathcal{O}(T^{-3})\,.
\end{equation}
This is (twice) the mismatch~\eqref{eq:mismatch}. It is also interesting to note that this CDD factor modifies both~\eqref{eq:2pishift-nomonodromy} and~\eqref{eq:2pishift-yesmonodromy} in such a way that both are left with a non-trivial monodromy, of the type
\begin{equation}
    \exp\left[\gamma_a(x_2^{+})-\gamma_a(x_2^{-})\right]\,.
\end{equation}
This monodromy can in principle be eliminated by modifying slightly the CDD phase~\eqref{eq:cddwithoutgamma} by adding a linear combination of $\gamma$'s,
\begin{equation}
\label{eq:cddwithgamma}
\begin{aligned}
&\theta^\cdd_{ab}(\tx_1^\pm,\tx_2^\pm)+\frac{i}{2}\left(\tg_a(\tx_1^-)-\tg_a(\tx_1^+)-\tg_a(\tx_2^-)+\tg_a(\tx_2^+) \right),\\
&\theta^\cdd_{a\bar{a}}(\tx_1^\pm,\tx_2^\pm)-\frac{i}{2}\left(\tg_a(\tx_1^-)-\tg_a(\tx_1^+)-\tg_{\bar{a}}(\tx_2^-)+\tg_{\bar{a}}(\tx_2^+) \right),
\end{aligned}
\end{equation}
One can check that this  would indeed make both~\eqref{eq:2pishift-nomonodromy} and~\eqref{eq:2pishift-yesmonodromy} hold with trivial monodromy --- something that might be desirable in principle. However, the linear combination breaks parity invariance in the mirror model, and it also  modifies the large-tension expansion of the dressing factors already at tree level.

We could not find any CDD factor which preserves the good properties of our dressing factors --- including the trivial monodromy in~\eqref{eq:2pishift-nomonodromy} --- and removes the monodromy in~\eqref{eq:2pishift-yesmonodromy}. One might still consider introducing a factor like~\eqref{eq:cddwithoutgamma}, which would improve the matching with some perturbative results. However, spoiling~\eqref{eq:2pishift-nomonodromy} seems undesirable. Even in the pure-RR model, a CDD factor like~\eqref{eq:cddwithoutgamma} is somewhat problematic as it introduces new branch points at $x=0$ and $x=\infty$, which were previously absent from the dressing factors.

\section{Conclusions}
\label{sec:conclusions}

The dressing factors for massive particles we proposed are smooth deformations of the ones for the pure-RR $AdS_3\times S^3\times T^4$ \cite{Frolov:2021fmj}. They satisfy the necessary physical properties and pass all available tests. Their interesting peculiarity and distinction from the pure-RR factors is that the starting point for solving the crossing equations is the mirror theory where the integration contours in the $\ka$-deformed BES phases are just straight lines that drastically simplifies many computations, and especially the analyses of CP invariance. In the first arXiv version of our letter~\cite{Frolov:2024pkz} we proposed a solution to the crossing equations which used string theory integration contours instead.
For the pure-RR case using the string or mirror contours gives the same dressing factor, as we detail in appendices~\ref{app:stringvsmirror} and~\ref{app:equivalence_pure_RR}.%
\footnote{Similar contours were also used in~\cite{Gromov:2009bc} to study the $AdS_5\times S^5$ dressing factors.}
In the mixed-flux case, we do not know whether the solution defined on the string contours matches with the one on the mirror contours presented here and in the second version of~\cite{Frolov:2024pkz}; it would be interesting to clarify this.

The next step is to find dressing factors for massless particles and $m$-particle bound states when $m$ is a multiple of~$k$.
 It seems that the dressing factors we proposed are valid for $k$-particle string bound states, and assuming relations \eqref{eq:2pishift-nomonodromy}, \eqref{eq:2pishift-nomonodromy2}, \eqref{eq:2pishift-yesmonodromy} continue to hold for $m=0, k$, one immediately finds the dressing factors involving massless string particles. It would then be necessary to check that the crossing equations and CP invariance for massless dressing factors are satisfied.
It may also provide a definite answer to a sign puzzle in the crossing equation for massless particles in the pure RR case which has been recently raised in~\cite{Ekhammar:2024kzp}.
In the mirror theory we expect that, similarly to the RR case  \cite{Frolov:2021fmj}, a massless particle with real momentum would have its $u$-rapidity on the mirror cuts. We do not expect any problem with $k$-particle mirror bound states because on the mirror $u$-plane the line of real momentum does not intersect any cut. 
  
Once all the dressing factors are known it should be straightforward to write the mirror TBA equations encoding the spectrum of mixed-flux $AdS_3\times S^3\times T^4$ superstring. 
We expect they would have the same form as in the pure RR case \cite{Frolov:2021bwp,Frolov:2023wji}. 
The TBA equations then should be analysed numerically for large $h$, and analytically (and numerically) for small $h$ and fixed $k$ where we expect to find a relation to spectrum of the Wess-Zumino-Witten models~\cite{Maldacena:2000hw}, whose TBA description is also known~\cite{Dei:2018mfl}. This may be particularly interesting for the case~$k=1$, whose dual theory is the symmetric-product orbifold CFT of a free theory~\cite{Eberhardt:2018ouy}, see~\cite{Seibold:2024qkh} for a recent review.

\section*{Acknowledgments}

We thank the participants of the Workshop \textit{Integrability in Low Supersymmetry Theories} in Filicudi in 2023 and in Trani in 2024 (Italy) for stimulating discussions that initiated and furthered this project. D.P.~thanks Bogdan Stefa\'nski for stimulating discussions during the 2024 workshop on `Gauge theories, supergravity and superstrings' in Benasque.
D.P.~and A.S.~are grateful to the Kavli Institute for
Theoretical Physics in Santa Barbara for the hospitality during the follow-on of the Integrable22 workshop,
where part of this work was carried out.
D.P.~and A.S.~acknowledge support from the EU - NextGenerationEU, program STARS@UNIPD, under project ``Exact-Holography'', and from the PRIN Project n.~2022ABPBEY. A.S.~also acknowledges support from the CARIPLO Foundation under grant n.~2022-1886, and from the CARIPARO Foundation Grant under grant n.~68079.
A.S.~also thanks the MATRIX Institute in Creswick \& Melbourne, Australia, for support through a MATRIX Simons fellowship in conjunction with the  program ``New Deformations of Quantum Field and Gravity Theories'', as well as the IAS in Princeton for hospitality. This work has received funding from the Deutsche Forschungsgemeinschaft (DFG, German Research Foundation) -- SFB-Gesch\"aftszeichen 1624 -- Projektnummer 506632645. 
S.F.~acknowledges support from the INFN under a Foreign Visiting Fellowship at the last stages of the project.

\appendix

\section{Definitions}
For the reader's convenience, we collect here the definitions introduced throughout the main text.

\paragraph{Parametrisations.}
The rapidity variables are 
\begin{equation}
u_a(x)=x+\frac{1}{x}- \frac{\kappa_a}{\pi}\,\ln x \,,
\end{equation}
where $\ka_{\L}=-\ka_{\R}=\frac{k}{h}$ (with $k=1,\,2, \dots$ and $h \ge0$).
The image of the main branch point on the $x$-plane is at
\begin{equation}
\xi_\L =\xi\,,\qquad \xi_\R={1\ov \xi}\,,\qquad  \xi\equiv {\ka\ov2\pi}+ \sqrt{1+{\ka^2\ov4\pi^2}}\,.
\end{equation}
The Zhukovski variables can be expressed in terms of momenta as
\begin{equation}
\label{eq:xpmofp_app}
\begin{aligned}
    x^{\pm m}_{\L} (p)&=& \frac{e^{\pm i p /2}}{2h\,\sin \tfrac{p}{2}} \Big(m+\tfrac{k}{2\pi}p+\sqrt{\big(m+\tfrac{k}{2\pi}p\big)^2+4h^2\sin^2\tfrac{p}{2}}\Bigg)\,,\\
    x^{\pm m}_{\R}(p) &=& \frac{e^{\pm i p /2}}{2h\,\sin \tfrac{p}{2}} \Big(m-\tfrac{k}{2\pi}p+\sqrt{\big(m-\tfrac{k}{2\pi}p\big)^2+4h^2\sin^2\tfrac{p}{2}}\Bigg)\,.
\end{aligned}
\end{equation}
whereas in terms of the rapidity we have
\begin{equation}
x_{a}^{\pm m}(u)=x_{a}(u\pm {i\ov h}m),\qquad
p_a = i\,(\ln x_a^{-m} - \ln x_a^{+m})\,,
\end{equation}
The string $\g$-rapidities are
\bal
x_a(\g)&={\xi_a+e^{\g} \ov 1 - \xi_a e^{\g} }\,,\qquad  
\g_a(x) = \ln{x - \xi_a\ov x\,\xi_a +1}\,,
 \eal
while the mirror ones are
\bal
\tx_a(\tg)&={\xi_a-i\,e^{\tg} \ov 1 + i\,\xi_a e^{\tg} }\,,
\qquad
\tg_a(\tx) &= \ln{\xi_a -\tx \ov \tx\,\xi_a +1}-\frac{i\pi}{2}\,.
 \eal

\paragraph{Dressing factors.}
The dressing factors are split into the product on an ``even'' and ``odd'' part. They are defined in the mirror region and then continued to the other relevant regions. The even part is itself given by the ratio of the BES and HL dressing factors $\Sigma^{\besratio}_{ab}= \Sigma^{\bes}_{ab}/\Sigma^{\hl}_{ab}$.
The BES phase is expressed in terms of the double integral
\begin{equation}
    \tPhi_{ab}^{\alpha \beta}(x_{1},x_{2}) 
= -  \lint_{{ \pa\cR_\alpha}} \frac{{\rm d} w_1}{2\pi i} \lint_{{ \pa\cR_\beta}} \frac{{\rm d} w_2}{2\pi i} {1\ov w_1-x_{1}}{1\ov w_2-x_{2}} K^\bes(u_a(w_1)-u_b(w_2))\,,
\end{equation}
where for the LL/RR phase we take the semi-sum of the $\pm\pm$ contours, and for the LR/RL phase we take the semi-sum of the $\pm\mp$ contours, and use the BES Kernel
\begin{equation}
    K^\bes(v)= i \log \frac{\Gamma \left(1+\frac{ih}{2}v \right)}{\Gamma \left(1-\frac{ih}{2}v \right)}\,.
\end{equation}
Expanding the BES Kernel at large-$h$ gives the AFS Kernel, the HL Kernel, and subleading terms, with
\begin{equation}
K_\eps^\afs(v) = -\frac{1}{2} v\big( \log (\eps -i v)+\log (\eps +i v)\big)  -  v \log \frac{h}{2 e},\quad
K_\eps^{\hl}(v)= \frac{\ln(\epsilon-iv)-\ln(\epsilon+iv)}{2i},
\end{equation}
where $\epsilon>0$ is a regulator. As we detail, it is possible to simplify these expressions and for the HL phase
\bal
\tPhi_{aa}^{\hl}(x_{1},x_{2}) =&- \frac{1}{4 \pi}   \lint_{\widetilde{ \rm cuts}} {\rm d} v \frac{\tx'_a(v)}{\tx_a(v) - x_{1}} \left( \log\left(\tx_a(v - i \eps)-x_{2} \right) - \log\left(\tx_a(v + i \eps)-\tx_{2} \right) \right)\\
&-{i\ov 4}\sgn\left(\Im(x_1)\right)\ln(-x_2) 
+{\pi\ov 8}\sgn\left(\Im(x_1)\right)\sgn\left(\Im(x_2)\right)\,,\\
\tPhi_{\bar a a}^{\hl}(x_{1}, x_{2})=&+\frac{1}{4 \pi}   \lint_{\widetilde{ \rm cuts}} {\rm d} v \frac{\tx'_{\bar{a}}(v)}{\tx_{\bar{a}}(v) - x_{1}} \left( \log\left(\frac{1}{\tx_{\bar{a}}(v-i \eps)}-x_{2} \right) - \log\left(\frac{1}{\tx_{\bar{a}}(v+ i \eps)}-x_{2} \right) \right)\\
&-{i\ov 4}\sgn\left(\Im(x_1)\right)\ln(-x_2) 
+{\pi\ov 8}\sgn\left(\Im(x_1)\right)\sgn\left(\Im(x_2)\right)\, .
\eal
The odd dressing factor is expressed in terms of the Barnes $G$-function with
\begin{equation}
    R (\g)= {G(1- \frac{\g}{2\pi i})\ov G(1+ \frac{\g}{2\pi i}) }\,,
\end{equation}
and it reads
\bal
\tPhi_{aa}^\barnes(\tx_{a1},\tx_{a2})=& +i \ln R(\tg_{a1}-\tg_{a2})\,,\\
\tPhi_{\bar{a}a}^\barnes(\tx_{\bar{a}1},\tx_{a2})=& -\frac{i}{2} \ln R(\tg_{\bar{a}1}-\tg_{a2}+i\pi)R(\tg_{\bar{a}1}-\tg_{a2}-i\pi)\,.
\eal
For any of these functions we consider the combination
\begin{equation}
\tilde{\theta}_{ab}^{\text{any}}(\tx^\pm_{a1},\tx^\pm_{b2})= 
\tPhi^{\text{any}}_{ab}(\tx_{a1}^+,\tx_{b2}^+)-\tPhi^{\text{any}}_{ab}(\tx_{a1}^+,\tx_{b2}^-)-\tPhi^{\text{any}}_{ab}(\tx_{a1}^-,\tx_{b2}^+)+\tPhi^{\text{any}}_{ab}(\tx_{a1}^-,\tx_{b2}^-)\,,
\end{equation}
where the phases are related to the dressing factors as
\bal
\Sigma^{\text{any}}_{ab}(\tx^\pm_{a1}, \tx^\pm_{b2})=\exp\left[{i\, \tilde{\theta}_{ab}^{\text{any}}(\tx^\pm_{a1}, \tx^\pm_{b2})}\right]\,.
\eal

\section{Improved BES as an integral around mirror cuts}
\label{app:stringvsmirror}

In this section, we show how to write the improved BES phase integrating over the mirror contours in the case $\ka=0$.
While this rewriting was considered previously~\cite{Gromov:2009bc}, we find it appropriate to present it here in some detail, as it provides our starting point for obtaining the $\ka$-deformed improved BES phases.
 
\paragraph{$\Phi$ functions.}

For $\ka=0$ and for points $\{x^\pm_1, x^\pm_2\}$ in the string region the string BES phase is given by the following combination of $\Phi$ functions\footnote{This definition also generalises to mirror $\tPhi$ functions and the HL functions as follows
\bal
\label{eq:app_varPhi_functions}
&\tilde\varPhi(x^\pm_1,x^\pm_2) \equiv \tilde\Phi(x^+_1,x^+_2)+\tilde\Phi(x^-_1,x^-_2)-\tilde\Phi(x^+_1,x^-_2)-\tilde\Phi(x^-_1,x^+_2)\,,\\
&\varPhi^\hl(x^\pm_1,x^\pm_2) \equiv \Phi^\hl(x^+_1,x^+_2)+\Phi^\hl(x^-_1,x^-_2)-\Phi^\hl(x^+_1,x^-_2)-\Phi^\hl(x^-_1,x^+_2)\,,\\
&\tilde\varPhi^\hl(x^\pm_1,x^\pm_2) \equiv \tilde\Phi^\hl(x^+_1,x^+_2)+\tilde\Phi^\hl(x^-_1,x^-_2)-\tilde\Phi^\hl(x^+_1,x^-_2)-\tilde\Phi^\hl(x^-_1,x^+_2)\,.
\eal
}
\bal
\label{eq:app_var_Phi_string}
\varPhi(x^\pm_1,x^\pm_2) \equiv \Phi(x^+_1,x^+_2)+\Phi(x^-_1,x^-_2)-\Phi(x^+_1,x^-_2)-\Phi(x^-_1,x^+_2)
\eal
with
\begin{equation}
\Phi(x_1,x_2)=\Phi_+(x_1,x_2)-\Phi_+(x_2,x_1)
\end{equation}
and
\bal
\Phi_+(x_1,x_2)&=i\oint\limits_{|w_1|=1}\frac{{\rm d}w_1}{2\pi i}\oint\limits_{|w_2|=1} \frac{{\rm
d}w_2}{2\pi i}\frac{1}{(w_1-x_1)(w_2-x_2)} \log \Gamma \bigl[1+\tfrac{ih}{
2}\big(u(w_1)-u(w_2)\bigl)\bigl]\,.
\eal
In the expression above $u$ is
\bal
\label{eq:Zhuk_map_k_eq_0}
u(x)=x+ \frac{1}{x}\,,
\eal
and the integrals are performed on the unit circle following the counterclockwise direction. We can equivalently define the integral in the $u$ plane as
\bal
\label{eq:app_Phip_Ram_Ram}
\Phi_+(x_1,x_2)=i\oint\limits_{\rm cut}\frac{{\rm d}v_1}{2\pi i}\oint\limits_{\rm cut} \frac{{\rm
d}v_2}{2\pi i}\frac{x'(v_1) x'(v_2)}{(x(v_1)-x_1)(x(v_2)-x_2)} \log \Gamma \bigl[1+\tfrac{ih}{
2}\big(v_1-v_2\bigl)\bigl],
\eal
where we integrate both $v_1$ and $v_2$ around the string theory cut $(-2, +2)$  counterclockwise. The functions $x(v_1)$ and $x(v_2)$ are obtained by inverting the Zhukovsky map~\eqref{eq:Zhuk_map_k_eq_0} and take values outside the unit circle.

While for points in the string region the BES phase is simply given by
\bal
\theta^\bes (x^\pm_1, x^\pm_2)=\varPhi(x^\pm_1, x^\pm_2),
\eal
when we move the points to the mirror region the BES phase is analytically continued as~\cite{Arutyunov:2009kf}
\bal
\label{eq:tild_theta_BES_keq0}
\theta^\bes(\tx^\pm_1, \tx^\pm_2)&=\varPhi(\tx^\pm_1, \tx^\pm_2)\\
&+\Psi(\tx^+_2, \tx^+_1)-\Psi(\tx^+_1, \tx^+_2)+\Psi(\tx^+_1, \tx^-_2)-\Psi(\tx^+_2, \tx^-_1)\\
&+ \frac{1}{i} \log \frac{\Gamma \left(1-\frac{i h}{2} \left( u(x^+_1) - u(x^+_2) \right) \right)}{\Gamma \left(1+\frac{i h}{2} \left( u(x^+_1) - u(x^+_2) \right) \right)}\,,
\eal
where
\bal
\Psi(x,y)=\Psi_+(x,y)- \Psi_-(x,y)
\eal
and
\bal
\Psi_\pm(x_1,x_2)=+i\int_{\rm cut} \frac{{\rm
d}v}{2\pi i}\,\frac{x'(v)}{x(v)-x_2} \log \Gamma\big[1\pm\tfrac{ih}{
2}\big(u_1-v\big)\big]\,,
\eal
where $u_i=u(x_i)$.
For later convenience, we also introduce the functions
\bal
\label{eq:app_tPhip_Ram_Ram}
\tPhi_+(x_1,x_2)
&=-i\int_{\widetilde{\rm cuts}}\frac{{\rm d}v_1}{2\pi i}\int_{\widetilde{\rm cuts}} \frac{{\rm
d}v_2}{2\pi i}\frac{\tx'(v_1) \tx'(v_2)}{(\tx(v_1)-x_1)(\tx(v_2)-x_2)} \log \Gamma \bigl[1+\tfrac{ih}{
2}\big(v_1-v_2\bigl)\bigl]\,,
\eal
\bal
\tilde{\Psi}_\pm(x_1,x_2)=-i\int_{\widetilde{\rm cuts}} \frac{{\rm
d}v}{2\pi i}\,\frac{\tx'(v)}{\tx(v)-x_2} \log \Gamma\big[1\pm\tfrac{ih}{
2}\big(u_1-v\big)\big]\,.
\eal
The integrals in the expressions above are now performed around the mirror theory cuts $(-\infty, -2)$ and $(+2,+\infty)$, and we move to the right on the lower edge of the cuts. The function $\tx(v)$ is the inverse of the Zhukovsky map~\eqref{eq:Zhuk_map_k_eq_0} evaluated in the mirror region.

In the mirror theory, it is convenient to write the phases in terms of the so-called improved BES dressing factor
\bal
\label{eq:improved_bes_in_mir_reg}
\tilde{\theta}^\bes(\tx^\pm_1, \tx^\pm_2)=\theta^\bes(\tx^\pm_1, \tx^\pm_2)+ \frac{1}{i} \log \left( \frac{1-\frac{1}{\tx^+_1 \tx^-_2}}{1-\frac{1}{\tx^-_1 \tx^+_2}} \right)\,,
\eal
which fuses well in the mirror region~\cite{Arutyunov:2009kf}.
In this appendix, we show that the expression in~\eqref{eq:improved_bes_in_mir_reg} can be equivalently written as
\bal
\label{eq:tild_theta_BES_keq0_mircont}
\tilde{\theta}^\bes(\tx^\pm_1, \tx^\pm_2)=\tilde\varPhi(\tx^\pm_1, \tx^\pm_2)
\eal
with $\tilde\varPhi$ given in~\eqref{eq:app_varPhi_functions} and
\bal
\tPhi(\tx_1, \tx_2)=\tPhi_+(x_1,x_2)-\tPhi_+(x_2,x_1)\,.
\eal

To obtain formula~\eqref{eq:tild_theta_BES_keq0_mircont} starting from~\eqref{eq:tild_theta_BES_keq0} we need to deform the integration contours around the string theory cut $(-2,+2)$ to contours around the mirror cuts $(-\infty, -2)$ and $(+2, +\infty)$.

\paragraph{\texorpdfstring{$v_2$}{v[2]}-integration contour.}

We start deforming the $v_2$ integration contour. 
We note that the function
\bal
\log \Gamma \bigl[1+\tfrac{ih}{2}\big(v_1-v_2\bigl)\bigl]
\eal
has branch points at 
\bal
v_2=v_1-\frac{2i}{h} n \quad n=1,\,2,\dots
\eal
with cuts in the $v_2$-plane running vertically to $-i \infty$.
For this reason, we need to deform the $v_2$ integration contour by moving it to the upper half of the complex plane (indeed a deformation to the lower half of the plane would produce an intersection with the branch cuts of the $\log \Gamma$ function). We deform the upper edge of the string cut so that it becomes the union of a semicircle of infinite radius lying in the upper half of the plane and the lower edges of the mirror cuts; then we move the lower edge of the string cut to the anti-string $v_2$-plane and then deform it there so that it becomes the upper semicircle of infinite radius and the upper edges of the mirror cuts. This preserves the original sign of the integral. 

Let us label $u_2=u(\tx_2)$. When we move the integration line on the lower edge of the string cut across the cut to the anti-string $v_2$ plane then 
\bal
x(v_2) \to \frac{1}{x(v_2)}=\tx(v_2) \,.
\eal
For $\Im(u_2)>0$ deforming this integration line in the upper half of the $v_2$ plane, we encounter a pole at $v_2=u_2$. Keeping the contribution of this pole into account the result of the deformation of the $v_2$ contour is
\bal
\label{eq:app_phip_one_cont_def}
\Phi_+(\tx_1,\tx_2)
&=i\lint_{\rm cut}\frac{{\rm d}v_1}{2\pi i}\lint_{\widetilde{\rm cuts}} \frac{{\rm
d}v_2}{2\pi i}\frac{x'(v_1) \tx'(v_2)}{(x(v_1)-\tx_1)(\tx(v_2)-\tx_2)} \log \Gamma \bigl[1+\tfrac{ih}{
2}\big(v_1-v_2\bigl)\bigl]
\\
&+ \theta\big(\Im(u_2)\big)\Psi_-(\tx_2,\tx_1)\,.
\eal

\paragraph{\texorpdfstring{$v_1$}{v[1]}-integration contour}

In the $v_1$ plane, the cuts of $\log \Gamma$ are directed to $+ i \infty$ and we must deform the contour in the lower half of the complex plane.
In this case, we move the upper edge of the string cut to the anti-string $v_1$-plane and then deform it so that it becomes the lower semicircle of infinite radius and the lower edges of the mirror cuts; then we deform the lower edge of the string cut so that it becomes the lower semicircle and the upper edges of the mirror cuts. This changes the original sign of the integral. If $\Im(u_1)<0$ we get an extra contribution from deforming the contour on the lower edge of the string cut due to the pole at $v_1=u_1$ on the string $v_1$-plane.  The result  of the deformation of both contours is then given by
\bal
\Phi_+(\tx_1,\tx_2)
&={1\ov i}\lint_{\widetilde{\rm cuts}}\frac{{\rm d}v_1}{2\pi i}\lint_{\widetilde{\rm cuts}} \frac{{\rm
d}v_2}{2\pi i}\frac{\tx'(v_1) \tx'(v_2)}{(\tx(v_1)-\tx_1)(\tx(v_2)-\tx_2)} \log \Gamma \bigl[1+\tfrac{ih}{
2}\big(v_1-v_2\bigl)\bigl]
\\
&+ \theta\big(-\Im(u_1)\big)\tilde\Psi_+(\tx_1,\tx_2)+ \theta\big(\Im(u_2)\big)\Psi_-(\tx_2,\tx_1)\,.
\eal
To relate $\tilde\Psi_+(\tx_1,\tx_2)$ to $\Psi_+(\tx_1,\tx_2)$ we deform the $v_2$ contour in $\Psi_+(\tx_1,\tx_2)$ in the same way as above and get 
\bal
\Psi_+(\tx_1,\tx_2)
&=-\tilde\Psi_+(\tx_1,\tx_2)+ \theta\big(\Im(u_2)\big)i\log \Gamma \bigl[1+\tfrac{ih}{
2}\big(u_1-u_2\bigl)\bigl]\,,
\eal
where $u_1 \equiv u(\tx_1)$ and $u_2 \equiv u(\tx_2)$.
Thus,
the final  result  of the deformation of the $v_1$ and $v_2$ contours is
\bal\label{Phiptx1tx2}
\Phi_+(\tx_1,\tx_2)
=&+\tilde{\Phi}_+(\tx_1,\tx_2)
- \theta\big(-\Im(u_1)\big)\Psi_+(\tx_1,\tx_2)
+ \theta\big(\Im(u_2)\big)\Psi_-(\tx_2,\tx_1)
\\
&+ \theta\big(-\Im(u_1)\big)\theta\big(\Im(u_2)\big)i\log \Gamma \bigl[1+\tfrac{ih}{
2}u_{12}\bigl]\,.
\eal

\paragraph{String $\Phi$ in term of mirror $\tilde\Phi$.}

Using~\eqref{Phiptx1tx2} we get
\bal\label{stringPhi1}
\varPhi(\tx_1^\pm,\tx_2^\pm)=&+\Phi(\tx_1^+,\tx_2^+)+\Phi(\tx_1^-,\tx_2^-)-\Phi(\tx_1^+,\tx_2^-)-\Phi(\tx_1^-,\tx_2^+)
\\
=&+\tilde\Phi(\tx_1^+,\tx_2^+)+\Psi_-(\tx_2^+,\tx_1^+)-\Psi_-(\tx_1^+,\tx_2^+)
\\
&+\tilde\Phi(\tx_1^-,\tx_2^-)-\Psi_+(\tx_1^-,\tx_2^-)+\Psi_+(\tx_2^-,\tx_1^-)
\\
&-\tilde\Phi(\tx_1^+,\tx_2^-)- \Psi_+(\tx_2^-,\tx_1^+)+\Psi_-(\tx_1^+,\tx_2^-)
+ i\log \Gamma \bigl[1+\tfrac{ih}{
2}\big(u_{21} -\tfrac{2i}{h}\big)\bigl]
\\
&-\tilde\Phi(\tx_1^-,\tx_2^+)+ \Psi_+(\tx_1^-,\tx_2^+)
-\Psi_-(\tx_2^+,\tx_1^-)
- i\log \Gamma \bigl[1+\tfrac{ih}{
2}\big(u_{12} -\tfrac{2i}{h}\big)\bigl]
\,.
\eal
The expression above can be written as
\bal\label{stringPhi3}
\varPhi(\tx_1^\pm, \tx_2^\pm)&=\tilde{\varPhi}(\tx_1^\pm, \tx_2^\pm)-{1\ov i}\log {\Gamma \bigl[1-\tfrac{ih}{
2}u_{12}\bigl]\ov \Gamma \bigl[1+\tfrac{ih}{
2}u_{12}\bigl]} +{1\ov i}\log{u_{12}-{2i\ov h}\ov u_{12}+{2i\ov h}}\\
&+\Psi(\tx_1^+,\tx_2^+)-\Psi(\tx_2^+,\tx_1^+) 
-\Psi(\tx_1^+,\tx_2^-)+\Psi(\tx_2^+,\tx_1^-)
\\
&+ \Psi_+(\tx_1^-,\tx_2^+)-\Psi_+(\tx_1^+,\tx_2^+)+\Psi_+(\tx_2^+,\tx_1^+)- \Psi_+(\tx_2^-,\tx_1^+)
\\
&+\Psi_+(\tx_1^+,\tx_2^-)
-\Psi_+(\tx_1^-,\tx_2^-)+\Psi_+(\tx_2^-,\tx_1^-)-\Psi_+(\tx_2^+,\tx_1^-) \,.
\eal
Next, we  compute the following difference
\bal
\Psi_+(\tx_1^-,\tx_2)
-\Psi_+(\tx_1^+,\tx_2)&=i\lint_{\rm cut} \frac{{\rm
d}v}{2\pi i}\,\frac{x'(v)}{x(v)-\tx_2} \log {\Gamma\big[1+\tfrac{ih}{
2}\big(u_1-\tfrac{i}{
h}-v\big)\big]\ov \Gamma\big[1+\tfrac{ih}{
2}\big(u_1+\tfrac{i}{
h}-v\big)\big]}
\\
&=i\lint_{\rm cut} \frac{{\rm
d}v}{2\pi i}\,\frac{x'(v)}{x(v)-\tx_2} \log \big[1+\tfrac{ih}{
2}\big(u_1+\tfrac{i}{h}-v\big)\big]
\\
&=i \log(\tx_1^--\tx_2) -i\theta(-\Im(u_2))\log \big[1+\tfrac{ih}{
2}\big(u_{12}+\tfrac{i}{h}\big)\big]\,.
\eal
The integral has been computed by closing a big circle at infinity surrounding the cut of the $\log$ and using Cauchy's theorem:
the first term in the last line of the expression above comes from the cut of $\log \big[1+\tfrac{ih}{
2}\big(u_1+\tfrac{i}{h}-v\big)\big]$ and the second one from the pole at $v=u_2$. There is also a singular contribution coming from closing the contour at infinity which cancels in the full phase.
Thus, we get
\bal\label{stringPhi4}
&+ \Psi_+(\tx_1^-,\tx_2^+)-\Psi_+(\tx_1^+,\tx_2^+)+\Psi_+(\tx_2^+,\tx_1^+)- \Psi_+(\tx_2^-,\tx_1^+)
\\
&+\Psi_+(\tx_1^+,\tx_2^-)
-\Psi_+(\tx_1^-,\tx_2^-)+\Psi_+(\tx_2^-,\tx_1^-)-\Psi_+(\tx_2^+,\tx_1^-)
\\
=&+i \log(\tx_1^--\tx_2^+)-i \log(\tx_1^--\tx_2^-) +i\log \big[1+\tfrac{ih}{
2}\big(u_{12}+\tfrac{2i}{h}\big)\big]
\\
&-i \log(\tx_2^--\tx_1^+)+i \log(\tx_2^--\tx_1^-) -i\log \big[1+\tfrac{ih}{
2}\big(u_{21}+\tfrac{2i}{h}\big)\big]
\\
&={1\ov i} \log{\tx_1^+-\tx_2^-\ov \tx_1^--\tx_2^+}
\,.
\eal
Plugging this result into~\eqref{stringPhi3} we obtain
\bal
\label{stringPhi5}
\varPhi(\tx^\pm_1,\tx^\pm_2)&=\tilde{\varPhi}(\tx_1^\pm, \tx_2^\pm)-{1\ov i}\log {\Gamma \bigl[1-\tfrac{ih}{
2}u_{12}\bigl]\ov \Gamma \bigl[1+\tfrac{ih}{
2}u_{12}\bigl]} -{1\ov i}\log{1-{1\ov \tx_1^+\tx_2^-}\ov 1-{1\ov \tx_1^-\tx_2^+}}\\
&+\Psi(\tx_1^+,\tx_2^+)-\Psi(\tx_2^+,\tx_1^+) 
-\Psi(\tx_1^+,\tx_2^-)+\Psi(\tx_2^+,\tx_1^-) \,.
\eal
Finally, substituting this formula into~\eqref{eq:tild_theta_BES_keq0}
we obtain
\bal
\theta^\bes(\tx^\pm_1, \tx^\pm_2)&=\tilde{\varPhi}(\tx_1^\pm, \tx_2^\pm)-{1\ov i}\log{1-{1\ov \tx_1^+\tx_2^-}\ov 1-{1\ov \tx_1^-\tx_2^+}} \,.
\eal
From this equality, we conclude that the expressions for the improved BES phase written in~\eqref{eq:improved_bes_in_mir_reg} and~\eqref{eq:tild_theta_BES_keq0_mircont} are equivalent.

\section{AFS and HL orders of mirror \texorpdfstring{$\tPhi$}{tPhi}-functions}
\label{app:mirPhiafs}

This appendix shows how to compute the AFS order of mirror  $\tPhi$-functions, which is the limit in which $h \gg 1$, $k \gg 1$ with $\ka=\frac{k}{h}$ fixed.
We perform the computation in the case in which both particles are in the mirror region. The two towers of branch points of  $K^\bes(v)$, located at
\bal
&v=  +\frac{2i}{h} n \,, \quad n=1, 2, \dots\,,\\
&v=  -\frac{2i}{h} n \,, \quad n=1, 2, \dots\,,\\
\eal
collapse to zero in the large $h$ limit trapping the integration contour in the $v$ plane from the opposite sides. We therefore introduce a regulator $\eps>0$ and consider the large-$h$ limit of the following regularised BES kernel
\bal
\label{eq:Kbes_v_mirror_reg}
K_\eps^\bes(v)= i \log \frac{\Gamma \left(1+\frac{ih}{2} (v -i\eps) \right)}{\Gamma \left(1-\frac{ih}{2} (v +i\eps) \right)}\,.
\eal
Thanks to the regulator $\eps$ the two towers of branch points are now shifted 
\bal
&v= +i \eps +\frac{2i}{h} n \,, \quad n=1, 2, \dots\,,\\
&v=  -i \eps -\frac{2i}{h} n \,, \quad n=1, 2, \dots\,,\\
\eal
and stay away from the integration contour when $h \to \infty$. The  large $h$ expansion of the regularised BES kernel~\eqref{eq:Kbes_v_mirror_reg} is given by
\bal\label{Kbeshlarge}
K_\eps^\bes(v)&= h\, K_\eps^\afs(v) +K_\eps^\hl(v) +\cO(\frac{1}{h})\,,
\eal
where
\begin{equation}
\label{eq:AFSkernel}
K_\eps^\afs(v) = -\frac{1}{2} v\big( \log (\eps -i v)+\log (\eps +i v)\big)  -  v \log \frac{h}{2 e}
\end{equation}
and
\begin{equation}
\label{eq:HLkernel}
    K_\eps^{\hl}(v)= \frac{\ln(\epsilon-iv)-\ln(\epsilon+iv)}{2i}\,.
\end{equation}
Both the AFS and HL kernels have branch cuts
\bal
&v=+i \eps +i t \,, \quad t \ge0\,,\\
&v=-i \eps -i t \,, \quad t \ge0\,,
\eal
which for $\ka<\ka_{\rm cr}$ do not intersect the integration contour. We perform the integration for $\eps$ finite and we send $\eps \to 0$ after integrating.

\subsection{The AFS order}
In order to find the AFS order of the phases, it is convenient to work with the mixed derivative
\bal
{\stackrel{\prime\prime}{\tPhi}}{}^{\afs}_{a b}(x,y) \equiv \frac{\partial^2}{\partial x \partial y} {\tPhi}{}^{\afs}_{a b}(x,y) \,.
\eal
After finding a result for ${\stackrel{\prime\prime}{\tPhi}}{}^{\afs}_{a b}(x, y)$ we will integrate it with respect to $x$ and $y$ and obtain ${\tPhi}{}^{\afs}_{a b}(x, y)$. 
A consequence of the fact we are working with the mixed derivative of the phase is that the result we find is correct up to certain single-valued functions of $x$ and $y$
\bal
{\tPhi}{}^{\afs}_{a b}(x,y) +f_1(x)+f_2(y)\,.
\eal
We stress that this ambiguities disappear in the full phase since the unknown functions $f_1$ and $f_2$ cancel in the combination
\bal
 {\tPhi}{}^{\afs}_{a b}(\tx^+_{a1},\tx^+_{b2})+ {\tPhi}{}^{\afs}_{a b}(\tx^-_{a1},\tx^-_{b2})- {\tPhi}{}^{\afs}_{a b}(\tx^+_{a1},\tx^-_{b2})- {\tPhi}{}^{\afs}_{a b}(\tx^-_{a1},\tx^+_{b2})\,.
\eal 

For $\tx_{1}$ and $\tx_{2}$ in the mirror region we find the following results for the mixed derivative of the phases
\bal 
\label{eq:afs_mixed_der}
{\stackrel{\prime\prime}{\tPhi}}{}^{\afs}_{a a}(\tx_{a1},\tx_{a2}) =&-\frac{1}{2}\frac{u_a'(\tx_{a1}) u_a'(\tx_{a2})}{u_1-u_2}+\frac{1}{4} \frac{u_a'(\tx_{a2})+u_a'(\tx_{a1})}{\tx_{a1} - \tx_{a2}}+\frac{\tx_{a1} - \tx_{a2}}{4 \tx^2_{a1} \tx^2_{a2}}\,,\\
{\stackrel{\prime\prime}{\tPhi}}{}^{\afs}_{a \bar a}(\tx_{a1},\tx_{\bar a2})=&\frac{1}{2 \tx_{a1} \tx_{\bar a2} \left(\tx_{a1} \tx_{\bar a2} -1 \right)} \left[ \tx_{a1} - \tx_{\bar a2} - \frac{\ka_a}{2 \pi} \left(\tx_{a1} \tx_{\bar a2} +1 \right) \right]\,,
\eal
where we define $u_1 \equiv u_\R(\tx_{\R1})$ and $u_2 \equiv u_\R(\tx_{\R2})$.
Integrating the functions above we get
\begin{equation}
\begin{aligned}
\tPhi_{aa}^{\afs}(\tx_1,\tx_2)
&=-\frac{\kappa_a}{4 \pi }\text{Li}_2\frac{\tx_1-\tx_2}{\tx_1} +\frac{\ln \tx_2}{2\tx_1}+\frac{u_{1}}{2}  \ln \frac{u_{1}-u_{2}}{\tx_1-\tx_2}   -   (1\leftrightarrow 2)
 \,,\\
\tPhi_{a\bar a}^{\afs}(\tx_1,\tx_2)
&= \frac{\kappa_a}{4\pi}\left[\ln \tx_1\ln \tx_2+2\text{Li}_2\frac{1}{\tx_1\tx_2}\right]\\
&\qquad+\frac{1}{2}
   \left[\frac{\ln \tx_2}{\tx_1}+\frac{\tx_1{}^2+1}{\tx_1}\ln\frac{\tx_1\tx_2-1}{\tx_1\tx_2}-(1\leftrightarrow2)\right].
\end{aligned}
\end{equation}
These primitives have been chosen in such a way that they have no branch cuts when $\tx_1$ and $\tx_2$ are both in the mirror region.

\subsubsection*{Computation of $\tPhi_{\R\R}^{\afs}$}
We show how to compute $\tPhi_{\R\R}^{\afs}$. The remaining cases can be found similarly.
We start with
\bal\label{eq:kernel_expansion_RR_afs_1_notbp}
{\stackrel{\prime\prime}{\tPhi}}{}^{--,\afs}_{\R\R}(\tx_1,\tx_2)=&-\frac{1}{2} \lint_{\widetilde{\rm cuts}} \frac{{\rm d}v_1}{2\pi i}\lint _{\widetilde{\rm cuts}}\frac{{\rm
d}v_2}{2\pi i}\frac{\tx'_\R(v_1) \tx'_\R(v_2)}{(\tx_\R(v_1)-\tx_{\R1})^2(\tx_\R(v_2)-\tx_{\R2})^2}
K^\afs_\eps(v_1 - v_2) \\
&- (\tx_{\R1} \leftrightarrow \tx_{\R2})
\eal
and
\bal\label{eq:kernel_expansion_RR_afs_2_notbp}
{\stackrel{\prime\prime}{\tPhi}}{}^{++,\afs}_{\R\R}(\tx_{\R1},\tx_{\R2})=&-\frac{1}{2} \lint_{\widetilde{\rm cuts}} \frac{{\rm d}v_1}{2\pi i}\lint _{\widetilde{\rm cuts}}\frac{{\rm
d}v_2}{2\pi i}\frac{\left(\frac{1}{\tx_\L(v_1)} \right)' \left(\frac{1}{\tx_\L(v_2)} \right)'}{(\frac{1}{\tx_\L(v_1)}-\tx_{\R1})^2(\frac{1}{\tx_\L(v_2)}-\tx_{\R2})^2}
K^\afs_\eps(v_1 - v_2) \\
&- (\tx_{\R1} \leftrightarrow \tx_{\R2})
  \,,
\eal
where $\tx_{\R1}$ and $\tx_{\R2}$ are both chosen to be in the mirror region: $\Im(\tx_{\R1})<0$, $\Im(\tx_{\R2})<0$. Integrating by parts we obtain
\bal\label{eq:kernel_expansion_RR_afs_1_bp}
&{\stackrel{\prime\prime}{\tPhi}}{}^{--,\afs}_{\R\R}(\tx_{\R1},\tx_{\R2})={\stackrel{\prime\prime}{\tPhi}}{}^{--,{\rm bdry}}_{\R\R}(\tx_{\R1},\tx_{\R2})+{\stackrel{\prime\prime}{\tPhi}}{}^{--,{\rm bulk}}_{\R\R}(\tx_{\R1},\tx_{\R2})\,,
\eal
\bal\label{eq:kernel_expansion_RR_afs_2_bp}
&{\stackrel{\prime\prime}{\tPhi}}{}^{++,\afs}_{\R\R}(\tx_{\R1},\tx_{\R2})={\stackrel{\prime\prime}{\tPhi}}{}^{++,{\rm bdry}}_{\R\R}(\tx_{\R1},\tx_{\R2})+{\stackrel{\prime\prime}{\tPhi}}{}^{++,{\rm bulk}}_{\R\R}(\tx_{\R1},\tx_{\R2})\,,
\eal
with the following bulk contributions
\bal\label{eq:kernel_expansion_RR_afs_1}
&{\stackrel{\prime\prime}{\tPhi}}{}^{--,{\rm bulk}}_{\R\R}(\tx_{\R1},\tx_{\R2})\\
&=-{1\ov4} \lint_{\widetilde{\rm cuts}} \frac{{\rm d}v_1}{2\pi i}\lint _{\widetilde{\rm cuts}}\frac{{\rm
d}v_2}{2\pi i}\frac{1}{(\tx_\R(v_1)-\tx_{\R1})(\tx_\R(v_2)-\tx_{\R2})}
\left({1\ov v_1-v_2-i\epsilon} +{1\ov v_1-v_2+i\epsilon}  \right) \\
&- (\tx_{\R1} \leftrightarrow \tx_{\R2})\,,
\eal
\bal\label{eq:kernel_expansion_RR_afs_2}
&{\stackrel{\prime\prime}{\tPhi}}{}^{++,{\rm bulk}}_{\R\R}(\tx_{\R1},\tx_{\R2})\\
&=-{1\ov4} \lint_{\widetilde{\rm cuts}} \frac{{\rm d}v_1}{2\pi i}\lint _{\widetilde{\rm cuts}}\frac{{\rm
d}v_2}{2\pi i}\frac{1}{(\frac{1}{\tx_\L(v_1)}-\tx_{\R1})(\frac{1}{\tx_\L(v_2)}-\tx_{\R2})}
\left({1\ov v_1-v_2-i\epsilon} +{1\ov v_1-v_2+i\epsilon}  \right) \\
&- (\tx_{\R1} \leftrightarrow \tx_{\R2})
  \,.
\eal
The boundary contributions ${\stackrel{\prime\prime}{\Phi}}{}^{--,{\rm bdy}}_{\R\R}(\tx_{\R1},\tx_{\R2})$ and ${\stackrel{\prime\prime}{\Phi}}{}^{++,{\rm bdy}}_{\R\R}(\tx_{\R1},\tx_{\R2})$ arising from the integration by parts are nonzero; they can be computed introducing a regulator $R$ and integrating around the mirror theory cut $(\ubr, +R)$ and around the cuts $(-\tR\pm i\ka, -\ubr \pm i\ka)$ where $\tR=\sqrt{R^2-\ka^2}$ (after integrating we take the limit $R \to \infty$).
The result for the boundary contribution is the following
\bal
\label{RR_afs_mirror_bdy}
{\stackrel{\prime\prime}{\tPhi}}{}^{{\rm bdry}}_{\R\R}(\tx_{\R1},\tx_{\R2})&=\frac{1}{2} \left({\stackrel{\prime\prime}{\Phi}}{}^{--, {\rm bdry}}_{\R\R}(\tx_{\R1},\tx_{\R2})+{\stackrel{\prime\prime}{\Phi}}{}^{++, {\rm bdry}}_{\R\R}(\tx_{\R1},\tx_{\R2}) \right)\\
&= \frac{1}{8} \left( \frac{1}{\tx_{\R1}} -  \frac{1}{\tx_{\R2}}+ \frac{1}{\tx_{\R1} \tx^2_{\R2}} -  \frac{1}{\tx^2_{\R1} \tx_{\R2}} \right)\,.
\eal
This result does not depend on $k$ and is therefore the same for all phases ${\stackrel{\prime\prime}{\tPhi}}{}^{{\rm bdry}}_{\R\R}$, ${\stackrel{\prime\prime}{\tPhi}}{}^{{\rm bdry}}_{\L\L}$, ${\stackrel{\prime\prime}{\tPhi}}{}^{{\rm bdry}}_{\L\R}$, ${\stackrel{\prime\prime}{\tPhi}}{}^{{\rm bdry}}_{\R\L}$. We omit the details related to the computation of this boundary contribution, while we will show below the computation for the main bulk terms ${\stackrel{\prime\prime}{\tPhi}}{}^{--,{\rm bulk}}_{\R\R}$ and ${\stackrel{\prime\prime}{\tPhi}}{}^{++,{\rm bulk}}_{\R\R}$ in full detail.

\paragraph{The bulk contribution ${\stackrel{\prime\prime}{\tPhi}}{}^{--, {\rm bulk}}_{\R\R}$.}
We start evaluating the expression in~\eqref{eq:kernel_expansion_RR_afs_1}. 
We integrate first wrt $v_2$. We close the contour by adding and subtracting a circle of large radius $R$ at infinity so that the result can be written as the difference between a closed integral (which can be computed by Cauchy's theorem) and the integral over the circle:
\bal 
\label{eq:AFS_split_into_Cv2poles_Cv2circle_mirror}
{\stackrel{\prime\prime}{\tPhi}}{}^{--,{\rm bulk}}_{\R\R}(\tx_1,\tx_2) =C^{v_2}_{\rm poles}-C^{v_2}_{\rm circle} \,.
\eal
 The poles inside the clockwise closed contour are 
\bal 
v_2=u_2\,,\quad v_2 = v_1\pm i\epsilon \,.
\eal
Then we obtain
\bal
C^{v_2}_{\rm poles} =A+B\,,
\eal
where
\bal
A={1\ov4} u'(\tx_{\R2}) \lint_{\widetilde{\rm cuts}} \frac{{\rm d}v_1}{2\pi i} \frac{1}{\tx_\R(v_1)-\tx_{\R1}}
\left({1\ov v_1-u_2-i\epsilon} +{1\ov v_1-u_2+i\epsilon}  \right) - (\tx_{\R1} \leftrightarrow \tx_{\R2}) \,,
\eal
\bal
\label{eq:B_contribution_phimm}
B=-{1\ov4} \lint_{\widetilde{\rm cuts}} \frac{{\rm d}v_1}{2\pi i} \frac{1}{\tx_\R(v_1)-\tx_{\R1}}
\left( \frac{1}{\tx_\R(v_1 + i\epsilon)-\tx_{\R2}}
+ \frac{1}{\tx_\R(v_1 - i\epsilon)-\tx_{\R2}}    \right) - (\tx_{\R1} \leftrightarrow \tx_{\R2}) \,.
\eal
Now we integrate wrt $v_1$. We follow the same procedure of before and we compute $A$ by adding and subtracting a big circle of radius $R$ at infinity so that we can write
\bal
A=A_{\rm poles}-A_{\rm circle}\,.
\eal
For $\Im(\tx_{1})<0$ and $\Im(\tx_{2})<0$ we find
\bal
A_{\rm poles}=-\frac{u_\R'(\tx_{\R1}) u_\R'(\tx_{\R2})}{u_1-u_2}+\frac{1}{2} \frac{u_\R'(\tx_{\R1})+u_\R'(\tx_{\R2})}{\tx_{\R1} - \tx_{\R2}} \,.
\eal
Using the asymptotic behaviour of $\tx_\R(v)$ (see~\eqref{eq:txR_large_u}), we find that only the upper half part of the circle, corresponding to a big arc starting from $-\tilde{R}-i \ka+i0$ and ending at $R+i 0$, contributes to $A_{\rm circle}$. In contrast, the half circle on the lower part of the complex plane is suppressed and we obtain
\bal
A_{\rm circle}= -\frac{1}{4} \frac{u_\R'(\tx_{\R1})}{\tx_{\R2}}-\frac{1}{4} \frac{u_\R'(\tx_{\R2})}{\tx_{\R1}} \,.
\eal
Then we have
\bal
\label{eq:result_Cv2_poles_keq0_mirror}
C^{v_2}_{\rm poles}=-\frac{u_{\R}'(\tx_{\R1}) u_{\R}'(\tx_{\R2})}{u_1-u_2}+\frac{1}{2} \frac{u_{\R}'(\tx_{\R2})+u_{\R}'(\tx_{\R1})}{\tx_{\R1} - \tx_{\R2}}-\frac{1}{4} \left(\frac{u_{\R}'(\tx_{\R2})}{\tx_{\R1}}-\frac{u_{\R}'(\tx_{\R1})}{\tx_{\R2}} \right)+B\,.
\eal
It does not appear that $B$ may be expressed it in terms of elementary functions of $\tx_1$ and $\tx_2$. Regardless, we will show in one moment that the same~$B$ term is also produced in the computation of ${\stackrel{\prime\prime}{\Phi}}{}^{++, {\rm bulk}}_{\R\R}$, with an opposite sign. As a consequence of this fact the sum of ${\stackrel{\prime\prime}{\Phi}}{}^{--, {\rm bulk}}_{\R\R}$ and ${\stackrel{\prime\prime}{\Phi}}{}^{++, {\rm bulk}}_{\R\R}$ takes a simple form.~\footnote{Having in mind that this expression is directly related to near-BMN expansion of the dressing factors, the cancellation of such unpleasant terms provides a further indication that the building blocks for the BES phase must be taken to be the half sum of the two contributions, as stressed in section~\ref{sec:proposal}.}.

Now we want to compute $C^{v_2}_{\rm circle}$. This term can be split into
\bal
\label{Cv2_circle_case_kequal0_first_definition}
C^{v_2}_{\rm circle}=C^{v_2}_{\rm lower \ arc}+C^{v_2}_{\rm upper \ arc}\,,
\eal
where $C^{v_2}_{\rm lower \ arc}$ and $C^{v_2}_{\rm upper \ arc}$ correspond to the contributions of the half circles on the lower and upper part of the complex plane. As before the two integration contours correspond to an arc starting from $-\tilde{R}-i \ka+i0$ and ending at $R+i 0$ (the upper half circle) and an arc starting from $R-i 0$ and ending at $-\tilde{R}-i \ka-i0$ (the lower half circle). These contributions are given by
\bal
C^{v_2}_{\rm lower \, arc}&=-{1\ov4} \lint_{\rm \widetilde{cuts}} \frac{{\rm d}v_1}{2\pi i}\lint_{{\rm  lower \, arc }}\frac{{\rm
d}v_2}{2\pi i}\frac{1}{(\tx_{\R}(v_1)-\tx_{\R1})(v_2-\tx_{\R2})}
\left({1\ov v_{12}-i\epsilon} +{1\ov v_{12}+i\epsilon}  \right) \\
&- (\tx_{\R1} \leftrightarrow \tx_{\R2})
  \,,
\eal
\bal
C^{v_2}_{\rm upper \, arc}&=-{1\ov4} \lint_{\rm \widetilde{cuts}} \frac{{\rm d}v_1}{2\pi i}\lint_{{\rm  upper \, arc }}\frac{{\rm
d}v_2}{2\pi i}\frac{1}{(\tx_{\R}(v_1)-\tx_{\R1})} \frac{1}{(\frac{1}{v_2}-\tx_{\R2})}
\left({1\ov v_{12}-i\epsilon} +{1\ov v_{12}+i\epsilon}  \right) \\
&- (\tx_{\R1} \leftrightarrow \tx_{\R2})
  \,.
\eal
We start computing $C^{v_2}_{\rm lower \ arc}$ integrating wrt $v_1$. We split this contribution into
\bal
C^{v_2}_{\rm lower \ arc}=C^{v_1}_{\rm poles}-C^{v_1}_{\rm circle}\,.
\eal
The poles contained in the contour are $v_1=u_1$ and $v_1=v_2+i \eps$ and we obtain
\bal
C^{v_1}_{\rm poles} =& + {1\ov4} u'(\tx_{\R1}) \lint_{{\rm  lower \ arc }}\frac{{\rm
d}v_2}{2\pi i}\frac{1}{v_2-\tx_{\R2}}
\left({1\ov u_1-v_2-i\epsilon} +{1\ov u_1-v_2+i\epsilon}  \right)\\
& + {1\ov4} \lint_{{\rm  lower \ arc }}\frac{{\rm
d}v_2}{2\pi i}\frac{1}{(\tx_{\R}(v_2+i \eps)-\tx_{\R1})(v_2-\tx_{\R2})}- (\tx_{\R1} \leftrightarrow \tx_{\R2})\,.
\eal
This contribution vansihes since on the lower arc it holds that $\tx_{\R}(v) \sim v$ and the radius of the arc is infinite.
The contribution on the circle is instead given by
\bal
C^{v_1}_{\rm circle}=&- {1\ov4} \lint_{\rm upper \ arc} \frac{{\rm d}v_1}{2\pi i}\lint_{{\rm  lower \ arc }}\frac{{\rm
d}v_2}{2\pi i}\frac{1}{(\frac{1}{v_1}-\tx_{\R1})(v_2-\tx_{\R2})}
\left({1\ov v_1-v_2-i\epsilon} +{1\ov v_1-v_2+i\epsilon}  \right)\\
&- (\tx_{\R1} \leftrightarrow \tx_{\R2})\,,
\eal
where we used the fact that the contribution on the lower arc of the circle is suppressed. The expression above can be integrated with respect to $v_1$ and returns
\bal
C^{v_1}_{\rm circle}=& +{1\ov4 \tx_{\R1}} \lint_{\rm upper \ arc} \frac{{\rm d}v_1}{2\pi i}\lint_{{\rm  lower \ arc }}\frac{{\rm
d}v_2}{2\pi i}\frac{1}{v_2} 
\left({1\ov v_1-v_2-i\epsilon} +{1\ov v_1-v_2+i\epsilon}  \right)- (\tx_{\R1} \leftrightarrow \tx_{\R2})\\
=& +{1\ov4 \tx_{\R1}}  \frac{1}{2\pi i}\lint_{{\rm  lower \ arc }}\frac{{\rm
d}v_2}{2\pi i}\frac{1}{v_2}
\left(2 \log \left( R-v_2 \right) - 2 \log \left( -R-v_2 \right) \right)- (\tx_{\R1} \leftrightarrow \tx_{\R2})\\
=& +{1\ov4 \tx_{\R1}}  \frac{1}{2\pi i}\int^{-\pi}_{0}\frac{{\rm
d} \phi }{2\pi}
\left(2 \log \left( 1-e^{i \phi} \right) - 2 \log \left( -1-e^{i \phi} \right) \right)- (\tx_{\R1} \leftrightarrow \tx_{\R2})\\
=&+ \frac{1}{16 \tx_{\R1}}- \frac{1}{16 \tx_{\R2}}\,,
\eal
where in the last row we parameterised $v_2=R e^{i \phi}$. Therefore we obtain
\bal
\label{CLower_arc_AFS_case_keq0}
C^{v_2}_{\rm lower \ arc}= -\frac{1}{16 \tx_{\R1}} + \frac{1}{16 \tx_{\R2}}\,.
\eal
$C^{v_2}_{\rm upper \ arc}$ can be computed similarly and returns
\bal
\label{CUpper_arc_AFS_case_keq0}
C^{v_2}_{\rm upper \ arc}=&-{1\ov 4\tx_{\R2}} u_{\R}'(\tx_{\R1})+{1\ov 4\tx_{\R1}} u_{\R}'(\tx_{\R2}) +\frac{1}{8 \tx_{\R1} \tx^2_{\R2}} - \frac{1}{8 \tx^2_{\R1} \tx_{\R2}}+\frac{1}{16 \tx_{\R2}}-\frac{1}{16 \tx_{\R1}}\,.
\eal
Finally combining~\eqref{eq:result_Cv2_poles_keq0_mirror}, \eqref{Cv2_circle_case_kequal0_first_definition}, \eqref{CLower_arc_AFS_case_keq0} and~\eqref{CUpper_arc_AFS_case_keq0} into~\eqref{eq:AFS_split_into_Cv2poles_Cv2circle_mirror} we obtain
\bal 
\label{mirror_AFS_expression_1final_keq0}
{\stackrel{\prime\prime}{\tPhi}}{}^{--, {\rm bulk}}_{\R\R}(\tx_{\R1},\tx_{\R2})=&-\frac{u_{\R}'(\tx_{\R1}) u_{\R}'(\tx_{\R2})}{u_1-u_2}+\frac{1}{2} \frac{u_{\R}'(\tx_{\R2})+u_{\R}'(\tx_{\R1})}{\tx_{\R1} - \tx_{\R2}}\\
&- \frac{3}{8} \left( \frac{u_{\R}'(\tx_{\R2})}{\tx_{\R1}}- \frac{u_{\R}'(\tx_{\R1})}{\tx_{\R2}} \right) +B\,.
\eal

\paragraph{The bulk contribution ${\stackrel{\prime\prime}{\tPhi}}{}^{++,{\rm bulk}}_{\R\R}$.}

Let now consider the expression in~\eqref{eq:kernel_expansion_RR_afs_2}.
We integrate first wrt $v_2$. We close the contour by a circle of large radius $R$ so that (as before) we can split the result into pole and circle contributions
\bal 
\label{eq:AFS_split_into_Iv2poles_Iv2circle_mirror}
{\stackrel{\prime\prime}{\tPhi}}{}^{++, {\rm bulk}}_{\R\R}(\tx_{\R1},\tx_{\R2}) =I^{v_2}_{\rm poles}-I^{v_2}_{\rm circle} \,.
\eal
 The poles inside the clockwise closed contour are 
\bal 
v_2 = v_1\pm i\epsilon\,.
\eal
Taking the residues we obtain
\bal
I^{v_2}_{\rm poles}&=-{1\ov4} \int_{\rm cuts} \frac{{\rm d}v}{2\pi i} \frac{1}{(\frac{1}{\tx_{\L}(v)}-\tx_{\R1})}  \left( \frac{1}{\frac{1}{\tx_{\L}(v+ i \eps)}-\tx_{\R2}} +\frac{1}{\frac{1}{\tx_{\L}(v- i \eps)}-\tx_{\R2}}  \right) - (\tx_{\R1} \leftrightarrow \tx_{\R2}) \,.
\eal
Using that for $r> \ubr$
\bal
\frac{1}{\tx_{\L}(r \pm i \eps)} = \tx_{\R}(r \mp i \eps)
\eal
and 
\bal
\frac{1}{\tx_{\L}(-r + i \ka \pm i \eps)} = \tx_{\R}(-r - i \ka \mp i \eps)
\eal
the expression above is equal to 
\bal
\label{Iv2_poles_mirror_k0}
I^{v_2}_{\rm poles}&={1\ov4} \int_{\rm cuts} \frac{{\rm d}v}{2\pi i} \frac{1}{(\tx_{\R}(v)-\tx_{\R1})}  \left( \frac{1}{\tx_{\R}(v+ i \eps)-\tx_{\R2}} +\frac{1}{\tx_{\R}(v- i \eps)-\tx_{\R2}}  \right)
\\
&- (\tx_{\R1} \leftrightarrow \tx_{\R2})=-B\,.
\eal
This is the opposite of the contribution in~\eqref{eq:B_contribution_phimm} we found in the computation of ${\stackrel{\prime\prime}{\tPhi}}{}^{--, {\rm bulk}}_{\R\R}$.

The contribution from the circle can instead be split into upper and lower arcs
\bal
I^{v_2}_{\rm circle}=I^{v_2}_{\rm upper \ arc}+I^{v_2}_{\rm lower \ arc} \,.
\eal
After a computation similar to the one done previously we obtain
\bal
I^{v_2}_{\rm circle}=-{1\ov8} \left( \frac{u_{\R}'(\tx_{\R2})}{\tx_{\R1}}- \frac{u_{\R}'(\tx_{\R1})}{\tx_{\R2}} \right)\,.
\eal
Then we obtain
\bal
\label{mirror_AFS_expression_2final_keq0}
{\stackrel{\prime\prime}{\tPhi}}{}^{++, {\rm bulk}}_{\R\R}(\tx_{\R1},\tx_{\R2}) =-B+{1\ov8} \left( \frac{u_{\R}'(\tx_{\R2})}{\tx_{\R1}}- \frac{u_{\R}'(\tx_{\R1})}{\tx_{\R2}} \right) \,.
\eal

Performing the half sum of~\eqref{mirror_AFS_expression_1final_keq0} and~\eqref{mirror_AFS_expression_2final_keq0} the non-analytic $B$ terms cancel and we obtain
\bal 
\label{RR_afs_mirror_bulk}
&{\stackrel{\prime\prime}{\tPhi}}{}^{{\rm bulk}}_{\R\R}(\tx_{\R1},\tx_{\R2})=\frac{1}{2} \left({\stackrel{\prime\prime}{\tPhi}}{}^{--, {\rm bulk}}_{\R\R}(\tx_{\R1},\tx_{\R2})+{\stackrel{\prime\prime}{\tPhi}}{}^{++, {\rm bulk}}_{\R\R}(\tx_{\R1},\tx_{\R2}) \right) \\
=&-\frac{1}{2}\frac{u_\R'(\tx_{\R1}) u_\R'(\tx_{\R2})}{u_1-u_2}+\frac{1}{4} \frac{u_\R'(\tx_{\R2})+u_\R'(\tx_{\R1})}{\tx_{\R1} - \tx_{\R2}}+\frac{1}{8} \left(\frac{u_\R'(\tx_{\R1})}{\tx_{\R2}}-\frac{u_\R'(\tx_{\R2})}{\tx_{\R1}} \right)\,.
\eal
Adding also the boundary contribution in~\eqref{RR_afs_mirror_bdy} we obtain
\bal
{\stackrel{\prime\prime}{\tPhi}}{}^{\afs}_{\R\R}(\tx_{\R1},\tx_{\R2}) =&-\frac{1}{2}\frac{u_\R'(\tx_{\R1}) u_\R'(\tx_{\R2})}{u_1-u_2}+\frac{1}{4} \frac{u_\R'(\tx_{\R2})+u_\R'(\tx_{\R1})}{\tx_{\R1} - \tx_{\R2}}+\frac{1}{8} \left(\frac{u_\R'(\tx_{\R1})}{\tx_{\R2}}-\frac{u_\R'(\tx_{\R2})}{\tx_{\R1}} \right)\\
&+ \frac{1}{8} \left( \frac{1}{\tx_{\R1}} -  \frac{1}{\tx_{\R2}}+ \frac{1}{\tx_{\R1} \tx^2_{\R2}} -  \frac{1}{\tx^2_{\R1} \tx_{\R2}} \right)\,,\\
\eal
which agrees with the first formula in~\eqref{eq:afs_mixed_der} for $a=\text{R}$.

\paragraph{Extending the results to arbitrary points.} In this section, we derived the AFS order of the phases when both points were in the mirror region, which is $\Im(x_1)<0$ and $\Im(x_2)<0$. The result can be easily generalised to the case in which $x_1$ and $x_2$ are arbitrary points in the complex plane. The computation can be performed similarly as before with the only caveat that now the poles can be inside or outside the integration contours depending on the signs of the imaginary parts of $x_1$ and $x_2$. For example, in the case in which the particles are of the same type we find
\bal
\label{eq:ddtPhiafsaa_arb_points}
{\stackrel{\prime\prime}{\tPhi}}{}_{aa}^{\afs}(x_1,x_2)=&-\frac{1}{2} \left(\theta\left(-\Im(x_1)\right) \theta\left(-\Im(x_2)\right) +\theta\left(\Im(x_1)\right) \theta\left(\Im(x_2)\right)\right) 
{u_a'(x_1)u_a'(x_2)\ov u_{1}-u_{2}} 
\\
&+{1\ov4} {u_a'(x_1)+ u_a'(x_2)\ov x_1- x_{2}}
+{x_1 - x_2 \ov 4x^2_1x_2^2}  
\,.
\eal
The formula above reduces to~\eqref{eq:afs_mixed_der} for $\Im(x_1)<0$ and $\Im(x_2)<0$. A similar formula can be derived for ${\stackrel{\prime\prime}{\tPhi}}{}_{\bar{a} a}^{\afs}(x_1,x_2)$.

\subsection{The HL order}\label{app:HL}

It is convenient to work out the HL order $\tPhi_{ab}^{\alpha \beta, \hl}(\tx_{a1},\tx_{b2})$ by using the $u$-plane. We start considering $\a=\b=-$ and $a=b$, in which case we have
\bal
\tPhi_{aa}^{--, \hl}(\tx_{a1},\tx_{a2}) 
=& -  \lint_{\widetilde{\rm cuts}} \frac{{\rm d} v_1}{2\pi i} \lint_{\widetilde{\rm cuts}} \frac{{\rm d} v_2}{2\pi i} {\tx'_a(v_1)\ov \tx_a(v_1)-\tx_{a1}}{\tx'_a(v_2)\ov \tx_a(v_2)-\tx_{a2}} K^\hl_\eps(v_1-v_2)\,,
\eal
where the regularised HL kernel is given in~\eqref{eq:HLkernel}. Integrating by parts with respect to $v_2$ we obtain
\bal
\label{eq:hl_aa_mm_mirror}
\tPhi&_{aa}^{--, \hl}(\tx_{a1},\tx_{a2}) = I^{\text{bdry}}(\tx_{a1},\tx_{a2}) + I^{\text{bulk}}(\tx_{a1},\tx_{a2})\,,
\eal
where 
\bal
I^{\text{bdry}}(\tx_{a1},\tx_{a2})=-  \lint_{\widetilde{\rm cuts}} \frac{{\rm d} v_1}{2\pi i} \lint_{\widetilde{\rm cuts}} \frac{{\rm d} v_2}{2\pi i} {\tx'_a(v_1)\ov \tx_a(v_1)-\tx_{a1}} \frac{\partial}{\partial v_2} \left[\log(\tx_a(v_2)-\tx_{a2})K^\hl_\eps(v_1-v_2) \right]\,.
\eal
and
\bal
I^{\text{bulk}}(\tx_{a1},\tx_{a2})=\frac{i}{8 \pi^2}  \lint_{\widetilde{\rm cuts}} {\rm d} v_1 \lint_{\widetilde{\rm cuts}} {\rm d} v_2 {\tx'_a(v_1)\ov \tx_a(v_1)-\tx_{a1}}  \log(\tx_a(v_2)-\tx_{a2})\left( \frac{1}{v_{12}- i \eps}- \frac{1}{v_{12}+ i \eps} \right) \,.
\eal
In the expressions above we defined $v_{12} \equiv v_1 - v_2 $ and used 
\bal
\frac{\partial}{\partial v} K^\hl_\eps(v)=\frac{i}{2} \left( \frac{1}{v- i \eps}- \frac{1}{v+ i \eps} \right) \,.
\eal
After a long computation it is possible to show that
\bal
I^{\text{bdry}}(\tx_{a1},\tx_{a2})=-{i\ov 4}\sgn\left(\Im(\tx_{a1})\right)\ln(-\tx_{a2}) 
+{\pi\ov 8}\sgn\left(\Im(\tx_{a1})\right)\sgn\left(\Im(\tx_{a2})\right)\,.
\eal
When both points are in the mirror region
\bal
I^{\text{bdry}}(\tx_{a1},\tx_{a2})={i\ov 4}\ln(-\tx_{a2}) +{\pi\ov 8}
\eal
and the boundary contribution cancel in the full phase since
\bal
I^{\text{bdry}}(\tx^+_{a1},\tx^+_{a2})+I^{\text{bdry}}(\tx^-_{a1},\tx^-_{a2})-I^{\text{bdry}}(\tx^+_{a1},\tx^-_{a2})-I^{\text{bdry}}(\tx^-_{a1},\tx^+_{a2})=0\,.
\eal
These boundary contributions are then completely irrelevant for the purpose of study crossing in the mirror region.
However they play a nontrivial role in the continuation of HL to the string region.

Now we perform the integral with respect to $v_1$ in~\eqref{eq:hl_aa_mm_mirror}. Note that we cannot send $\eps \to 0$ since there are two poles at $v_2= v_1 \pm i \eps$ trapping the contour in the limit. Due to this fact while the integrand is zero in the limit, this is not the case for the integrated function. To evaluate the integral we deform the $v_1$ contour in such a way as to surround the pole $v_2+ i \eps$ from above and the pole $v_2 - i\eps$ from below;
by doing this operation we need of course to subtract the residues of the two poles. Then we obtain
\bal
I^{\text{bulk}}(\tx_{a1},\tx_{a2})&=\frac{i}{8 \pi^2}  \lint_{\rm{def. \, cont.}} {\rm d} v_1 \lint_{\widetilde{\rm cuts}} {\rm d} v_2 {\tx'_a(v_1)\ov \tx_a(v_1)-\tx_{a1}}  \log(\tx_a(v_2)-\tx_{a2}) \frac{2i\eps}{v_{12}^2+ \eps^2}\\
&-\frac{i}{8 \pi^2} (2 \pi i)  \lint_{\widetilde{\rm cuts}} {\rm d} v_2 {\tx'_a(v_2+ i \eps)\ov \tx_a(v_2+ i \eps)-\tx_{a1}}  \log(\tx_a(v_2)-\tx_{a2})\\
&+\frac{i}{8 \pi^2} (2 \pi i)  \lint_{\widetilde{\rm cuts}} {\rm d} v_2 {\tx'_a(v_2- i \eps)\ov \tx_a(v_2- i \eps)-\tx_{a1}}  \log(\tx_a(v_2)-\tx_{a2}) \,.
\eal
By `def. cont.' in the first line of the expression above we mean the integration contour around the cuts of $\tx_a$ deformed so that $v_2 - i \eps$ and $v_2 + i \eps$ are both between the integration lines on the upper and lower edge of the cut. Then in the limit $\eps \to 0$ the first line of the expression above vanishes and we remain with
\bal
I^{\text{bulk}}(\tx_{a1},\tx_{a2})=&\frac{1}{4 \pi} \lint_{\widetilde{\rm cuts}} {\rm d} v_2 \left( {\tx'_a(v_2+ i \eps)\ov \tx_a(v_2+ i \eps)-\tx_{a1}} -  {\tx'_a(v_2- i \eps)\ov \tx_a(v_2- i \eps)-\tx_{a1}} \right)  \log(\tx_a(v_2)-\tx_{a2})\,.
\eal
Using that for any functions $F$ and $G$
\bal
& \int_{{\rm cuts}\, {\rm of}\, \tx_a(v)}  \frac{{\rm
d}v}{2\pi i} \left(F(\tx_a(v+i\eps))-F(\tx_a(v-i\eps))\right)G(\tx_a(v))
\\
&= \int_{{\rm cuts}\, {\rm of}\, \tx_a(v)}  \frac{{\rm
d}v}{2\pi i} F(\tx_a(v))\left(G(\tx_a(v+i \eps))-G(\tx_a(v-i\eps))\right)\,,
\eal
we can write the result as the following single integral
\bal
\tPhi_{aa}^{--, \hl}(\tx_{a1},\tx_{a2}) &= I^{\text{bdry}}(\tx_{a1},\tx_{a2})\\
& - \frac{1}{4 \pi}   \lint_{\widetilde{ \rm cuts}} {\rm d} v \frac{\tx'_a(v)}{\tx_a(v) - \tx_{a1}} \left( \log\left(\tx_a(v - i \eps)-\tx_{a2} \right) - \log\left(\tx_a(v + i \eps)-\tx_{a2} \right) \right)\,,
\eal
Repeating the same analysis for the remaining phases we obtain (up to the same boundary contribution of before, cancelling in the full phase):
\bal
\Phi&_{aa}^{++, \hl}(x_{a1},x_{a2}) 
\\
&=-   \frac{1}{4 \pi}   \lint_{\widetilde{ \rm cuts}} {\rm d} v \frac{\left(\frac{1}{\tx_{\bar{a}}(v)} \right)'}{\frac{1}{\tx_{\bar{a}}(v)} - \tx_{a1}} \left( \log\left(\frac{1}{\tx_{\bar{a}}(v-i \eps)} -\tx_{a2} \right) - \log\left(\frac{1}{\tx_{\bar{a}}(v+ i \eps)}-\tx_{a2} \right) \right)\,,
\eal
\bal
\Phi&_{\bar a a}^{-+, \hl}(\tx_{\bar a 1},\tx_{ a 2}) 
\\
&= \frac{1}{4 \pi}   \lint_{\widetilde{ \rm cuts}} {\rm d} v \frac{\tx'_{\bar{a}}(v)}{\tx_{\bar{a}}(v) - \tx_{\bar a1}} \left( \log\left(\frac{1}{\tx_{\bar{a}}(v-i \eps)}-\tx_{a2} \right) - \log\left(\frac{1}{\tx_{\bar{a}}(v+ i \eps)}-\tx_{a2} \right) \right)
\,,
\\
\Phi&_{\bar a a}^{+-, \hl}(x_{\bar a 1},x_{ a 2}) 
= \frac{1}{4 \pi}   \lint_{\widetilde{ \rm cuts}} {\rm d} v \frac{\left(\frac{1}{\tx_{a}(v)} \right)'}{\frac{1}{\tx_{a}(v)} - \tx_{\bar a1}} \left( \log\left(\tx_{a}(v-i \eps) -\tx_{a2} \right) - \log\left(\tx_{a}(v+i \eps)-\tx_{a2} \right) \right) \,.
\eal

By connecting the Zhukovsy variables on the different edges of the cuts\footnote{As usual the relations are $
\tx_{a}(r+ i \eps)=\frac{1}{\tx_{\bar{a}}(r- i \eps)}\,,$ and $\tx_{a}(-r+ i \ka_a + i \eps)=\frac{1}{\tx_{\bar{a}}(-r + i \ka_{\bar{a}}- i \eps)}\,,
$
with $r > \nu$.} it is not difficult to show that
\bal
\tPhi_{aa}^{--, \hl}(\tx_{a1},\tx_{a2})&=\tPhi_{aa}^{++, \hl}(\tx_{a1},\tx_{a2}) = I^\hl_{aa}(\tx_{a1},\tx_{a2})+{i\ov 4}\ln(-\tx_{a2}) +{\pi\ov 8}\,,\\
\tPhi_{\bar{a} a}^{-+, \hl}(\tx_{\bar{a} 1},\tx_{a2})&=\tPhi_{\bar{a} a}^{+-, \hl}(\tx_{\bar{a} 1},\tx_{a2}) = I^\hl_{\bar a a}(\tx_{\bar{a}1},\tx_{a2})+{i\ov 4}\ln(-\tx_{a2}) +{\pi\ov 8} \,,
\eal
where
\bal
\label{eq:app_I_HL}
I^\hl_{aa}(x_{1},x_{2}) &\equiv - \frac{1}{4 \pi}   \lint_{\widetilde{ \rm cuts}} {\rm d} v \frac{\tx'_a(v)}{\tx_a(v) - x_{1}} \left( \log\left(\tx_a(v - i \eps)-x_{2} \right) - \log\left(\tx_a(v + i \eps)-x_{2} \right) \right)\,,\\
I^\hl_{\bar a a}(x_{1},x_{2})&\equiv +\frac{1}{4 \pi}   \lint_{\widetilde{ \rm cuts}} {\rm d} v \frac{\tx'_{\bar{a}}(v)}{\tx_{\bar{a}}(v) - x_{1}} \left( \log\left(\frac{1}{\tx_{\bar{a}}(v-i \eps)}-x_{2} \right) - \log\left(\frac{1}{\tx_{\bar{a}}(v+ i \eps)}-x_{2} \right) \right) \, .
\eal

For arbitrary points $x_1$ and $x_2$ in the complex plane then (taking into account also the boundary contributions) it holds that
\bal
\label{eq:PhiHL_as_functions_of_IHL1}
\tPhi_{aa}^{\hl}(x_{1},x_{2})&= \frac{1}{2} \left( \tPhi_{aa}^{--, \hl}(x_{1},x_{2}) + \tPhi_{aa}^{++, \hl}(x_{1},x_{2}) \right)\\
&= I^\hl_{aa}(x_{1},x_{2})-{i\ov 4}\sgn\left(\Im(x_1)\right)\ln(-x_2) 
+{\pi\ov 8}\sgn\left(\Im(x_1)\right)\sgn\left(\Im(x_2)\right)\,,
\eal
\bal
\label{eq:PhiHL_as_functions_of_IHL2}
\tPhi_{\bar a a}^{\hl}(x_{1}, x_{2}) &= \frac{1}{2} \left( \tPhi_{\bar aa}^{-+, \hl}(x_{1},x_{2}) + \tPhi_{\bar aa}^{+-, \hl}(x_{1},x_{2}) \right)\\
&=  I^\hl_{\bar a a}(x_{1},x_{2})-{i\ov 4}\sgn\left(\Im(x_1)\right)\ln(-x_2) 
+{\pi\ov 8}\sgn\left(\Im(x_1)\right)\sgn\left(\Im(x_2)\right)\,.
\eal

\section{Discontinuities}
\label{app:disc}

\subsection{Discontinuities of \texorpdfstring{$\tPhi$}{tPhi}-functions}

The continuation of $\tPhi^{\a \b}_{ab} (x,y)$ from the region $\{ \Im(x)<0, \Im(y)<0\}$ to the region $\{\Im(x)>0, \Im(y)<0\}$ generates the $\tPsi$ function defined in~\eqref{PsiBESkamirror}:
\bal
\tPhi^{\a \b}_{ab} (x,y) \to \tPhi^{\a \b}_{ab} (x,y) - \tPsi^{\b}_{b} (x,y) \,.
\eal
Similarly, the continuation in the second variable from the region $\{ \Im(x)<0, \Im(y)<0\}$ to the region $\{\Im(x)<0, \Im(y)>0\}$ is
\bal
\tPhi^{\a \b}_{ab} (x,y) \to \tPhi^{\a \b}_{ab} (x,y)+\tPsi^{\a}_{a} (y,x) \,.
\eal
The continuation in both variables ($\{ \Im(x)<0, \Im(y)<0\} \to \{ \Im(x)>0, \Im(y)>0\}$) is obtained by combining the expressions above and considering additional contributions from the discontinuities of $\tPsi$ functions:
\bal
\tPhi^{\a \b}_{ab} (x,y) &\to \tPhi^{\a \b}_{ab} (x,y) - \tPsi^{\b}_{b} (x,y)+\tPsi^{\a}_{a} (y,x)+\text{Discontinuities of $\tPsi$}\,.
\eal
We discuss the additional discontinuities of $\tPsi$ functions in the next subsection.

\subsection{Discontinuities of  \texorpdfstring{$\tPsi$}{tPsi}-functions}
\label{app:tPsi_disc}

In section~\ref{sec:proposal} we introduced the functions
\bal
\tPsi_{b}^{-}(x,y)&=-\lint_{\pa\cR_-}\frac{{\rm
d}w}{2\pi i}\,\frac{K^\bes(u_a(x)-u_b(w))}{w-y} =-\lint_{\widetilde{\rm cuts}}\frac{{\rm
d}v}{2\pi i}\,\frac{\tx'_{b}(v)}{\tx_{b}(v)-y} K^\bes(u_a(x)-v) \,,
\eal
\bal
\tPsi_{b}^{+}(x,y)&=-\lint_{\pa\cR_+}\frac{{\rm
d}w_2}{2\pi i}\,\frac{K^\bes(u_a(x)-u_b(w_2))}{w_2-y}=\lint_{\widetilde{\rm cuts}}\frac{{\rm
d}v}{2\pi i}\,\frac{ \left( \frac{1}{\tx_{\bar{b}} (v)} \right)'}{\frac{1}{\tx_{\bar{b}} (v)}-y} K^\bes(u_a(x)-v)\,,
\eal
as well as their combination~\eqref{eq:tPsi_as_half_sum}. Let us describe the analytic properties of these functions in some detail.

The function $\tPsi_{b}^{\beta}(x,y)$ ($\beta=\pm$) is originally defined for $\Im(y)<0$, i.e. when the second entry belongs to the mirror region. While continuing $y$ to the upper half of the complex plane a singularity approaches the integration line from below and the integrand of the $\tPsi$ function develops a pole when $\Im(y)=0$.
If we want to continue $y$ to the region $\Im(y)>0$ passing through the semi-line $(0, +\infty)$ (if we pass through $(-\infty,0)$ we must be careful to extra $\log$ discontinuities however we never consider this path for crossing) we must keep into account the residue of this pole and the $\tPsi$ function is continued as follows
\bal
\tPsi_{b}^{\beta}(x, y) \to \tPsi_{b}^{\beta}(x, y) + K^\bes(u_a(x)-u_b(y)) \,, \qquad \beta= \pm \,.
\eal

Additionally to these poles, we also need to consider branch points of the BES kernel $K^\bes(u_a(x)-v)$, whose cuts run vertically in the $v$ plane. Extra discontinuities of $\tPsi$ must be taken into account if one or more of these branch points crosses the integration line.  Let us show how these additional contributions arise. First, we integrate by parts and get 
\bal
&\tPsi_{b}^{-}(x,y)=+{h\ov 2}\lint_{\widetilde{\rm cuts}}\frac{{\rm
d }v}{2\pi i} \log(\tx_b(v)-y) \,  \left[\psi\Big(1+\tfrac{ih}{
2}\big(u_a(x)-v \big)\Big)+ \psi\Big(1-\tfrac{ih}{
2}\big(u_a(x)-v\big)\Big)\right]\,,\\
&\tPsi_{b}^{+}(x,y)=-{h\ov 2}\lint_{\widetilde{\rm cuts}}\frac{{\rm
d }v}{2\pi i} \log(\frac{1}{\tx_{\bar{b}}(v)}-y) \,  \left[\psi\Big(1+\tfrac{ih}{
2}\big(u_a(x)-v \big)\Big)+ \psi\Big(1-\tfrac{ih}{
2}\big(u_a(x)-v\big)\Big)\right]\,,
\eal 
where $\psi$ is the Digamma function.
Taking into account that for any positive integer $n$
\bal
\psi(z-n+1) = -{1\ov z} + \text{ regular terms}\,,\quad n=1,2, \dots
\eal
one sees that the integrand has poles in the $v$-plane located at 
\bal
\label{eq:poles_psi_function}
&v_*^{(n)}=u_a(x)-\frac{2i}{h} n  \,,\quad n= 1,2, \dots\,,\\
&v_{**}^{(n)}=u_a(x)+\frac{2i}{h} n \,,\quad n=1, 2, \dots\,.
\eal
These poles of $\psi$ functions correspond to the branch points of the BES kernel.
While moving $x$ in the complex plane (or if we prefer moving $u_a(x)$ in the $u$ plane) some of the poles of $\psi$ functions may cross the integration contour. For example, this is the case any time
\bal
u_a(x)= \nu +t \pm \frac{2i}{h}n \,, \qquad n=1, 2,\dots \,,\quad t \ge0\,,
\eal
in which case the poles in~\eqref{eq:poles_psi_function} overlap the integration contours around the main mirror cut $(\nu, + \infty)$.
The same problem happens in $\tPsi^{-}_{b}$ if 
\bal
u_a(x)= -\nu+ i \ka_b -t \pm \frac{2i}{h}n \,, \qquad n=1, 2,\dots \,, \quad t \ge0\,,
\eal
and in $\tPsi^{+}_{b}$ if 
\bal
u_a(x)= -\nu + i \ka_{\bar{b}} -t \pm \frac{2i}{h}n \,, \qquad n=1, 2,\dots \,, \quad t \ge0\,.
\eal
Instead of considering all possible continuations of the $\tPsi$ functions (that are infinitely many), we just focus on the one entering the crossing equations for mirror kinematics. 

\paragraph{$\tPsi$ continuation for mirror crossing.}
Let us start with
\bal
\tPsi^{\beta}_{b}(\frac{1}{{\tx}^-_{\bar{a}1}},\tx_{b2})
\eal
entering equation~\eqref{eq:tchi_crossing_step1} and suppose we want to continue $\tx^+_{a1}$ from the mirror to the anti-mirror region. Following the previous steps we have that
\bal
\tPsi_{b}^{-}(\frac{1}{\tx^-_{\bar{a}1}},\tx_{b2})=&+{h\ov 2}\lint_{\widetilde{\rm cuts}}\frac{{\rm
d }v}{2\pi i} \log(\tx_b(v)-\tx_{b2})\\
&\left[\psi\Big(1+\tfrac{ih}{
2}\big(u_1 - \frac{i}{h}-v \big)\Big)+ \psi\Big(1-\tfrac{ih}{
2}\big(u_1 - \frac{i}{h}-v\big)\Big)\right]
\eal
and
\bal
\tPsi_{b}^{+}(\frac{1}{\tx^-_{\bar{a}1}},\tx_{b2})=&-{h\ov 2}\lint_{\widetilde{\rm cuts}}\frac{{\rm
d }v}{2\pi i} \log(\frac{1}{\tx_{\bar{b}}(v)}-\tx_{b2})\\ &\left[\psi\Big(1+\tfrac{ih}{
2}\big(u_1 - \frac{i}{h}-v \big)\Big)+ \psi\Big(1-\tfrac{ih}{
2}\big(u_1 - \frac{i}{h}-v\big)\Big)\right]\,,
\eal 
In the expressions above we used that
\bal
u_a(\frac{1}{\tx^-_{\bar{a}1}})=u_{\bar{a}}(\tx^-_{\bar{a}1})=u_1 - \frac{i}{h} \,.
\eal
The poles of the $\psi$ functions are then located at  
\bal
&v_*^{(n)}=u_1-\frac{2i}{h} n- \frac{i}{h}  \,,\quad n= 1,2, \dots\,,\\
&v_{**}^{(n)}=u_1+\frac{2i}{h} n- \frac{i}{h} \,,\quad n=1, 2, \dots\,.
\eal
These poles are not problematic as far as they are away from the integration contour.
When $\tx^+_{a1}=\tx_a(u_1+\frac{i}{h})$ enters the anti-mirror region then $u_1+\frac{i}{h}$ crosses the cut $(\ubr, +\infty)$ from above; at the point at which it crosses the cut then the pole 
$$
v_{**}^{(1)}=u_1+\frac{i}{h}
$$ 
crosses the integration line on the upper edge of $(\ubr, +\infty)$ from above and then the integration line on the lower edge of $(\ubr, +\infty)$ also from above. The $\tPsi$ function must be continued by picking up the two residues associated with the pole that crossed the contour two times.
We obtain
\bal
\label{eq:tPsi_continuation_crossing_2}
\tPsi_{b}^{\beta}(\frac{1}{\tx^-_{\bar{a}1}},\tx_{b2}) &\to \tPsi_{b}^{\beta}(\frac{1}{\tx^-_{\bar{a}1}},\tx_{b2})-\frac{1}{i} \log \left( \frac{\frac{1}{\tx_{\bar{b}}(u_1+\frac{i}{h})}-\tx_{b2}}{\tx_b(u_1+\frac{i}{h})-\tx_{b2}} \right)\\
&= \tPsi_{b}^{\beta}(\frac{1}{\tx^-_{\bar{a}1}},\tx_{b2})-\frac{1}{i} \log \left( \frac{\frac{1}{\tx^+_{\bar{b}1}}-\tx_{b2}}{\tx^+_b-\tx_{b2}} \right)\,.
\eal

\subsection{Discontinuities of the HL integral}
\label{app:discontinuities_HL}

The expressions for the building blocks of HL provided in~\eqref{eq:PhiHL_as_functions_of_IHL1} and~\eqref{eq:PhiHL_as_functions_of_IHL2} are good for the analytic continuation of the first variable and can be directly used to study the crossing equations for the mirror theory. However, due to the logs in~\eqref{eq:app_I_HL} they are not suitable for the continuation of the second variable. To continue the second variable we should first use the antisymmetry properties of the phases
\bal
\label{eq:app_antisym_HL}
\Phi_{aa}^{\hl}(x_{1},x_{2})=-\Phi_{aa}^{\hl}(x_{2},x_{1})\, , \qquad \Phi_{\bar{a} a}^{\hl}(x_{1},x_{2})=-\Phi_{a\bar{a}}^{\hl}(x_{2},x_{1})\,,
\eal
and then perform the continuation in the second variable. By doing so the boundary contributions play a non-trivial role in the continuation of the HL phase to the string region.
Keeping into account the antisymmetry of HL, we can continue this phase in both variables from the mirror region to any other region using the discontinuity properties of the functions $I^\hl_{aa}$ and $I^\hl_{\bar{a} a}$. We summarise the discontinuities of $I^\hl_{aa}$ and $I^\hl_{\bar{a} a}$ below.

\paragraph{Continuation of $u_1$ across the main mirror cut.}
Let us start with both the first and second particles in the mirror region and continue the first variable across the main mirror cut $(\ubr, +\infty)$ from above. We obtain
\bal
\label{eq:IHL_continuation_properties_main_above}
I_{aa}^{\hl}(\tx_{a1},\tx_{a2}) &\to I_{aa}^{\hl}(\frac{1}{\tx_{\bar a1}},\tx_{a2}) +\frac{1}{2i} \log \left( \frac{\frac{1}{\tx_{\bar a 1}} - \tx_{a 2}}{\tx_{a 1} - \tx_{a 2}} \right)\,,\\
I_{\bar a a}^{\hl}(\tx_{ \bar{a} 1},\tx_{ a 2}) &\to I_{\bar a a}^{\hl}(\frac{1}{\tx_{ a 1}},\tx_{ a 2})   +\frac{1}{2i} \log \left( \frac{\frac{1}{\tx_{\bar a 1}} - \tx_{a 2}}{\tx_{a 1} - \tx_{a 2}} \right)\
\eal

Doing the continuation across the cut $(\ubr, +\infty)$ from below we obtain instead 
\bal
\label{eq:IHL_continuation_properties_main_below}
I_{aa}^{\hl}(\tx_{a1},\tx_{a2}) &\to I_{aa}^{\hl}(\frac{1}{\tx_{\bar a1}},\tx_{a2}) -\frac{1}{2i} \log \left( \frac{\frac{1}{\tx_{\bar a 1}} - \tx_{a 2}}{\tx_{a 1} - \tx_{a 2}} \right)\,,\\
I_{\bar a a}^{\hl}(\tx_{ \bar{a} 1},\tx_{ a 2}) &\to I_{\bar a a}^{\hl}(\frac{1}{\tx_{ a 1}},\tx_{ a 2})   -\frac{1}{2i} \log \left( \frac{\frac{1}{\tx_{\bar a 1}} - \tx_{a 2}}{\tx_{a 1} - \tx_{a 2}} \right)\
\eal

\paragraph{Continuation of $u_1$ across the $\ka$ cuts.}

We recall the convention that $\ka_\L=-\ka_\R=\frac{k}{h}$. Crossing the cut $(-\infty+ i \ka_{a}, -\ubr+ i \ka_{a})$ from above with $u_1$ we obtain
\bal
\label{eq:IaaHL_continuation_properties_ka_above}
I_{aa}^{\hl}(\tx_{a1},\tx_{a2}) \to I_{aa}^{\hl}(\frac{1}{\tx_{\bar a}(u_1+2i \ka_{\bar{a}})},\tx_{a2}) +\frac{1}{2i} \log \left( \frac{\frac{1}{\tx_{\bar a}(u_1+2i \ka_{\bar{a}})} - \tx_{a 2}}{\tx_{a 1} - \tx_{a 2}} \right)\,.
\eal
If we cross the cut $(-\infty+ i \ka_{\bar{a}}, -\ubr+ i \ka_{\bar{a}})$ from above we obtain
\bal
\label{eq:IbaaHL_continuation_properties_ka_above}
I_{\bar a a}^{\hl}(\tx_{\bar a 1},\tx_{ a 2}) \to I_{\bar a a}^{\hl}(\frac{1}{\tx_{a}(u_1 + 2 i \ka_a)},\tx_{ a 2}) + \frac{1}{2i} \log \left( \frac{\frac{1}{\tx_{\bar a 1}} - \tx_{a2}}{\tx_a(u_1 + 2i \ka_a) - \tx_{a2}}  \right)\,.
\eal

Crossing the same cuts from below we have instead
\bal
\label{eq:IHL_continuation_properties_ka_below}
I_{aa}^{\hl}(\tx_{a1},\tx_{a2}) &\to I_{aa}^{\hl}(\frac{1}{\tx_{\bar a}(u_1+2i \ka_{\bar{a}})},\tx_{a2}) -\frac{1}{2i} \log \left( \frac{\frac{1}{\tx_{\bar a}(u_1+2i \ka_{\bar{a}})} - \tx_{a 2}}{\tx_{a 1} - \tx_{a 2}} \right)\,,\\
I_{\bar a a}^{\hl}(\tx_{\bar a 1},\tx_{ a 2}) &\to I_{\bar a a}^{\hl}(\frac{1}{\tx_{ a}(u_1+2 i \ka_a)},\tx_{ a 2}) - \frac{1}{2i} \log \left( \frac{\frac{1}{\tx_{\bar a 1}} - \tx_{a2}}{\tx_a(u_1 + 2i \ka_a) - \tx_{a2}}  \right)\,.
\eal
The analytic continuation in the second variable is obtained by first using~\eqref{eq:app_antisym_HL} to write 
$I_{aa}^{\hl}(\tx_{a1},\tx_{a2})$ and $I_{\bar{a} a}^{\hl}(\tx_{\bar{a}1},\tx_{a2})$ as a functions of $I_{aa}^{\hl}(\tx_{a2},\tx_{a1})$ and $I_{a \bar{a}}^{\hl}(\tx_{a2},\tx_{\bar{a}1})$ and then continuing the first variable.

\section{Identities}
\subsection{Identities of  \texorpdfstring{$\tPhi$}{tPhi}-functions}

We collect some identities of $\tPhi$ functions necessary to check different properties in the mirror and string regions. First of all it is easy to check that
\bal
\left( \tPhi^{\a \, \b}_{a b} (x, y) \right)^*= \tPhi^{-\a \, -\b}_{a b} (x^*, y^*)
\eal
and therefore, with the choice~\eqref{Phiabara}, it holds
\bal
\label{eq:app_tPhi_compl_conj}
\tPhi_{ab}(x,y)^*=\tPhi_{ab}(x^*,y^*) \,.
\eal
This property is useful to show that unitarity is satisfied in the string model.

$\tPhi$ functions also satisfy the following additional relations. Each relation is useful to check a particular property, that we write before the relation.
\begin{enumerate}
\item Braiding unitarity
    \bal
\tPhi^{\a \b}_{ab}(x,y)+\tPhi^{\b \a}_{ba}(y,x)=0 \,.
\eal

\item Crossing in the mirror and string models
\bal
\tPhi^{\a \b}_{ab}(\frac{1}{x}, y)+\tPhi^{-\a \, \b}_{\bar a b}(x, y)=\tPhi^{-\a \, \b}_{\bar a b}(0, y)\,.
\eal 

\item Parity in the mirror model
\bal
\tPhi^{\a  \a}_{aa} (- \frac{1}{\tx},- \frac{1}{\ty})+ \tPhi^{\a \a}_{aa} (\tx,\ty)&=+\tPhi^{\a \a}_{aa} (0,\ty)+\tPhi^{\a \a}_{aa} (\tx,0)-\tPhi^{\a \a}_{aa} (0,0) \,,\\
\tPhi^{\a , -\a}_{\bar aa} (- \frac{1}{\tx},- \frac{1}{\ty})+ \tPhi^{\a, -\a}_{\bar aa} (\tx,\ty)&=+\tPhi^{\a,  -\a}_{\bar aa} (0,\ty)+\tPhi^{\a,  -\a}_{\bar aa} (\tx,0)-\tPhi^{\a,  -\a}_{\bar aa} (0,0) \,.
\eal

\end{enumerate}

\subsection{Identities for  \texorpdfstring{$\tPsi$}{tPsi}-functions}\label{app:psi-identities}

Similarly to the $\tPhi$ functions, the $\tPsi$ functions also satisfy simple identities under complex conjugation. In this case, we have
\bal
\label{eq:app_tPsi_compl_conj}
\left(\tPsi_b^{\b}(x,y) \right)^*= - \tPsi_b^{-\b}(x^*,y^*) \implies \left(\tPsi_b(x,y) \right)^*= - \tPsi_b(x^*,y^*)\,.
\eal

Additionally to this simple identity, necessary to prove the unitarity of the string model, one can derive and check numerically several other identities necessary to prove all the remaining symmetries of the theory. 
Let
\bal\la{eq:Dab_definition}
\Delta_{b}^{\b}(u_1\pm {i\ov h}m_1,\tx_{b 2}^{\pm m_2})\equiv &+\tPsi_{b}^{\b}(u_1 +{i\ov h}m_1,\tx_{b 2}^{+ m_2})-\tPsi_{b}^{\b}(u_1 +{i\ov h}m_1,\tx_{b 2}^{- m_2})
\\
&-\tPsi_{b}^{\b}(u_1 -{i\ov h}m_1,\tx_{b 2}^{+ m_2})+\tPsi_{b}^{\b}(u_1 -{i\ov h}m_1,\tx_{b 2}^{- m_2})\,,
\eal
where  $\tx_{b 2}^{\pm m_2} = \tx_{b}(u_2\pm {i\ov h}m_2)$, $u_1,u_2\in\bR$. Then, by using
\bal
\label{eq:KBESshiftid}
K^\bes(v-{i\ov h}n)-K^\bes(v+{i\ov h}n) =i \sum _{j=1}^{n} \left(\log (-\frac{i h
   v}{2}+j-\frac{n}{2})+\log (\frac{i h v}{2}+j-\frac{n}{2})\right)\,,
   \eal
we get
\bal\label{eq:Deltaepsm_2mm}
\exp\big(i\,\Delta_{b}^{-}&(u_1\pm {i\ov h}m_1,\tx_{b 2}^{\pm m_2})\big) = S^{m_1m_2}(u_{12})
\left(\frac{\tx_{b 2}^{+ m_2} }{\tx_{b 2}^{- m_2} }\right)^{m_1}
   \\
   &\times \frac{ (\tx_{b 1}^{-m_1}-\tx_{b 2}^{- m_2}) (\tx_{b 1}^{+m_1}-\tx_{b 2}^{- m_2}) }{
   (\tx_{b 1}^{-m_1}-\tx_{b 2}^{+ m_2}) (\tx_{b 1}^{+m_1}-\tx_{b 2}^{+ m_2}) } \prod_{j=1}^{m_1-1}
 \left( \frac{ \tx_{b 1}^{+(m_1-2j)}-\tx_{b 2}^{- m_2} }{
\tx_{b 1}^{+(m_1-2j)}-\tx_{b 2}^{+ m_2} }\right)^2
 \,,
\eal
\bal\la{eq:Deltaepsp_2mm}
\exp\big(i\,\Delta_{b}^{+}(u_1\pm {i\ov h}m_1,\tx_{b 2}^{\pm m_2})\big) =& 
\left(\frac{\tx_{b 2}^{- m_2} }{\tx_{b 2}^{+ m_2} }\right)^{m_1}
 \frac{
   (\tx_{\bar b 1}^{-m_1}\tx_{b 2}^{+ m_2}-1) (\tx_{\bar b 1}^{+m_1}\tx_{b 2}^{+ m_2}-1) }{ (\tx_{\bar b 1}^{-m_1}\tx_{b 2}^{- m_2}-1) (\tx_{\bar b 1}^{+m_1}\tx_{b 2}^{- m_2}-1) } 
      \\
   &\times\prod_{j=1}^{m_1-1}
 \left( \frac{
\tx_{\bar b 1}^{+(m_1-2j)}\tx_{b 2}^{+ m_2} -1}{ \tx_{\bar b 1}^{+(m_1-2j)}\tx_{b 2}^{- m_2}-1 }\right)^2
 \,,
\eal
where
\bal
S^{m_1m_2}(u)&=\frac{\left(u-\frac{i
   \left(m_1+m_2\right)}{h}\right) \left(u-\frac{i
   \left(m_2-m_1\right)}{h}\right)}{\left(u+\frac{
   i \left(m_1+m_2\right)}{h}\right)
   \left(u+\frac{i \left(m_2-m_1\right)}{h}\right)} \prod_{j=1}^{m_1-1} \left(\frac{u-\frac{i \left(2
   j-m_1+m_2\right)}{h}}{u+\frac{i \left(2 j-m_1+m_2\right)}{h}}\right)^2
   \\
   &=\prod _{j=1}^{m_1} \frac{\left(u+\frac{i \left(2j-m_1-m_2\right)}{h}\right) \left(u-\frac{i\left(2j-m_1+m_2\right)}{h}\right)}{\left(u-\frac{i \left(2j-m_1-m_2\right)}{h}\right)\left(u+\frac{i\left(2 j-m_1+m_2\right)}{h}\right)}\,.
\eal
We also need identities for the cases where $x_2^+$ is on the anti-mirror plane, or both $x_2^+$ and $x_2^-$ are on the anti-mirror plane.
Let
\bal
\label{eq:app_delta_string}
\Delta_{b}^{\b}(u_1\pm {i\ov h}m_1,{1\ov\tx_{\bar b 2}^{+ m_2}},\tx_{b 2}^{- m_2})\equiv &+\tPsi_{b}^{\b}(u_1 +{i\ov h}m_1,{1\ov\tx_{\bar b 2}^{+ m_2}})-\tPsi_{b}^{\b}(u_1 +{i\ov h}m_1,\tx_{b 2}^{- m_2})
\\
&-\tPsi_{b}^{\b}(u_1 -{i\ov h}m_1,{1\ov\tx_{\bar b 2}^{+ m_2}})+\tPsi_{b}^{\b}(u_1 -{i\ov h}m_1,\tx_{b 2}^{- m_2})\,,
\eal
\bal
\Delta_{b}^{\b}(u_1\pm {i\ov h}m_1,{1\ov\tx_{\bar b 2}^{\pm m_2}})\equiv &+\tPsi_{b}^{\b}(u_1 +{i\ov h}m_1,{1\ov\tx_{\bar b 2}^{+ m_2}})-\tPsi_{b}^{\b}(u_1 +{i\ov h}m_1,{1\ov\tx_{\bar b 2}^{- m_2}})
\\
&-\tPsi_{b}^{\b}(u_1 -{i\ov h}m_1,{1\ov\tx_{\bar b 2}^{+ m_2}})+\tPsi_{b}^{\b}(u_1 -{i\ov h}m_1,{1\ov\tx_{\bar b 2}^{- m_2}})\,.
\eal
Then, we get
\bal
\label{eq:Deltaepsm_2ms}
\exp\big(i\,\Delta_{b}^{ -}&(u_1\pm {i\ov h}m_1,{1\ov\tx_{\bar b 2}^{+ m_2}},\tx_{b 2}^{- m_2})\big) =\frac{\left(\frac{h}{2}\right)^{-2 m_1}}{\prod
   _{j=1}^{m_1} \left(u_{12}+\frac{i \left(2
   j-m_1+m_2\right)}{h}\right) \left(u_{12}+\frac{i \left(-2
   j+m_1+m_2\right)}{h}\right)}
   \\
   &\times \left(\frac{\tx_{\bar b 2}^{+ m_2} }{\tx_{b 2}^{- m_2} }\right)^{m_1}\frac{ (\tx_{b 1}^{-m_1}-\tx_{b 2}^{- m_2}) (\tx_{b 1}^{+m_1}-\tx_{b 2}^{- m_2}) }{
   (\tx_{b 1}^{-m_1}\tx_{\bar b 2}^{+ m_2}-1) (\tx_{b 1}^{+m_1}\tx_{\bar b 2}^{+ m_2}-1)} \prod_{j=1}^{m_1-1}
 \left( \frac{ \tx_{b 1}^{+(m_1-2j)}-\tx_{b 2}^{- m_2} }{
\tx_{b 1}^{+(m_1-2j)}\tx_{\bar b 2}^{+ m_2} -1}\right)^2
 \,,
\eal
\bal
\label{eq:Deltaepsp_2ms}
\exp\big(i\,\Delta_{b}^{+}&(u_1\pm {i\ov h}m_1,{1\ov\tx_{\bar b 2}^{+ m_2}},\tx_{b 2}^{- m_2})\big) =\frac{\left(\frac{h}{2}\right)^{-2 m_1}}{\prod
   _{j=1}^{m_1} \left(u_{12}+\frac{i \left(2
   j-m_1-m_2\right)}{h}\right) \left(u_{12}+\frac{i \left(-2
   j+m_1-m_2\right)}{h}\right)}
   \\
   &\times  
\left(\frac{\tx_{b 2}^{- m_2} }{\tx_{\bar b 2}^{+ m_2} }\right)^{m_1}
 \frac{
   (\tx_{\bar b 1}^{-m_1}-\tx_{\bar b 2}^{+ m_2}) (\tx_{\bar b 1}^{+m_1}-\tx_{\bar b 2}^{+ m_2}) }{ (\tx_{\bar b 1}^{-m_1}\tx_{b 2}^{- m_2}-1) (\tx_{\bar b 1}^{+m_1}\tx_{b 2}^{- m_2}-1) } 
\prod_{j=1}^{m_1-1}
 \left( \frac{
\tx_{\bar b 1}^{+(m_1-2j)}-\tx_{\bar b 2}^{+ m_2} }{ \tx_{\bar b 1}^{+(m_1-2j)}\tx_{b 2}^{- m_2} -1}\right)^2
 \,,
\eal
\bal
\exp\big(i\,\Delta_{b}^{-}&(u_1\pm {i\ov h}m_1,{1\ov\tx_{\bar b 2}^{\pm m_2}})\big) =\left(\frac{\tx_{\bar b 2}^{+ m_2} }{\tx_{\bar b 2}^{- m_2} }\right)^{m_1}
 \frac{
   (\tx_{ b 1}^{-m_1}\tx_{\bar b 2}^{- m_2}-1) (\tx_{ b 1}^{+m_1}\tx_{\bar b 2}^{- m_2}-1) }{ (\tx_{ b 1}^{-m_1}\tx_{\bar b 2}^{+ m_2}-1) (\tx_{ b 1}^{+m_1}\tx_{\bar b 2}^{+ m_2}-1) } 
      \\
   &\times\prod_{j=1}^{m_1-1}
 \left( \frac{
\tx_{ b 1}^{+(m_1-2j)}\tx_{\bar b 2}^{- m_2} -1}{ \tx_{ b 1}^{+(m_1-2j)}\tx_{\bar b 2}^{+ m_2}-1 }\right)^2
 \,,
\eal

\bal
\exp\big(i\,\Delta_{b}^{+}&(u_1\pm {i\ov h}m_1,{1\ov\tx_{\bar b 2}^{\pm m_2}})\big) ={1\ov S^{m_1m_2}(u_{12})}
\left(\frac{\tx_{\bar b 2}^{- m_2} }{\tx_{\bar b 2}^{+ m_2} }\right)^{m_1}
   \\
   &\times \frac{ (\tx_{\bar b 1}^{-m_1}-\tx_{\bar b 2}^{-+m_2}) (\tx_{\bar b 1}^{+m_1}-\tx_{\bar b 2}^{+ m_2}) }{
   (\tx_{\bar b 1}^{-m_1}-\tx_{\bar b 2}^{- m_2}) (\tx_{\bar b 1}^{+m_1}-\tx_{\bar b 2}^{- m_2}) } \prod_{j=1}^{m_1-1}
 \left( \frac{ \tx_{\bar b 1}^{+(m_1-2j)}-\tx_{\bar b 2}^{+ m_2} }{
\tx_{\bar b 1}^{+(m_1-2j)}-\tx_{\bar b 2}^{- m_2} }\right)^2
 \,.
\eal
These identities can be used to check crossing relations for mirror and string bound state S-matrices, and to prove the CP invariance of the string model.

\section{Solutions to crossing for bound states}

The dressing factors for the bound states are obtained from the ones of fundamental particles through fusion and must satisfy certain crossing equations.
This appendix shows that the bound states dressing factors obtained from fusion satisfy the associated crossing equations in the mirror and string model. 
This is a nontrivial consistency check of our proposal for the dressing factors.

\subsection{Crossing equations for mirror bound states}

In mirror theory, the bound states are composed of multiple particles of type $Z$ or multiple particles of type $\bar{Z}$. 
As already mentioned, the constituents of a $m$-particle mirror bound state satisfy
\bal
\tx_{a}^{+m}(u) =\tx_{a1}^+ ,\quad  \tx_{a1}^- =\tx_{a2}^+ ,\quad   \quad\ldots\,,\quad    \tx_{a(m-1)}^- =\tx_{am}^+ , \quad \tx_{am}^- =\tx_{a}^{-m}(u) .
\eal
The rapidities of the particles composing the bound state are then
\bal
u_j=u+\frac{i}{h} (m+1) - \frac{2i}{h} j \,, \qquad j=1, \, 2,\, \dots,\, m\,,
\eal
where $u$ is the centre of the bound state. The S-matrix elements for the bound states are then obtained by fusing the elements $S^{11}_{\bar{Z} \bar{Z}}$, $S^{11}_{Z \bar{Z}}$, $S^{11}_{\bar{Z} Z}$ and~$S^{11}_{Z Z}$ for fundamental particles with rapidities chosen as above. From these elements we can read the normalisation for the different sectors (Right-Right, Left-Right, Right-Left and Left-Left) of the bound states. Applying fusion on the elements in~\eqref{eq:massivenormmir} we obtain
\bal
\mathbf{S}\,\big|Y^{m_1}_{1} Y^{m_2}_{2}\big\rangle=&
\frac{\tx^{+m_1}_{\L1}}{\tx^{-m_1}_{\L1}} \ \frac{\tx^{-m_2}_{\L2}}{\tx^{+m_2}_{\L2}} \ \left(\frac{\tx^{-m_1}_{\L1} - \tx^{+m_2}_{\L2}}{\tx^{+m_1}_{\L1} - \tx^{-m_2}_{\L2}} \right)^2\prod^{m_2-1}_{j=1} \left(\frac{u_{12} + \frac{i}{h} (m_1-m_2+2 j)}{u_{12} - \frac{i}{h} (m_1+m_2-2 j)} \right)^2 \\
& \frac{u_{12} + \frac{i}{h} (m_1 + m_2)}{u_{12} - \frac{i}{h} (m_1 + m_2)} \ \frac{u_{12} + \frac{i}{h} (m_1 - m_2)}{u_{12} - \frac{i}{h} (m_1 - m_2)} \ \big(\Sigma^{m_1 m_2}_{\L\L}\big)^{-2}\,
\big|Y^{m_1}_{1}Y^{m_2}_{2}\big\rangle \,,\\
\\
 \mathbf{S}\,\big|Y^{m_1}_{1}\bar{Z}^{m_2}_{2}\big\rangle=&
     {\tx_{\R2}^{-m_2}\ov \tx_{\R2}^{+m_2}}
    \frac{1-\frac{1}{\tx_{\L1}^{-m_1}\tx_{\R2}^{-m_2}}}{1-\frac{1}{\tx_{\L1}^{+m_1}x_{\R2}^{+m_2}}}
    \frac{1-\frac{1}{\tx_{\L1}^{+m_1}\tx_{\R2}^{-m_2}}}{1-\frac{1}{\tx_{\L1}^{-m_1}\tx_{\R2}^{+m_2}}}\big(\Sigma^{m_1 m_2}_{\L\R}\big)^{-2}\,
    \big|Y^{m_1}_{1}\bar{Z}^{m_2}_{2}\big\rangle \, ,\\
    \\
      \mathbf{S}\,\big|\bar{Z}^{m_1}_{1}Y^{m_2}_{2}\big\rangle=&
 {\tx_{\R1}^{+m_1}\ov \tx_{\R1}^{-m_1}} 
   \frac{1-\frac{1}{\tx_{\R1}^{+m_1}\tx_{\L2}^{+m_2}}}{1-\frac{1}{\tx_{\R1}^{-m_1}\tx_{\L2}^{-m_2}}}
    \frac{1-\frac{1}{\tx_{\R1}^{+m_1}\tx_{\L2}^{-m_2}}}{1-\frac{1}{\tx_{\R1}^{-m_1}\tx_{\L2}^{+m_2}}}\big(\Sigma^{m_1 m_2}_{\R\L}\big)^{-2}\,
    \big|\bar{Z}^{m_1}_{1}Y^{m_2}_{2}\big\rangle \,,\\
    \\
    \mathbf{S}\,\big|\bar{Z}^{m_1}_{1}\bar{Z}^{m_2}_{2}\big\rangle=&
\prod^{m_2-1}_{j=1} \left(\frac{u_{12} + \frac{i}{h} (m_1-m_2+2 j)}{u_{12} - \frac{i}{h} (m_1+m_2-2 j)} \right)^2 \\
& \frac{u_{12} + \frac{i}{h} (m_1 + m_2)}{u_{12} - \frac{i}{h} (m_1 + m_2)} \ \frac{u_{12} + \frac{i}{h} (m_1 - m_2)}{u_{12} - \frac{i}{h} (m_1 - m_2)} \ \big(\Sigma^{m_1 m_2}_{\R\R}\big)^{-2}\,
\big|\bar{Z}^{m_1}_{1}\bar{Z}^{m_2}_{2}\big\rangle \,.
\eal

The crossing equations for the mirror bound-states dressing factors are easily obtainied by fusing the crossing equations~\eqref{eq:mirror_crossing_eq} for fundamental mirror particles.
By doing so we obtain
\bal
\label{crossing_eq_bound_state_mirror_dfactors1}
&\big(\Sigma^{m_1 \, m_2}_{aa}(u_1,u_2)\big)^{2} \big(\Sigma^{m_1 \, m_2}_{\bar{a} a}(\bar u_1,u_2)\big)^{2}=
\frac{(\tx_{a1}^{-m_1} - \tx_{a2}^{+m_2})(\tx_{a1}^{+m_1} - \tx_{a2}^{-m_2})}{(\tx_{a1}^{-m_1} - \tx_{a2}^{-m_2})(\tx_{a1}^{+m_1} - \tx_{a2}^{+m_2})}\\
&\prod^{m_2-1}_{j=1} \left(\frac{u_{12} + \frac{i}{h} (m_1-m_2+2 j)}{u_{12} - \frac{i}{h} (m_1+m_2-2 j)} \right)^2 \ \frac{u_{12} + \frac{i}{h} (m_1 + m_2)}{u_{12} - \frac{i}{h} (m_1 + m_2)} \ \frac{u_{12} + \frac{i}{h} (m_1 - m_2)}{u_{12} - \frac{i}{h} (m_1 - m_2)}\,,
\eal
\bal
\label{crossing_eq_bound_state_mirror_dfactors2}
 &\big(\Sigma^{m_1 \, m_2}_{aa}(\bar u_1,u_2)\big)^{2} \big(\Sigma^{m_1 \, m_2}_{\bar{a} a}( u_1,u_2)\big)^{2}=
\frac{\big(1-\frac{1}{\tx^{+m_1}_{\bar{a}1}\tx^{+m_2}_{a2}}\big)\big(1-\frac{1}{\tx^{-m_1}_{\bar{a}1}\tx^{-m_2}_{a2}}\big)}{\big(1-\frac{1}{\tx^{+m_1}_{\bar{a}1}\tx^{-m_2}_{a2}}\big)\big(1-\frac{1}{\tx^{-m_1}_{\bar{a}1}\tx^{+m_2}_{a2}}\big)}\\
 &\prod^{m_2-1}_{j=1} \left(\frac{u_{12} + \frac{i}{h} (m_1-m_2+2 j)}{u_{12} - \frac{i}{h} (m_1+m_2-2 j)} \right)^2 \ \frac{u_{12} + \frac{i}{h} (m_1 + m_2)}{u_{12} - \frac{i}{h} (m_1 + m_2)} \ \frac{u_{12} + \frac{i}{h} (m_1 - m_2)}{u_{12} - \frac{i}{h} (m_1 - m_2)}\,.
\eal

\subsection{Solving crossing for mirror bound states}

Using the notation of section~\ref{mirrorPhi} the analytic continuation of the mirror bound states to the anti-mirror region is obtained as follows.
For a bound state of type $a$ ($a=\text{L}, \, \text{R}$) composed of $m_1$ particles we first move $\tx_{a}^{-m_1}$ and then $\tx_{a}^{+m_1}$ to the antimirror region through the interval $(0, \xbr_a)$ (we recall that $\xbr_\L=\xbr$ and $\xbr_\R=\frac{1}{\xbr}$). In the $u$ plane this corresponds to crossing first the main cut of $\tx^{-m_1}_a(u)$ from above and then the main cut of $\tx^{+m_1}_a(u)$ also from above.

\paragraph{Improved BES for mirror bound states and its continuation.}
Continuing $\tx^{-m_1}_{a1}$ to the anti mirror region we obtain
\bal
&\tchi^{\a \b}_{ab}(\frac{1}{\tx^{-m_1}_{\bar{a}1}},\tx_{b2})=\tPhi^{\a \b}_{ab}(\frac{1}{\tx^{-m_1}_{\bar{a}1}},\tx_{b2})-\tPsi^{\b}_{b}(\bar{\tx}^{-m_1}_{a1},\tx_{b2})\,,\\
&\tchi^{\a \b}_{ab}(\tx^{+m_1}_{a1},\tx_{b2})=\tPhi^{\a \b}_{ab}(\tx^{+m_1}_{a1},\tx_{b2})\,.
\eal
We repeat the same procedure for $\tx^{+m_1}_{a1}$. Analogously to what happens for fundamental mirror particle in this second procedure the function $\tPsi^{\b}_{b}(\frac{1}{\tx^{-m_1}_{\bar{a}1}},\tx_{b2})$ generates additional terms. To see this, we first integrate by parts and get (up to boundary terms irrelevant for the continuation) the following expression  
\bal
&\tPsi_{b}^{\b}(\frac{1}{\tx^{-m_1}_{\bar{a} 1}},\tx_{b2})={h\ov 2}\int_{\pa\cR_\b}\frac{{\rm
d }w}{2\pi i} \log(w-\tx_{b2})\, {du_b(w)\ov dw}
\\
&~~~~~~~\times \left[\psi\Big(1+\tfrac{ih}{
2}\big(u_1 - \frac{i}{h} m_1-u_b(w)\big)\Big)+ \psi\Big(1-\tfrac{ih}{
2}\big(u_1 - \frac{i}{h} m_1-u_b(w)\big)\Big)\right]\,,~~~
\eal 
where as usual we defined
\bal
u_a(\frac{1}{\tx^{-m_1}_{\bar{a}1}})=u_1 - \frac{i}{h} m_1 \,.
\eal

We start considering the case $\b=-$, for which we have
\bal
&\tPsi_{b}^{\b}(\frac{1}{\tx^{-m_1}_{\bar{a} 1}},\tx_{b2})={h\ov 2}\int_{\widetilde{\rm cuts}}\frac{{\rm
d }v}{2\pi i} \log(\tx_b(v)-\tx_{b2})\,
\\
&~~~~~~~\times \left[\psi\Big(1+\tfrac{ih}{
2}\big(u_1- \frac{i}{h}m_1-v \big)\Big)+ \psi\Big(1-\tfrac{ih}{
2}\big(u_1- \frac{i}{h}m_1-v\big)\Big)\right]\,,~~~ 
\eal 
The $\psi$ functions in the integrand have poles located at
\bal
&v_*^{(n)}=u_1-\frac{2i}{h} n- \frac{i}{h}m_1  \,,\quad n= 1,2, \dots\,,\\
&v_{**}^{(n)}=u_1+\frac{2i}{h} n- \frac{i}{h}m_1 \,,\quad n=1, 2, \dots\,.
\eal
When $\tx^{-m_1}_{a1}$ enters the anti-mirror region then $u_1$ takes the value $u_1=t+\frac{i}{h} m_1$, with $t> \ubr$.
This coresponds to have $\tx^{-m_1}_{a1}$ evaluated on the main mirror cut. At this point we move $t$ down on a vertical line $t \to t-\frac{2i}{h} m_1$; in this manner (after the shift) we are on the main mirror cut of $\tx^{+m_1}_{a1}$. Performing this shift the poles $v_{**}^{(1)}$, $v_{**}^{(2)}$, $\dots$, $v_{**}^{(m_1)}$ cross the integration line $(\ubr, +\infty)$ from above and we need to pick up the associated residues. In the end we obtain
\bal
&\tchi^{\a \b}_{ab}(\frac{1}{\tx^{-m_1}_{\bar a1}},\tx_{b2})=\tPhi^{\a \b}_{ab}(\frac{1}{\tx^{-m_1}_{\bar a1}},\tx_{b2})-\tPsi^{\b}_{b}(\frac{1}{\tx^{-m_1}_{\bar a1}},\tx_{b2})\\
&-\frac{1}{i} \log \left(\frac{\frac{1}{\tx^{+m_1}_{\bar{b}}}-\tx_{b2}}{\tx^{+m_1}_{b} -\tx_{b2}} \right) -\frac{1}{i} \sum_{n=1}^{m_1-1}\log \left(\frac{\frac{1}{\tx_{\bar{b}} (u_1 + \frac{i}{h} (2n -m_1))}-\tx_{b2}}{\tx_{b}  (u_1 + \frac{i}{h} (2n -m_1))-\tx_{b2}} \right)\,,\\
&\tchi^{\a \b}_{ab}(\frac{1}{\tx^{+m_1}_{\bar a1}},\tx_{b2})=\tPhi^{\a \b}_{ab}(\frac{1}{\tx^{+m_1}_{\bar a1}},\tx_{b2})-\tPsi^{\b}_{b}(\frac{1}{\tx^{+m_1}_{\bar a1}},\tx_{b2})\,,
\eal
where as usual we define
 \bal
 & \tx_{c 1}^{\pm m_1} = \tx_c(u_1\pm {i\ov h} m_1)\,.
 \eal
The computation is similar for $\b=+$; the only difference is the change of parameterisation to $w=\bar{\tx}_b(v)=\frac{1}{\tx_{\bar b}(v)}$) but the final result does not change.

\paragraph{BES crossing equations for mirror bound states.} 
Using the analytic continuation just described, the crossing equations for the improved BES of mirror bound states take the following form
 \bal\la{creqthmir_bound}
 &\tilde{\theta}_{ab}^{\a \b}({1\ov \tx_{\bar a 1}^{\pm m_1}}, \tx_{b2}^{\pm m_2}) + \tilde{\theta}_{\bar a b}^{-\a \, \b}( \tx_{\bar a 1}^{\pm m_1}, \tx_{b2}^{\pm m_2})=\\
 &\tPsi_{b}^{\b}({1\ov \tx_{\bar a 1}^{+m_1}}, \tx_{b2}^{-m_2})  -
\tPsi_{b}^{\b}({1\ov \tx_{\bar a 1}^{-m_1}}, \tx_{b2}^{-m_2}) +
\tPsi_{b}^{\b}({1\ov \tx_{\bar a 1}^{-m_1}}, \tx_{b2}^{+m_2})- \tPsi_{b}^{\b}({1\ov \tx_{\bar a 1}^{+m_1}}, \tx_{b2}^{+m_2}) 
\\
&-\frac{1}{i} \log \left(\frac{\frac{1}{\tx^{+m_1}_{\bar{b}} }-\tx^{-m_2}_{b2}}{\tx^{+m_1}_{b}  -\tx^{-m_2}_{b2}} \right)+\frac{1}{i} \log \left(\frac{\frac{1}{\tx^{+m_1}_{\bar{b}} }-\tx^{+m_2}_{b2}}{\tx^{+m_1}_{b}  -\tx^{+m_2}_{b2}} \right)\\
&-\frac{1}{i} \sum_{n=1}^{m_1-1}\log \frac{\frac{1}{\tx_{\bar{b}} (u_1 + \frac{i}{h} (2n -m_1))}-\tx^{-m_2}_{b2}}{\tx_{b}  (u_1 + \frac{i}{h} (2n -m_1))-\tx^{-m_2}_{b2}}
+\frac{1}{i} \sum_{n=1}^{m_1-1}\log \frac{\frac{1}{\tx_{\bar{b}} (u_1 + \frac{i}{h} (2n -m_1))}-\tx^{+m_2}_{b2}}{\tx_{b}  (u_1 + \frac{i}{h} (2n -m_1))-\tx^{+m_2}_{b2}}\,.
\eal
The second line of the expression above is equal to $\Delta_{b}^{\b}(u_1\pm {i\ov h}m_1,\tx_{b 2}^{\pm m_2})$ defined in~\eqref{eq:Dab_definition}. 
Using the relations~\eqref{eq:ttheta_definition_mir_fund} and the results in appendix~\ref{app:psi-identities} then we obtain the following crossing equations for the improved BES phase of mirror bound states
  \bal
  \la{eq:app_bes_bstates_mirror_crossing}
&2\tilde{\theta}^\bes_{\bar a a}({1\ov \tx_{a 1}^{\pm m_1}}, \tx_{a 2}^{\pm m_2})+2\tilde{\theta}^\bes_{a a}( \tx_{a 1}^{\pm m_1}, \tx_{a 2}^{\pm m_2}) =2\tilde{\theta}^\bes_{a a}({1\ov \tx_{\bar a 1}^\pm}, \tx_{a 2}^\pm) + 2\tilde{\theta}^\bes_{\bar a b}( \tx_{\bar a 1}^\pm, \tx_{a 2}^\pm)=\\
&+{ 1\ov i}\log \left( \frac{\tx_{a 1}^{+m_1}-\tx_{a 2}^{-m_2}}{{ \tx_{a1}^{+m_1}}-\tx_{a2}^{+m_2}} \ \frac{\tx_{a 1}^{-m_1}-\tx_{a2}^{+m_2}}{{ \tx_{a1}^{-m_1}}-\tx_{a2}^{-m_2}} \ \frac{1-{1\ov \tx_{\bar a 1}^{-m_1} \tx_{a2}^{-m_2}}} {1-{1\ov \tx_{\bar a 1}^{-m_1}\tx_{a2}^{+m_2}}} \ \frac{1-{1\ov \tx_{\bar a 1}^{+m_1}\tx_{a2}^{+m_2}}}{1-{1\ov \tx_{\bar a 1}^{+m_1}\tx_{a2}^{-m_2}}} \right)\\
&+\frac{1}{i} \log \left( \frac{u_{12} + \frac{i}{h} (m_1+m_2)}{u_{12} - \frac{i}{h} (m_1+m_2)} \ \frac{u_{12} + \frac{i}{h} (m_1-m_2)}{u_{12} - \frac{i}{h} (m_1-m_2)} \prod^{m_2-1}_{j=1}\left( \frac{u_{12} + \frac{i}{h} (m_1-m_2 + 2j) }{u_{12} - \frac{i}{h} (m_1+m_2-2 j)} \right)^2 \right)\,.
\eal
 
\paragraph{HL crossing equations for mirror bound states.}
The derivation of the crossing equations for the HL phase is identical to the one for fundamental particles; the equations are the following
 \bal
   \la{eq:app_hl_bstates_mirror_crossing}
&2\tilde{\theta}^{\hl}_{aa}(\frac{1}{\tx^{\pm m_1}_{\bar a1}},\tx^{\pm m_2}_{a2}) + 2\tilde{\theta}^{\hl}_{\bar a a}(\tx^{\pm m_1}_{\bar a1},\tx^{\pm m_2}_{a2})=2\tilde{\theta}^{\hl}_{aa}(\tx^{\pm m_1}_{ a1},\tx^{\pm m_2}_{a2}) + 2\tilde{\theta}^{\hl}_{\bar a a}(\frac{1}{\tx^{\pm m_1}_{a1}},\tx^{\pm m_2}_{a2})\\
&={1\ov  i}  \log \left( \frac{\frac{1}{\tx^{+m_1}_{\bar a 1}} - \tx^{+m_2}_{a 2}}{\tx^{+m_1}_{ a 1} - \tx^{+m_2}_{a 2}}  \frac{\frac{1}{\tx^{-m_1}_{\bar a 1}} - \tx^{-m_2}_{a 2}}{\tx^{-m_1}_{ a 1} - \tx^{-m_2}_{ a 2}}  \frac{\tx^{+m_1}_{ a 1} - \tx^{-m_2}_{a 2}}{\frac{1}{\tx^{+m_1}_{\bar a 1}} - \tx^{-m_2}_{a 2}}    \frac{\tx^{-m_1}_{ a 1} - \tx^{+m_2}_{a 2}}{\frac{1}{\tx^{-m_1}_{\bar a 1}} - \tx^{+m_2}_{a 2}}     \right)\,.
\eal

 \paragraph{Crossing equations for the full mirror phase.}
If we combine~\eqref{eq:app_bes_bstates_mirror_crossing} and~\eqref{eq:app_hl_bstates_mirror_crossing} then we get the following crossing equation for the even part of the dressing factors
\bal
\la{eq:crossing_eq_ratio_BES_HL_bound}
& \left( \Sigma^{\besratio}_{\bar{a} a}(\frac{1}{x^{\pm m_1}_{a1}}, x^{\pm m_2}_{a2})  \Sigma^{\besratio}_{aa}(x^{\pm m_1}_{a1}, x^{\pm m_2}_{a2})  \right)^2=\left(\Sigma^{\besratio}_{aa}(\frac{1}{x^{\pm m_1}_{\bar{a}1}}, x^{\pm m_2}_{a2}) \Sigma^{\besratio}_{\bar{a} a}(x^{\pm m_1}_{\bar{a} 1}, x^{\pm m_2}_{a2})  \right)^2\\
&= \frac{u_{12} + \frac{i}{h} (m_1+m_2)}{u_{12} - \frac{i}{h} (m_1+m_2)} \ \frac{u_{12} + \frac{i}{h} (m_1-m_2)}{u_{12} - \frac{i}{h} (m_1-m_2)} \prod^{m_2-1}_{j=1}\left( \frac{u_{12} + \frac{i}{h} (m_1-m_2 + 2j) }{u_{12} - \frac{i}{h} (m_1+m_2-2 j)} \right)^2 \,,
\eal
where as usual we defined
\bal
\la{eq:even_dressing_mir_bound}
\Sigma^{\besratio}_{ab}(\tx^{\pm m_1}_{a1}, \tx^{\pm m_2}_{b2})=\exp \left[ i \tilde{\theta}_{ab}^{\bes}(\tx^{\pm m_1}_{a1}, \tx^{\pm m_2}_{b2}) - i \tilde{\theta}_{ab}^{\hl}(\tx^{\pm m_1}_{a1}, \tx^{\pm m_2}_{b2}) \right]\,.
\eal

Minimal solutions to~(\eqref{crossing_eq_bound_state_mirror_dfactors1}, \eqref{crossing_eq_bound_state_mirror_dfactors2}) are then provided by
\bal
&\left(\Sigma^{m_1 m_2}_{a a} (u_1, u_2)\right)^{-2}= \left( \Sigma^{\besratio}_{a a} (x^{\pm m_1}_{a1}, x^{\pm m_2}_{a2}) \right)^{-2}\frac{R^2(\tg^{-m_1 \, -m_2}_{aa}) R^2(\tg^{+m_1 \, +m_2}_{aa})}{R^2(\tg^{-m_1 \, +m_2}_{aa}) R^2(\tg^{+m_1 \, -m_2}_{aa}) } \,,
\eal
and
\bal
&\left(\Sigma^{m_1 m_2}_{a \bar a} (u_1, u_2)\right)^{-2}=\left(\Sigma^{\besratio}_{a \bar a} (x^{\pm m_1}_{a1}, x^{\pm m_2}_{\bar{a}2}) \right)^{-2}\\
& \frac{R(\tg^{-m_1 \, +m_2}_{a \bar a}+ i \pi) R(\tg^{-m_1 \, +m_2}_{a \bar a}- i \pi) R(\tg^{+m_1 \, -m_2}_{a\bar a}+ i \pi)R(\tg^{+m_1 \, -m_2}_{a \bar a}- i \pi)}{R(\tg^{-m_1 \, -m_2}_{a  \bar a}+ i \pi) R(\tg^{-m_1 \, -m_2}_{a  \bar a}- i \pi) R(\tg^{+m_1 \, +m_2}_{a  \bar a}+ i \pi) R(\tg^{+m_1 \, +m_2}_{a  \bar a}- i \pi)} \,,
\eal
in agreement with what we obtain from fusion. Note that the odd part of the dressing factors (composed of the $R$ functions) satisfies crossing in the same way as for fundamental particles and we do not repeat its study here.  

\subsection{Crossing equations for string bound states}
\label{app:crossing-string-bs}
Differently from mirror theory, in string theory the bound states are obtained by fusing particles of type $Y$, or particles of type $\bar{Y}$.
The constituents of a string theory $m$-particle bound states must be chosen to satisfy the condition 
\bal
\la{eq:boundstatecond_stringth}
x_{a}^{-m}(u) =x_{a1}^- \,,\quad  x_{a1}^+ =x_{a2}^- \,,\quad    x_{a2}^+ =x_{a3}^- \,,\quad\ldots\,,\quad    x_{a(m-1)}^+ =x_{am}^- \,, \quad x_{am}^+ =x_{a}^{+m}(u)\,,
\eal
which can be realised by the following set of rapidities
\bal
u_j=u-\frac{i}{h} (m+1) + \frac{2i}{h} j \,, \qquad j=1, \, 2,\, \dots,\, m\,.
\eal
As before, $u$ is the centre of the bound state. The S-matrix elements for the string bound states are obtained by fusing the elements $S^{11}_{\bar{Y} \bar{Y}}$, $S^{11}_{Y \bar{Y}}$, $S^{11}_{\bar{Y} Y}$ and~$S^{11}_{Y Y}$ of the constituents particles, with rapidities chosen as above. Note that this differs from mirror theory, where we fused particles of type $Z$ and $\bar{Z}$.
These elements allow us to read the bound states' normalisation for the different string theory sectors (Right-Right, Left-Right, Right-Left and Left-Left). The normalisation for the S-matrices of string-bound states is provided in appendix~\ref{app:string-bound-states}.

Crossing equations for the string theory bound states are easily obtained by first continuing the crossing equations in~\eqref{eq:mirror_crossing_eq} to the string region (this is obtained by just replacing $\tx_{a} \to x_a$) and then fusing these equations. After fusion we obtain the following equations for the string bound states
\bal
\label{eq:string_boundstates_crossing_eq1}
&\big(\Sigma^{m_1 \, m_2}_{aa}(u_1,u_2)\big)^{2} \big(\Sigma^{m_1 \, m_2}_{\bar{a} a}(\bar u_1,u_2)\big)^{2}=
\frac{(x_{a1}^{-m_1} - x_{a2}^{+m_2})(x_{a1}^{+m_1} - x_{a2}^{-m_2})}{(x_{a1}^{-m_1} - x_{a2}^{-m_2})(x_{a1}^{+m_1} - x_{a2}^{+m_2})}\\
 &\prod^{m_2-1}_{j=1} \left(\frac{u_{12} + {i\ov h} (m_1-m_2+2 j)}{u_{12} - {i\ov h} (m_1+m_2-2 j)} \right)^2 \left(\frac{u_{12}+ \frac{i}{h} (m_1+m_2)}{u_{12}- \frac{i}{h} (m_1+m_2)} \right) \, \left(\frac{u_{12}+ \frac{i}{h} (m_1-m_2)}{u_{12}- \frac{i}{h} (m_1-m_2)} \right)\,,
\eal
\bal
\label{eq:string_boundstates_crossing_eq2}
&\big(\Sigma^{m_1 \, m_2}_{aa}(\bar u_1,u_2)\big)^{2} \big(\Sigma^{m_1 \, m_2}_{\bar{a} a}( u_1,u_2)\big)^{2}=
\frac{\big(1-\frac{1}{x^{+m_1}_{\bar{a}1}x^{+m_2}_{a2}}\big)\big(1-\frac{1}{x^{-m_1}_{\bar{a}1}x^{-m_2}_{a2}}\big)}{\big(1-\frac{1}{x^{+m_1}_{\bar{a}1}x^{-m_2}_{a2}}\big)\big(1-\frac{1}{x^{-m_1}_{\bar{a}1}x^{+m_2}_{a2}}\big)}\\
& \prod^{m_2-1}_{j=1} \left(\frac{u_{12} + {i\ov h} (m_1-m_2+2 j)}{u_{12} - {i\ov h} (m_1+m_2-2 j)} \right)^2 \left(\frac{u_{12}+ \frac{i}{h} (m_1+m_2)}{u_{12}- \frac{i}{h} (m_1+m_2)} \right) \, \left(\frac{u_{12}+ \frac{i}{h} (m_1-m_2)}{u_{12}- \frac{i}{h} (m_1-m_2)} \right)\,,
\eal
which are just a trivial continuation of the equations for mirror bound states~(\eqref{crossing_eq_bound_state_mirror_dfactors1}, \eqref{crossing_eq_bound_state_mirror_dfactors2}) to the string region.

\subsection{Solving crossing for string bound states}

We choose the following solution for the constituents $x_{aj}^\pm$ ($j=1, \dots, m$) solving the bound state condition~\eqref{eq:boundstatecond_stringth}
\bal
\la{eq:cond_st_boundstate}
x_{aj}^- &=  \tx_a\big(u -{i\ov h}(m+2-2j)\big) \,,\quad x_{aj}^+ =  \tx_a\big(u -{i\ov h}(m-2j)\big)\,,\quad j=1,\ldots, m-1 \,,
\\
x_{am}^- &=  \tx_a\big(u +{i\ov h}(m-2)\big) \,,\quad x_{am}^{+} = x_a\big(u+{i\ov h}m\big) = {1\ov \tx_{\bar a}\big(u+{i\ov h}m\big)}\,,
\eal
which is particularly convenient for the fusion of the BES phase. For $u \in \mathbb{R}$ the energy and momentum of the bound states are real since $(x^+_{am})^*=x^-_{a1}$. It is possible to show that S-matrix elements do not depend on the choice of the constituents; we refer to appendix~\ref{app:string-bound-states} for a proof of this fact.

Let us analyse the scattering of two string bound states of quantum numbers $m_1$ and $m_2$.
With the condition above all the BES phases for the constituents, but the ones containing $x_{am_1}^{+}$ and $x_{am_2}^{+}$, do not require to be modified by $\tPsi$ functions
and after fusion we obtain the following improved BES phases for the bound states 
\bal
\label{eq:thetaBES_in_string_region}
&\tilde{\theta}_{a b}^{\bes}(x^{\pm m_1}_{a1}, x^{\pm m_2}_{b2})=\\
&+\tPhi_{ab}(x^{+m_1}_{a1},x^{+m_2}_{b2})+\tPhi_{ab}(x^{-m_1}_{a1},x^{-m_2}_{b2})-\tPhi_{ab}(x^{+m_1}_{a1},x^{-m_2}_{b2})-\tPhi_{ab}(x^{-m_1}_{a1},x^{+m_2}_{b2})\\
&+\tPsi_{a}(x^{+m_2}_{b2}, x^{+m_1}_{a1})-\tPsi_{b}(x^{+m_1}_{a1}, x^{+m_2}_{b2})+\tPsi_{b}(x^{+m_1}_{a1}, x^{-m_2}_{b2})-\tPsi_{a}(x^{+m_2}_{b2}, x^{-m_1}_{a1})\\
& -K^\bes(u_{12}+{i\ov h}(m_1-m_2))\,.
\eal

\paragraph{Analytic continuation to anti-string region.}

As for fundamental string particles we move $x^{+m_1}_{a1}$ to the anti-string region first. We start with 
\bal
x_{a}(u_1+\frac{i}{h} m_1)=\frac{1}{\tx_{a}(u_1+\frac{i}{h} m_1)}\, \qquad u_1 \in \mathbb{R}\,.
\eal 
Then in order we cross the mirror cut $(\ubr,+\infty)$ from above (in the $x$ plane this corresponds to crossing the line $(\xbr_{a}, +\infty)$ from above) and then the string theory cut $(-\infty, \ubr)$ from below (this corresponds to enter the deformed unit circle from below). Then we repeat a similar procedure for $x^{-m_1}_{a1}$, crossing before the string theory main cut from below and then the mirror theory cut from above.

In doing the continuation of $x^{+m_1}_{a1}$ nontrivial contributions are generated by the $\tPsi$ functions. In particular by $\tPsi_{b}^{\b}(x^{+m_1}_{a1},x_{2})$ in~\eqref{eq:thetaBES_in_string_region}. We focus on the case $\b=-$, where we have
\bal
&\tPsi_{b}^{-}(x^{+m_1}_{a1},x_{2})={h\ov 2}\lint_{\widetilde{\rm cuts}}\frac{{\rm
d }v}{2\pi i} \log(\tx_b(v)-x_{2})\,
\\
&~~~~~~~\times \left[\psi\Big(1+\tfrac{ih}{
2}\big(u_1+ \frac{i}{h}m_1-v \big)\Big)+ \psi\Big(1-\tfrac{ih}{
2}\big(u_1+ \frac{i}{h}m_1-v\big)\Big)\right]\,,~~~ 
\eal 
The $\psi$ functions in the integrand have poles located at
\bal
&v_*^{(j)}=u_1-\frac{2i}{h} j+ \frac{i}{h}m_1  \,,\quad n= 1,2, \dots\,,\\
&v_{**}^{(j)}=u_1+\frac{2i}{h} j+ \frac{i}{h}m_1 \,,\quad n=1, 2, \dots\,.
\eal
We start with $u_1 \in \mathbb{R}$. Then we shift $u_1\to u_1 -\frac{i}{h}m_1$ to reach the mirror cut of $\tx^{+m_1}_{a1}$. We make this operation in such a way that the poles $v_{*}^{(j)}$, with $j=1,\, 2,\, \dots, \, m_1-1$ cross the integration line on the upper edge of the main mirror cut from above and we obtain
\bal
\tPsi_{b}^{-}(x^{+m_1}_{a1},x_{2}) &\to \tPsi_{b}^{-}(x^{+m_1}_{a1},x_{2})+i \sum^{m_1-1}_{j=1}\log \left( \frac{\frac{1}{\tx_{\bar{b}}(u_1 - \frac{2i}{h}j+ \frac{i}{h}m_1)} - x_2}{\tx_{b}(u_1 - \frac{2i}{h}j+ \frac{i}{h}m_1) - x_2} \right)\\
&=\tPsi_{b}^{-}(x^{+m_1}_{a1},x_{2})+i \sum^{m_1-1}_{j=1}\log \left( \frac{\frac{1}{\tx^{+(m_1-2j)}_{\bar{b}1}} - x_2}{\tx^{+(m_1-2j)}_{b1} - x_2} \right)\,.
\eal
The computation for $\tPsi_{b}^{+}(x^{+m_1}_{a1},x_{2})$ is identical, with the only caveat that we must use the parameterisation $w=\frac{1}{\tx_{\bar{b}}(v)}$ and add a minus sign due to the fact that we integrate around the cuts in the opposite direction. Then for $\b=+$ we obtain
\bal
\tPsi_{b}^{+}(x^{+m_1}_{a1},x_{2}) \to \tPsi_{b}^{+}(x^{+m_1}_{a1},x_{2})+i \sum^{m_1-1}_{j=1}\log \left( \frac{\frac{1}{\tx^{+(m_1-2j)}_{\bar{b}1}} - x_2}{\tx^{+(m_1-2j)}_{b1} - x_2} \right)\,.
\eal
Apart from these additional residues, the continuation of $x_{a1}^{+m_1}$ to the anti-string region is just the opposite of what we did to go from the mirror to the string region with $\tx^{+m_1}_{a1}$ and we just need to remove from~\eqref{eq:thetaBES_in_string_region}  the terms arising from the previous continuation
\bal
&\tilde{\theta}_{a b}^{\bes}(\frac{1}{x^{+m_1}_{\bar{a} 1}},x^{-m_1}_{a1}; x^{\pm m_2}_{b2})=\\
&+\tPhi_{ab}(\frac{1}{x^{+m_1}_{\bar{a} 1}},x^{+m_2}_{b2})+\tPhi_{ab}(x^{-m_1}_{a1},x^{-m_2}_{b2})-\tPhi_{ab}(\frac{1}{x^{+m_1}_{\bar{a}1}},x^{-m_2}_{b2})-\tPhi_{ab}(x^{-m_1}_{a1},x^{+m_2}_{b2})\\
&+\tPsi_{a}(x^{+m_2}_{b2}, \frac{1}{x^{+m_1}_{\bar{a}1}})-\tPsi_{a}(x^{+m_2}_{b2}, x^{-m_1}_{a1})\\
&+\frac{i}{2} \sum^{m_1-1}_{j=1}\log \left( \frac{\frac{1}{\tx^{+(m_1-2j)}_{\bar{b}1}} - x^{-m_2}_{b2}}{\tx^{+(m_1-2j)}_{b1} - x^{-m_2}_{b2}} \right)^2- \frac{i}{2} \sum^{m_1-1}_{j=1}\log \left( \frac{\frac{1}{\tx^{+(m_1-2j)}_{\bar{b}1}} - x^{+m_2}_{b2}}{\tx^{+(m_1-2j)}_{b1} - x^{+m_2}_{b2}} \right)^2
\eal

Now we repeat the same procedure for $x^{-m_1}_{a1}$, where we enter first the string theory cut from below and then the mirror theory cut from above. The continuation of the $\tPhi$ and $\tPsi$ functions is given by
\bal
\tPhi_{ab}(x^{-m_1}_{a1},\tx_{2}) \to \tPhi_{ab}(\frac{1}{x^{-m_1}_{\bar{a} 1}},\tx_{2}) - \tPsi_{b}(\frac{1}{x^{-m_1}_{\bar{a} 1}},\tx_{2})\,,
\eal
\bal
\tPsi_{a}(x^{+m_2}_{b2}, x^{-m_1}_{a1}) &\to \tPsi_{a}(x^{+m_2}_{b2}, \frac{1}{x^{-m_1}_{\bar{a}1}}) -K^\bes(u_{12}-{i\ov h}(m_1+m_2))
\eal
Then the continuation of the improved BES phase for bound states to the anti-string region is given by
\bal
\label{eq:continuation_BES_antistring_boundstates}
&\tilde{\theta}_{a b}^{\bes}(\frac{1}{x^{\pm m_1}_{\bar{a} 1}}, x^{\pm m_2}_{b2})=\\
&+\tPhi_{ab}(\frac{1}{x^{+m_1}_{\bar{a} 1}},x^{+m_2}_{b2})+\tPhi_{ab}(\frac{1}{\tx^{-m_1}_{\bar{a} 1}},x^{-m_2}_{b2})-\tPhi_{ab}(\frac{1}{x^{+m_1}_{\bar{a}1}},x^{-m_2}_{b2})-\tPhi_{ab}(\frac{1}{x^{-m_1}_{\bar{a} 1}},x^{+m_2}_{b2})\\
&+\tPsi_{a}(x^{+m_2}_{b2}, \frac{1}{x^{+m_1}_{\bar{a}1}})-\tPsi_{a}(x^{+m_2}_{b2}, \frac{1}{x^{-m_1}_{\bar{a} 1}})- \tPsi_{b}(\frac{1}{x^{-m_1}_{\bar{a} 1}},x^{-m_2}_{b2})+ \tPsi_{b}(\frac{1}{x^{-m_1}_{\bar{a} 1}},x^{+m_2}_{b2})\\
&+\frac{i}{2} \sum^{m_1-1}_{j=1}\log \left( \frac{\frac{1}{\tx^{+(m_1-2j)}_{\bar{b}1}} - x^{-m_2}_{b2}}{\tx^{+(m_1-2j)}_{b1} - x^{-m_2}_{b2}} \right)^2-\frac{i}{2} \sum^{m_1-1}_{j=1}\log \left( \frac{\frac{1}{\tx^{+(m_1-2j)}_{\bar{b}1}} - x^{+m_2}_{b2}}{\tx^{+(m_1-2j)}_{b1} - x^{+m_2}_{b2}} \right)^2\\
& +K^\bes(u_{12}-{i\ov h}(m_1+m_2))\,.
\eal

\paragraph{BES crossing equations for string bound states.}
Combining~\eqref{eq:thetaBES_in_string_region} and~\eqref{eq:continuation_BES_antistring_boundstates} we obtain the following crossing equations for the BES phases
\bal
&2\tilde{\theta}_{a b}^{\bes}(\frac{1}{x^{\pm m_1}_{\bar{a} 1}}, x^{\pm m_2}_{b2})+2\tilde{\theta}_{\bar{a} b}^{\bes}(x^{\pm m_1}_{\bar{a}1}, x^{\pm m_2}_{b2})=\\
&+\tPsi_{b}(x^{+m_1}_{\bar{a} 1}, x^{-m_2}_{b2})- \tPsi_{b}(\frac{1}{x^{-m_1}_{\bar{a} 1}},x^{-m_2}_{b2}) +\tPsi_{ b}(\frac{1}{x^{-m_1}_{\bar{a} 1}},x^{+m_2}_{b2}) -\tPsi_{ b}(x^{+m_1}_{\bar{a} 1}, x^{+m_2}_{b2}) \\
&+\tPsi_{a}(x^{+m_2}_{b2}, \frac{1}{x^{+m_1}_{\bar{a}1}}) +\tPsi_{\bar{a}}(x^{+m_2}_{b2}, x^{+m_1}_{\bar{a} 1})-\tPsi_{a}(x^{+m_2}_{b2}, \frac{1}{x^{-m_1}_{\bar{a} 1}})-\tPsi_{\bar{a}}(x^{+m_2}_{b2}, x^{-m_1}_{\bar{a} 1})\\
&+\frac{1}{i}  \log \prod^{m_1-1}_{j=1} \left( \frac{\tx^{+(m_1-2j)}_{b1} - x^{-m_2}_{b2}}{\tx^{+(m_1-2j)}_{b1} - x^{+m_2}_{b2}} \right)^2 +\frac{1}{i}  \log \prod^{m_1-1}_{j=1} \left( \frac{1-\tx^{+(m_1-2j)}_{\bar{b}1}x^{+m_2}_{b2}}{1-\tx^{+(m_1-2j)}_{\bar{b}1}x^{-m_2}_{b2}} \right)^2 \\
& -2K^\bes(u_{12}+{i\ov h}(m_1-m_2)) +2K^\bes(u_{12}-{i\ov h}(m_1+m_2))
\eal
It is easy to show that the third row of the expression above cancels while the second row is equal to
\bal
-\frac{1}{2}\Delta_{b}^{-}(u_1\pm {i\ov h}m_1,x_{b 2}^{\pm m_2})-\frac{1}{2}\Delta_{b}^{+}(u_1\pm {i\ov h}m_1,x_{b 2}^{\pm m_2})\,,
\eal
whose result is provided in appendix~\ref{app:psi-identities}.
Then after some manipulations it is possible to show that the crossing equations above become
\bal
\la{eq:crossingBESstringboundstates}
&2\tilde{\theta}_{a b}^{\bes}(\frac{1}{x^{\pm m_1}_{\bar{a} 1}}, x^{\pm m_2}_{b2})+2\tilde{\theta}_{\bar{a} b}^{\bes}(x^{\pm m_1}_{\bar{a}1}, x^{\pm m_2}_{b2})=\\
&+ \frac{1}{i}\log \left[ \frac{\left( u_{12}+\frac{i}{h} (m_1+m_2) \right) \left( u_{12}+\frac{i}{h} (m_1-m_2) \right) }{\left( u_{12}-\frac{i}{h} (m_1 + m_2)  \right)  \left( u_{12}-\frac{i}{h} (m_1 - m_2)  \right)}   \prod^{m_2-1}_{j=1}\left( \frac{u_{12} + \frac{i}{h} (m_1-m_2 )+ \frac{2i}{h}j }{u_{12} - \frac{i}{h} (m_1+m_2) + \frac{2i}{h}j} \right)^2\right]\\
&-\frac{1}{i}\log \left[ \frac{\left(x^{+m_1}_{b1} -x^{+m_2}_{b2} \right) \left(x^{-m_1}_{b1} -x^{-m_2}_{b2} \right) \left(1-\frac{1}{x^{+m_1}_{\bar{b} 1} x^{-m_2}_{b2}} \right) \left(1-\frac{1}{x^{-m_1}_{\bar{b} 1} x^{+m_2}_{b2}} \right)}{\left(x^{+m_1}_{b1} -x^{-m_2}_{b2} \right) \left(x^{-m_1}_{b1} -x^{+m_2}_{b2} \right) \left(1-\frac{1}{x^{+m_1}_{\bar{b} 1} x^{+m_2}_{b2}} \right) \left(1-\frac{1}{x^{-m_1}_{\bar{b} 1} x^{-m_2}_{b2}} \right)} \right]\,.
\eal
Note that the nontrivial dependence on the bound state constituents cancels between the terms arising by the analytic continuation and the expressions for $\Delta_{ab}^{~ \pm}(u_1\pm {i\ov h}m_1,x_{b 2}^{\pm m_2})$. As a consequence of this fact the crossing equations for the improved BES phases are independent of the constituent particles of the bound states.

\paragraph{Crossing equations for the full string phase.}
The HL dressing factors for bound states are the same as the ones for fundamental string particles (see~\eqref{eq:newHL_string_aa} and~\eqref{eq:newHL_string_baa}) after replacing $x^\pm_{a} \to x^{\pm m}_{a}$. 
The crossing equations for HL are also equal to those of the fundamental string particles in~\eqref{eq:crossing_string_HL} after the same replacement. 
Then defining 
\bal
\la{eq:even_dressing_string_bound}
\Sigma^{\besratio}_{ab}(x^{\pm m_1}_{a1}, x^{\pm m_2}_{b2})=\exp \left[ i \tilde{\theta}_{ab}^{\bes}(x^{\pm m_1}_{a1}, x^{\pm m_2}_{b2}) - i \tilde{\theta}_{ab}^{\hl}(x^{\pm m_1}_{a1}, x^{\pm m_2}_{b2}) \right]\,.
\eal
we see that the dressing factors
\bal
&\left(\Sigma^{m_1 m_2}_{a a} (u_1, u_2)\right)^{-2}= \left( \Sigma^{\besratio}_{a a} (x^{\pm m_1}_{a1}, x^{\pm m_2}_{a2}) \right)^{-2}\frac{R^2(\g^{-m_1 \, -m_2}_{aa}) R^2(\g^{+m_1 \, +m_2}_{aa})}{R^2(\g^{-m_1 \, +m_2}_{aa}) R^2(\g^{+m_1 \, -m_2}_{aa}) } \,,
\eal
and
\bal
&\left(\Sigma^{m_1 m_2}_{a \bar a} (u_1, u_2)\right)^{-2}=\left(\Sigma^{\besratio}_{a \bar a} (x^{\pm m_1}_{a1}, x^{\pm m_2}_{\bar{a}2}) \right)^{-2}\\
& \frac{R(\g^{-m_1 \, +m_2}_{a \bar a}+ i \pi) R(\g^{-m_1 \, +m_2}_{a \bar a}- i \pi) R(\g^{+m_1 \, -m_2}_{a\bar a}+ i \pi)R(\tg^{+m_1 \, -m_2}_{a \bar a}- i \pi)}{R(\g^{-m_1 \, -m_2}_{a  \bar a}+ i \pi) R(\g^{-m_1 \, -m_2}_{a  \bar a}- i \pi) R(\g^{+m_1 \, +m_2}_{a  \bar a}+ i \pi) R(\g^{+m_1 \, +m_2}_{a  \bar a}- i \pi)} \,,
\eal
(obtained from fusion) solve the crossing equations for string bound states~(\eqref{eq:string_boundstates_crossing_eq1}, \eqref{eq:string_boundstates_crossing_eq2}).
Again the odd part of the phases satisfies crossing in the same way as for fundamental particles; therefore we do not repeat its study here.

\section{ String bound-state S-matrix elements }
\label{app:string-bound-states}
In this appendix we use fusion to  find the normalisation for the string bound-state S-matrix elements used in the paper, and then we show that the S-matrix elements are independent of the constituent particles of the bound states. 

\subsection{ String bound-state S-matrix elements normalisation}

Let us recall that an $m$-particle string bound state satisfies the condition $x_{a j}^+ = x_{a j+1}^-$, $j=1,\ldots, m-1$, and we use the solution~\eqref{eq:cond_st_boundstate}.
In what follows we are going to use the following notation for an $m_\b$-particle string bound state constituent with rapidity $u_\b$
\bal
 x_{a \b j}^+ =\tx_a(u_\b - {i\ov h}(m_\b-2j))\,,\quad  \b=1,2\,.
\eal
By using  fusion, we find the following normalisation for the string bound-state S-matrix elements used in the paper
\bal\la{eq:SbYbY}
& \mathbf{S}\,\big|\bar Y_{1}^{m_1} \bar Y_{2}^{m_2}\big\rangle=S_{\bar Y \bar Y}^{m_1m_2}(p_{1}, p_{2}) 
   \,
    \big|\bar Y_{1}^{m_1} \bar Y_{2}^{m_2}\big\rangle\,,\quad 
    S_{\bar Y  \bar Y}^{m_1m_2}(p_{1}, p_{2}) = A_{\bar Y  \bar Y}^{m_1m_2}(p_{1}, p_{2})\ \big(\Sigma^{m_1 m_2}_{\R\R}\big)^{-2}\,,
    \\
&A_{\bar Y  \bar Y}^{m_1m_2}(p_{1}, p_{2})
=
    \left({x_{\R1}^{+m_1}\ov x_{\R1}^{-m_1}}\right)^{m_2}   \left({x_{\R2}^{-m_2}\ov x_{\R2}^{+m_2}}\right)^{m_1} \left(\frac{x^{- m_1}_{\R1} - x^{+ m_2}_{\R2}}{x^{+ m_1}_{\R1} - x^{- m_2}_{\R2}} \right)^2  \prod^{m_1-1}_{j=1} \left(\frac{x^{+}_{\R1 j} - x^{+m_2}_{\R 2}}{x^{+}_{\R1 j} - x^{-m_2}_{\R2}} \right)^2
 \\
 & \hspace{2.7cm} \times  \prod^{m_2-1}_{j=1} \left(\frac{x^{- m_1}_{\R1} - x^{+}_{\R2 j}}{x^{+ m_1}_{\R1} - x^{+}_{\R2 j}} \right)^2 \left(\frac{u_{12}+ \frac{i}{h} (m_1-m_2)}{u_{12}- \frac{i}{h} (m_1-m_2)} \right) \, \left(\frac{u_{12}+ \frac{i}{h} (m_1+m_2)}{u_{12}- \frac{i}{h} (m_1+m_2)} \right) 
  \\
  & \hspace{2.7cm}  \times\prod^{m_2-1}_{j=1} \left(\frac{u_{12} + {i\ov h} (m_1+m_2-2 j)}{u_{12} - {i\ov h} (m_1+m_2-2 j)} \right)^2 \,,\eal
\bal\la{eq:SbYZ}
& \mathbf{S}\,\big|\bar Y_{1}^{m_1} Z_{2}^{m_2}\big\rangle=S_{\bar Y Z}^{m_1m_2}(p_{1}, p_{2}) 
   \,
    \big|\bar Y_{1}^{m_1} Z_{2}^{m_2}\big\rangle\,,\quad
    S_{\bar Y  Z}^{m_1m_2}(p_{1}, p_{2}) = A_{\bar Y  Z}^{m_1m_2}(p_{1}, p_{2})\ \big(\Sigma^{m_1 m_2}_{\R\L}\big)^{-2}\,,
    \\
&A_{\bar Y  Z}^{m_1m_2}(p_{1}, p_{2})
= \left({x_{\R 1}^{-m_1}\ov x_{\R 1}^{+m_1}}\right)^{m_2-1}
      \left({x_{\L 2}^{+m_2}\ov x_{\L 2}^{-m_2}}\right)^{m_1} \ \frac{{x^{-m_1}_{\R 1} x^{-m_2}_{\L 2}}-1}{{x^{+m_1}_{\R 1} x^{+m_2}_{\L 2}}-1} \ \frac{{x^{+m_1}_{\R 1} x^{-m_2}_{\L 2}}-1}{{x^{-m_1}_{\R 1} x^{+m_2}_{\L 2}}-1} \\
      & \hspace{2.7cm}  \times\prod_{j=1}^{m_1-1}  \left(\frac{{x^+_{\R1  j} x^{-m_2}_{\L 2}}-1}{{x^+_{\R1 j} x^{+m_2}_{\L 2}}-1}\right)^2 \  \prod^{m_2-1}_{j=1}\left( \frac{{x^{+m_1}_{\R 1} x^{+}_{\L2 j}}-1}{{x^{-m_1}_{\R 1} x^{+}_{\L2 j}}-1} \right)^2 
      \,,
      \eal
\bal\la{eq:SbZY}
&\mathbf{S}\,\big| \bar{Z}_{1}^{m_1}{Y}_{2}^{m_2}\big\rangle =S_{ \bar Z  Y}^{m_1,m_2}(p_{1}, p_{2})\big| \bar{Z}_{1}^{m_1}{Y}_{2}^{m_2}\big\rangle \,,\quad S_{ \bar Z  Y}^{m_1,m_2}(p_{1}, p_{2})=A_{ \bar Z  Y}^{m_1,m_2}(p_{1}, p_{2})\ \big(\Sigma^{m_1, m_2}_{\R\L}\big)^{-2}\,,
\\
 &A_{ \bar Z  Y}^{m_1,m_2}(p_{1}, p_{2})=  \left({x_{\R1}^{-m_1}\ov x_{\R1}^{+m_1}}\right)^{m_2} \left({x_{\L2}^{+m_2}\ov x_{\L2}^{-m_2}}\right)^{m_1-1}  \frac{{x^{+m_1}_{\R 1} x^{+m_2}_{\L2}}-1}{{x^{-m_1}_{\R 1}x^{-m_2}_{\L2}}-1} \ \frac{{x^{+m_1}_{\R 1} x^{-m_2}_{\L2}}-1}{{x^{-m_1}_{\R 1}x^{+m_2}_{\L2}}-1}\\
      & \hspace{2.7cm}  \times \prod^{m_1-1}_{j=1}\left( \frac{{x^{+}_{\R 1j}x^{-m_2}_{\L2}}-1}{{x^{+}_{\R 1j} x^{+m_2}_{\L2}}-1} \right)^2 \ \prod_{j=1}^{m_2-1}  \left(\frac{{x^{+m_1}_{\R 1}x^+_{\L 2j}}-1}{{x^{-m_1}_{\R 1}x^+_{\L 2j}}-1}\right)^2 
      \,,
\eal
\bal\la{eq:SbZbZ}
&\mathbf{S}\,\big| \bar Z_1^{m_1}\bar{Z}_2^{m_2}\big\rangle =S_{ \bar Z \bar Z }^{m_1,m_2}(p_{1}, p_{2})\,\big|  \bar Z_1^{m_1}\bar{Z}_2^{m_2}\big\rangle\,,\quad S_{ \bar Z \bar Z }^{m_1,m_2}(p_{1}, p_{2})=A_{ \bar Z \bar Z }^{m_1,m_2}(p_{1}, p_{2})\ \big(\Sigma^{m_1, m_2}_{\R\R}\big)^{-2}\,,
\\
& A_{ \bar Z  \bar Z}^{m_1,m_2}(p_{1}, p_{2})=  \left({x_{\R1}^{+m_1}\ov x_{\R1}^{-m_1}}\right)^{m_2-1}   \left({x_{\R2}^{-m_2}\ov x_{\R2}^{+m_2}}\right)^{m_1-1} 
\\
& \hspace{2.7cm}  \times\prod^{m_1-1}_{j=1} \left(\frac{x^{+}_{\R 1j} - x^{+m_2}_{\R 2}}{x^{+}_{\R 1j} - x^{-m_2}_{\R 2}} \right)^2
\  \prod^{m_2-1}_{j=1} \left(\frac{x^{- m_1}_{\R1} - x^{+}_{\R 2j}}{x^{+ m_1}_{\R1} - x^{+}_{\R 2j}} \right)^2 
\\
& \hspace{2.cm}  \times \frac{u_{12}+ \frac{i}{h} (m_1-m_2)}{u_{12}- \frac{i}{h} (m_1-m_2)} \ \frac{u_{12}+ \frac{i}{h} (m_1+m_2)}{u_{12}- \frac{i}{h} (m_1+m_2)}   \prod^{m_2-1}_{j=1} \left(\frac{u_{12} + {i\ov h} (m_1+m_2-2 j)}{u_{12} - {i\ov h} (m_1+m_2-2 j)} \right)^2  
      \,,
\eal
and the other four S-matrix elements are restored by using the left-right symmetry.

\subsection{S-matrix elements independence of the constituent particles}

One can think that the S-matrix elements depend on the constituent particles. It is not the case, and here we follow \cite{Arutyunov:2009kf}, and 
 show that the S-matrix elements only depend on the bound state rapidities $x_{a\b}^{\pm m_\b}$. 
Due to the left-right symmetry it is sufficient to consider the S-matrix elements $S_{Y Y}^{m_1,m_2}(u_{1},u_{2})$ and $S_{Y \bar Z}^{m_1,m_2}(u_{1},u_{2})$.

\subsubsection*{Computation of $S_{Y Y}^{m_1m_2}(u_{1},u_{2})$}

The S-matrix element $S_{Y Y}^{m_1,m_2}(u_{1},u_{2})$ is given by \eqref{eq:SbYbY} with the replacement R$\to$L.
We want to show that
\bal
  \prod^{m_1-1}_{j=1} \left(\frac{x^{+}_{\L 1j} - x^{+m_2}_{\L 2}}{x^{+}_{\L 1j} - x^{-m_2}_{\L 2}} \right)^2
\  \prod^{m_2-1}_{j=1} \left(\frac{x^{- m_1}_{\L1} - x^{+}_{\L 2j}}{x^{+ m_1}_{\L1} - x^{+}_{\L2 j}} \right)^2 \, e^{-2i(\tilde{\theta}^{\bes}_{\L\L}(x_{\L1}^{\pm m_1},x_{\L 2}^{\pm m_2})-\tilde{\theta}^{\hl}_{\L\L}(x_{\L1}^{\pm m_1},x_{\L 2}^{\pm m_2}))}
\eal
depends only on $x^{\pm m_j}_{\L j}$. We begin with $2\tilde{\theta}^{\bes}_{\L\L}(x_{\L1}^{\pm m_1},x_{\L 2}^{\pm m_2})$, see \eqref{eq:thetaBES_in_string_region}, and rewrite the combination of $\tPsi$ functions in it as follows
\bal
&-2\tilde\Psi_{\L}(x_{\L1}^{+m_1},x_{\L 2}^{+m_2})+2\tilde\Psi_{\L}(x_{\L1}^{+m_1},x_{\L 2}^{-m_2})+2\tilde\Psi_{\L}(x_{\L 2}^{+m_2},x_{\L1}^{+m_1})-2\tilde\Psi_{\L}(x_{\L 2}^{+m_2},x_{\L 1}^{-m_1})
\\
=&-\tilde\Psi_{\L}^{+}(x_{\L1}^{+m_1},x_{\L 2}^{+m_2})+\tilde\Psi^{+}_{\L}(x_{\L1}^{+m_1},x_{\L 2}^{-m_2})+\tilde\Psi^{+}_{\L}(x_{\L 2}^{+m_2},x_{\L1}^{+m_1})-\tilde\Psi^{+}_{\L}(x_{\L 2}^{+m_2},x_{\L 1}^{-m_1})
\\
&-\tilde\Psi_{\L}^{-}(x_{\L1}^{-m_1},x_{\L 2}^{+m_2})+\tilde\Psi^{-}_{\L}(x_{\L1}^{-m_1},x_{\L 2}^{-m_2})+\tilde\Psi^{-}_{\L}(x_{\L 2}^{-m_2},x_{\L1}^{+m_1})-\tilde\Psi^{-}_{\L}(x_{\L 2}^{-m_2},x_{\L 1}^{-m_1})
\\
& -\Delta_{\L}^{-}(u_1\pm {i\ov h}m_1,x_{\L 2}^{+ m_2},x_{\L 2}^{- m_2})+\Delta_{\L}^{-}(u_2\pm {i\ov h}m_2,x_{\L 1}^{+ m_1},x_{\L 1}^{- m_1})
\,.
\eal
By using  identity \eqref{eq:Deltaepsm_2ms} for $\tPsi$ functions, we find
\bal
\label{eq:app_DeltaL_ind_const}
e&^{i\Delta_{\L}^{-}(u_1\pm {i\ov h}m_1,x_{\L 2}^{+ m_2},x_{\L 2}^{- m_2})-i\Delta_{\L}^{-}(u_2\pm {i\ov h}m_2,x_{\L 1}^{+ m_1},x_{\L 1}^{- m_1})}
\\
&=\left(\frac{h}{2}\right)^{+2 m_2-2 m_1}
\frac{\prod_{j=1}^{m_2} \left(u_{12}-\frac{i \left(2 j-m_2+m_1\right)}{h}\right) \left(u_{12}-\frac{i \left(-2j+m_1+m_2\right)}{h}\right) }{\prod_{j=1}^{m_1} \left(u_{12}+\frac{i \left(2 j-m_1+m_2\right)}{h}\right) \left(u_{12}+\frac{i \left(-2j+m_1+m_2\right)}{h}\right)}
   \\
   &\times \left(\frac{x_{\L 1}^{- m_1} }{x_{\L 1}^{+ m_1} }\right)^{m_2}\left(\frac{x_{\L 2}^{+ m_2} }{x_{\L 2}^{- m_2} }\right)^{m_1}\frac{ (x_{\L 1}^{+m_1}-x_{\L 2}^{- m_2}) (\tx_{\L 1}^{+m_1}-x_{\L 2}^{- m_2}) }{
   (x_{\L 1}^{-m_1}-x_{\L 2}^{+ m_2}) (\tx_{\L 1}^{+m_1}-x_{\L  2}^{+ m_2})} \frac{ (\tx_{\L 2}^{+m_2}-x_{\L 1}^{+ m_1}) }{
    (\tx_{\L 2}^{+m_2}-x_{\L  1}^{- m_1})} 
   \\
   &\times \prod_{j=1}^{m_1-1}
 \left( \frac{ \tx_{\L 1}^{+(m_1-2j)}-x_{\L 2}^{- m_2} }{
\tx_{\L 1}^{+(m_1-2j)}-x_{\L  2}^{+ m_2} }\right)^2\prod_{j=1}^{m_2-1}
 \left( \frac{ \tx_{\L 2}^{+(m_2-2j)}-x_{\L 1}^{+ m_1} }{
\tx_{\L 2}^{+(m_2-2j)}-x_{\L  1}^{- m_1} }\right)^2\,.
\eal
The last line cancels the corresponding contribution in the S-matrix element. 
Then, we also have in $e^{2i\tilde\theta^\hl(x_{1}^{\pm m_1},x_{2}^{\pm m_2}) }$ (see~\eqref{eq:newHL_string_aa})
\bal
-\frac{x_{\L 1}^{+m_1}}{x_{\L 1}^{-m_1}}   
 \frac{x_{\L 2}^{-m_2}}{x_{\L 2}^{+m_2}}\frac{x_{\L 1}^{+m_1} -x_{\L 2}^{-m_2}}{x_{\L 1}^{-m_1}-x_{\L 2}^{+m_2} }  
\frac{x_{\L 1}^{-m_1}-{\tx^{+m_2}_{\L 2}} }{x_{\L 1}^{+m_1}-{\tx^{+m_2}_{\L 2}} }  
 \frac{{\tx^{+m_1}_{\L 1}} -x_{\L 2}^{+m_2}}{{\tx^{+m_1}_{\L 1}} -x_{\L 2}^{-m_2}}\,.
\eal
We see that the terms with $\tx_{\L j}^{+ m_j}$ cancel out, and indeed 
\begin{multline}
  \prod^{m_1-1}_{j_1=1} \left(\frac{x^{+}_{\L j_1} - x^{+m_2}_{\L 2}}{x^{+}_{\L j_1} - x^{-m_2}_{\L 2}} \right)^2
\  \prod^{m_2-1}_{j_2=1} \left(\frac{x^{- m_1}_{\L1} - x^{+}_{\L j_2}}{x^{+ m_1}_{\L1} - x^{+}_{\L j_2}} \right)^2\\
\times \exp \left[ -2i \left(\tilde{\theta}^{\bes}_{\L\L}(x^{\pm m_1}_{\L1} , x^{\pm m_2}_{\L2})-\tilde{\theta}^{\hl}_{\L\L}(x^{\pm m_1}_{\L1} , x^{\pm m_2}_{\L2}) \right) \right]
\end{multline}
depends only on $x^{\pm m_j}_{\L j}$.

It is possible to show that the terms dependent on $u$ in~\eqref{eq:app_DeltaL_ind_const} partially simplify with $K^\bes(u_{12}+{i\ov h}(m_1-m_2))$ in~\eqref{eq:thetaBES_in_string_region}. Combining all the terms, we get 
\bal
\label{eq:app_SYY_indep}
&S_{Y Y}^{m_1m_2}(u_{1},u_{2})=\left(\frac{x_{\L 1}^{+m_1}}{x_{\L 1}^{-m_1}}\right)^{m_2}\left(\frac{x_{\L 2}^{-m_2}}{x_{\L 2}^{+m_2}}\right)^{m_1}
 \\
 &\times
 \frac{u_{12}- \frac{i}{h} (m_1-m_2)}{u_{12}+ \frac{i}{h} (m_1-m_2)} 
  \frac{u_{12}- \frac{i}{h} (m_1+m_2)}{u_{12}+ \frac{i}{h} (m_1+m_2)}  
 \prod^{m_2-1}_{j=1} \left(\frac{u_{12} + {i\ov h} (2j-m_1-m_2 )}{u_{12} - {i\ov h} (2j-m_1-m_2 )} \right)^2\, \big(\sigma_{\L\L}^{m_1m_2}\big)^{-2}
 \,,
    \eal
 where $\sigma_{\L\L}^{m_1m_2}$ is  the string dressing factor
\bal\la{eq:sigmaLL}
\sigma_{\L\L}^{m_1m_2}(u_1,u_2)^{-2}&= \left(-1\right)^{m_1-m_2+1} 
  \left(\frac{x_{\L 1}^{-m_1}}{x_{\L 1}^{+m_1}}\right)^{m_2-1}
  \left( \frac{x_{\L 2}^{+m_2}}{x_{\L 2}^{-m_2}}\right)^{m_1-1}
  \frac{u_{12}+ \frac{i}{h} (m_1+m_2)}{u_{12}- \frac{i}{h} (m_1+m_2)} 
  \\
  &\times \left({ \Gamma\big[{m_1+m_2\ov2}-\tfrac{ih}{2}u_{12}\big]\ov \Gamma\big[{m_1+m_2\ov2}+\tfrac{ih}{2}u_{12}\big]}\right)^2
   \, \big(\Xi^{m_1 m_2}_{\L\L}(u_1,u_2)\big)^{-2}\,,
\eal
and
\bal
\big(\Xi^{m_1 m_2}_{\L\L}(u_1,u_2)\big)^{-2}=\big(\Sigma^\barnes_{\L\L}(\g_{\L1}^{\pm m_1},\g_{\L2}^{\pm m_2})\big)^{-2} \, e^{-2i\delta^{m_1 m_2}_{\L\L}}\,,
\eal
\bal
&\delta^{m_1 m_2}_{\L\L}= +\tilde\Phi_{\L\L}(x_{\L1}^{+m_1},x_{\L 2}^{+m_2})+ \tilde\Phi_{\L\L}(x_{\L1}^{-m_1},x_{\L 2}^{-m_2})- \tilde\Phi_{\L\L}(x_{\L1}^{+m_1},x_{\L 2}^{-m_2}) - \tilde\Phi_{\L\L}(x_{\L1}^{-m_1},x_{\L 2}^{+m_2})
\\
&\hspace{1.3cm}-\tilde\Phi_{\L\L}^\hl(x_{\L1}^{+m_1},x_{\L 2}^{+m_2})- \tilde\Phi_{\L\L}^\hl(x_{\L1}^{-m_1},x_{\L 2}^{-m_2})+\tilde\Phi_{\L\L}^\hl(x_{\L1}^{+m_1},x_{\L 2}^{-m_2}) + \tilde\Phi_{\L\L}^\hl(x_{\L1}^{-m_1},x_{\L 2}^{+m_2})
\\
&\hspace{1.3cm} -{\tilde\Psi_{\L}^{+}(x_{\L1}^{+m_1},x_{\L 2}^{+m_2})\ov2}-{\tilde\Psi_{\L}^{-}(x_{\L1}^{-m_1},x_{\L 2}^{+m_2})\ov2}+{\tilde\Psi^{+}_{\L}(x_{\L1}^{+m_1},x_{\L 2}^{-m_2})\ov2}+{\tilde\Psi^{-}_{\L}(x_{\L1}^{-m_1},x_{\L 2}^{-m_2})\ov2}
\\
&\hspace{1.3cm}+{\tilde\Psi^{+}_{\L}(x_{\L 2}^{+m_2},x_{\L1}^{+m_1})\ov2}+{\tilde\Psi^{-}_{\L}(x_{\L 2}^{-m_2},x_{\L1}^{+m_1})\ov2}-{\tilde\Psi^{+}_{\L}(x_{\L 2}^{+m_2},x_{\L 1}^{-m_1})\ov2}
-{\tilde\Psi^{-}_{\L}(x_{\L 2}^{-m_2},x_{\L 1}^{-m_1})\ov2}
\,.
\eal
Using the expressions above the S-matrix can also be written in the more compact notation
\begin{multline}
S_{Y Y}^{m_1m_2}(u_{1},u_{2})=\frac{x_{\L 1}^{+m_1}}{x_{\L 1}^{-m_1}} \, \frac{x_{\L 2}^{-m_2}}{x_{\L 2}^{+m_2}} \, (-1)^{m_1-m_2-1}\\
\times \frac{u_{12} + \frac{i}{h}(m_1 - m_2) }{u_{12} - \frac{i}{h}(m_1 - m_2)} \left({ \Gamma\big[{m_1-m_2\ov2}-\tfrac{ih}{2}u_{12}\big]\ov \Gamma\big[{m_1-m_2\ov2}+\tfrac{ih}{2}u_{12}\big]}\right)^2
    \big(\Xi^{m_1 m_2}_{\L\L}(u_1,u_2)\big)^{-2}\,,
\end{multline}
which makes it obvious that the S-matrix element satisfies both braiding and physical unitarity.

\subsubsection*{Computation of $S_{Y \bar Z}^{m_1,m_2}(u_{1},u_{2})$ }

The S-matrix element $S_{Y \bar Z}^{m_1,m_2}(u_{1},u_{2})$ is given by \eqref{eq:SbYZ} with the replacement R$\leftrightarrow$L.
Let us show that
\begin{multline}
\label{eq:app_starting_prod_const_LR}
 \prod_{j=1}^{m_1-1}  \left(\frac{{x^+_{\L 1j} x^{-m_2}_{\R 2}}-1}{{x^+_{\L 1j} x^{+m_2}_{\R 2}}-1}\right)^2 \  \prod^{m_2-1}_{j=1}\left( \frac{{x^{+m_1}_{\L1} x^{+}_{\R 2j}}-1}{{x^{-m_1}_{\L1} x^{+}_{\R 2j}}-1} \right)^2\\
\times \exp \left[ -2i \left(\tilde{\theta}^{\bes}_{\L\R}(x^{\pm m_1}_{\L1} , x^{\pm m_2}_{\R2})-\tilde{\theta}^{\hl}_{\L\R}(x^{\pm m_1}_{\L1} , x^{\pm m_2}_{\R2}) \right) \right]
\end{multline}
depends only on $x^{\pm m_1}_{\L 1}$ and $x^{\pm m_2}_{\R 2}$.
As before, we start from~\eqref{eq:thetaBES_in_string_region} and rewrite the combination of $\tPsi$ in it as follows
\bal
&-2\tilde\Psi_{\R}(x_{\L1}^{+m_1},x_{\R 2}^{+m_2})+2\tilde\Psi_{\R}(x_{\L1}^{+m_1},x_{\R 2}^{-m_2})+2\tilde\Psi_{\L}(x_{\R 2}^{+m_2},x_{\L1}^{+m_1})-2\tilde\Psi_{\L}(x_{\R 2}^{+m_2},x_{\L 1}^{-m_1})
\\
=&-\tilde\Psi_{\R}^{-}(x_{\L1}^{+m_1},x_{\R 2}^{+m_2})+\tilde\Psi^{-}_{\R}(x_{\L1}^{+m_1},x_{\R 2}^{-m_2})+\tilde\Psi^{-}_{\L}(x_{\R 2}^{+m_2},x_{\L1}^{+m_1})-\tilde\Psi^{-}_{\L}(x_{\R 2}^{+m_2},x_{\L 1}^{-m_1})
\\
&-\tilde\Psi_{\R}^{+}(x_{\L1}^{-m_1},x_{\R 2}^{+m_2})+\tilde\Psi^{+}_{\R}(x_{\L1}^{-m_1},x_{\R 2}^{-m_2})+\tilde\Psi^{+}_{\L}(x_{\R 2}^{-m_2},x_{\L1}^{+m_1})-\tilde\Psi^{+}_{\L}(x_{\R 2}^{-m_2},x_{\L 1}^{-m_1})
\\
& -\Delta_{\R}^{+}(u_1\pm {i\ov h}m_1,x_{\R 2}^{+ m_2},x_{\R 2}^{- m_2})+\Delta_{\L}^{+}(u_2\pm {i\ov h}m_2,x_{\L 1}^{+ m_1},x_{\L 1}^{- m_1})
\,.
\eal
Using identity~\eqref{eq:Deltaepsp_2ms}, we obtain
\bal
e&^{i\Delta_{\R}^{+}(u_1\pm {i\ov h}m_1,x_{\R 2}^{+ m_2},x_{\R 2}^{- m_2})-i\Delta_{\L}^{+}(u_2\pm {i\ov h}m_2,x_{\L 1}^{+ m_1},x_{\L 1}^{- m_1})}
\\
&=\left(\frac{h}{2}\right)^{+2 m_2-2 m_1}
\frac{\prod_{j=1}^{m_2} \left(u_{12}-\frac{i \left(2 j-m_2-m_1\right)}{h}\right) \left(u_{12}-\frac{i \left(-2j-m_1+m_2\right)}{h}\right) }{\prod_{j=1}^{m_1} \left(u_{12}+\frac{i \left(2 j-m_1-m_2\right)}{h}\right) \left(u_{12}+\frac{i \left(-2j+m_1-m_2\right)}{h}\right)}
   \\
   &\times \left(\frac{x_{\L 1}^{+ m_1} }{x_{\L 1}^{- m_1} }\right)^{m_2}\left(\frac{x_{\R 2}^{- m_2} }{x_{\R 2}^{+ m_2} }\right)^{m_1}
   \frac{ (x_{\L 1}^{-m_1}x_{\R 2}^{+ m_2}-1) (\tx_{\L 1}^{+m_1}x_{\R 2}^{+ m_2}-1) }{
 (x_{\L 1}^{+ m_1}x_{\R 2}^{-m_2}-1) (\tx_{\L 1}^{+m_1}x_{\R  2}^{- m_2}-1)} 
    \frac{ (\tx_{\R 2}^{+m_2}x_{\L  1}^{- m_1}-1)}{ (\tx_{\R 2}^{+m_2}x_{\L 1}^{+ m_1}-1) }
   \\
   &\times \prod_{j=1}^{m_1-1}
 \left( \frac{ \tx_{\L 1}^{+(m_1-2j)}x_{\R 2}^{+ m_2} -1}{
\tx_{\L 1}^{+(m_1-2j)}x_{\R  2}^{- m_2}-1 }\right)^2 \prod_{j=1}^{m_2-1}
 \left( \frac{\tx_{\R 2}^{+(m_2-2j)}x_{\L  1}^{- m_1}-1 }{ \tx_{\R 2}^{+(m_2-2j)}x_{\L 1}^{+ m_1} -1}\right)^2\,.
\eal
The last line cancels the same contribution in~\eqref{eq:app_starting_prod_const_LR}. 
Taking into account the contribution
\bal
-\frac{x_{\L 1}^{+m_1}}{x_{\L 1}^{-m_1}}   
 \frac{x_{\R 2}^{-m_2}}{x_{\R 2}^{+m_2}}
\frac{ x_{\L1}^{-m_1} x_{\R2}^{+m_2}-1}{x_{\L1}^{+m_1} x_{\R2}^{-m_2}-1} 
\frac{ x_{\L1}^{+m_1}\tx_{\R2}^{+m_2}-1}{x_{\L1}^{-m_1} \tx_{\R2}^{+m_2}-1} 
 \frac{ \tx_{\L1}^{+m_1}x_{\R2}^{-m_2} -1}{ \tx_{\L1}^{+m_1}x_{\R2}^{+m_2}-1}
\eal
from $e^{2i\tilde\theta^\hl_{\L\R}(x_{\L1}^{\pm m_1},x_{\R2}^{\pm m_2}) }$,
we see that the terms with $\tx_{\L 1}^{+ m_1}$ and $\tx_{\R 2}^{+ m_2}$ cancel out.
Finally 
\begin{multline}
     \prod_{j=1}^{m_1-1}
 \left( \frac{ \tx_{\L 1}^{+(m_1-2j)}x_{\R 2}^{+ m_2} -1}{
\tx_{\L 1}^{+(m_1-2j)}x_{\R  2}^{- m_2}-1 }\right)^2 \prod_{j=1}^{m_2-1}
 \left( \frac{\tx_{\R 2}^{+(m_2-2j)}x_{\L  1}^{- m_1}-1 }{ \tx_{\R 2}^{+(m_2-2j)}x_{\L 1}^{+ m_1} -1}\right)^2\\
 \times \exp \left[ -2i \left(\tilde{\theta}^{\bes}_{\L\R}(x^{\pm m_1}_{\L1} , x^{\pm m_2}_{\R2})-\tilde{\theta}^{\hl}_{\L\R}(x^{\pm m_1}_{\L1} , x^{\pm m_2}_{\R2}) \right) \right]
\end{multline}
depends only on $x_{\L 1}^{\pm m_1}$ and $x_{\R 2}^{\pm m_2}$.

Combining the $u$-dependent terms in  $\exp(-2i\tilde{\theta}^{\bes}_{\L\R} (x^{\pm m_1}_{\L1}, x^{\pm m_2}_{\L2}))$ and collecting all the terms we obtain
 \bal
 \label{eq:app_SYbZ_indep}
&S_{Y \bar Z}^{m_1,m_2}(u_{1},u_{2})=
  \left(\frac{x_{\L 1}^{+m_1}}{x_{\L 1}^{-m_1}}\right)^{m_2-1}
  \left( \frac{x_{\R 2}^{-m_2}}{x_{\R 2}^{+m_2}}\right)^{m_1}\ \frac{1-{1\ov x^{-m_1}_{\L1} x^{-m_2}_{\R 2}}}{1-{1\ov x^{+m_1}_{\L1} x^{+m_2}_{\R 2}}} \ \frac{1-{1\ov x^{-m_1}_{\L1} x^{+m_2}_{\R 2}}}{1-{1\ov x^{+m_1}_{\L1} x^{-m_2}_{\R 2}}}\,  \big(\sigma_{\L\R}^{m_1m_2}\big)^{-2}
 \,,
    \eal    
where
$ \big(\sigma_{\L\R}^{m_1m_2}\big)^{-2}$  is the string dressing factor
\bal\la{eq:sigmaLR}
\sigma_{\L\R}^{m_1m_2}(u_1,u_2)^{-2}&= \left(-1\right)^{m_1-m_2+1} 
  \left(\frac{x_{\L 1}^{-m_1}}{x_{\L 1}^{+m_1}}\right)^{m_2-1}
  \left( \frac{x_{\R 2}^{+m_2}}{x_{\R 2}^{-m_2}}\right)^{m_1-1}
  \frac{u_{12}+ \frac{i}{h} (m_1+m_2)}{u_{12}- \frac{i}{h} (m_1+m_2)} 
  \\
  &\times \left({ \Gamma\big[{m_1+m_2\ov2}-\tfrac{ih}{2}u_{12}\big]\ov \Gamma\big[{m_1+m_2\ov2}+\tfrac{ih}{2}u_{12}\big]}\right)^2
   \, \big(\Xi^{m_1 m_2}_{\L\R}(u_1,u_2)\big)^{-2}\,,
\eal
and
\bal
  \big(\Xi^{m_1 m_2}_{\L\R}(u_1,u_2)\big)^{-2}= \big(\Sigma^\barnes_{\L\R}(\g_{\L1}^{\pm m_1},\g_{\R2}^{\pm m_2})\big)^{-2} \,  e^{-2i\delta^{m_1 m_2}_{\L\R}}\,,
\eal
\bal
&\delta^{m_1 m_2}_{\L\R}= +\tilde\Phi_{\L\R}(x_{\L1}^{+m_1},x_{\R 2}^{+m_2})+ \tilde\Phi_{\L\R}(x_{\L1}^{-m_1},x_{\R 2}^{-m_2})- \tilde\Phi_{\L\R}(x_{\L1}^{+m_1},x_{\R 2}^{-m_2}) - \tilde\Phi_{\L\R}(x_{\L1}^{-m_1},x_{\R 2}^{+m_2})
\\
&\hspace{1.3cm}-\tilde\Phi_{\L\R}^\hl(x_{\L1}^{+m_1},x_{\R 2}^{+m_2})- \tilde\Phi_{\L\R}^\hl(x_{\L1}^{-m_1},x_{\R 2}^{-m_2})+\tilde\Phi_{\L\R}^\hl(x_{\L1}^{+m_1},x_{\R 2}^{-m_2}) + \tilde\Phi_{\L\R}^\hl(x_{\L1}^{-m_1},x_{\R 2}^{+m_2})
\\
&\hspace{1.3cm}-{\tilde\Psi_{\R}^{-}(x_{\L1}^{+m_1},x_{\R 2}^{+m_2})\ov2}+{\tilde\Psi^{-}_{\R}(x_{\L1}^{+m_1},x_{\R 2}^{-m_2})\ov2}+{\tilde\Psi^{-}_{\L}(x_{\R 2}^{+m_2},x_{\L1}^{+m_1})\ov2}-{\tilde\Psi^{-}_{\L}(x_{\R 2}^{+m_2},x_{\L 1}^{-m_1})\ov2}
\\
&\hspace{1.3cm}-{\tilde\Psi_{\R}^{+}(x_{\L1}^{-m_1},x_{\R 2}^{+m_2})\ov2}+{\tilde\Psi^{+}_{\R}(x_{\L1}^{-m_1},x_{\R 2}^{-m_2})\ov2}+{\tilde\Psi^{+}_{\L}(x_{\R 2}^{-m_2},x_{\L1}^{+m_1})\ov2}-{\tilde\Psi^{+}_{\L}(x_{\R 2}^{-m_2},x_{\L 1}^{-m_1})\ov2}\,.
\eal
This form makes it obvious that the S-matrix element satisfies the physical unitarity. 

Formulas~\eqref{eq:sigmaLL} and~\eqref{eq:sigmaLR} are valid for $-m_j/h<\Im(u_j)<m_j/h$, and therefore $|\Im(u_{12})|<(m_1+m_2)/h$. The double poles in the strip are the expected ones for these S-matrix elements.
The cuts of $\tPsi$'s in the strips $-m_j/h<\Im(u_j)<m_j/h$ cancel out, and therefore the S-matrix elements $S^{m_1, m_2}_{Y Y}$ and $S^{m_1, m_2}_{Y \bar{Z}}$ are meromorphic functions there.

\subsection{ \texorpdfstring{$\s_{ab}^{m_1m_2}$}{sabm1m2}  in the Ramond-Ramond case}
\label{app:equivalence_pure_RR}

Here we show that in the RR case the string dressing factors $\s_{ab}^{m_1m_2}$ coincide with the ones proposed in \cite{Frolov:2021fmj}. 
In fact since the odd factors obviously reduce for $k=0$ to the ones  in \cite{Frolov:2021fmj}, it is sufficient to show that the ratio of the HL and BES string phases is equal to  \eqref{eq:sigmaLL} (or  \eqref{eq:sigmaLR}) without the odd factor in $\Xi^{m_1 m_2}_{\L\L}$ (or in $\Xi^{m_1 m_2}_{\L\R}$). 

\subsubsection*{$\Phi$ function}

For $k=0$ there is no difference between $x_\L$ and $x_\R$, and we have the usual Zhukovsky variables 
\bal
u(x)=x+ \frac{1}{x}\quad \Leftrightarrow \quad x=x(u)\quad \text{or}\quad x=\tx(u) \,.
\eal
Then, the string $\Phi$  and the mirror $\tPhi$ functions can be defined as 
\begin{equation}
\Phi(x_1,x_2)=\Phi_+(x_1,x_2)-\Phi_+(x_2,x_1)\,,\quad \tPhi(x_1,x_2)=\tPhi_+(x_1,x_2)-\tPhi_+(x_2,x_1)\,,
\end{equation}
where we used the definitions in~\eqref{eq:app_Phip_Ram_Ram} and~\eqref{eq:app_tPhip_Ram_Ram}. We now assume that both $x_1$ and $x_2$ are outside the unit circle, and therefore can be written as $x_i=x(u_i)$, and first deform the $v_2$-integration contour, and then the $v_1$ contour. The contour deformation is the same as the one already described in appendix~\ref{app:stringvsmirror}; the only difference is that now (compared to appendix~\ref{app:stringvsmirror}) the points are in the string region (instead of being in the mirror region) and some residues generated by performing the contour deformation have an opposite sign.

Let us deform the $v_2$ contour first, which must be done by moving it to the upper half of the complex plane (see discussion in appendix~\ref{app:stringvsmirror}).
If $\Im(u_2)>0$,  we have an extra contribution from deforming the upper edge of the string cut due to the pole at $v_2=u_2$ on the string $v_2$-plane. We recall that in appendix~\ref{app:stringvsmirror} the extra contribution was coming from the deformation of the contour on the lower edge of the cut and the residue is the opposite of the one in~\eqref{eq:app_phip_one_cont_def}.  Then the result of the deformation of the $v_2$ contour is
\bal
\Phi_+(x_1,x_2)
&=i\oint\limits_{\rm cut}\frac{{\rm d}v_1}{2\pi i}\lint_{\widetilde{\rm cuts}} \frac{{\rm
d}v_2}{2\pi i}\frac{x'(v_1) \tx'(v_2)}{(x(v_1)-x_1)(\tx(v_2)-x_2)} \log \Gamma \bigl[1+\tfrac{ih}{
2}\big(v_1-v_2\bigl)\bigl]
\\
&- \theta\big(\Im(u_2)\big)\Psi_-(x_2,x_1)\,.
\eal

The deformation of the $v_1$ contour is identical to the one already discussed in appendix~\ref{app:stringvsmirror} and must be done in the lower half of the complex plane. If $\Im(u_1)<0$ we need to cross a pole and we get
\bal
\Phi_+(x_1,x_2)
&={1\ov i}\lint_{\widetilde{\rm cuts}}\frac{{\rm d}v_1}{2\pi i}\lint_{\widetilde{\rm cuts}} \frac{{\rm
d}v_2}{2\pi i}\frac{\tx'(v_1) \tx'(v_2)}{(\tx(v_1)-x_1)(\tx(v_2)-x_2)} \log \Gamma \bigl[1+\tfrac{ih}{
2}\big(v_1-v_2\bigl)\bigl]
\\
&+\theta\big(-\Im(u_1)\big)\tilde\Psi_+(x_1,x_2)- \theta\big(\Im(u_2)\big)\Psi_-(x_2,x_1)\,.
\eal
To relate $\Psi_-(x_2,x_1)$ to $\tPsi_-(x_2,x_1)$ we deform the $v_1$ contour in $\Psi_-(x_2,x_1)$ in the same way as above and get 
\bal
\Psi_-(x_1,x_2)
&=\tilde\Psi_-(x_1,x_2)- \theta\big(-\Im(u_1)\big)i\log \Gamma \bigl[1+\tfrac{ih}{
2}\big(u_1-u_2\bigl)\bigl]\,.
\eal
Thus,
the final  result  of the deformation of the $v_1$ and $v_2$ contours is
\bal\la{Phipx1x2}
\Phi_+(x_1,x_2)
=&+\tilde{\Phi}_+(x_1,x_2)+\theta\big(-\Im(u_1)\big)\tilde\Psi_+(x_1,x_2)- \theta\big(\Im(u_2)\big)\tPsi_-(x_2,x_1)
\\
&+ \theta\big(-\Im(u_1)\big)\theta\big(\Im(u_2)\big)i\log \Gamma \bigl[1+\tfrac{ih}{
2}u_{12}\bigl]\,.
\eal

\subsubsection*{String $\Phi$ in term of mirror $\tilde\Phi$}

We use \eqref{Phipx1x2} and setting $x_1^\pm = x(u_1\pm {i m_1\ov h})$, $x_2^\pm = x(u_2\pm {i m_2\ov h})$ (see also~\eqref{eq:app_var_Phi_string}), we get
\bal\la{strPhi1}
\varPhi&(x_1^\pm,x_2^\pm)=+\Phi(x_1^+,x_2^+)+\Phi(x_1^-,x_2^-)-\Phi(x_1^+,x_2^-)-\Phi(x_1^-,x_2^+)
\\
 =&+\tilde\Phi(x_1^+,x_2^+)-\tPsi_-(x_2^+,x_1^+)+\tPsi_-(x_1^+,x_2^+)
\\
&+\tilde\Phi(x_1^-,x_2^-)+\tPsi_+(x_1^-,x_2^-)-\tPsi_+(x_2^-,x_1^-)
\\
&-\tilde\Phi(x_1^+,x_2^-)+ \tPsi_+(x_2^-,x_1^+)-\tPsi_-(x_1^+,x_2^-)
+ i\log \Gamma \bigl[1+\tfrac{ih}{
2}\big(u_{21} -\tfrac{i}{h}(m_1+m_2)\big)\bigl]
\\
&-\tilde\Phi(x_1^-,x_2^+)- \tPsi_+(x_1^-,x_2^+)
+\tPsi_-(x_2^+,x_1^-)
- i\log \Gamma \bigl[1+\tfrac{ih}{
2}\big(u_{12} -\tfrac{i}{h}(m_1+m_2)\big)\bigl]
\,.
\eal
This formula can be cast in the form
\bal\la{strPhi3}
\varPhi&(x_1^\pm,x_2^\pm)=+\tilde\varPhi(x_1^\pm,x_2^\pm)-{\tPsi(x_1^+,x_2^+)\ov2}+{\tPsi(x_2^+,x_1^+)\ov2}+{\tPsi(x_1^-,x_2^-)\ov2}-{\tPsi(x_2^-,x_1^-)\ov2}
\\
&
+{\tPsi(x_1^+,x_2^-)\ov2}+{\tPsi(x_2^-,x_1^+)\ov2}-{\tPsi(x_1^-,x_2^+)\ov2}-{\tPsi(x_2^+,x_1^-)\ov2}
\\
&
+{\Delta_+(u_1\pm\tfrac{i m_1}{h},x_2^\pm)+\Delta_-(u_1\pm\tfrac{i m_1}{h},x_2^\pm) \ov2}
-{\Delta_+(u_2\pm\tfrac{i m_2}{h},x_1^\pm)+\Delta_-(u_2\pm\tfrac{i m_2}{h},x_1^\pm) \ov2}
\\
&+ i\log \left(-\frac{u_{12}+ \frac{i}{h} (m_1+m_2)}{u_{12}- \frac{i}{h} (m_1+m_2)} \,{ \Gamma\big[{m_1+m_2\ov2}-\tfrac{ih}{2}u_{12}\big]\ov \Gamma\big[{m_1+m_2\ov2}+\tfrac{ih}{2}u_{12}\big]}\right)
\,,
\eal
where 
\bal
\Delta_\pm(u_1\pm\tfrac{i m_1}{h},x_2^\pm) = \tPsi_\pm(x_1^+,x_2^+)-\tPsi_\pm(x_1^-,x_2^+)-\tPsi_\pm(x_1^+,x_2^-)+\tPsi_\pm(x_1^-,x_2^-)\,.
\eal
It is easy to show that
\bal
\Delta_+(u_1\pm\tfrac{i m_1}{h},x_2^\pm)=-{1\ov i}\sum_{j=1}^{m_1} \left(\log \big[j-{m_1+m_2\ov2}+\tfrac{ih}{
2}u_{12}\big]-\log {\tx_1^{m_1-2j}-x_2^-\ov \tx_1^{m_1-2j} - x_2^+}\right)\,,
\eal
\bal
\Delta_-(u_1\pm\tfrac{i m_1}{h},x_2^\pm)
=+{1\ov i}\sum_{j=1}^{m_1}\left( \log \big[j-{m_1-m_2\ov2}-\tfrac{ih}{
2}u_{12}\big]-\log {\tx_1^{m_1-2j}-x_2^-\ov \tx_1^{m_1-2j}- x_2^+}\left(x_2^+\ov x_2^-\right)^{m_1}\right)\,,
\eal
By using the formulae,  we get
\bal\la{strPhi5}
&2 \, \varPhi(x_1^\pm,x_2^\pm)=2 \, \tilde\varPhi(x_1^\pm,x_2^\pm)-{\tPsi(x_1^+,x_2^+)}+{\tPsi(x_2^+,x_1^+)}+{\tPsi(x_1^-,x_2^-)}-{\tPsi(x_2^-,x_1^-)}
\\
&
+{\tPsi(x_1^+,x_2^-)}+{\tPsi(x_2^-,x_1^+)}-{\tPsi(x_1^-,x_2^+)}-{\tPsi(x_2^+,x_1^-)}
\\
&+ i\log\left(-1\right)^{m_1-m_2}\left(x_1^-\ov x_1^+\right)^{m_2} \left(x_2^+\ov x_2^-\right)^{m_1} \frac{u_{12}+ \frac{i}{h} (m_1+m_2)}{u_{12}- \frac{i}{h} (m_1+m_2)} \,\left({ \Gamma\big[{m_1+m_2\ov2}-\tfrac{ih}{2}u_{12}\big]\ov \Gamma\big[{m_1+m_2\ov2}+\tfrac{ih}{2}u_{12}\big]}\right)^2
\\
&+{1\ov i}\log {\tx_1^{+m_1}- x_2^+\ov \tx_1^{+m_1}-x_2^-} {\tx_1^{-m_1}-x_2^-\ov \tx_1^{-m_1}- x_2^+}
{\tx_2^{+m_2}-x_1^-\ov \tx_2^{+m_2}- x_1^+} {\tx_2^{-m_2}- x_1^+\ov \tx_2^{-m_2}-x_1^-}
\,.
\eal

\subsubsection*{$\Phi^\hl$ function}

Similarly, we  define
\begin{equation}
\Phi^\hl(x_1,x_2)=\Phi^\hl_+(x_1,x_2)-\Phi^\hl_+(x_2,x_1)
\end{equation}
with
\bal
\Phi^\hl_+(x_1,x_2)=\oint\limits_{\rm cut}\frac{{\rm d}v_1}{2\pi i}\oint\limits_{\rm cut} \frac{{\rm
d}v_2}{2\pi i}\frac{x'(v_1) x'(v_2)}{(x(v_1)-x_1)(x(v_2)-x_2)}\frac{i}{2}\log \big(\eps+i v_{12}\big)\,,
\eal
and introduce the functions
\bal
\Psi^\hl_\pm(x_1,x_2)=+\oint\limits_{\rm cut} \frac{{\rm
d}v}{2\pi i}\,\frac{x'(v)}{x(v)-x_2}\frac{i}{2}\log \big(\eps \pm i (u_1-v)\big)\,,
\eal
\bal
\tilde{\Psi}^\hl_\pm(x_1,x_2)=-\lint_{\widetilde{\rm cuts}} \frac{{\rm
d}v}{2\pi i}\,\frac{\tx'(v)}{\tx(v)-x_2} \frac{i}{2}\log \big(\eps \pm i (u_1-v)\big)\,.
\eal
As before we deform the $v_2$ contour first, and then the $v_1$ contour.
We note that the function
\bal
\log \big(\eps+i v_{12}\big)
\eal
has a branch point at 
\bal
v_2=v_1- i\eps
\eal
with a cut in the $v_2$-plane running vertically to $-i \infty$. Thus, we deform the integration contours as for $\Phi$. The  final  result  of the deformation of the $v_1$ and $v_2$ contours is
\bal\la{Phihlpx1x2}
\Phi^\hl_+(x_1,x_2)
=&+\tilde{\Phi}^\hl_+(x_1,x_2)+\theta\big(-\Im(u_1)\big)\tilde\Psi^\hl_+(x_1,x_2)- \theta\big(\Im(u_2)\big)\tPsi^\hl_-(x_2,x_1)
\\
&+ \theta\big(-\Im(u_1)\big)\theta\big(\Im(u_2)\big)\frac{i}{2}\log \big(\eps+i u_{12}\big)\,.
\eal

\subsubsection*{String $\Phi^\hl$ in term of mirror $\tilde\Phi^\hl$}

We use \eqref{Phihlpx1x2} and recalling~\eqref{eq:app_varPhi_functions} we get ($x_1^\pm = x(u_1\pm {i m_1\ov h})$, $x_2^\pm = x(u_2\pm {i m_2\ov h})$)
\bal\la{strPhihl2}
\varPhi^\hl(x_1^\pm,x_2^\pm)=&+\tilde\varPhi^\hl(x_1^\pm,x_2^\pm)
+\Delta^\hl(u_1\pm\tfrac{i m_1}{h},x_2^\pm)
-\Delta^\hl(u_2\pm\tfrac{i m_2}{h},x_1^\pm)
\\
&-  \frac{1}{2i}\log \big(-i ( u_{12} +\tfrac{i}{h}(m_1+m_2))\big)
+ \frac{1}{2i}\log \big(+i (u_{12} -\tfrac{i}{h}(m_1+m_2))\big)
\,,
\eal
where 
\bal
\Delta^\hl(u_1\pm\tfrac{i m_1}{h},x_2^\pm) = \tPsi^\hl_-(x_1^+,x_2^+)-\tPsi^\hl_-(x_1^+,x_2^-)-\tPsi^\hl_+(x_1^-,x_2^+)+\tPsi^\hl_+(x_1^-,x_2^-)\,.
\eal
Next, we  find
\bal
\Delta^\hl(u_1\pm\tfrac{i m_1}{h},&x_2^\pm)={1\ov 2i}\int_{\tilde{\rm cuts}} \frac{{\rm
d}v}{2\pi i}\left(\frac{\tx'(v)}{\tx(v)-x_2^-}-\frac{\tx'(v)}{\tx(v)-x_2^+}\right) \log { \eps+i\big(u_1-\tfrac{i m_1}{h}-v\big)\ov \eps-i\big(u_1+\tfrac{i m_1}{h}-v\big)}
\\
&={i\ov 2}\log {+i\big(u_{12}-\tfrac{i (m_1+m_2)}{h}\big)\ov -i\big(u_{12}+\tfrac{i (m_1-m_2)}{h}\big)}+{1\ov 2i} \log {\tx_1^{-m_1}-x_2^-\ov \tx_1^{-m_1}- x_2^+} {\tx_1^{+m_1}-x_2^+\ov \tx_1^{+m_1}- x_2^-}-{1\ov 2i}\log{x_2^+\ov x_2^-}\,,
\eal
and therefore the ratio of the HL and BES string factors is equal to
\bal\la{strPhihl6}
e^{-i(2\varPhi(x_1^\pm,x_2^\pm)
-2\varPhi^\hl(x_1^\pm,x_2^\pm))}&=e^{-2i\delta^{m_1 m_2}}
\left(-1\right)^{m_1-m_2+1} 
  \left(\frac{x_{ 1}^{-m_1}}{x_{ 1}^{+m_1}}\right)^{m_2-1}
  \left( \frac{x_{ 2}^{+m_2}}{x_{ 2}^{-m_2}}\right)^{m_1-1}
\\
&
\times  \frac{u_{12}+ \frac{i}{h} (m_1+m_2)}{u_{12}- \frac{i}{h} (m_1+m_2)} 
  \left({ \Gamma\big[{m_1+m_2\ov2}-\tfrac{ih}{2}u_{12}\big]\ov \Gamma\big[{m_1+m_2\ov2}+\tfrac{ih}{2}u_{12}\big]}\right)^2
\,.
\eal
This is exactly the string dressing factor \eqref{eq:sigmaLL} without the odd factor.

\section{Continuation to  the region  \texorpdfstring{$-2\pi <p<0$}{m2pip0}}\la{app:pto2pimp}

Here we calculate the diagonal S-matrix elements  $S_{\bar Y\bar Y}^{m_1m_2}(p_1-2\pi,p_2)$, $S_{\bar Y Z}^{m_1m_2}(p_1-2\pi,p_2)$,  $S_{YY}^{m_1m_2}(p_1-2\pi,p_2)$, $S_{Y\bar Z}^{m_1m_2}(p_1-2\pi,p_2)$
 by performing the two steps of the 
 analytic continuation  to the region $-2\pi <p<0$.

\subsection{Calculating  \texorpdfstring{$S_{\bar Y \bar Y}^{m_1m_2}(p_{1}-2\pi, p_{2})$}{sbybym1m2} }

The right-right S-matrix element $S_{\bar Y \bar Y}^{m_1m_2}(p_1,p_2)$ is given by \eqref{eq:SbYbY}.
The factor $A_{\bar Y  \bar Y}^{m_1m_2}(p_{1},p_{2})$ after the two steps $x_{\R1}^{\pm m_1}\to x_{\R1}^{\pm (m_1+k)}$, $u_1\to u_1-{i\ov h}k$ becomes
\bal
 A_{\bar Y  \bar Y}^{m_1m_2}&(p_{1},p_{2})
  \to A_{\bar Y  \bar Y}^{m_1m_2}(p_{1}-2\pi,p_{2})
 =
     \left({x_{\R1}^{+(m_1+k)}\ov x_{\R1}^{-(m_1+k)}}\right)^{m_2}   \left({x_{\R2}^{-m_2}\ov x_{\R2}^{+m_2}}\right)^{m_1} \left(\frac{x^{- (m_1+k)}_{\R1} - x^{+ m_2}_{\R2}}{x^{+(m_1+k)}_{\R1} - x^{- m_2}_{\R2}} \right)^2
      \\
  &\times  \prod^{m_1-1}_{j=1} \left(\frac{x^{+}_{\R1 j} - x^{+m_2}_{\R 2}}{x^{+}_{\R1 j} - x^{-m_2}_{\R2}} \right)^2
  \prod^{m_2-1}_{j=1} \left(\frac{x^{- (m_1+k)}_{\R1} - x^{+}_{\R2 j}}{x^{+(m_1+k)}_{\R1} - x^{+}_{\R2 j}} \right)^2 \left(\frac{u_{12}+ \frac{i}{h} (m_1-m_2-k)}{u_{12}- \frac{i}{h} (m_1-m_2+k)} \right) \, 
   \\
  &\times\left(\frac{u_{12}+ \frac{i}{h} (m_1+m_2-k)}{u_{12}- \frac{i}{h} (m_1+m_2+k)} \right) 
  \prod^{m_2-1}_{j=1} \left(\frac{u_{12} + {i\ov h} (m_1+m_2-2 j-k)}{u_{12} - {i\ov h} (m_1+m_2-2 j+k)} \right)^2 
  \,,
\eal
 where
\bal
x_{\R 1 j}^+ =\tx_\R(u_1 - {i\ov h}(k+m_1-2j))\,,\quad  x_{\R2 j}^+ =\tx_\R(u_2 - {i\ov h}(m_2-2j))\,.
\eal

Next, the first step of
the analytic continuation moves the first particle to the mirror region and replaces $x_{\R 1}^{\pm m_1}\to \tx_{\R 1}^{\pm m_1}$, 
$\g_{\R 1}^{\pm m_1}\to \tg_{\R 1}^{\pm m_1}$, and $\Sigma^{\bes}_{\R\R}$ and $\Sigma^{\hl}_{\R\R}$ are  given by
\bal\la{BESRRmirrorstring}
\tilde\theta_{\R\R}^\bes( \tx_{\R 1}^{\pm m_1},x^{\pm m_2}_{\R 2})  &=  \tilde\Phi_{\R\R}( \tx_{\R 1}^{+ m_1},x^{+m_2}_{\R 2})+ \tilde\Phi_{\R\R}( \tx_{\R 1}^{- m_1},x^{-m_2}_{\R 2})- \tilde\Phi_{\R\R}( \tx_{\R 1}^{+ m_1},x^{-m_2}_{\R 2}) 
\\
&- \tilde\Phi_{\R\R}( \tx_{\R 1}^{- m_1},x^{+m_2}_{\R 2})
+\tilde\Psi_{\R}(x^{+m_2}_{\R 2}, \tx_{\R 1}^{+ m_1})-\tilde\Psi_{\R}(x^{+m_2}_{\R 2}, \tx_{\R 1}^{- m_1})
\,,
\eal
\bal\la{HRRLmirrorstring}
\tilde\theta_{\R\R}^\hl( \tx_{\R 1}^{\pm m_1},x^{\pm m_2}_{\R 2})  &=  \tilde\Phi_{\R\R}^\hl( \tx_{\R 1}^{+ m_1},x^{+m_2}_{\R 2})+ \tilde\Phi_{\R\R}^\hl( \tx_{\R 1}^{- m_1},x^{-m_2}_{\R 2})- \tilde\Phi_{\R\R}^\hl( \tx_{\R 1}^{+ m_1},x^{-m_2}_{\R 2}) 
\\
&- \tilde\Phi_{\R\R}^\hl( \tx_{\R 1}^{- m_1},x^{+m_2}_{\R 2})+{1\ov 2i}\log \frac{\tx^{+m_1}_{\R1}}{\tx^{-m_1}_{\R1} } \frac{\tx^{+m_1}_{\R1}-x^{+m_2}_{\R 2}}{\tx^{+m_1}_{\R1}-{\tx^{+m_2}_{\R 2}} }  \frac{\tx^{-m_1}_{\R1}-{\tx^{+m_2}_{\R 2}} } {\tx^{-m_1}_{\R1}-x^{+m_2}_{\R 2} }
\,.
\eal
Then, we need to analytically continue $\tPhi_{\R\R}( \tx_{\R 1}^{\pm m_1},x^{\pm m_2}_{\R 2}) $ and $\tPhi_{\R\R}^\hl( \tx_{\R 1}^{\pm m_1},x^{\pm m_2}_{\R 2})$ in $\tx_{\R1}^{+m_1 }$ through the negative semi-axis to the upper half-plane.

\subsubsection*{Continuation of $\tPhi_{\R\R}$ }

We first shift $u_1$ variable $u_1\to u_1-{i\ov h}k$ so that 
$\tx_{\R 1}^{\pm m_1}\to\tx_\R(u_1\pm{i\ov h}m_1-{i\ov h}k)=\tx_{\R 1}^{\pm m_1-k}$, and then analytically continue  $\tx_{\R 1}^{+m_1-k }$ to $1/\tx_{\L 1}^{+ (m_1+k)}=x_{\R 1}^{+ (m_1+k)}$, $u_1\in\bR$. We get
\bal
\tPhi_{\R\R}^{--}( \tx_{\R 1}^{+ m_1},x^{\pm m_2}_{\R 2})\to &+\tPhi_{\R\R}^{--}({x_{\R 1}^{+ (m_1+k)}},x_{\R 2}^{\pm m_2}) - \tPsi_{\R}^{-}(u_1+{i\ov h}(m_1-k),x_{\R 2}^{\pm m_2})\,,
\\
\tPhi_{\R\R}^{++}( \tx_{\R 1}^{+ m_1},x^{\pm m_2}_{\R 2})\to &+\tPhi_{\R\R}^{++}({ x_{\R 1}^{+ (m_1+k)}},x_{\R 2}^{\pm m_2 }) - \tPsi_{\R}^{+}(u_1+{i\ov h}(m_1+k),x_{\R 2}^{\pm  m_2})\,,
\eal
\bal
\tilde\Psi^{-}_{\R}(u_{2} + {i\ov h}m_2, \tx_{\R 1}^{+ m_1})\to & \  \tilde\Psi^{-}_{\R}(u_{2} + {i\ov h}m_2, x_{\R 1}^{+ (m_1+k)})-K^\bes(u_{12} +{i\ov h}(m_1-m_2)-{i\ov h}k)\,,
\\
\tilde\Psi^{+}_{\R}(u_{2} + {i\ov h}m_2, \tx_{\R 1}^{+ m_1})\to & \ \tilde\Psi^{+}_{\R}(u_{2} + {i\ov h}m_2, x_{\R 1}^{+ (m_1+k)})-K^\bes(u_{12}+{i\ov h}(m_1-m_2) +{i\ov h}k)\,.
\eal
Note that we do not cross any cut of $\tPsi$-functions.

Summing up the $\tPhi$ functions, we get
\bal
2\tilde\theta_{\R\R}^\bes&(\tx_{\R 1}^{\pm m_1},x_{\R 2}^{\pm m_2})\to 
2\tilde\theta_{\R\R}^\bes({ x_{\R 1}^{\pm (m_1+k)}},{x_{\R 2}^{\pm m_2 }})
+\Delta_{\R}^{-}(u_1+{i\ov h}m_1\pm {i\ov h}k,x_{\R 2}^{\pm m_2})
\\
&-K^\bes(u_{12}+{i\ov h}(m_1-m_2) -{i\ov h}k)+K^\bes(u_{12} +{i\ov h}(m_1-m_2)+{i\ov h}k)
\,.
\eal
By using  identity \eqref{eq:Deltaepsm_2ms} for $\tPsi$ functions, 
and the identity
\bal\la{eq:KbesmKbes}
\exp[i(K^\bes&(u -{i\ov h}n)-K^\bes(u+{i\ov h}n))]
={\left(\frac{h}{2}\right)^{-2 n}\ov \prod
   _{j=1}^{n} \left(u+\frac{i \left(2
   j-n\right)}{h}\right) \left(u+\frac{i \left(-2
   j+n\right)}{h}\right)}
\,,
\eal
we get 
\bal\la{eqqq1a}
&e^{-2i\tilde\theta_{\R\R}^\bes(\tx_{\R 1}^{\pm m_1},x_{\R 2}^{\pm m_2})}\to 
e^{-2i\tilde\theta_{\R\R}^\bes({ x_{\R 1}^{\pm (m_1+k)}},{x_{\R 2}^{\pm m_2 }})}
 \\
&\times \prod_{j=1}^k{\big(u_{12}+{i\ov h}(m_1+m_2-k+2j)\big)\big(u_{12}+{i\ov h}(m_1+m_2+k-2j)\big)\ov
 \big(u_{12}+{i\ov h}(m_1-m_2-k+2j)\big)\big(u_{12}+{i\ov h}(m_1-m_2+k-2j)\big) }
 \\
&\times
\left(\frac{x_{\R 2}^{- m_2} }{x_{\R 2}^{+ m_2} }\right)^{k}
 \frac{ (\tx_{\R 1}^{-(k-m_1)}-x_{\R 2}^{- m_2}) (\tx_{\R1}^{+(k+m_1)}-x_{\R 2}^{+ m_2}) } {
  (\tx_{\R 1}^{-(k-m_1)}-x_{\R 2}^{+ m_2})(\tx_{\R 1}^{+(k+m_1)}-x_{\R 2}^{- m_2}) }
\prod_{j=m_1}^{m_1+k-1}
 \left( 
 \frac{ \tx_{\R1}^{2j-m_1-k}-x_{\R 2}^{+ m_2} } {
\tx_{\R1}^{2j-m_1-k}-x_{\R 2}^{- m_2}}\right)^2
 \,.
\eal
One can then show that the product of the factor $A_{\bar Y  \bar Y}^{m_1,m_2}(p_{1}-2\pi,p_{2})$ and the factor in \eqref{eqqq1a}  is equal to the factor $A_{\bar Y  \bar Y}^{m_1+k,m_2}(p_{1},p_{2})$ in $S_{\bar Y  \bar Y}^{m_1+k,m_2}(p_{1},p_{2})$ multiplied by the following factor
\bal\la{eqqq2a}
 \frac{ (\tx_{\R 1}^{-(k-m_1)}-x_{\R 2}^{- m_2}) (\tx_{\R1}^{+(k+m_1)}-x_{\R 2}^{+ m_2}) } {
  (\tx_{\R 1}^{-(k-m_1)}-x_{\R 2}^{+ m_2})(\tx_{\R 1}^{+(k+m_1)}-x_{\R 2}^{- m_2}) }\,.
  \eal
  
\subsubsection*{Continuation of $\tPhi_{\R\R}^\hl$}

Next, we analytically continue $\tPhi_{\R\R}^\hl$ in $\tx_{\R 1}^{+ m_1}\to x_{\R 1}^{+ (m_1+k)}$, 
and  get
\bal
\tPhi_{\R\R}^{\hl}(\tx_{\R1}^{+ m_1},x_{\R 2}^{\pm m_2})\to &+\tPhi_{\R\R}^{\hl}( x_{\R 1}^{+ (k+m_1)},x_{\R 2}^{\pm m_2})+  {1\ov 2i}\log  \frac{\tx_{\R1}^{-(k-m_1)} - x_{\R 2}^{\pm m_2}}{{ x_{\R 1}^{+(k+m_1)}} - x_{\R 2}^{\pm m_2}}  {1\ov x_{\R 2}^{\pm m_2}} + {\pi\ov2}\,.
\eal
Summing up all the terms, we get
\bal
e^{2i\tilde\theta_{\R\R}^{\hl}(x_{\R 1}^{\pm m_1},x_{\R 2}^{\pm m_2})}\to & \ 
e^{2i\tilde\theta_{\R\R}^{\hl}(x_{\R 1}^{\pm(m_1+k)},x_{\R 2}^{\pm m_2})}
\\
&
 \times 
  \frac{{\tx^{+(m_1+k)}_{\R 1}} -x_{\R2}^{-m_2}}{{\tx^{+(m_1+k)}_{\R 1}} -x_{\R2}^{+m_2}}\ 
 \frac{\tx_{\R 1}^{-(k-m_1)} - x_{\R 2}^{+m_2}}{\tx_{\R 1}^{-(k-m_1)} - x_{\R 2}^{-m_2}}
\,,
\eal
where $\tilde\theta_{\R\R}^{\hl}(x_{\R 1}^{\pm(m_1+k)},x_{\R 2}^{\pm m_2})$ is the string HL phase.

The term on the second line cancels the remaining term in \eqref{eqqq2a}, and since the odd dressing factor in $\Sigma^{m_1,m_2}_{\R\R} (p_1, p_2)$ becomes the odd factor in $\Sigma^{m_1+k,m_2}_{\R\R} (p_1, p_2)$, we conclude that 
\bal\la{Sbybyp1m2pip2Sbybyp1p2}
 S_{ \bar Y \bar Y}^{m_1,m_2}(p_{1}-2\pi,p_{2}) = S_{ \bar Y \bar Y}^{m_1+k,m_2}(p_{1},p_{2}) \,.
\eal

\subsection{Calculating   \texorpdfstring{$S_{\bar Y Z}^{m_1m_2}(p_{1}-2\pi, p_{2})$}{sbyzm1m2} }

The right-left S-matrix element $S_{\bar Y Z}^{m_1m_2}(p_1,p_2)$ is given by
\eqref{eq:SbYZ}.
The factor $A_{\bar Y  Z}^{m_1m_2}(p_{1},p_{2})$ after the two steps $x_{\R1}^{\pm m_1}\to x_{\R1}^{\pm (m_1+k)}$ becomes
\bal
 A_{\bar Y  Z}^{m_1m_2}(p_{1},p_{2})
  &\to A_{\bar Y  Z}^{m_1m_2}(p_{1}-2\pi,p_{2})=
   \left({x_{\R 1}^{-(k+m_1)}\ov x_{\R 1}^{+(k+m_1)}}\right)^{m_2-1}
      \left({x_{\L 2}^{+m_2}\ov x_{\L 2}^{-m_2}}\right)^{m_1} \ \frac{{x^{-(k+m_1)}_{\R 1} x^{-m_2}_{\L 2}}-1}{{x^{+(k+m_1)}_{\R 1} x^{+m_2}_{\L 2}}-1} 
      \\  
        &\times\frac{{x^{+(k+m_1)}_{\R 1} x^{-m_2}_{\L 2}}-1}{{x^{-(k+m_1)}_{\R 1} x^{+m_2}_{\L 2}}-1} 
    \prod_{j=1}^{m_1-1}  \left(\frac{{x^+_{\R1  j} x^{-m_2}_{\L 2}}-1}{{x^+_{\R 1j} x^{+m_2}_{\L 2}}-1}\right)^2 \  \prod^{m_2-1}_{j=1}\left( \frac{{x^{+(k+m_1)}_{\R 1} x^{+}_{\L 2j}}-1}{{x^{-(k+m_1)}_{\R 1} x^{+}_{\L 2j}}-1} \right)^2\,,
\eal
 where
\bal
x_{\R 1 j}^+ =\tx_\R(u_1 - {i\ov h}(k+m_1-2j))\,,\quad  x_{\L2 j}^+ =\tx_\L(u_2 - {i\ov h}(m_2-2j))\,.
\eal

Next, the first step of
the analytic continuation moves the first particle to the mirror region and replaces $x_{\R 1}^{\pm m_1}\to \tx_{\R 1}^{\pm m_1}$, 
$\g_{\R 1}^{\pm m_1}\to \tg_{\R 1}^{\pm m_1}$, and $\Sigma^{\bes}_{\R\L}$ and $\Sigma^{\hl}_{\R\L}$ are  given by
\bal\la{BESRLmirrorstring}
\tilde\theta_{\R\L}^\bes( \tx_{\R 1}^{\pm m_1},x^{\pm m_2}_{\L 2})  &=  \tilde\Phi_{\R\L}( \tx_{\R 1}^{+ m_1},x^{+m_2}_{\L 2})+ \tilde\Phi_{\R\L}( \tx_{\R 1}^{- m_1},x^{-m_2}_{\L 2})- \tilde\Phi_{\R\L}( \tx_{\R 1}^{+ m_1},x^{-m_2}_{\L 2}) 
\\
&- \tilde\Phi_{\R\L}( \tx_{\R 1}^{- m_1},x^{+m_2}_{\L 2})
+\tilde\Psi_{\R}(x^{+m_2}_{\L 2}, \tx_{\R 1}^{+ m_1})-\tilde\Psi_{\R}(x^{+m_2}_{\L 2}, \tx_{\R 1}^{- m_1})
\,,
\eal
\bal\la{HLRLmirrorstring}
\tilde\theta_{\R\L}^\hl( \tx_{\R 1}^{\pm m_1},x^{\pm m_2}_{\L 2})  &=  \tilde\Phi_{\R\L}^\hl( \tx_{\R 1}^{+ m_1},x^{+m_2}_{\L 2})+ \tilde\Phi_{\R\L}^\hl( \tx_{\R 1}^{- m_1},x^{-m_2}_{\L 2})- \tilde\Phi_{\R\L}^\hl( \tx_{\R 1}^{+ m_1},x^{-m_2}_{\L 2}) 
\\
&- \tilde\Phi_{\R\L}^\hl( \tx_{\R 1}^{- m_1},x^{+m_2}_{\L 2})+{1\ov 2i}\log \frac{\tx^{+m_1}_{\R1}}{\tx^{-m_1}_{\R1} } \frac{\tx^{+m_1}_{\R1}-x^{+m_2}_{\R 2}}{\tx^{+m_1}_{\R1}{x^{+m_2}_{\L 2}}-1 }  \frac{\tx^{-m_1}_{\R1}{x^{+m_2}_{\L 2}}-1 } {\tx^{-m_1}_{\R1}-x^{+m_2}_{\R 2} }
\,.
\eal
Then, we need to analytically continue $\tPhi_{\R\L}( \tx_{\R 1}^{\pm m_1},x^{\pm m_2}_{\L 2}) $ and $\tPhi_{\R\L}^\hl( \tx_{\R 1}^{\pm m_1},x^{\pm m_2}_{\L 2}) $ in $\tx_{\R1}^{+m_1 }$ through the negative semi-axis to the upper half-plane.

\subsubsection*{Continuation of $\tPhi_{\R\L}$ }

We first shift $u_1$ variable $u_1\to u_1-{i\ov h}k$ so that 
$\tx_{\R 1}^{\pm m_1}\to\tx_\R(u_1\pm{i\ov h}m_1-{i\ov h}k)=\tx_{\R 1}^{\pm m_1-k}$, and then analytically continue  $\tx_{\R 1}^{+m_1-k }$ to $1/\tx_{\L 1}^{+ (m_1+k)}=x_{\R 1}^{+ (m_1+k)}$, $u_1\in\bR$. We get
\bal
\tPhi_{\R\L}^{-+}( \tx_{\R 1}^{+ m_1},x^{\pm m_2}_{\L 2})\to &+\tPhi_{\R\L}^{-+}({x_{\R 1}^{+ (m_1+k)}},x_{\L 2}^{\pm m_2}) - \tPsi_{\L}^{+}(u_1+{i\ov h}(m_1-k),x_{\L 2}^{\pm m_2})\,,
\\
\tPhi_{\R\L}^{+-}( \tx_{\R 1}^{+ m_1},x^{\pm m_2}_{\L 2})\to &+\tPhi_{\R\L}^{+-}({ x_{\R 1}^{+ (m_1+k)}},x_{\L 2}^{\pm m_2 }) - \tPsi_{\L}^{-}(u_1+{i\ov h}(m_1+k),x_{\L 2}^{\pm  m_2})\,,
\eal
\bal
\tilde\Psi^{-}_{\R}(u_{2} + {i\ov h}m_2, \tx_{\R 1}^{+ m_1})\to & \  \tilde\Psi^{-}_{\R}(u_{2} + {i\ov h}m_2, x_{\R 1}^{+ (m_1+k)})-K^\bes(u_{12} +{i\ov h}(m_1-m_2)-{i\ov h}k)\,,
\\
\tilde\Psi^{+}_{\R}(u_{2} + {i\ov h}m_2, \tx_{\R 1}^{+ m_1})\to & \ \tilde\Psi^{+}_{\R}(u_{2} + {i\ov h}m_2, x_{\R 1}^{+ (m_1+k)})-K^\bes(u_{12}+{i\ov h}(m_1-m_2) +{i\ov h}k)\,.
\eal
Note that we do not cross any cut of $\tPsi$-functions.

Summing up the $\tPhi$ functions, we get
\bal
2\tilde\theta_{\R\L}^\bes&(\tx_{\R 1}^{\pm m_1},x_{\L 2}^{\pm m_2})\to 
2\tilde\theta_{\R\L}^\bes({ x_{\R 1}^{\pm (m_1+k)}},{x_{\L 2}^{\pm m_2 }})
+\Delta_{\L}^{+}(u_1+{i\ov h}m_1\pm {i\ov h}k,x_{\L 2}^{\pm m_2})
\\
&-K^\bes(u_{12}+{i\ov h}(m_1-m_2) -{i\ov h}k)+K^\bes(u_{12} +{i\ov h}(m_1-m_2)+{i\ov h}k)
\,.
\eal
By using  identity \eqref{eq:Deltaepsp_2ms} for $\tPsi$ functions, 
and  identity \eqref{eq:KbesmKbes} for $K^\bes$,
we get 
\bal\la{eqqq1}
&e^{-2i\tilde\theta_{\R\L}^\bes(\tx_{\R 1}^{\pm m_1},x_{\L 2}^{\pm m_2})}\to
e^{-2i\tilde\theta_{\R\L}^\bes({ x_{\R 1}^{\pm (m_1+k)}},{x_{\L 2}^{\pm m_2 }})}
 \\
&\times
\left(\frac{x_{\L 2}^{+ m_2} }{x_{\L 2}^{- m_2} }\right)^{k}
 \frac{ (\tx_{\R 1}^{-(k-m_1)}x_{\L 2}^{+ m_2}-1) (x_{\L 1}^{+(k+m_1)}-x_{\L 2}^{- m_2}) } {
  (\tx_{\R 1}^{-(k-m_1)}x_{\L 2}^{- m_2}-1)(x_{\L 1}^{+(k+m_1)}-x_{\L 2}^{+ m_2}) }
\prod_{j=m_1}^{m_1+k-1}
 \left( 
 \frac{  \tx_{\R1}^{2j-m_1-k}x_{\L 2}^{- m_2}-1 } {
 \tx_{\R1}^{2j-m_1-k}x_{\L 2}^{+ m_2}-1}\right)^2
 \,.
\eal
Thus, the factor $A_{\bar Y  Z}^{m_1,m_2}(p_{1}-2\pi,p_{2})$ multiplied by the factor in \eqref{eqqq1}  becomes equal the factor $A_{\bar Y  Z}^{m_1+k,m_2}(p_{1},p_{2})$ in $S_{\bar Y  Z}^{m_1+k,m_2}(p_{1},p_{2})$ times the following factor
\bal\la{eqqq2}
 \frac{ (\tx_{\R 1}^{-(k-m_1)}x_{\L 2}^{+ m_2}-1) (x_{\L 1}^{+(k+m_1)}-x_{\L 2}^{- m_2}) } {
  (\tx_{\R 1}^{-(k-m_1)}x_{\L 2}^{- m_2}-1)(x_{\L 1}^{+(k+m_1)}-x_{\L 2}^{+ m_2}) }\,.
  \eal
  
\subsubsection*{Continuation of $\tPhi_{\R\L}^\hl$}

Next, we analytically continue $\tPhi_{\R\L}^\hl$ in $\tx_{\R 1}^{+ m_1}\to x_{\R 1}^{+ (m_1+k)}$, 
and  get
\bal
\tPhi_{\R\L}^{\hl}(\tx_{\R1}^{+ m_1},x_{\L 2}^{\pm m_2})\to &+\tPhi_{\R\L}^{\hl}( x_{\R 1}^{+ (k+m_1)},x_{\L 2}^{\pm m_2})+  {1\ov 2i}\log  \frac{\tx_{\L 1}^{+(k+m_1)} - x_{\L 2}^{\pm m_2}}{{1\ov \tx_{\R 1}^{-(k-m_1)}} - x_{\L 2}^{\pm m_2}}   {1\ov x_{\L 2}^{\pm m_2}} \pm {\pi\ov4}\,.
\eal
Summing up all the terms, we get
\bal
e^{2i\tilde\theta_{\R\L}^{\hl}(x_{\R 1}^{\pm m_1},x_{\L 2}^{\pm m_2})}\to & \ 
e^{2i\tilde\theta_{\R\L}^{\hl}(x_{\R 1}^{\pm(m_1+k)},x_{\L 2}^{\pm m_2})}
\\
&
 \times 
  \frac{x_{\L1}^{+(k+m_1)} -x_{\L2}^{+m_2}}{x_{\L1}^{+(k+m_1)} -x_{\L2}^{-m_2}}
\frac{{ \tx_{\R 1}^{-(k-m_1)}} x_{\L 2}^{- m_2}-1}{{ \tx_{\R 1}^{-(k-m_1)}}x_{\L 2}^{+ m_2}-1} 
\,,
\eal
where $\tilde\theta_{\R\L}^{\hl}(x_{\R 1}^{\pm(m_1+k)},x_{\L 2}^{\pm m_2})$ is the string HL phase.

The term on the second line cancels the remaining term in \eqref{eqqq2}, and since the odd dressing factor in $\Sigma^{m_1m_2}_{\R\L} (p_1, p_2)$ becomes the odd factor in $\Sigma^{m_1+k,m_2}_{\R\L} (p_1, p_2)$, we conclude that 
\bal\la{Sbyzp1m2pip2Sbyzp1p2}
 S_{ \bar Y Z}^{m_1,m_2}(p_{1}-2\pi,p_{2}) = S_{ \bar Y Z}^{m_1+k,m_2}(p_{1},p_{2}) \,.
\eal

\subsection{Calculating \texorpdfstring{$S_{ Y Y}^{m_1m_2}(p_{1}-2\pi, p_{2})\,,\, m_1=1,2,\ldots, k-1$ }{syym1m2a} }

The left-left S-matrix element $S_{Y Y}^{m_1m_2}(p_1,p_2)$ is given by \eqref{eq:SbYbY} with the replacement R$\to$L.
We want to relate it to $S_{ \bar Z  Y}^{k-m_1,m_2}(p_{1}, p_{2})$ in \eqref{eq:SbZY}.
The factor $A_{ Y  Y}^{m_1m_2}(p_{1},p_{2})$ after the two steps $x_{\L1}^{\pm m_1}\to 1/x_{\R1}^{\mp (k-m_1)}$, $u_1\to u_1+{i\ov h}k$ becomes
\bal
 A&_{YY}^{m_1m_2}(p_{1},p_{2})
  \to A_{Y  Y}^{m_1m_2}(p_{1}-2\pi,p_{2})=
  \left({x_{\R1}^{-(k-m_1)}\ov x_{\R 1}^{+(k-m_1)}}\right)^{m_2}   \left({x_{\L2}^{-m_2}\ov x_{\L2}^{+m_2}}\right)^{m_1}
     \\
  &\times    \left(\frac{{ x^{+ (k-m_1)}_{\R1} }x^{+ m_2}_{\L2}-1}{{ x^{-(k-m_1)}_{\R1} }x^{- m_2}_{\L2}-1} \right)^2 
    \prod^{m_1-1}_{j=1} \left(\frac{x^{+}_{\L 1j} - x^{+m_2}_{\L 2}}{x^{+}_{\L 1j} - x^{-m_2}_{\L 2}} \right)^2
\  \prod^{m_2-1}_{j=1} \left(\frac{{ x^{+(k- m_1)}_{\R1}} x^{+}_{\L 2j}-1}{{ x^{-(k-m_1)}_{\R1}}  x^{+}_{\L 2j}-1} \right)^2 
\\
    &\times \frac{u_{12}+ \frac{i}{h} (m_1-m_2+k)}{u_{12}- \frac{i}{h} (m_1-m_2-k)}  \, \frac{u_{12}+ \frac{i}{h} (m_1+m_2+k)}{u_{12}- \frac{i}{h} (m_1+m_2-k)}  \prod^{m_2-1}_{j=1} \left(\frac{u_{12} - {i\ov h} (2j-k-m_1-m_2 )}{u_{12} + {i\ov h} (2j+k-m_1-m_2 )} \right)^2
    \,,
\eal
where
\bal
x_{\L 1 j}^+ =\tx_\L(u_1 - {i\ov h}(m_1-k-2j))\,,\quad  x_{\L2 j}^+ =\tx_\L(u_2 - {i\ov h}(m_2-2j))\,.
\eal

Next, the first step of
the analytic continuation moves the first particle to the mirror region and replaces $x_{\L 1}^{\pm m_1}\to \tx_{\L 1}^{\pm m_1}$, 
$\g_{\L 1}^{\pm m_1}\to \tg_{\L 1}^{\pm m_1}$, and $\Sigma^{\bes}_{\L\L}$ and $\Sigma^{\hl}_{\L\L}$ are  given by
\bal\la{BESLLmirrorstring}
\tilde\theta_{\L\L}^\bes( \tx_{\L 1}^{\pm m_1},x^{\pm m_2}_{\L 2})  &=  \tilde\Phi_{\L\L}( \tx_{\L 1}^{+ m_1},x^{+m_2}_{\L 2})+ \tilde\Phi_{\L\L}( \tx_{\L 1}^{- m_1},x^{-m_2}_{\L 2})- \tilde\Phi_{\L\L}( \tx_{\L 1}^{+ m_1},x^{-m_2}_{\L 2}) 
\\
&- \tilde\Phi_{\L\L}( \tx_{\L 1}^{- m_1},x^{+m_2}_{\L 2})
+\tilde\Psi_{\L}(x^{+m_2}_{\L 2}, \tx_{\L 1}^{+ m_1})-\tilde\Psi_{\L}(x^{+m_2}_{\L 2}, \tx_{\L 1}^{- m_1})
\,,
\eal
\bal\la{HLLLmirrorstring}
\tilde\theta_{\L\L}^\hl( \tx_{\L 1}^{\pm m_1},x^{\pm m_2}_{\L 2})  &=  \tilde\Phi_{\L\L}^\hl( \tx_{\L 1}^{+ m_1},x^{+m_2}_{\L 2})+ \tilde\Phi_{\L\L}^\hl( \tx_{\L 1}^{- m_1},x^{-m_2}_{\L 2})- \tilde\Phi_{\L\L}^\hl( \tx_{\L 1}^{+ m_1},x^{-m_2}_{\L 2}) 
\\
&- \tilde\Phi_{\L\L}^\hl( \tx_{\L 1}^{- m_1},x^{+m_2}_{\L 2})
+{1\ov 2i}\log \frac{\tx^{+m_1}_{\L 1}}{\tx^{-m_1}_{\L 1} } \frac{\tx^{+m_1}_{\L 1}-x^{+m_2}_{\L 2}}{\tx^{+m_1}_{\L 1}-{\tx^{+m_2}_{\L 2}} }  \frac{\tx^{-m_1}_{\L 1}-{\tx^{+m_2}_{\L 2}} } {\tx^{-m_1}_{\L 1}-x^{+m_2}_{\L 2} }\,.
\eal

Then, we need to analytically continue $\tPhi_{\L\L}( \tx_{\L 1}^{\pm m_1},x^{\pm m_2}_{\L 2}) $ and $\tPhi_{\L\L}^\hl( \tx_{\L 1}^{\pm m_1},x^{\pm m_2}_{\L 2}) $ in $\tx_{\L 1}^{+m_1 }$ through the negative semi-axis to the upper half-plane.

\subsubsection*{Continuation of $\tPhi_{\L\L}$ }

We first shift $u_1$ variable $u_1\to u_1+{i\ov h}k$ so that 
$\tx_{\L 1}^{\pm m_1}\to\tx_\L(u_1\pm{i\ov h}m_1+{i\ov h}k)=\tx_{\L 1}^{\pm m_1+k}$, and then analytically continue  $\tx_{\L 1}^{m_1+k }$ to $1/\tx_{\R 1}^{- (k-m_1)}=1/x_{\R 1}^{-(k-m_1)}$, $u_1\in\bR$. We get
\bal
\tPhi_{\L\L}^{--}( \tx_{\L 1}^{+ m_1},x^{\pm m_2}_{\L 2})\to &+\tPhi_{\L\L}^{--}({1\ov x_{\R 1}^{- (k-m_1)}},x_{\L 2}^{\pm m_2}) - \tPsi_{\L}^{-}(u_1+{i\ov h}(m_1+k),x_{\L 2}^{\pm m_2})\,,
\\
\tPhi_{\L\L}^{++}( \tx_{\L 1}^{+ m_1},x^{\pm m_2}_{\L 2})\to &+\tPhi_{\L\L}^{++}({1\ov x_{\R 1}^{- (k-m_1)}},x_{\L 2}^{\pm m_2 }) - \tPsi_{\L}^{+}(u_1+{i\ov h}(m_1-k),x_{\L 2}^{\pm  m_2})\,,
\eal
\bal
\tilde\Psi^{-}_{\L}(u_{2} + {i\ov h}m_2, \tx_{\L 1}^{+ m_1})\to & \ \tilde\Psi^{-}_{\L}(u_{2} + {i\ov h}m_2, {1\ov x_{\R 1}^{- (k-m_1)}})-K^\bes(u_{12}+{i\ov h}(m_1-m_2) +{i\ov h}k)\,,
\\
\tilde\Psi^{+}_{\L}(u_{2} + {i\ov h}m_2, \tx_{\L 1}^{+ m_1})\to & \  \tilde\Psi^{+}_{\L}(u_{2} + {i\ov h}m_2, {1\ov x_{\R 1}^{- (k-m_1)}})-K^\bes(u_{12} +{i\ov h}(m_1-m_2)-{i\ov h}k)\,.
\eal
Note that we do not cross any cut of $\tPsi$-functions.

Summing up the $\tPhi$ functions, we get
\bal
2\tilde\theta_{\L\L}^\bes&(\tx_{\L 1}^{\pm},x_{\L 2}^{\pm})\to 
2\tilde\theta_{\R\L}^{k-m_1,m_2}({ x_{\R 1}^{\pm (k-m_1)}},{x_{\L 2}^{\pm m_2 }})-\Delta_{\L}^{-}(u_1+{i\ov h}k\pm {i\ov h}m_1,x_{\L 2}^{\pm m_2})
\\
&+\Delta_{\L}^{+}(u_1\pm{i\ov h}(k-m_1),x_{\L 2}^{\pm m_2})+2K^\bes(u_{12}+{i\ov h}(k-m_1-m_2))
\\
&-K^\bes(u_{12}+{i\ov h}(m_1-m_2) -{i\ov h}k)-K^\bes(u_{12} +{i\ov h}(m_1-m_2)+{i\ov h}k)
\,.
\eal
By using  identities \eqref{eq:Deltaepsm_2ms}, \eqref{eq:Deltaepsp_2ms} for $\tPsi$ functions, 
and  identity \eqref{eq:KbesmKbes} for $K^\bes$,
we get 
\bal\la{eqqq1b}
e^{-2i\tilde\theta_{\L\L}^\bes(\tx_{\L 1}^{\pm},x_{\L 2}^{\pm})}&A_{ Y  Y}^{m_1,m_2}(p_{1},p_{2})\to
e^{-2i\tilde\theta_{\R\L}^\bes({ x_{\R 1}^{\pm (k-m_1)}},{x_{\L 2}^{\pm m_2 }})} A_{\bar  Z Y}^{k-m_1,m_2}(p_{1},p_{2})
 \\
&\times
\frac{x_{\L 2}^{+ m_2} }{x_{\L 2}^{- m_2} }
 \frac{ (\tx_{\R 1}^{+(k-m_1)}x_{\L 2}^{- m_2}-1) (x_{\R 1}^{+(k+m_1)}x_{\L 2}^{- m_2}-1) } {
  (\tx_{\R 1}^{+(k-m_1)}x_{\L 2}^{+ m_2}-1)(x_{\R 1}^{+(k+m_1)}x_{\L 2}^{+ m_2}-1) }
 \,.
\eal
  
\subsubsection*{Continuation of $\tPhi_{\L\L}^\hl$}

Next, we analytically continue $\tPhi_{\L\L}^\hl$ in $\tx_{\L 1}^{+ m_1}\to 1/x_{\R 1}^{- (k-m_1)}$, 
and  get
\bal
\tPhi_{\L\L}^{\hl}(\tx_{\L1}^{+ m_1},x_{\L 2}^{\pm m_2})\to &+\tPhi_{\L\L}^{\hl}( {1\ov x_{\R 1}^{-(k-m_1)}},x_{\L 2}^{\pm m_2})+  {1\ov 2i}\log  \frac{\tx_{\R 1}^{+(k+m_1)} - x_{\L 2}^{\pm m_2}}{{1\ov x_{\R 1}^{-(k-m_1)}} - x_{\L 2}^{\pm m_2}} {1\ov x_{\L 2}^{\pm m_2}} \pm {\pi\ov4}\,.
\eal
Summing up all the terms, we get
\bal
&e^{2i\tilde\theta_{\L\L}^{\hl}(x_{\L 1}^{\pm m_1},x_{\L 2}^{\pm m_2})}\to  \ 
e^{2i\tilde\theta_{\R\L}^{\hl }(x_{\R 1}^{\pm(k-m_1)},x_{\L 2}^{\pm m_2})}
\\
&
 \times 
\frac{ x_{\R 1}^{-(k-m_1)}x_{\L 2}^{-m_2} - 1}{ x_{\R 1}^{+(k-m_1)}x^{+m_2}_{\L2} -1}
\frac{ x_{\R1}^{+(k-m_1)}x_{\L2}^{-m_2} -1}{x_{\R1}^{-(k-m_1)}x_{\L2}^{+m_2} -1} 
\frac{ \tx_{\R1}^{+(k-m_1)}x_{\L2}^{+m_2} -1}{\tx_{\R1}^{+(k-m_1)}x_{\L2}^{-m_2} -1} 
  \frac{ x_{\R 1}^{+(k+m_1)} x_{\L 2}^{+m_2}- 1}{ x_{\R 1}^{+(k+m_1)} x_{\L 2}^{-m_2}- 1}
\,,
\eal
where $\tilde\theta_{\R\L}^{\hl}(x_{\R 1}^{\pm(k-m_1)},x_{\L 2}^{\pm m_2})$ is the string HL phase.

Thus,
\bal\la{eqqq2b}
&e^{-2i\tilde\theta_{\L\L}^\bes(\tx_{\L 1}^{\pm},x_{\L 2}^{\pm})+2i\tilde\theta_{\L\L}^{\hl}(x_{\L 1}^{\pm m_1},x_{\L 2}^{\pm m_2})}A_{ Y  Y}^{m_1,m_2}(p_{1},p_{2})
\\
&\to
e^{-2i\tilde\theta_{\R\L}^\bes({ x_{\R 1}^{\pm (k-m_1)}},{x_{\L 2}^{\pm m_2 }})+2i\tilde\theta_{\R\L}^{\hl }(x_{\R 1}^{\pm(k-m_1)},x_{\L 2}^{\pm m_2})} A_{\bar  Z Y}^{k-m_1,m_2}(p_{1},p_{2})
 \\
&\times
\frac{x_{\L 2}^{+ m_2} }{x_{\L 2}^{- m_2} }
\frac{ x_{\R 1}^{-(k-m_1)}x_{\L 2}^{-m_2} - 1}{ x_{\R 1}^{+(k-m_1)}x^{+m_2}_{\L2} -1}
\frac{ x_{\R1}^{+(k-m_1)}x_{\L2}^{-m_2} -1}{x_{\R1}^{-(k-m_1)}x_{\L2}^{+m_2} -1} 
 \,.
\eal
Finally, we take into account the contribution of the odd factor.
Since we do not cross the cut of $\g^\pm_\L$-rapidities between $(-1/\xi,\xi)$ the  $\g^\pm_\L$-rapidities transform as
\bal
\g_\L(x_{\L 1}^{\pm m_1}) \to \g_\L(1/x_{\R 1}^{\mp (k-m_1)}) &=  \g_{\R 1}^{\mp(k-m_1)} \pm i\pi
\,.
\eal
Thus,
the odd factor in $\Sigma^{\barnes}_{\L\L} (\g_{\L 1}^{\pm m_1},\g_{\L 2}^{\pm m_2})$ becomes
\bal\la{eqqq3b}
&\Sigma^{\barnes}_{\L\L} (\g_{\L 1}^{\pm m_1},\g_{\L 2}^{\pm m_2})
\to
\Sigma^{\barnes}_{\R\L} (\g_{\R 1}^{\pm (k-m_1)},\g_{\L 2}^{\pm m_2})
 {\cosh{\g_{\R\L}^{+(k-m_1),+m_2}\ov2}\ov \cosh{\g_{\R\L}^{+(k-m_1),-m_2}\ov2}}{\cosh{\g_{\R\L}^{-(k-m_1),+m_2}\ov2}\ov \cosh{\g_{\R\L}^{-(k-m_1),-m_2}\ov2}} 
\,.
\eal
Multiplying \eqref{eqqq2b} and \eqref{eqqq3b}, we
find that under the second step of the analytic continuation 
\bal\la{Syyp1m2pip2Sbzyp1p2}
&S^{m_1 \, m_2}_{Y Y} (p_1-2 \pi, p_2) = S^{k-m_1 \, m_2}_{\bar{Z} Y} (p_1, p_2)\, {\a_\L( x^{-m_2}_{\L 2}) \ov \a_\L( x^{+m_2}_{\L 2}) } \,,
\eal
where $ \alpha_a(x)$ is defined in \eqref{eq:defalpha}.

\subsection{Calculating  \texorpdfstring{$S_{ Y Y}^{m_1m_2}(p_{1}-2\pi, p_{2})\,,\, m_1>k$ }{syym1m2b} }

The factor $A_{ Y  Y}^{m_1m_2}(p_{1},p_{2})$ after the two steps $x_{\L1}^{\pm m_1}\to x_{\L1}^{\pm (m_1-k)}$, $u_1\to u_1+{i\ov h}k$ becomes
\bal
 A&_{YY}^{m_1m_2}(p_{1},p_{2})
  \to A_{Y  Y}^{m_1m_2}(p_{1}-2\pi,p_{2})=
 \left({x_{\L1}^{+ (m_1-k)}\ov x_{\L1}^{- (m_1-k)}}\right)^{m_2}   \left({x_{\L2}^{-m_2}\ov x_{\L2}^{+m_2}}\right)^{m_1} \left(\frac{x^{-  (m_1-k)}_{\L1} - x^{+ m_2}_{\L2}}{x^{+  (m_1-k)}_{\L1} - x^{- m_2}_{\L2}} \right)^2 \\
    &\times \frac{u_{12}+ \frac{i}{h} (m_1-m_2+k)}{u_{12}- \frac{i}{h} (m_1-m_2-k)}  \, \frac{u_{12}+ \frac{i}{h} (m_1+m_2+k)}{u_{12}- \frac{i}{h} (m_1+m_2-k)}  \prod^{m_2-1}_{j=1} \left(\frac{u_{12} - {i\ov h} (2j-m_1-m_2 -k)}{u_{12} + {i\ov h} (2j-m_1-m_2 +k)} \right)^2
      \\
  &\times 
    \prod^{m_1-1}_{j=1} \left(\frac{x^{+}_{\L1 j} - x^{+m_2}_{\L 2}}{x^{+}_{\L 1j} - x^{-m_2}_{\L 2}} \right)^2
\  \prod^{m_2-1}_{j=1} \left(\frac{x^{-  (m_1-k)}_{\L1} - x^{+}_{\L 2j}}{x^{+  (m_1-k)}_{\L1} - x^{+}_{\L 2j}} \right)^2 
    \,,
\eal
where
\bal
x_{\L 1 j}^+ =\tx_\L(u_1 - {i\ov h}(m_1-k-2j))\,,\quad  x_{\L2 j}^+ =\tx_\L(u_2 - {i\ov h}(m_2-2j))\,.
\eal

After the first step of
the analytic continuation  $\Sigma^{\bes}_{\L\L}$ and $\Sigma^{\hl}_{\L\L}$ are still given by \eqref{BESLLmirrorstring}
 and \eqref{HLLLmirrorstring}.
Then, we need to analytically continue $\tPhi_{\L\L}( \tx_{\L 1}^{\pm m_1},x^{\pm m_2}_{\L 2}) $ and $\tPhi_{\L\L}^\hl( \tx_{\L 1}^{\pm m_1},x^{\pm m_2}_{\L 2}) $ in $\tx_{\L 1}^{+m_1 }$ through the negative semi-axis to the upper half-plane.

\subsubsection*{Continuation of $\tPhi_{\L\L}$ }

We first shift $u_1$ variable $u_1\to u_1+{i\ov h}k$ so that 
$\tx_{\L 1}^{\pm m_1}\to\tx_\L(u_1\pm{i\ov h}m_1+{i\ov h}k)=\tx_{\L 1}^{\pm m_1+k}$, and then analytically continue  $\tx_{\L 1}^{m_1+k }$ to $1/\tx_{\R 1}^{m_1-k}=x_{\L 1}^{+(m_1-k)}$, $u_1\in\bR$. We get
\bal
\tPhi_{\L\L}^{--}( \tx_{\L 1}^{+ m_1},x^{\pm m_2}_{\L 2})\to &+\tPhi_{\L\L}^{--}({ x_{\L 1}^{+(m_1-k)}},x_{\L 2}^{\pm m_2}) - \tPsi_{\L}^{-}(u_1+{i\ov h}(m_1+k),x_{\L 2}^{\pm m_2})\,,
\\
\tPhi_{\L\L}^{++}( \tx_{\L 1}^{+ m_1},x^{\pm m_2}_{\L 2})\to &+\tPhi_{\L\L}^{++}({ x_{\L 1}^{+(m_1-k)}},x_{\L 2}^{\pm m_2 }) - \tPsi_{\L}^{+}(u_1+{i\ov h}(m_1-k),x_{\L 2}^{\pm  m_2})\,,
\eal
\bal
\tilde\Psi^{-}_{\L}(u_{2} + {i\ov h}m_2, \tx_{\L 1}^{+ m_1})\to & \ \tilde\Psi^{-}_{\L}(u_{2} + {i\ov h}m_2, { x_{\L 1}^{+(m_1-k)}})-K^\bes(u_{12}+{i\ov h}(m_1-m_2) +{i\ov h}k)\,,
\\
\tilde\Psi^{+}_{\L}(u_{2} + {i\ov h}m_2, \tx_{\L 1}^{+ m_1})\to & \  \tilde\Psi^{+}_{\L}(u_{2} + {i\ov h}m_2, { x_{\L 1}^{+(m_1-k)}})-K^\bes(u_{12} +{i\ov h}(m_1-m_2)-{i\ov h}k)\,.
\eal
Note that we do not cross any cut of $\tPsi$-functions.

Summing up the $\tPhi$ functions, we get
\bal
2\tilde\theta_{\L\L}^\bes&(\tx_{\L 1}^{\pm m_1},x_{\L 2}^{\pm m_2})\to 
2\tilde\theta_{\L\L}^{k-m_1,m_2}({ x_{\L 1}^{\pm (k-m_1)}},{x_{\L 2}^{\pm m_2 }})-\Delta_{\L}^{-}(u_1+{i\ov h}m_1\pm {i\ov h}k,x_{\L 2}^{\pm m_2})
\\
&+K^\bes(u_{12}+{i\ov h}(m_1-m_2) -{i\ov h}k)-K^\bes(u_{12} +{i\ov h}(m_1-m_2)+{i\ov h}k)
\,.
\eal
By using  identity \eqref{eq:Deltaepsm_2ms} for $\tPsi$ functions, 
and  identity \eqref{eq:KbesmKbes} for $K^\bes$,
we get 
\bal\la{eqqq1c}
e^{-2i\tilde\theta_{\L\L}^\bes(\tx_{\L 1}^{\pm m_1},x_{\L 2}^{\pm m_2})}&A_{ Y  Y}^{m_1,m_2}(p_{1},p_{2})\to
e^{-2i\tilde\theta_{\L\L}^\bes({ x_{\L 1}^{\pm (m_1-k)}},{x_{\L 2}^{\pm m_2 }})} A_{ Y Y}^{m_1-k,m_2}(p_{1},p_{2})
 \\
&\times
 \frac{ (\tx_{\L 1}^{+(m_1-k)}-x_{\L 2}^{+m_2}) (\tx_{\L 1}^{+(m_1+k)}-x_{\L 2}^{- m_2}) } {
 (\tx_{\L 1}^{+(m_1-k)}-x_{\L 2}^{-m_2}) (\tx_{\L 1}^{+(m_1+k)}-x_{\L 2}^{+ m_2}) }
 \,.
\eal
  
\subsubsection*{Continuation of $\tPhi_{\L\L}^\hl$}

Next, we analytically continue $\tPhi_{\L\L}^\hl$ in $\tx_{\L 1}^{+ m_1}\to x_{\L 1}^{+ (m_1-k)}$, 
and  get
\bal
\tPhi_{\L\L}^{\hl}(\tx_{\L1}^{+ m_1},x_{\L 2}^{\pm m_2})\to &+\tPhi_{\L\L}^{\hl}( { x_{\L 1}^{+(m_1-k)}},x_{\L 2}^{\pm m_2})+  {1\ov 2i}\log  \frac{\tx_{\L 1}^{+(k+m_1)} - x_{\L 2}^{\pm m_2}}{{x_{\L 1}^{+(m_1-k)}} - x_{\L 2}^{\pm m_2}}{1\ov   x_{\L 2}^{\pm m_2}} \pm {\pi\ov4}\,.
\eal
Summing up all the terms, we get
\bal
&e^{2i\tilde\theta_{\L\L}^{\hl}(x_{\L 1}^{\pm m_1},x_{\L 2}^{\pm m_2})}\to  \ 
e^{2i\tilde\theta_{\L\L}^{\hl }(x_{\L 1}^{\pm(m_1-k)},x_{\L 2}^{\pm m_2})}
\\
&
 \times 
 \frac {(\tx_{\L 1}^{+(m_1-k)}-x_{\L 2}^{-m_2}) (\tx_{\L 1}^{+(m_1+k)}-x_{\L 2}^{+ m_2}) }
 { (\tx_{\L 1}^{+(m_1-k)}-x_{\L 2}^{+m_2}) (\tx_{\L 1}^{+(m_1+k)}-x_{\L 2}^{- m_2}) }
\,,
\eal
where $\tilde\theta_{\R\L}^{\hl}(x_{\R 1}^{\pm(k-m_1)},x_{\L 2}^{\pm m_2})$ is the string HL phase.

The term on the second line cancels the remaining term in \eqref{eqqq1c}, and since the odd dressing factor in $\Sigma^{m_1,m_2}_{\L\L} (p_1, p_2)$ becomes the odd factor in $\Sigma^{m_1-k,m_2}_{\L\L} (p_1, p_2)$, we conclude that 
\bal
 S_{ Y  Y}^{m_1,m_2}(p_{1}-2\pi,p_{2}) = S_{  Y Y}^{m_1-k,m_2}(p_{1},p_{2}) \,, \quad m_1>k\,.
\eal
Shifting $p_1$ by $2\pi$, and $m_1\to m_1+k$ we can also write the relation in the form
\bal\la{Syyp1m2pip2Syyp1p2}
 S_{ Y  Y}^{m_1,m_2}(p_{1}+2\pi,p_{2}) = S_{  Y Y}^{m_1+k,m_2}(p_{1},p_{2}) \,, \quad m_1>0\,.
\eal

\subsection{Calculating \texorpdfstring{$S_{ Y \bar Z}^{m_1m_2}(p_{1}-2\pi, p_{2})\,,\, m_1=1,2,\ldots, k-1$ }{sybzm1m2a}  }

The left-right S-matrix element $S_{Y \bar Z}^{m_1m_2}(p_1,p_2)$ is given by
\eqref{eq:SbYZ} with the replacement R$\leftrightarrow$L.
We want to relate it to $S_{ \bar Z \bar Z }^{k-m_1,m_2}(p_{1}, p_{2})$, see \eqref{eq:SbZbZ}.
The factor $A_{Y\bar Z}^{m_1m_2}(p_{1},p_{2})$ after the two steps $x_{\L1}^{\pm m_1}\to 1/x_{\R1}^{\mp (k-m_1)}$, $u_1\to u_1+{i\ov h}k$ becomes
\bal
 A&_{Y\bar Z}^{m_1m_2}(p_{1},p_{2})
  \to A_{Y  \bar Z}^{m_1m_2}(p_{1}-2\pi,p_{2})=
 \left({x_{\R1}^{+(k-m_1)}\ov x_{\R1}^{-(k-m_1)}}\right)^{m_2-1}
      \left({x_{\R2}^{+m_2}\ov x_{\R2}^{-m_2}}\right)^{m_1} \ \frac{{x^{+(k-m_1)}_{\R1} -x^{-m_2}_{\R 2}}}{{x^{-(k-m_1)}_{\R1} -x^{+m_2}_{\R 2}}}  \\
      &\times \frac{{x^{-(k-m_1)}_{\R1} -x^{-m_2}_{\R 2}}}{{x^{+(k-m_1)}_{\R1} -x^{+m_2}_{\R 2}}}\prod_{j=1}^{m_1-1}  \left(\frac{{x^+_{\L 1j} x^{-m_2}_{\R 2}}-1}{{x^+_{\L 1j} x^{+m_2}_{\R 2}}-1}\right)^2 \  \prod^{m_2-1}_{j=1}\left( \frac{{x^{-(k-m_1)}_{\R1} -x^{+}_{\R 2j}}}{{x^{+(k-m_1)}_{\R1} -x^{+}_{\R 2j}}} \right)^2
    \,,
\eal
where
\bal
x_{\L 1 j}^+ =\tx_\L(u_1 - {i\ov h}(m_1-k-2j))\,,\quad  x_{\R2 j}^+ =\tx_\R(u_2 - {i\ov h}(m_2-2j))\,.
\eal

Next, the first step of
the analytic continuation moves the first particle to the mirror region and replaces $x_{\L 1}^{\pm m_1}\to \tx_{\L 1}^{\pm m_1}$, 
$\g_{\L 1}^{\pm m_1}\to \tg_{\L 1}^{\pm m_1}$, and $\Sigma^{\bes}_{\L\R}$ and $\Sigma^{\hl}_{\L\R}$ are  given by
\bal\la{BESLRmirrorstring}
\tilde\theta_{\L\R}^\bes( \tx_{\L 1}^{\pm m_1},x^{\pm m_2}_{\R 2})  &=  \tilde\Phi_{\L\R}( \tx_{\L 1}^{+ m_1},x^{+m_2}_{\R 2})+ \tilde\Phi_{\L\R}( \tx_{\L 1}^{- m_1},x^{-m_2}_{\R 2})- \tilde\Phi_{\L\R}( \tx_{\L 1}^{+ m_1},x^{-m_2}_{\R 2}) 
\\
&- \tilde\Phi_{\L\R}( \tx_{\L 1}^{- m_1},x^{+m_2}_{\R 2})
+\tilde\Psi_{\L}(x^{+m_2}_{\R 2}, \tx_{\L 1}^{+ m_1})-\tilde\Psi_{\L}(x^{+m_2}_{\R 2}, \tx_{\L 1}^{- m_1})
\,,
\eal
\bal\la{HLLRmirrorstring}
\tilde\theta_{\L\R}^\hl( \tx_{\L 1}^{\pm m_1},x^{\pm m_2}_{\R 2})  &=  \tilde\Phi_{\L\R}^\hl( \tx_{\L 1}^{+ m_1},x^{+m_2}_{\R 2})+ \tilde\Phi_{\L\R}^\hl( \tx_{\L 1}^{- m_1},x^{-m_2}_{\R 2})- \tilde\Phi_{\L\R}^\hl( \tx_{\L 1}^{+ m_1},x^{-m_2}_{\R 2}) 
\\
&- \tilde\Phi_{\L\R}^\hl( \tx_{\L 1}^{- m_1},x^{+m_2}_{\R 2})
+{1\ov 2i}\log \frac{\tx^{+m_1}_{\L 1}}{\tx^{-m_1}_{\L 1} }\frac{\tx^{+m_1}_{\L 1}-x^{+m_2}_{\L 2}}{\tx^{+m_1}_{\L 1}{x^{+m_2}_{\R 2}}-1 }  \frac{\tx^{-m_1}_{\L 1}{x^{+m_2}_{\R 2}}-1 } {\tx^{-m_1}_{\L 1}-x^{+m_2}_{\L 2} }\,.
\eal

Then, we need to analytically continue $\tPhi_{\L\R}( \tx_{\L 1}^{\pm m_1},x^{\pm m_2}_{\R 2}) $ and $\tPhi_{\L\R}^\hl( \tx_{\L 1}^{\pm m_1},x^{\pm m_2}_{\R 2}) $ in $\tx_{\L 1}^{+m_1 }$ through the negative semi-axis to the upper half-plane.

\subsubsection*{Continuation of $\tPhi_{\L\R}$ }

We first shift $u_1$ variable $u_1\to u_1+{i\ov h}k$ so that 
$\tx_{\L 1}^{\pm m_1}\to\tx_\L(u_1\pm{i\ov h}m_1+{i\ov h}k)=\tx_{\L 1}^{\pm m_1+k}$, and then analytically continue  $\tx_{\L 1}^{m_1+k }$ to $1/\tx_{\R 1}^{- (k-m_1)}=1/x_{\R 1}^{-(k-m_1)}$, $u_1\in\bR$. We get
\bal
\tPhi_{\L\R}^{-+}( \tx_{\L 1}^{+ m_1},x^{\pm m_2}_{\R 2})\to &+\tPhi_{\L\R}^{-+}({1\ov x_{\R 1}^{- (k-m_1)}},x_{\R 2}^{\pm m_2}) - \tPsi_{\R}^{+}(u_1+{i\ov h}(m_1+k),x_{\R 2}^{\pm m_2})\,,
\\
\tPhi_{\L\R}^{+-}( \tx_{\L 1}^{+ m_1},x^{\pm m_2}_{\L 2})\to &+\tPhi_{\L\R}^{+-}({1\ov x_{\R 1}^{- (k-m_1)}},x_{\L 2}^{\pm m_2 }) - \tPsi_{\R}^{-}(u_1+{i\ov h}(m_1-k),x_{\R 2}^{\pm  m_2})\,,
\eal
\bal
\tilde\Psi^{-}_{\L}(u_{2} + {i\ov h}m_2, \tx_{\L 1}^{+ m_1})\to & \ \tilde\Psi^{-}_{\L}(u_{2} + {i\ov h}m_2, {1\ov x_{\R 1}^{- (k-m_1)}})-K^\bes(u_{12}+{i\ov h}(m_1-m_2) +{i\ov h}k)\,,
\\
\tilde\Psi^{+}_{\L}(u_{2} + {i\ov h}m_2, \tx_{\L 1}^{+ m_1})\to & \  \tilde\Psi^{+}_{\L}(u_{2} + {i\ov h}m_2, {1\ov x_{\R 1}^{- (k-m_1)}})-K^\bes(u_{12} +{i\ov h}(m_1-m_2)-{i\ov h}k)\,.
\eal
Note that  we do not cross any cut of $\tPsi$-functions.

Summing up the $\tPhi$ functions, we get
\bal
2\tilde\theta_{\L\R}^\bes&(\tx_{\L 1}^{\pm m_1},x_{\R 2}^{\pm m_2})\to 
2\tilde\theta_{\R\R}^{k-m_1,m_2}({ x_{\R 1}^{\pm (k-m_1)}},{x_{\R 2}^{\pm m_2 }})-\Delta_{\R}^{+}(u_1+{i\ov h}k\pm {i\ov h}m_1,x_{\R 2}^{\pm m_2})
\\
&+\Delta_{\R}^{-}(u_1\pm{i\ov h}(k-m_1),x_{\R 2}^{\pm m_2})+2K^\bes(u_{12}+{i\ov h}(k-m_1-m_2))
\\
&-K^\bes(u_{12}+{i\ov h}(m_1-m_2) -{i\ov h}k)-K^\bes(u_{12} +{i\ov h}(m_1-m_2)+{i\ov h}k)
\,.
\eal
By using  identities \eqref{eq:Deltaepsm_2ms}, \eqref{eq:Deltaepsp_2ms} for $\tPsi$ functions, 
and  identity \eqref{eq:KbesmKbes} for $K^\bes$,
we get 
\bal\la{eqqq1d0}
e^{-2i\tilde\theta_{\L\R}^\bes(\tx_{\L 1}^{\pm m_1},x_{\R 2}^{\pm m_2})}&A_{ Y \bar Z}^{m_1,m_2}(p_{1},p_{2})\to
e^{-2i\tilde\theta_{\R\R}^\bes({ x_{\R 1}^{\pm (k-m_1)}},{x_{\R 2}^{\pm m_2 }})} A_{\bar  Z Y}^{k-m_1,m_2}(p_{1},p_{2})
 \\
&\times
\frac{x_{\R 2}^{- m_2} }{x_{\R 2}^{+m_2} }
 \frac{ (\tx_{\R 1}^{+(k-m_1)}-x_{\R 2}^{+m_2}) (x_{\R 1}^{+(k+m_1)}-x_{\R 2}^{+m_2}) } {
  (\tx_{\R 1}^{+(k-m_1)}-x_{\R2}^{- m_2})(x_{\R 1}^{+(k+m_1)}-x_{\R 2}^{- m_2}) }
 \,.
\eal
  
\subsubsection*{Continuation of $\tPhi_{\L\R}^\hl$}

Next, we analytically continue $\tPhi_{\L\R}^\hl$ in $\tx_{\L 1}^{+ m_1}\to 1/x_{\R 1}^{- (k-m_1)}$, 
and  get
\bal
\tPhi_{\L\R}^{\hl}(\tx_{\L1}^{+ m_1},x_{\R 2}^{\pm m_2})\to &+\tPhi_{\R\R}^{\hl}( {1\ov x_{\R 1}^{-(k-m_1)}},x_{\R 2}^{\pm m_2})+  {1\ov 2i}\log  \frac{x_{\R 1}^{-(k-m_1)} - x_{\R 2}^{\pm m_2}}{{ x_{\R 1}^{+(k+m_1)}} - x_{\R 2}^{\pm m_2}} {1\ov x_{\R 2}^{\pm m_2}} \pm {\pi\ov4}\,.
\eal
Summing up all the terms, we get
\bal
&e^{2i\tilde\theta_{\L\R}^{\hl}(x_{\L 1}^{\pm m_1},x_{\R 2}^{\pm m_2})}\to  \ 
e^{2i\tilde\theta_{\R\R}^{\hl }(x_{\R 1}^{\pm(k-m_1)},x_{\R 2}^{\pm m_2})}
\\
&
 \times 
\frac{ x_{\R 1}^{-(k-m_1)}-x_{\R 2}^{+m_2} }{ x_{\R 1}^{+(k-m_1)}-x^{-m_2}_{\R 2} }
\frac{ x_{\R1}^{+(k-m_1)}-x_{\R 2}^{+m_2} }{ x_{\R1 }^{-(k-m_1)}-x_{\R 2}^{-m_2} } 
\frac{ \tx_{\R1}^{+(k-m_1)}-x_{\R 2}^{-m_2} }{\tx_{\R1}^{+(k-m_1)} -x_{\R 2}^{+m_2} } 
  \frac{ x_{\R 1}^{+(k+m_1)} -x_{\R 2}^{-m_2}}{ x_{\R 1}^{+(k+m_1)} -x_{\R 2}^{+m_2}}
\,,
\eal
where $\tilde\theta_{\R\R}^{\hl}(x_{\R 1}^{\pm(k-m_1)},x_{\R 2}^{\pm m_2})$ is the string HL phase.

Thus,
\bal\la{eqqq2d}
&e^{-2i\tilde\theta_{\L\R}^\bes(\tx_{\L 1}^{\pm},x_{\R 2}^{\pm})+2i\tilde\theta_{\L\R}^{\hl}(x_{\L 1}^{\pm m_1},x_{\R 2}^{\pm m_2})}A_{ Y  \bar Z}^{m_1,m_2}(p_{1},p_{2})
\\
&\to
e^{-2i\tilde\theta_{\R\R}^\bes({ x_{\R 1}^{\pm (k-m_1)}},{x_{\R 2}^{\pm m_2 }})+2i\tilde\theta_{\R\R}^{\hl }(x_{\R 1}^{\pm(k-m_1)},x_{\R 2}^{\pm m_2})} A_{\bar  Z \bar Z}^{k-m_1,m_2}(p_{1},p_{2})
 \\
&\times
\frac{x_{\R 2}^{- m_2} }{x_{\R 2}^{+ m_2} }
\frac{ x_{\R 1}^{-(k-m_1)}-x_{\R 2}^{+m_2} }{ x_{\R 1}^{+(k-m_1)}-x^{-m_2}_{\R 2} }
\frac{ x_{\R1}^{+(k-m_1)}-x_{\R 2}^{+m_2} }{ x_{\R1 }^{-(k-m_1)}-x_{\R 2}^{-m_2} } 
 \,.
\eal
Finally, we take into account the contribution of the odd factor.
Since we do not cross the cut of $\g^\pm_\L$-rapidities between $(-1/\xi,\xi)$ the  $\g^\pm_\L$-rapidities transform as
\bal
\g_\L(x_{\L 1}^{\pm m_1}) \to \g_\L(1/x_{\R 1}^{\mp (k-m_1)}) &=  \g_{\R 1}^{\mp(k-m_1)} \pm i\pi
\,.
\eal
Thus,
the odd factor in $\Sigma^{\barnes}_{\L\R} (\g_{\L 1}^{\pm m_1},\g_{\R 2}^{\pm m_2})$ becomes
\bal\la{eqqq3d}
&\Sigma^{\barnes}_{\L\R} (\g_{\L 1}^{\pm m_1},\g_{\R 2}^{\pm m_2})
\to
\Sigma^{\barnes}_{\R\R} (\g_{\R 1}^{\pm (k-m_1)},\g_{\R 2}^{\pm m_2})
 {\sinh{\g_{\R\R}^{-(k-m_1),-m_2}\ov2}\ov \sinh{\g_{\R\R}^{+(k-m_1),+m_2}\ov2}}{\sinh{\g_{\R\R}^{+(k-m_1),-m_2}\ov2}\ov \sinh{\g_{\R\R}^{-(k-m_1),+m_2}\ov2}} 
\,.
\eal
Multiplying \eqref{eqqq2d} and \eqref{eqqq3d}, we
find that after the second step of the analytic continuation 
\bal\la{Sybzp1m2pip2Sbzbzp1p2}
S^{m_1 \, m_2}_{Y \bar{Z}} (p_1-2 \pi, p_2) =  S^{k-m_1 \, m_2}_{\bar{Z} \bar{Z}} (p_1, p_2)\,{\a_\R( x^{+m_2}_{\R 2}) \ov \a_\R( x^{-m_2}_{\R 2}) }  \,,
\eal
where $ \alpha_a(x)$ is defined in \eqref{eq:defalpha}.

\subsection{Calculating \texorpdfstring{$S_{ Y \bar Z}^{m_1m_2}(p_{1}-2\pi, p_{2})\,,\, m_1>k$ }{sybzm1m2b}  }

The factor $A_{ Y  \bar Z}^{m_1m_2}(p_{1},p_{2})$ after the two steps $x_{\L1}^{\pm m_1}\to x_{\L1}^{\pm (m_1-k)}$, $u_1\to u_1+{i\ov h}k$ becomes
\bal
 A&_{Y\bar Z}^{m_1m_2}(p_{1},p_{2})
  \to A_{Y \bar Z}^{m_1m_2}(p_{1}-2\pi,p_{2})=
 \left({x_{\L1}^{- (m_1-k)}\ov x_{\L1}^{+ (m_1-k)}}\right)^{m_2-1}
      \left({x_{\R2}^{+m_2}\ov x_{\R2}^{-m_2}}\right)^{m_1} \ \frac{{x^{- (m_1-k)}_{\L1} x^{-m_2}_{\R 2}}-1}{{x^{+ (m_1-k)}_{\L1} x^{+m_2}_{\R 2}}-1}  \\
      &\times\frac{{x^{+ (m_1-k)}_{\L1} x^{-m_2}_{\R 2}}-1}{{x^{- (m_1-k)}_{\L1} x^{+m_2}_{\R 2}}-1}\prod_{j=1}^{m_1-1}  \left(\frac{{x^+_{\L 1j} x^{-m_2}_{\R 2}}-1}{{x^+_{\L 1j} x^{+m_2}_{\R 2}}-1}\right)^2 \  \prod^{m_2-1}_{j=1}\left( \frac{{x^{+ (m_1-k)}_{\L1} x^{+}_{\R 2j}}-1}{{x^{- (m_1-k)}_{\L1} x^{+}_{\R 2j}}-1} \right)^2
    \,,
\eal
where
\bal
x_{\L 1 j}^+ =\tx_\L(u_1 - {i\ov h}(m_1-k-2j))\,,\quad  x_{\R 2 j}^+ =\tx_\R(u_2 - {i\ov h}(m_2-2j))\,.
\eal

After the first step of
the analytic continuation  $\Sigma^{\bes}_{\L\R}$ and $\Sigma^{\hl}_{\L\R}$ are still given by \eqref{BESLRmirrorstring}
 and \eqref{HLLRmirrorstring}.
Then, we need to analytically continue $\tPhi_{\L\R}( \tx_{\L 1}^{\pm m_1},x^{\pm m_2}_{\R 2}) $ and $\tPhi_{\L\R}^\hl( \tx_{\L 1}^{\pm m_1},x^{\pm m_2}_{\R 2}) $ in $\tx_{\L 1}^{+m_1 }$ through the negative semi-axis to the upper half-plane.

\subsubsection*{Continuation of $\tPhi_{\L\R}$ }

We first shift $u_1$ variable $u_1\to u_1+{i\ov h}k$ so that 
$\tx_{\L 1}^{\pm m_1}\to\tx_\L(u_1\pm{i\ov h}m_1+{i\ov h}k)=\tx_{\L 1}^{\pm m_1+k}$, and then analytically continue  $\tx_{\L 1}^{m_1+k }$ to $1/\tx_{\R 1}^{m_1-k}=x_{\L 1}^{+(m_1-k)}$, $u_1\in\bR$. We get
\bal
\tPhi_{\L\R}^{-+}( \tx_{\L 1}^{+ m_1},x^{\pm m_2}_{\R 2})\to &+\tPhi_{\L\R}^{-+}({ x_{\L 1}^{+(m_1-k)}},x_{\R 2}^{\pm m_2}) - \tPsi_{\R}^{+}(u_1+{i\ov h}(m_1+k),x_{\R 2}^{\pm m_2})\,,
\\
\tPhi_{\L\R}^{+-}( \tx_{\L 1}^{+ m_1},x^{\pm m_2}_{\R 2})\to &+\tPhi_{\L\L}^{+-}({ x_{\L 1}^{+(m_1-k)}},x_{\R 2}^{\pm m_2 }) - \tPsi_{\R}^{-}(u_1+{i\ov h}(m_1-k),x_{\R 2}^{\pm  m_2})\,,
\eal
\bal
\tilde\Psi^{-}_{\L}(u_{2} + {i\ov h}m_2, \tx_{\L 1}^{+ m_1})\to & \ \tilde\Psi^{-}_{\L}(u_{2} + {i\ov h}m_2, { x_{\L 1}^{+(m_1-k)}})-K^\bes(u_{12}+{i\ov h}(m_1-m_2) +{i\ov h}k)\,,
\\
\tilde\Psi^{+}_{\L}(u_{2} + {i\ov h}m_2, \tx_{\L 1}^{+ m_1})\to & \  \tilde\Psi^{+}_{\L}(u_{2} + {i\ov h}m_2, { x_{\L 1}^{+(m_1-k)}})-K^\bes(u_{12} +{i\ov h}(m_1-m_2)-{i\ov h}k)\,.
\eal
Note that  we do not cross any cut of $\tPsi$-functions.

Summing up the $\tPhi$ functions, we get
\bal
2\tilde\theta_{\L\R}^\bes&(\tx_{\L 1}^{\pm},x_{\R 2}^{\pm})\to 
2\tilde\theta_{\L\R}^\bes({ x_{\L 1}^{\pm (m_1-k)}},{x_{\R 2}^{\pm m_2 }})-\Delta_{\L\R}^{-+}(u_1+{i\ov h}m_1\pm {i\ov h}k,x_{\R 2}^{\pm m_2})
\\
&+K^\bes(u_{12}+{i\ov h}(m_1-m_2) -{i\ov h}k)-K^\bes(u_{12} +{i\ov h}(m_1-m_2)+{i\ov h}k)
\,.
\eal
By using  identity \eqref{eq:Deltaepsm_2ms} for $\tPsi$ functions, 
and  identity \eqref{eq:KbesmKbes} for $K^\bes$,
we get 
\bal\la{eqqq1d}
e^{-2i\tilde\theta_{\L\R}^\bes(\tx_{\L 1}^{\pm},x_{\R 2}^{\pm})}&A_{ Y  \bar Z}^{m_1,m_2}(p_{1},p_{2})\to
e^{-2i\tilde\theta_{\L\R}^\bes({ x_{\L 1}^{\pm (m_1-k)}},{x_{\R 2}^{\pm m_2 }})} A_{ Y \bar Z}^{m_1-k,m_2}(p_{1},p_{2})
 \\
&\times
 \frac{ (x_{\R 1}^{+(m_1-k)}-x_{\R 2}^{-m_2}) (x_{\R 1}^{+(m_1+k)}-x_{\R 2}^{+ m_2}) } {
 (x_{\R 1}^{+(m_1-k)}-x_{\R 2}^{+m_2}) (x_{\R 1}^{+(m_1+k)}-x_{\R 2}^{- m_2}) }
 \,.
\eal
  
\subsubsection*{Continuation of $\tPhi_{\L\R}^\hl$}

Next, we analytically continue $\tPhi_{\L\R}^\hl$ in $\tx_{\L 1}^{+ m_1}\to x_{\L 1}^{+ (m_1-k)}$, 
and  get
\bal
\tPhi_{\L\R}^{\hl}(\tx_{\L1}^{+ m_1},x_{\R 2}^{\pm m_2})\to &+\tPhi_{\L\R}^{\hl}( { x_{\L 1}^{+(m_1-k)}},x_{\R 2}^{\pm m_2})+  {1\ov 2i}\log  \frac{\tx_{\R 1}^{+(m_1-k)} - x_{\R 2}^{\pm m_2}}{{x_{\R 1}^{+(m_1+k)}} - x_{\R 2}^{\pm m_2}} {1\ov x_{\R 2}^{\pm m_2}} \pm {\pi\ov4}\,.
\eal
Summing up all the terms, we get
\bal
&e^{2i\tilde\theta_{\L\R}^{\hl}(x_{\L 1}^{\pm m_1},x_{\R 2}^{\pm m_2})}\to  \ 
e^{2i\tilde\theta_{\L\R}^{\hl }(x_{\L 1}^{\pm(m_1-k)},x_{\R 2}^{\pm m_2})}
\\
&
 \times 
 \frac{
 (x_{\R 1}^{+(m_1-k)}-x_{\R 2}^{+m_2}) (x_{\R 1}^{+(m_1+k)}-x_{\R 2}^{- m_2}) }{ (x_{\R 1}^{+(m_1-k)}-x_{\R 2}^{-m_2}) (x_{\R 1}^{+(m_1+k)}-x_{\R 2}^{+ m_2}) } 
\,,
\eal
where $\tilde\theta_{\L\R}^{\hl}(x_{\R 1}^{\pm(m_1-k)},x_{\R 2}^{\pm m_2})$ is the string HL phase.

The term on the second line cancels the remaining term in \eqref{eqqq1d}, and since the odd dressing factor in $\Sigma^{m_1,m_2}_{\L\R} (p_1, p_2)$ becomes the odd factor in $\Sigma^{m_1-k,m_2}_{\L\R} (p_1, p_2)$, we conclude that 
\bal\la{Sybzp1m2pip2Sybzp1p2a}
 S_{ Y  \bar Z}^{m_1,m_2}(p_{1}-2\pi,p_{2}) = S_{  Y \bar Z}^{m_1-k,m_2}(p_{1},p_{2}) \,, \quad m_1>k\,.
\eal
Shifting $p_1$ by $2\pi$, and $m_1\to m_1+k$ we can also write the relation in the form
\bal\la{Sybzp1m2pip2Sybzp1p2b}
 S_{ Y  \bar Z}^{m_1,m_2}(p_{1}+2\pi,p_{2}) = S_{  Y \bar Z}^{m_1+k,m_2}(p_{1},p_{2}) \,, \quad m_1>0\,.
\eal

\section{CP invariance relations}\la{app:CP}

\subsection{Proof of \texorpdfstring{$S_{\bar Y\bar Y}^{m_1m_2}(-p_{1},-p_{2})=S_{ Y Y}^{m_2m_1}(p_{2},p_{1})$ }{sbybyyym1m2} }

To prove the CP invariance condition $S_{\bar Y\bar Y}^{m_1m_2}(-p_{1},-p_{2})=S_{Y Y}^{m_2m_1}(p_{2},p_{1})$ we analytically continue  the equation \eqref{Sbybyp1m2pip2Sbybyp1p2} in $p_2$ to $p_2-2\pi$, and then replace $p_i\to 2\pi - p_i$. Then,  \eqref{Sbybyp1m2pip2Sbybyp1p2}  takes the form
\bal\la{Sbybymp1mp2Sbybyp1p2}
 S_{ \bar Y \bar Y}^{m_1,m_2}(-p_{1},-p_{2}) = S_{ \bar Y \bar Y}^{m_1+k,m_2+k}(2\pi-p_{1},2\pi-p_{2}) \,.
\eal
Since on the $u$-plane the transformation $p\to 2\pi-p$ corresponds to $u\to-u$ (see equation~\eqref{eq:secCP_pm2p_to_mp}), to prove the CP invariance condition we need to show that
\bal
S_{ \bar Y \bar Y}^{m_1+k,m_2+k}(-u_{1},-u_{2}) = S_{ Y Y}^{m_2m_1}(u_{2},u_{1})\,.
\eal
Taking into account that under the reflection transformation $u\to -u$ the right Zhukovsky variables transform as
\bal
x_{\R}^{\pm (k+m)} \ \xrightarrow{u\to-u} \ -x_{\L}^{\mp m}\,,
\eal
\bal
x_{\R j}^+= \tx_\R(u - {i\ov h}(k+m-2j))\ \xrightarrow{u\to-u} 
\  -{1\ov \tx_\R(u+ {i\ov h}m-{2i\ov h}j) }=-{1\ov \tx_\R^{m-2j }}\,,
\eal
and using \eqref{eq:SbYbY}, we get
\bal
&S_{\bar Y\bar Y}^{k+m_1,k+m_2}(-u_{1},-u_{2}) = A_{\bar Y\bar Y}^{k+m_1,k+m_2}(-u_{1},-u_{2}) \big(\Sigma^{k+m_1,k+m_2}_{\R\R}( -x_{\L1}^{\mp m_1},  -x_{\L2}^{\mp m_2})\big)^{-2}\,,
\\
&A_{\bar Y\bar Y}^{k+m_1,k+m_2}(-u_{1},-u_{2})= \left( {x_{\L1}^{-m_1}\ov x_{\L1}^{+m_1}} \right)^{k+m_2} \left( {x_{\L2}^{+m_2}\ov x_{\L2}^{-m_2}}\right)^{k+m_1} \left(\frac{x^{+m_1}_{\L1} - x^{- m_2}_{\L2}}{x^{- m_1}_{\L1} - x^{+ m_2}_{\L2}} \right)^2
\\
&\times
 \prod^{k+m_1-1}_{j=1} \left(\frac{{ \tx_{\R1}^{m_1-2j}} x_{\L2}^{-m_2}-1}{{\tx_{\R1}^{m_1-2j}} x_{\L2}^{+m_2}-1 } \right)^2  
 \prod^{k+m_2-1}_{j=1} \left(\frac{ x_{\L1}^{+m_1}\tx_{\R2}^{m_2-2j }-1}{x_{\L1}^{-m_1} \tx_{\R2}^{m_2-2j }-1} \right)^2
 \\
 &\times   \frac{u_{12}- \frac{i}{h} (m_1-m_2)}{u_{12}+ \frac{i}{h} (m_1-m_2)}\ \frac{u_{12}- \frac{i}{h} (2k+m_1+m_2)}{u_{12}+ \frac{i}{h} (2k+m_1+m_2)}  
  \prod^{k+m_2-1}_{j=1} \left(\frac{u_{12} - {i\ov h} (m_1-m_2+2 j)}{u_{12} + {i\ov h} (m_1-m_2+2 j)} \right)^2
\,.
\eal
Next, we need to relate
$\Sigma^{k+m_1,k+m_2}_{\R\R}( -x_{\L1}^{\mp m_1},  -x_{\L2}^{\mp m_2})$ to $\Sigma^{m_2m_1}_{\L\L}( x_{\L2}^{\pm m_2},  x_{\L1}^{\pm m_1})$.
We have ($\veps=\pm)$
\bal
\tilde\Phi_{aa}^{\veps\veps}(x_1,x_2)&=-\lint_{\pa\cR_\veps}\frac{{\rm d}w_1}{2\pi i}\lint_{\pa\cR_\veps} \frac{{\rm
d}w_2}{2\pi i}\frac{1}{(w_1-x_1)(w_2-x_2)} K^\bes\big(u_a(w_1)-u_a(w_2)\big)\,,
\eal
\bal
\tilde\Psi_{a}^{\veps}(u_1,x_2)&=-\lint_{\pa\cR_\veps} \frac{{\rm
d}w_2}{2\pi i}\frac{1}{w_2-x_2}K^\bes\big(u_1-u_a(w_2)\big)\,.
\eal
Since
\bal
 u_a(-w)=-u_{\bar a}(w) +i\ka_a\,\sgn(\Im(w))\,,
\eal
we get
\bal
\tilde\Phi_{aa}^{\veps\veps}(-{ x_1},-{x_2})&=-\tilde\Phi_{\bar a\bar a}^{-\veps-\veps}(x_1,x_2)\,,
\eal
and
\bal
\tilde\Psi_{a}^{\veps}(-u_1,-x_2)&=\tPsi_{\bar a}^{ -\veps}\big(u_1+i\ka_{\bar a}\,\sgn(\veps),x_2\big)\,.
\eal
By using the formulae, we get
\bal\la{thetarrthetall}
2\tilde\theta_{\R\R}^\bes(-x_{\L1}^{\mp m_1}&,-x_{\L2}^{\mp m_2})  =2\tilde\theta_{\L\L}^\bes(x_{\L2}^{\pm m_2},x_{\L1 }^{\pm m_1}) 
\\
&-\Delta_{\L}^{-}(u_1\pm{i\ov h}m_1,x_{\L2}^{\pm m_2})-\Delta_{\L}^{+}(u_1-{i\ov h}k\pm {i\ov h}(m_1+k),x_{\L2}^{\pm m_2})
\\
&+\Delta_{\L}^{-}(u_2\pm{i\ov h}m_2,x_{\L1}^{\pm m_1})+\Delta_{\L}^{+}(u_2-{i\ov h}k\pm {i\ov h}(m_2+k),x_{\L1}^{\pm m_1})
\\
&
-2K^\bes\big(u_{12}+{i\ov h}(m_1-m_2)\big)+2K^\bes\big(u_{12}-{i\ov h}(m_1-m_2)\big)
\,.
\eal
Applying  identities \eqref{eq:Deltaepsm_2ms}, \eqref{eq:Deltaepsp_2ms} for $\tPsi$ functions, 
and  identity \eqref{eq:KbesmKbes} for $K^\bes$,
we get 
\bal
&A_{\bar Y\bar Y}^{k+m_1,k+m_2}(-u_{1},-u_{2}) e^{-2i\theta^\bes_{\R\R}( -x_{\L 1}^{\mp m_1},  -x_{\L2}^{\mp m_2})} =A_{ Y Y}^{m_2,m_1}(u_{2},u_{1}) e^{-2i\tilde{\theta}^\bes_{\L\L}( x_{\L2}^{\pm m_2},  x_{\L1}^{\pm m_1})} 
\\
&\times 
\frac{x_{\L 1}^{+m_1}-\tx^{+m_2}_{\L 2}} {x_{\L 1}^{-m_1}-\tx^{+m_2}_{\L 2} } 
 \frac{\tx^{+m_1}_{\L 1} -x_{\L 2}^{-m_2}}{\tx^{+m_1}_{\L 1} -x_{\L 2}^{+m_2}}
 \frac{x_{\L1}^{-m_1} x_{\R2}^{-m_2-2k}-1} {x_{\L1}^{+m_1}x_{\R2}^{-m_2-2k}-1 } 
 \frac{ x_{\R1}^{-m_1-2k}x_{\L2}^{+m_2}-1} {x_{\R1}^{-m_1-2k}x_{\L2}^{-m_2}-1}\,.
\eal
It is then easy to check that 
\bal
 e^{2i \tilde{\theta}^\hl_{\R\R}( -x_{\L1}^{\mp m_1},  -x_{\L2}^{\mp m_2})}& \frac{x_{\L 1}^{+m_1}-\tx^{+m_2}_{\L 2}} {x_{\L 1}^{-m_1}-\tx^{+m_2}_{\L 2} } 
 \frac{\tx^{+m_1}_{\L 1} -x_{\L 2}^{-m_2}}{\tx^{+m_1}_{\L 1} -x_{\L 2}^{+m_2}}
 \frac{x_{\L1}^{-m_1} x_{\R2}^{-m_2-2k}-1} {x_{\L1}^{+m_1}x_{\R2}^{-m_2-2k}-1 } 
 \frac{ x_{\R1}^{-m_1-2k}x_{\L2}^{+m_2}-1} {x_{\R1}^{-m_1-2k}x_{\L2}^{-m_2}-1} 
 \\
 &= e^{2i \tilde{\theta}^\hl_{\L\L}( x_{\L2}^{\pm m_2},  x_{\L1}^{\pm m_1})} \,,
\eal
and 
\bal
 e^{2i \tilde{\theta}^\barnes_{\R\R}( -x_{\L1}^{\mp m_1},  -x_{\L2}^{\mp m_2})}= e^{2i \tilde{\theta}^\barnes_{\L\L}( x_{\L2}^{\pm m_2},  x_{\L1}^{\pm m_1})} \,,
\eal
and therefore 
\bal
S_{\bar Y\bar Y}^{k+m_1,k+m_2}(-u_{1},-u_{2})=S_{ Y Y}^{m_2m_1}(u_{2},u_{1})\quad \Leftrightarrow\quad  S_{\bar Y\bar Y}^{k+m_1,k+m_2}(-u_{1},-u_{2})S_{ Y Y}^{m_1m_2}(u_{1},u_{2})=1\,.
\eal
This completes the proof of the CP invariance condition $S_{\bar Y\bar Y}^{m_1m_2}(-p_{1},-p_{2})=S_{Y Y}^{m_2m_1}(p_{2},p_{1})$.

\subsection{Proof of \texorpdfstring{$S_{\bar Y Z}^{m_1m_2}(-p_{1},p_{2})=S_{ \bar Z Y}^{m_2m_1}(-p_{2},p_{1})$ }{sbyzbzym1m2}}

To prove the CP invariance condition $S_{\bar Y Z}^{m_1m_2}(-p_{1},p_{2})=S_{ \bar Z Y}^{m_2m_1}(-p_{2},p_{1})$ 
we  first replace $p_2\to 2\pi - p_2$. Then,  the CP invariance condition  becomes
\bal
S_{\bar Y Z}^{m_1m_2}(-p_{1},2\pi - p_{2})=S_{ \bar Z Y}^{m_2m_1}(p_{2}-2\pi ,p_{1}) \,.
 \eal
Next, we 
replace $p_1\to 2\pi - p_1$, and cast \eqref{Sbybyp1m2pip2Sbybyp1p2}  in the form
\bal\la{Sbyzp1p2Sbyzp1p2a}
 S_{ \bar Y Z}^{m_1,m_2}(-p_{1},2\pi-p_{2}) = S_{ \bar Y Z}^{m_1+k,m_2}(2\pi-p_{1},2\pi-p_{2}) \,.
 \eal
Since $S_{ \bar Z Y}^{m_2m_1}$ satisfies an equation similar to  \eqref{Sbybyp1m2pip2Sbybyp1p2}  
\bal\la{Sbyzp1p2Sbyzp1p2b}
   S_{ \bar Z Y}^{m_1,m_2}(p_{2}-2\pi,p_{1}) = S_{ \bar Z Y}^{m_2+k,m_1}(p_{2},p_{1}) \,,
\eal
we see that the CP invariance condition becomes equivalent to the relation
\bal\la{Sbyzp1p2Szbyp1p2}
S_{ \bar Y Z}^{m_1+k,m_2}(2\pi-p_{1},2\pi-p_{2}) =  S_{ \bar Z Y}^{m_2+k,m_1}(p_{2},p_{1})\,,
\eal
or, by using the $u$-plane rapidities
\bal
S_{ \bar Y Z}^{m_1+k,m_2}(-u_{1},-u_{2}) =  S_{ \bar Z Y}^{m_2+k,m_1}(u_{2},u_{1})\ \  \Leftrightarrow\ \  S_{ \bar Y Z}^{m_1+k,m_2}(-u_{1},-u_{2})S_{Y \bar Z }^{m_1,m_2+k}(u_{1},u_{2})=1.
\eal
Taking into account that under the reflection transformation $u\to -u$ the right and left Zhukovsky variables transform as
\bal
x_{\R}^{\pm (k+m)} \ \xrightarrow{u\to-u} \ -x_{\L}^{\mp m}\,,
\eal
\bal
x_{\R j}^+= \tx_\R(u - {i\ov h}(k+m-2j))\ \xrightarrow{u\to-u} 
\  -{1\ov \tx_\R(u+ {i\ov h}m-{2i\ov h}j) }=-{1\ov \tx_\R^{m-2j }}\,,
\eal
\bal
x_{\L}^{\pm m}  \ \xrightarrow{u\to-u} \  -x_{\R}^{\mp (k+m)}\,,
\eal
\bal
x_{\L j}^+ =\tx_\L(u - {i\ov h}(m-2j)) \ \xrightarrow{u\to-u} \  -{1\ov \tx_\L(u + {i\ov h}(k+m-2j))} =  -{1\ov \tx_\L^{k+m-2j}}\,.
\eal
and using \eqref{eq:SbYZ}, we get
\bal
&S_{\bar Y Z}^{k+m_1,m_2}(-u_{1},-u_{2}) = A_{\bar Y Z}^{k+m_1,m_2}(-u_{1},-u_{2}) \big(\Sigma^{k+m_1,k+m_2}_{\R\L}( -x_{\L1}^{\mp m_1},  -x_{\R2}^{\mp (m_2+k)})\big)^{-2}\,,
\\
&A_{\bar Y Z}^{k+m_1,m_2}(-u_{1},-u_{2})=  \left({x_{\L 1}^{+m_1}\ov x_{\L 1}^{-m_1}}\right)^{m_2-1}
      \left({x_{\R 2}^{-(k+m_2)}\ov x_{\R 2}^{+(k+m_2)}}\right)^{k+m_1} \ \frac{{x^{+m_1}_{\L 1} x^{+(k+m_2)}_{\R 2}}-1}{{x^{-m_1}_{\L 1} x^{-(k+m_2)}_{\R 2}}-1} 
      \\
      &\times\frac{{x^{-m_1}_{\L 1} x^{+(k+m_2)}_{\R 2}}-1}{{x^{+m_1}_{\L 1} x^{-(k+m_2)}_{\R 2}}-1} 
      \prod_{j=1}^{k+m_1-1}  \left(\frac{{\tx^{m_1-2j}_{\R 1} -x^{+(k+m_2)}_{\R 2}}}{{\tx^{m_1-2j}_{\R 1} -x^{-(k+m_2)}_{\R 2}}}\right)^2 \  
      \prod^{m_2-1}_{j=1}\left( \frac{{x^{-m_1}_{\L 1} -\tx^{k+m_2-2j}_{\L 2}}}{{x^{+m_1}_{\L 1} -\tx^{k+m_2-2j}_{\L 2}}}\right)^2 
\,.
\eal
Thus,
\bal\la{AybzmmAyzb}
A&_{\bar Y  Z}^{k+m_1,m_2}(-u_{1},-u_{2})A_{Y \bar Z}^{m_1,k+m_2}(u_{1},u_{2})
\\
&=
      \left({x_{\R 2}^{-(k+m_2)}\ov x_{\R 2}^{+(k+m_2)}}\right)^{k} \prod_{j=1}^{m_1-1}  \left(\frac{{\tx^{m_1-2j}_{\L1} x^{-(k+m_2)}_{\R 2}}-1}{{\tx^{m_1-2j}_{\L1} x^{+(k+m_2)}_{\R 2}}-1}\right)^2\ \prod_{j=1}^{k+m_1-1}  \left(\frac{{\tx^{m_1-2j}_{\R 1} -x^{+(k+m_2)}_{\R 2}}}{{\tx^{m_1-2j}_{\R 1} -x^{-(k+m_2)}_{\R 2}}}\right)^2 
 \\
  &\times \left({x_{\L 1}^{-m_1}\ov x_{\L 1}^{+m_1}}\right)^{k}     \prod^{m_2-1}_{j=1}\left( \frac{{x^{-m_1}_{\L 1} -\tx^{k+m_2-2j}_{\L 2}}}{{x^{+m_1}_{\L 1} -\tx^{k+m_2-2j}_{\L 2}}}\right)^2 \  \prod^{k+m_2-1}_{j=1}\left( \frac{{x^{+m_1}_{\L1} \tx^{k+m_2-2j}_{\R2}}-1}{{x^{-m_1}_{\L1} \tx^{k+m_2-2j}_{\R2}}-1} \right)^2 
      \,.
      \eal
Next, we need to compute
\bal
\big(\Sigma^{k+m_1, m_2}_{\R\L}(-u_1,-u_2)\Sigma^{m_1,k+m_2}_{\L\R}(u_{1},u_{2})\big)^{2} \,.
\eal
Since 
\bal
\g_a\big(- x\big) = - \g_{\bar a}\big(x\big) \,,
\eal
the product of odd factors is equal to 1.

Then, by using 
\bal
\tilde\Phi_{a\bar a}^{+\veps,-\veps}(-{ x_1},-{x_2})&=-\tilde\Phi_{\bar a a}^{-\veps,+\veps}(x_1,x_2)\,,
\\
\tilde\Psi_{\bar a}^{-\veps}(-u_1,-x_2)&=\tilde\Psi_{ a}^{\veps}(u_1-i\ka_a \sgn(\veps),x_2)\,,
\eal
 we get
\bal
2\tilde\theta_{\R\L}^\bes&(-x_{\L1}^{\mp m_1},-x_{\R2}^{\mp (k+m_2)})  +2\tilde\theta_{\L\R}^\bes(x_{\L1}^{\pm m_1},x_{\R2}^{\pm(k+m_2)}) =  
\\
&-\Delta_{\R}^{+}\big(u_1\pm {i\ov h}m_1,x_{\R2}^{\pm(k+m_2)}\big) 
- \Delta_{\R}^{-}\big(u_1- {i\ov h}k \pm{i\ov h}(k+m_1),x_{\R2}^{\pm(k+m_2)}\big) 
\\
&+\Delta_{\L}^{+}\big(u_2\pm {i\ov h}(k+m_2),x_{\L1}^{\pm m_1}\big) 
+\Delta_{\L}^{-}\big(u_2+{i\ov h}k \pm {i\ov h}m_2,x_{\L1}^{\pm m_1}\big)
\\
&-2K^\bes\big(u_{12}+{i\ov h}(m_1-m_2-k)\big)+2K^\bes\big(u_{12}-{i\ov h}(m_1-m_2+k)\big)
\,.
\eal
Applying  identities \eqref{eq:Deltaepsm_2ms}, \eqref{eq:Deltaepsp_2ms} for $\tPsi$ functions, 
and  identity \eqref{eq:KbesmKbes} for $K^\bes$,
we find
 \bal
\exp&\Big[2i\big(\tilde\theta_{\R\L}^\bes(-x_{\L1}^{\mp m_1},-x_{\R2}^{\mp (k+m_2)})  +\tilde\theta_{\L\R}^\bes(x_{\L1}^{\pm m_1},x_{\R2}^{\pm(k+m_2)})\big)\Big] 
\\
&=
      \left({x_{\R 2}^{-(k+m_2)}\ov x_{\R 2}^{+(k+m_2)}}\right)^{k} \prod_{j=1}^{m_1-1}  \left(\frac{{\tx^{m_1-2j}_{\L1} x^{-(k+m_2)}_{\R 2}}-1}{{\tx^{m_1-2j}_{\L1} x^{+(k+m_2)}_{\R 2}}-1}\right)^2\ \prod_{j=1}^{k+m_1-1}  \left(\frac{{\tx^{m_1-2j}_{\R 1} -x^{+(k+m_2)}_{\R 2}}}{{\tx^{m_1-2j}_{\R 1} -x^{-(k+m_2)}_{\R 2}}}\right)^2 
 \\
  &\times \left({x_{\L 1}^{-m_1}\ov x_{\L 1}^{+m_1}}\right)^{k}     \prod^{m_2-1}_{j=1}\left( \frac{{x^{-m_1}_{\L 1} -\tx^{k+m_2-2j}_{\L 2}}}{{x^{+m_1}_{\L 1} -\tx^{k+m_2-2j}_{\L 2}}}\right)^2 \  \prod^{k+m_2-1}_{j=1}\left( \frac{{x^{+m_1}_{\L1} \tx^{k+m_2-2j}_{\R2}}-1}{{x^{-m_1}_{\L1} \tx^{k+m_2-2j}_{\R2}}-1} \right)^2 
\\
&\times   \frac{  \tx_{\L 1}^{+m_1}x_{\R 2}^{- (k+m_2)}-1 } {
  \tx_{\L 1}^{+m_1}x_{\R 2}^{+(k+ m_2)}-1}  \
  \frac{\tx_{\R 1}^{-(2k+m_1)}-x_{\R 2}^{+ (k+m_2)} } { \tx_{\R 1}^{-(2k+m_1)}-x_{\R 2}^{- (k+m_2)}  }\
  \frac{
   x_{\L1}^{+ m_1}\tx_{\R 2}^{+ (k+m_2)} -1 }{ x_{\L1}^{- m_1}\tx_{\R 2}^{+ (k+m_2)}-1 } \
   \frac{ x_{\L1}^{- m_1} -\tx_{\L 2}^{k-m_2} }{
   x_{\L1}^{+ m_1}-\tx_{\L 2}^{k-m_2}}\,.
\eal
Thus,
\bal\la{AybzmmAyzbbes}
&A_{\bar Y  Z}^{k+m_1,m_2}(-u_{1},-u_{2})A_{Y \bar Z}^{m_1,k+m_2}(u_{1},u_{2})e^{-2i\big(\tilde\theta_{\R\L}^\bes(-x_{\L1}^{\mp m_1},-x_{\R2}^{\mp (k+m_2)})  +\tilde\theta_{\L\R}^\bes(x_{\L1}^{\pm m_1},x_{\R2}^{\pm(k+m_2)})\big)}
\\
&=  
\frac{\tx_{\L 1}^{+m_1}x_{\R 2}^{+(k+ m_2)}-1}{  \tx_{\L 1}^{+m_1}x_{\R 2}^{- (k+m_2)}-1 }  \
  \frac{ \tx_{\R 1}^{-(2k+m_1)}-x_{\R 2}^{- (k+m_2)}  }{\tx_{\R 1}^{-(2k+m_1)}-x_{\R 2}^{+ (k+m_2)} } \
  \frac{ x_{\L1}^{- m_1}\tx_{\R 2}^{+ (k+m_2)}-1 }{ x_{\L1}^{+ m_1}\tx_{\R 2}^{+ (k+m_2)} -1 } \
   \frac{ x_{\L1}^{+ m_1}-\tx_{\L 2}^{k-m_2}}{ x_{\L1}^{- m_1} -\tx_{\L 2}^{k-m_2} }
      \,.
      \eal
Finally, it is easy to check that the contribution of the HL phases is given by
\bal
\exp&\Big[2i\big(\tilde\theta_{\R\L}^\hl(-x_{\L1}^{\mp m_1},-x_{\R2}^{\mp (k+m_2)})  +\tilde\theta_{\L\R}^\hl(x_{\L1}^{\pm m_1},x_{\R2}^{\pm(k+m_2)})\big)\Big] 
\\
&=
 \frac{  \tx_{\L 1}^{+m_1}x_{\R 2}^{- (k+m_2)}-1 } {
  \tx_{\L 1}^{+m_1}x_{\R 2}^{+(k+ m_2)}-1}  \
  \frac{\tx_{\R 1}^{-(2k+m_1)}-x_{\R 2}^{+ (k+m_2)} } { \tx_{\R 1}^{-(2k+m_1)}-x_{\R 2}^{- (k+m_2)}  }\
  \frac{
   x_{\L1}^{+ m_1}\tx_{\R 2}^{+ (k+m_2)} -1 }{ x_{\L1}^{- m_1}\tx_{\R 2}^{+ (k+m_2)}-1 } \
   \frac{ x_{\L1}^{- m_1} -\tx_{\L 2}^{k-m_2} }{
   x_{\L1}^{+ m_1}-\tx_{\L 2}^{k-m_2}}\,,
\eal
 and therefore
\bal
S_{\bar Y  Z}^{k+m_1,m_2}&(-u_{1},-u_{2})S_{Y \bar Z}^{m_1,k+m_2}(u_{1},u_{2})=1\,.
\eal
This completes the proof of the CP invariance condition $S_{\bar Y Z}^{m_1m_2}(-p_{1},p_{2})=S_{ \bar Z Y}^{m_2m_1}(-p_{2},p_{1})$. 

\subsection{List of relations for \texorpdfstring{$S_{AB}(-u_1,-u_2)$}{sAB}}

In this appendix we list various useful relations between S-matrix elements related by the reflection transformation $u\to -u$

\bal
\label{eq:listofrelationsminusu}
S_{ \bar Y \bar Y}^{m_1+k,m_2+k}(-u_{1},-u_{2}) &= S_{ Y Y}^{m_2m_1}(u_{2},u_{1})\,,
\\
S_{\bar Y  Z}^{k+m_1,m_2}(-u_{1},-u_{2})&=S_{ \bar Z  Y}^{k+m_2,m_1}(u_{2},u_{1})\,,
\\
S_{\bar Y\bar Y}^{k+m_1,k-m_2}(-u_{1},-u_{2})&=S_{  \bar Z Y}^{m_2m_1}(u_{2},u_{1})\, {\a_\L( x^{-m_1}_{\L 1}) \ov \a_\L( x^{+m_1}_{\L 1}) }
\\
S_{\bar Y\bar Y}^{k-m_1,k-m_2}(-u_{1},-u_{2})&={\a_\R( x^{-m_2}_{\R 2}) \ov \a_\R( x^{+m_2}_{\R 2}) } \, S_{ \bar Z\bar Z}^{m_2m_1}(u_{2},u_{1})\, {\a_\R( x^{+m_1}_{\R 1}) \ov \a_\R( x^{-m_1}_{\R 1}) }
 \,,
\eal
where $ \alpha_a(x)$ is defined in \eqref{eq:defalpha}.

\section{Large-tension expansion}
\label{app:large-tension}

In this appendix we consider the large-tension limit of the dressing factors. Recall that $\ka$ and $h$ are related to the string tension $T$ and the continuous parameter $q \in [0,1]$ interpolating between the pure R-R and pure NS-NS backgrounds as follows
\bal
\ka = \frac{2 \pi  q}{\sqrt{1-q^2}}\,,\qquad h=\sqrt{1-q^2} T\,,
\eal
where $T \gg1$. The worldsheet theory is perturbative in this regime and we can split the S-matrix into a free and interacting part as
\bal
\mathbf{S}=\boldsymbol{1}+i \mathbf{T} \,,
\eal
In the rest of this appendix we compute certain elements of the operator $\frac{1}{i} \log \mathbf{S}$ to one-loop order in the large string tension.

\subsection{Tree-level mirror S-matrix}

Let us start calculating the tree-level S-matrix element for right mirror fundamental particles in the near-BMN limit. Before the limit, this element is given by (see~\eqref{eq:massivenormmir})
\bal
\la{eq:app_largeTmirror}
\mathbf{S}\,\big|\bar{Z}_{u_1}\bar{Z}_{u_2}\big\rangle&=&
    \frac{u_\R(\tx^+_{\R1})-u_\R(\tx^-_{\R2})}{u_\R(\tx^-_{\R1})-u_\R(\tx^+_{\R2})} \, \big(\Sigma^{11}_{\R\R}\big)^{-2}\,
    \big|\bar{Z}_{u_1}\bar{Z}_{u_2}\big\rangle\,.
    \eal
We set $\tx_{\R i}^\pm=\tx_\R(u_i\pm {i\ov h})$.
To find the tree-level {\bf T}-matrix we expand $\tx_{\R i}^\pm$ in powers of $1/h$ keeping $u_i$ and $\ka$ fixed
\bal
\la{eq:app_x_u_expansion}
\tx^\pm_{\R i} = \tx_{\R}(u_i \pm \frac{i}{h}) =\tx_{\R i}\pm \frac{i}{h}\frac{1}{u'_{\R}(\tx_{i})}+ \frac{1}{2h^2} \frac{u''_\R(\tx_i)}{(u'_\R(\tx_i))^3}+ \mathcal{O}(h^{-3})
\eal
with $\tx_{\R i} \equiv \tx_\R(u_i)$. This expansion can be straightforwardly related to the $T\gg1$ expansion at~$q$ fixed.
If we take $T \gg1$ with $q$ fixed then the difference between the BES and HL phases returns
\bal
\tilde{\theta}_{\R\R}^\bes(\tx^\pm_{\R 1}, \tx^\pm_{\R 2})-\tilde{\theta}_{\R\R}^\hl(\tx^\pm_{\R 1}, \tx^\pm_{\R 2}) = \sqrt{1-q^2}\,T\, \tilde{\theta}_{\R\R}^\afs(\tx^\pm_{\R 1}, \tx^\pm_{\R 2})+ \mathcal{O}(T^{-3})
\eal
which can be seen using the form of the dressing factors~\eqref{eq:fourPhidecomp} and recalling the expansion~\eqref{eq:BES-expansion}. The subleading terms matter at two loops in perturbation theory.
In the near-BMN limit we have, using again~\eqref{eq:fourPhidecomp},
\bal\la{eq:thetaafs_mir_sec_der}
&\tilde{\theta}_{\R\R}^\afs(\tx^\pm_{\R 1}, \tx^\pm_{\R 2})=- \frac{4}{(1-q^2)T^2} \frac{1}{u'_\R(\tx_{\R 1}) u'_\R(\tx_{\R 2})}  {\stackrel{\prime\prime}{\tPhi}}{}^{\afs}_{\R \R}(\tx_{\R1},\tx_{\R2}) +\mathcal{O}(T^{-4})
\eal
and substituting the first formula of~\eqref{eq:afs_mixed_der} into the expression above we obtain
\bal
2\sqrt{1-q^2}T\tilde{\theta}_{\R\R}^\afs(\tx^\pm_{\R 1}, \tx^\pm_{\R 2})=\frac{1}{\sqrt{1-q^2}T}f_{u}(\tx_{\R 1}, \tx_{\R 2})+\frac{1}{\sqrt{1-q^2}T}f_{x}(\tx_{\R 1}, \tx_{\R 2})+
\mathcal{O}(T^{-3})
\eal
where
\bal
f_{u}(\tx_{\R 1}, \tx_{\R 2}) \equiv  \frac{4}{u_\R(\tx_{\R 1})-u_\R(\tx_{\R 2})}\,,
\eal
while the remaining term is a rational function of $\tilde{x}_{\R j}$,
\bal
\label{eq:fx_defined}
f_{x}(\tx_{\R1}, \tx_{\R2})\equiv& - 2\frac{u'_\R(\tx_{\R1})+u'_\R(\tx_{\R2})}{u'_\R(\tx_{\R1}) u'_\R(\tx_{\R2})} \frac{1}{\tx_{\R1} - \tx_{\R2}} + \frac{1}{\tx_{\R1} u'_\R(\tx_{\R1})} - \frac{1}{\tx_{\R2} u'_\R(\tx_{\R2})}\\
&- \frac{1}{u'_\R(\tx_{\R1}) u'_\R(\tx_{\R2})}\left( \frac{1}{\tx_{\R1}} -  \frac{1}{\tx_{\R2}}+ \frac{1}{\tx_{\R1} \tx^2_{\R2}} -  \frac{1}{\tx^2_{\R1} \tx_{\R2}} \right)\,.
\eal
The normalisation factor in front of~\eqref{eq:app_largeTmirror} gives
\bal
\frac{1}{i} \log \frac{u_\R(\tx^+_{\R1})-u_\R(\tx^-_{\R2})}{u_\R(\tx^-_{\R1})-u_\R(\tx^+_{\R2})} = \frac{1}{\sqrt{1-q^2}T} \frac{4}{u_{\R}(\tx_{\R1}) - u_{\R}(\tx_{\R2})}+\mathcal{O}(T^{-3})\,.
\eal
This cancels $f_u$, which is necessary to match with perturbation theory, lest we encountered terms which are logarithmic in $\tx_j$ (and therefore in $\tilde{p}_j$) at the tree level.
The odd-part of the dressing factor does not contribute at the tree level, hence
\bal
\la{eq:tree_S_matrix_bZbZ_mirror1}
T^{11}_{\bar{Z} \bar{Z}} =\frac{1}{i} \log S^{11}_{\bar{Z} \bar{Z}}  =-\frac{1}{\sqrt{1-q^2}T}f_{x}(\tx_{\R1}, \tx_{\R2})+\mathcal{O}(T^{-2})\,.
\eal
We can express the result as a function of the energies and momenta by using the explicit functionality of $u_\R$ in~\eqref{eq:app_x_u_expansion}; then we find 
\bal
\tx_{\R i}^\pm = \tx_{\R i} \pm\frac{i \tx_{\R i}^2}{T \left(\sqrt{1-q^2}
   \left(\tx_{\R i}^2-1\right)+2 q \tx_{\R i}\right)} +
   \mathcal{O}(T^{-2})\,.
\eal
Similarly
\bal
\label{eq:mirror_constraints_xi}
{ \tilde\omega_{\R i}\ov T} &= \ln \tx_{\R i}^- -\ln \tx_{\R i}^+  \,,\quad \tilde\omega_{\R i}=-\frac{2 i \tx_{\R i}}{\sqrt{1-q^2}
   \left(\tx_{\R i}^2-1\right)+2 q \tx_{\R i}}+\mathcal{O}(T^{-2}) \,,
\\
\tp_{\R i} &= {h\ov2}\left( \tx_{\R i}^- -{1\ov \tx_{\R i}^-} - \tx_{\R i}^+ +{1\ov \tx_{\R i}^+} \right) = -\frac{i \sqrt{1-q^2}
   \left(\tx_{\R i}^2+1\right)}{\sqrt{1-q^2}
   \left(\tx_{\R i}^2-1\right)+2 q \tx_{\R i}}+\mathcal{O}(T^{-2}) \,.
\eal
Expressing $ \tx_{\R i}$ in terms of $\tp_{\R i}$, we get
\bal
\tx_{\R i} = \frac{\sqrt{{\tp_{\R i}}^2-q^2+1}-{\tp_{\R i}}
  \, q}{({\tp_{\R i}}+i) \sqrt{1-q^2}}+\mathcal{O}(T^{-2})\,,
\eal
where the square root branch is chosen from the condition $\Im (\tx_{\R i} )<0$. Expressing $\tE_{\R i} $ in terms of $\tp_{\R i}$, we obtain the dispersion relation of the mirror theory in the BMN limit
\bal
 \tilde\omega_{\R i}=-i q+\sqrt{{\tp_{\R i}}^2-q^2+1}\,.
\eal
Plugging the values of $\tx_{\R 1}$ and $\tx_{\R 2}$ into~\eqref{eq:tree_S_matrix_bZbZ_mirror1} we obtain the following tree-level expression for the T-matrix element
\bal
\la{eq:mirror_T_matrix_RR_tree}
&T^{11}_{\bar{Z} \bar{Z}}=\frac{1}{T} \, \tilde\omega_{\R 1} \tilde\omega_{\R 2} \, \frac{\tp_{\R 1} + \tp_{\R 2}}{\tilde\omega_{\R 1}-\tilde\omega_{\R 2}}+\mathcal{O}(T^{-2})\,,
\eal
which matches the result of~\cite{Baglioni:2023zsf} in the gauge $a=0$, as expected.
We repeated this computation for the remaining elements $T^{11}_{Y \bar{Z}}$, $T^{11}_{\bar{Z} Y}$ and $T^{11}_{Y Y}$. In all cases we found exact agreement with the results of~\cite{Baglioni:2023zsf}.

\subsection{String S-matrix to one-loop}
\label{app:String_S_matrix_one_loop}

We now derive the S-matrix for the scattering of right excitations to the order $T^{-2}$. We perform the derivation for the scattering of right particles but a similar argument can be applied to all other cases.
Let us consider two points $y_1$ and $y_2$ in the mirror region, i.e. $\Im(y_1)<0$ and  $\Im(y_2)<0$.
The analytic continuation of  $\tPhi_{\R\R}(y_1,y_2)$-functions in $y_1$ (or $y_2$) to the upper half-plane produces  $\tPsi_{\R\R}$-functions which can be used to check, e.g.\ the crossing equations. There are situations, however, where it is more beneficial to define the analytic continuation through an integration contours deformation; an example is the large tension limit considered in this appendix. Note indeed that if we leave the contour undeformed then the two points $x^+_{\R i}$ and $x^-_{\R i}$ for $u_i \in \mathbb{R}$ are evaluated on the opposite sides of the contour and in the large tension limit they trap the contour approaching it from opposite directions. 
If (as before) we expand
\bal
x^\pm_{\R i}= x_\R(u_i \pm \frac{i}{h})= x_{\R i} \pm \frac{i}{h} \frac{1}{u'_{\R}(x_{\R i})} + \mathcal{O}(h^{-2})\,,
\eal
with $x_{\R i} \equiv x_\R(u_i)$, then the $\tPhi$ functions are discontinuous at the point $x_{\R i}$ around which we are expanding since this point is exactly on the integration contour. 
Due to this fact we cannot perform the expansion of the phase as we did in~\eqref{eq:thetaafs_mir_sec_der}.
It is then simpler, while moving $x^+_{\R1}$ and $x^+_{\R2}$ to the upper-half plane, to deform the integration contour so that both $x^+_{\R1}$ and $x^+_{\R2}$ stay below it, see LHS of Figure~\ref{fig:continuation} in the main text.

Let us explain in some detail how the contour deformation can be performed.
The analytic continuation to the string region requires crossing the main mirror cut in the $u$-plane from below, and therefore the positive real semi-line in the $x$-plane on the right of $\xbr_\R$. Let us denote the integration contours along the lower and upper edges of the main cut in the $u$ plane by $C_{v_1}^-$ and $C_{v_1}^+$, respectively. The integrand of  $\tPhi_{\R\R}(y_1,y_2)$ may have a pole at $v_1=u_\R(y_1)$, and therefore one needs to deform the $v_1$-integration contour to avoid it as $u_\R(y_1)$ approaches the lower edge of the main cut. 
This can be done by deforming  a part of $C_{v_1}^-$ so that the deformed contour $C_{v_1}^{-\rm def}$ would always be above $u_\R(y_1)$. It requires the deformed part of the contour to be moved to the anti-mirror plane. This however  creates a problem with the vertical cuts of the BES kernel because they can intersect the deformed contour. To resolve the problem we have to deform  $C_{v_1}^+$ so that it would have the same shape as $C_{v_1}^{-\rm def}$ but the deformed contour $C_{v_1}^{+\rm def}$ remains on the original mirror plane. We also have to do the same deformation of the integration contours in $v_2$ variable: $C_{v_2}^\pm\to C_{v_2}^{\pm\rm def}$. 

Through this deformation, even if we moved the points $\tx^+_{\R1}$ and $\tx^+_{\R2}$ to the string region, they remain on the same side of the contour. Then the even part of the dressing factor, together with the normalisation factor, is trivially continued as follows
\bal
f_{x}(\tx_{\R1}, \tx_{\R2}) \to f_{x}(x_{\R1}, x_{\R2}) \,,
\eal
where the function $f_x$ we defined in~\eqref{eq:fx_defined}. The result as a function of $x_{\R1}$ and $x_{\R2}$ does not change compared to before. However, if we want to express $x_{\R1}$ and $x_{\R2}$ in terms of the energies and momenta of the string particles we need now to solve 

\bal
\label{eq:string_constraints_xi}
-i{ p_{\R i}\ov T} &= \ln x_{\R i}^- -\ln x_{\R i}^+  \,,\quad -i p_{\R i}=-\frac{2 i x_{\R i}}{\sqrt{1-q^2}
   \left(x_{\R i}^2-1\right)+2 q x_{\R i}}+\mathcal{O}(T^{-2}) \,,
\\
-i \omega_{\R i} &= {h\ov2}\left( x_{\R i}^- -{1\ov x_{\R i}^-} - x_{\R i}^+ +{1\ov x_{\R i}^+} \right) = -\frac{i \sqrt{1-q^2}
   \left(x_{\R i}^2+1\right)}{\sqrt{1-q^2}
   \left(x_{\R i}^2-1\right)+2 q x_{\R i}}+\mathcal{O}(T^{-2}) \,.
\eal
To solve the constraints above we require that for $\omega_{\R i}>0$ then $x_{\R i}>0$. On this branch, we can express $x_{\R i}$ as a function of the momentum as
\bal
\label{eq:app_xRi}
x_{\R i} = \frac{1 - q \, p_{\R i} + \omega_{\R i}}{p_{\R i} \, \sqrt{1-q^2}}
\eal
with the string dispersion relation given by
\bal
\omega_{\R i}=\sqrt{p^2_{\R i}-2 q p_{\R i}+1}\,.
\eal
We notice that the constraints~\eqref{eq:mirror_constraints_xi} and~\eqref{eq:string_constraints_xi} are the same under the map $\tp_{\R i} \to -i \omega_{\R i} $, $\tilde \omega_{\R i} \to -i p_{\R i} $ and it is possible to show that the branches for the mirror and string solutions also map one into the other. The T-matrix element for string right particles is then the same as the one given in~\eqref{eq:mirror_T_matrix_RR_tree} after mapping $\tp_{\R i} \to -i \omega_{\R i} $, $\tilde \omega_{\R i} \to -i p_{\R i} $ and we can write
\bal
\la{eq:string_T_string_RR_tree}
&T^{11}_{\bar{Z} \bar{Z}}=-\frac{1}{T} \, p_{\R 1} p_{\R 2} \, \frac{\omega_{\R 1} + \omega_{\R 2}}{p_{\R 1}-p_{\R 2}}+\mathcal{O}(T^{-2})\,.
\eal

\paragraph{One-loop.}

For string excitations one-loop results are available in the literature and is therefore important to expand our phases to the next order in $1/T$.
For the scattering of $\bar{Z}$ excitations the only one-loop contribution comes from the odd part of the dressing factor, since in the even-part the BES and HL cancel in the expansion.
Parameterising
\bal
\label{eq:grpm_expansion_pos_mom}
\g^\pm_{\R i} = \g_{\R i} \pm \frac{i}{2T} p^2_{\R i}+ \mathcal{O}(\frac{1}{T^2}) \,, \qquad \g_{\R i}= \ln \frac{x_{\R i} \xbr -1}{x_{\R i} + \xbr}\,,
\eal
we obtain
\bal
&\frac{1}{i} \log \frac{R^2(\g^+_{\R1}-\g^+_{\R2}) R^2(\g^-_{\R1}-\g^-_{\R2})}{R^2(\g^+_{\R1}-\g^-_{\R2}) R^2(\g^-_{\R1}-\g^+_{\R2})}=-\frac{2}{T^2} p_{\R 1} p_{\R 2} {\stackrel{\prime\prime}{\tPhi}}{}^{\rm odd}_{\R\R}(x_{\R 1},x_{\R 2})+\mathcal{O}(T^{-3})=\\
&-\frac{1}{T^2} \frac{x_{\R 1} x_{\R 2}}{\pi (x_{\R1} - x_{\R2})^2} p_{\R1} p_{\R2} \ln \left(\frac{\sqrt{1-q} x_{\R1} + \sqrt{1+q}}{\sqrt{1+q} x_{\R1} - \sqrt{1-q}} \ \frac{\sqrt{1+q} x_{\R2} - \sqrt{1-q}}{\sqrt{1-q} x_{\R2} + \sqrt{1+q}} \right)\\
&+\frac{1}{T^2} \, \frac{p^2_{\R1} p^2_{\R2}}{2 \pi} \, \frac{\omega_{\R1} + \omega_{\R2}}{p_{\R1} - p_{\R2}} +\mathcal{O}(T^{-3})\,. 
\eal
The expansion of the S-matrix element, to one-loop order in perturbation theory is then given by
\bal
\frac{1}{i} \log S^{11}_{\bar{Z} \bar{Z}}&=-\frac{1}{T} \, p_{\R 1} p_{\R 2} \, \frac{\omega_{\R 1} + \omega_{\R 2}}{p_{\R 1}-p_{\R 2}}\\
&-\frac{1}{T^2} \frac{x_{\R1} x_{\R2}}{\pi (x_{\R1} - x_{\R2})^2} p_{\R1} p_{\R2} \ln \left(\frac{\sqrt{1-q} x_{\R1} + \sqrt{1+q}}{\sqrt{1+q} x_{\R1} - \sqrt{1-q}} \ \frac{\sqrt{1+q} x_{\R2} - \sqrt{1-q}}{\sqrt{1-q} x_{\R2} + \sqrt{1+q}} \right)\\
&+\frac{1}{T^2} \, \frac{p^2_{\R1} p^2_{\R2}}{2 \pi} \, \frac{\omega_{\R1} + \omega_{\R2}}{p_{\R1} - p_{\R2}}+\mathcal{O}(T^{-3})\,.
\eal
At one-loop the expression above differs from the results in the literature~\cite{Babichenko:2014yaa,Stepanchuk:2014kza} by a rational factor which can be removed by a local counterterm in the Lagrangian. This is the same type of mismatch that was observed in the Ramond-Ramond case~\cite{Frolov:2021fmj}. In particular if we take the difference between our perturbative expansion and the result of~\cite{Bianchi:2014rfa} (see equation (4.15) of that paper) we obtain
\bal
\label{eq:app_mismatch_BH}
\frac{1}{i} \log S^{11}_{\bar{Z} \bar{Z}}-\frac{1}{i} \log S^{11, {\rm BH}}_{\bar{Z} \bar{Z}} = \frac{p_{\R1} p_{\R2}}{4 \pi T^2}  (\omega_{\R1} p_{\R2} - \omega_{\R2} p_{\R1}) \,.
\eal

The computations for the S-matrix elements of the remaining sectors (LL, LR and RL) are analogous. For left particles, we need to expand
\bal
x^\pm_{\L i}= x_\L(u_i \pm \frac{i}{h})= x_{\L i} \pm \frac{i}{h} \frac{1}{u'_{\L}(x_{\L i})} + \mathcal{O}(h^{-2})\,,
\eal
where $x_{\L i} \equiv x_\L(u_i)$. In this case, repeating a similar study to the one above we obtain
\bal
\label{eq:app_xLi}
x_{\L i} = \frac{1 + q \, p_{\L i} + \omega_{\L i}}{p_{\L i} \, \sqrt{1-q^2}} \, , \qquad \omega_{\L i}=\sqrt{p^2_{\L i}+2 q p_{\L i}+1}\,.
\eal
and
\bal
\label{eq:glpm_expansion_pos_mom}
\g^\pm_{\L i} = \g_{\L i} \pm \frac{i}{2T} p^2_{\L i}+ \mathcal{O}(\frac{1}{T^2}) \,, \qquad \g_{\L i}= \ln \frac{x_{\L i} - \xbr}{x_{\L i} \xbr + 1 }\,.
\eal
The computations are very similar and we do not report them here.
In all cases, we find the same type of mismatch~\eqref{eq:app_mismatch_BH}, up to replacing R $\to$ L if the associated scattered particle is of left type. We remark that this mismatch is not cut constructable.

\subsection{BMN limit for negative momenta in string theory}\la{app:bmn_neg_p_new}

Our way of describing the dressing factors treats differently the momentum region $0<p<2\pi$ and any other. Vice-versa, in the near-BMN there is no distinction between positive and negative momentum (for massive particles). Hence it is an important test of our construction to reproduce the results of the BMN expansion at negative momentum. We begin by recalling that S-matrix elements at negative momentum are related to those at positive momentum as in appendix~\ref{app:pto2pimp}. In particular, note that
\begin{equation}
\label{eq:negativeprelations_new}
 S_{Z \bar{Y}}(-p_1, q_2 ) =  S^{k-1, 1}_{\bar{Y} \bar{Y}}(2 \pi -p_1, p_2)\, \frac{\alpha_{\R}(x^+_{\R2})}{\alpha_{\R}(x^-_{\R2})},\quad
 S_{\bar{Y} \bar{Y}}(-p_1, q_2 ) = 
 S^{k+1, 1}_{\bar{Y} \bar{Y}}(2 \pi -p_1, p_2),
\end{equation}
where $\alpha_{\R}(x)$ is defined in~\eqref{eq:defalpha}.
We can therefore obtain the negative-momentum processes by computing the scattering of a (string) bound-state of $k-1$ or $k+1$ right particles, when the momentum is close to (but smaller than) $2\pi$. 
An $m$-particle string bound state must satisfy
\begin{equation}
x_\R^{-m}\equiv x_{\R1}^-,\quad
x_{\R1}^+=x_{\R2}^-,\quad
x_{\R2}^+=x_{\R3}^-,\quad\dots
\quad x_\R^{+m}\equiv x_{\R m}^+\,,
\end{equation}
which we solve by setting the condition~\eqref{eq:cond_st_boundstate} for the bound state constituents (we recall anyway that the result for the S~matrix elements is independent of the constituents, as shown in appendix~\ref{app:string-bound-states}).

\paragraph{Scattering of $Z$ with $\bar{Y}$, obtained from $m=k-1$ bound state scattering.}

We start considering the BES dressing factor for the scattering of a bound state of $k-1$ right particles and a fundamental right particle. After choosing the constituents of the bound state as in~\eqref{eq:cond_st_boundstate} and leaving the second particle in the mirror region, we get
\begin{equation}
\label{eq:BESstringQboundstate_new}
\begin{aligned}
&\tilde{\theta}_{\R \R}^{\bes}(x^{\pm (k-1)}_{\R1}, \tx^{\pm }_{\R2})=\\
&+\tPhi_{\R\R}(x^{+(k-1)}_{\R1},\tx^{+}_{\R2})+\tPhi_{\R\R}(x^{-(k-1)}_{\R1},\tx^{-}_{\R2})-\tPhi_{\R\R}(x^{+(k-1)}_{\R1},\tx^{-}_{\R2})-\tPhi_{\R\R}(x^{-(k-1)}_{\R1},\tx^{+}_{\R2})\\
&-\tPsi_{\R}(u_1+\frac{i}{h}(k-1), \tx^{+}_{\R2})+\tPsi_{\R}(u_1+\frac{i}{h} (k-1), \tx^{-}_{\R2})\,.
\end{aligned}
\end{equation}
This expression is equivalent to~\eqref{eq:thetaBES_in_string_region} when the second particle is left in the mirror region. For the reason already explained in section~\ref{app:String_S_matrix_one_loop} we leave both $\tx^{-}_{\R2}$ and $\tx^{+}_{\R2}$ in the mirror region and continue them to the string region at the end of the computation. Instead, we take the variables $x^{\pm(k-1)}_{\R1}$ in the string region.

We consider $u_1< -\ubr$, so that the momentum of the bound state is close to and smaller than~$2\pi$. In particular, this means that the point $x_{\R1}^{- (k-1)}$ lies in the intersection of the mirror and string region, close to the negative-$x$ half-line (see Figure~\ref{fig:rightplanes} in the main text). For such a point we can write
\bal
\tPhi_{\R\R}(x_{\R1}^{- (k-1)},\tx^\pm_{\R2})&=\tPhi^{\cup}_{\R\R}(x_{\R1}^{-(k-1)},\tx^\pm_{\R2})\\
&-\frac{1}{2}\tPsi^{+}_{\R}(u_1-\frac{i}{h} (k-1)+ 2 i \ka,\tx^\pm_{\R2})- \frac{1}{2}\tPsi^{-}_{\R}(u_1-\frac{i}{h} (k-1),\tx_{\R2})\,,
\eal
where $\tPhi^{\cup}_{\R\R}(x_{\R1}^{-(k-1)},\tx^\pm_{\R2})$ is defined so that the integration contour in the first variable runs just below $x_{\R1}^{-(k-1)}$. When we perform the contour deformation on $\tPhi^{++}_{\R\R}$ we need to enter the cut of the $\log$ in the $x$-plane and therefore the argument of $\tPsi^{+}_{\R}$ is translated by $2 i \ka$.
Substituting the expression above into~\eqref{eq:BESstringQboundstate_new} we can write
\begin{equation}
\label{eq:app_tildeThetaBES_bneg1}
\begin{aligned}
2\tilde{\theta}_{\R\R}^{\bes}(x^{\pm (k-1)}_{\R}, \tx^{\pm }_{\R2})&=2\tilde{\varPhi}_{\R\R}^{\cup}(x^{\pm(k-1)}_{\R},\tx^{\pm}_{\R2})\\
&+ \Delta^+_\R (u_1 + \frac{i}{h} k \pm \frac{i}{h}, \tx^\pm_{\R2}) -  \Delta^-_\R (u_1 \pm \frac{i}{h} (k-1), \tx^\pm_{\R2})\,,
\end{aligned}
\end{equation}
where we used~\eqref{eq:Dab_definition} and defined
\begin{multline}
\label{eq:varPhiU_appendix}
\tilde{\varPhi}_{\R\R}^{\cup}(x^{\pm(k-1)}_{\R},\tx^{\pm}_{\R2}) \equiv \tPhi_{\R\R}(x^{+(k-1)}_{\R},\tx^{+}_{\R2})+\tPhi_{\R\R}^{\cup}(x^{-(k-1)}_{\R},\tx^{-}_{\R2})\\
-\tPhi_{\R\R}(x^{+(k-1)}_{\R},\tx^{-}_{\R2})-\tPhi_{\R\R}^{\cup}(x^{-(k-1)}_{\R},\tx^{+}_{\R2})\,.
\end{multline}
Using the identities~(\eqref{eq:Deltaepsm_2mm}, \eqref{eq:Deltaepsp_2mm}) the expression~\eqref{eq:app_tildeThetaBES_bneg1} becomes
\bal
&\exp \left[ -2 i \tilde{\theta}_{\R\R}^{\bes}(x^{\pm (k-1)}_{\R}, \tx^{\pm }_{\R2}) \right] = \frac{\exp \left[ -2 i \tilde{\varPhi}_{\R\R}^{\cup}(x^{\pm(k-1)}_{\R},\tx^{\pm}_{\R2})  \right]}{A^{k-1, 1}_{\bar{Y} \bar{Y}} (u_1, u_2) } \frac{x^{+ (k-1)}_{\R1}}{x^{- (k-1)}_{\R1}} \, \frac{\tx^{+}_{\R2}}{\tx^{-}_{\R2}} \, \frac{(x^{-(k-1)}_{\R1} - \tx^{+}_{\R2})^2}{(x^{+(k-1)}_{\R1} - \tx^{-}_{\R2})^2}  \\
& \left(\frac{\tx^{+(k-1)}_{\L1} \tx^{-}_{\R2} -1}{\tx^{+(k-1)}_{\L1} \tx^{+}_{\R2} -1} \right) \, \left(\frac{\tx^{+(k+1)}_{\L1} \tx^{-}_{\R2} -1}{\tx^{+(k+1)}_{\L1} \tx^{+}_{\R2} -1} \right)  \, \frac{\tx^{-(k-1)}_{\R1} - \tx^{-}_{\R2}}{\tx^{-(k-1)}_{\R1} - \tx^{+}_{\R2}} \, \frac{\tx^{+(k-1)}_{\R1} - \tx^{-}_{\R2}}{\tx^{+(k-1)}_{\R1} - \tx^{+}_{\R2}} \,,
\eal
with $A^{k-1, 1}_{\bar{Y} \bar{Y}} (u_1, u_2)$ given in~\eqref{eq:SbYbY}.

Next, the continuation of the HL phase is performed in the same way, using the formulae of appendix~\ref{app:discontinuities_HL}.
Here it is important to note that $x^{-(k-1)}_{\R1}$ lies close to the real line between $-\xi$ and~$0$ (this did not matter for the BES computation), corresponding to approaching the cut with imaginary part $\Im[u]=-\kappa$ from above. We find
\begin{equation}
\label{eq:resultHLsimpl2pi_new}
\begin{aligned}
&\exp \left[ 2 i \tilde{\theta}^{\hl}_{\R \R}(x^{\pm (k-1)}_{ a},\tx^{\pm}_{ a2}) \right] = \exp \left[2 i \tilde{\varPhi}_{\R\R}^{\hl, \, \cup}(x^{\pm(k-1)}_{\R},\tx^{\pm}_{\R2}) \right]\\
& \times \frac{x_{\R 1}^{+(k-1)} - \tx_{\R 2}^-}{x_{\R 1}^{+(k-1)} - \tx_{\R 2}^+} \ \frac{x_{\R1}^{-(k-1)} - \tx_{\R 2}^-}{x_{\R 1}^{-(k-1)} - \tx_{\R 2}^+} \ \frac{\frac{1}{x_{\L1}^{+(k-1)}} - \tx_{\R 2}^+}{\frac{1}{x_{\L1}^{+(k-1)}} - \tx_{\R 2}^-} \ \frac{x_{\R1}^{+(k+1)} - \tx_{\R 2}^+}{x_{\R 1}^{+(k+1)} - \tx_{\R 2}^-}\,,
\end{aligned}
\end{equation}
where we defined $\tilde{\varPhi}_{\R\R}^{\hl, \, \cup}$ as in~\eqref{eq:varPhiU_appendix}, but for the HL phase.
We can then write the even part of the dressing factor, given by the ratio of BES and HL, as
\bal
\label{eq:app_Sigma_even_neg_mom}
\left( \Sigma^\besratio(x^{\pm(k-1)}_{\R1}, \tx^{\pm}_{\R2}) \right)^{-2}&=\frac{\exp \left[ -2 i \tilde{\varPhi}_{\R\R}^{\cup}(x^{\pm(k-1)}_{\R},\tx^{\pm}_{\R2}) +2 i \tilde{\varPhi}_{\R\R}^{\hl, \, \cup}(x^{\pm(k-1)}_{\R},\tx^{\pm}_{\R2})  \right]}{A^{k-1, 1}_{\bar{Y} \bar{Y}} (u_1, u_2) }\\
&\times \frac{x^{+(k-1)}_{\R1}}{x^{-(k-1)}_{\R1}} \ \frac{\tx^{+}_{\R2}}{\tx^{-}_{\R2}} \ \left( \frac{x^{-(k-1)}_{\R1} - \tx^-_{\R2}}{x^{+(k-1)}_{\R1} - \tx^+_{\R2}} \right)^2 \,.
\eal

The last piece of the dressing factor that we may worry about is the odd one; in this case, since it can be directly and straightforwardly evaluated in terms of Barnes functions, there is no need for any manipulation (though we will later need to take a little care when performing the expansion).
Using the first relation in~\eqref{eq:negativeprelations_new}, and putting together the results for the even dressing factor~\eqref{eq:app_Sigma_even_neg_mom} and the normalisation~\eqref{eq:SbYbY} we obtain simply
\begin{equation}
\label{eq:app_SZbarY_neg_mom}
\begin{aligned}
&\frac{1}{i} \log  S_{Z \bar{Y}}(-p_1/T, p_2/T )=\\
&-2  \tilde{\varPhi}_{\R\R}^{\cup}(x^{\pm(k-1)}_{\R1},\tx^{\pm}_{\R2}) +2  \tilde{\varPhi}_{\R\R}^{\hl, \, \cup}(x^{\pm(k-1)}_{\R1},\tx^{\pm}_{\R2})-2\tilde{\varPhi}_{\R\R}^\barnes(x^{+(k-1)}_{\R1},\tx^{+}_{\R2})\\
&-i\log
\frac{x_{\R1}^{+(k-1)}}{x_{\R1}^{-(k-1)}}\frac{\tx_{\R2}^+}{\tx_{\R2}^-}
\frac{(x^{-(k-1)}_{\R1}-\tx_{\R2}^-)^2}{(x_{\R1}^{+(k-1)} - \tx_{\R 2}^+)^2}\frac{\alpha_{\R}(\tx^+_{\R2})}{\alpha_{\R}(\tx^-_{\R2})}
\,.
\end{aligned}
\end{equation}
All the terms in this expression may be evaluated in a straightforward way, similarly to what we did in the previous subsection. Let us only highlight where the two computations differ.
The main difference is that after the contour deformation the variables $x^{\pm (k-1)}_{\R1}$ in $\tPhi^{\bes}$ (and $\tPhi^{\hl}$) are both evaluated \textit{above} the integration contour and $x^{\pm}_{\R2}$ are both evaluated below the contour. Recalling~\eqref{eq:xpm-BMN-pis2pi} we want to evaluate the phases at the points
\begin{equation}
\label{eq:xpm-BMN-pis2pi_app_new}
\begin{aligned}
x^{\pm(k-1)}_{\R}\left(2\pi-\frac{p_1}{T}\right)&=&\frac{1}{x_{\R}(p_1)}\left(-1\pm \frac{ip_1}{2T}\right)+O(T^{-2}) \, ,\\
x^{\pm}_{\R}\left(\frac{p_2}{T}\right)&=&x_{\R}(p_2)\left(+1\pm \frac{ip_2}{2T}\right)+O(T^{-2}) \, .
\end{aligned}
\end{equation}
Because the integrands with the deformed contour are regular as we cross the $x$ axis (even for negative~$x$), we can perform the computation by assuming that $\check{x}_1=-1/x_{\R}(p_1)$ lies well inside the \textit{antimirror} region, while $\tilde{x}_2=x_{\R}(p_2)$ lies well inside in the mirror region (like in the previous subsection). Because of that, we will need the double derivatives of the dressing factors with one argument in the antimirror region and the other in the mirror region. 
Moreover, like before we are dealing with a combination of $\tPhi^{\bes}$ and $\tPhi^{\hl}$ functions which can be evaluated in terms of $\tPhi^{\afs}$, while the HL order cancels and the higher orders contribute from two loops. Hence, to deal with the ``even'' part of the phase we only need~\eqref{eq:doublederivativeAFSmixed}, that is
\begin{equation}
\frac{\partial^2\tPhi^\afs_{\R\R}(\check{x}_1,\tx_2)}{\partial \check{x}_1\partial \tx_2}=
\frac{1}{4}\frac{u_{\R1}+u_{\R2}}{\check{x}_1-\tx_2}+\frac{\check{x}_1-\tx_2}{4(\check{x}_1)^2(\tx_2)^2}\,.
\end{equation}
The expansion of the odd part of the phases can be performed straightforwardly, having care to note that (as already remarked in the main text)
\begin{equation}
\label{eq:expansiongammakminus1}
\begin{aligned}
\gamma_{\R}^{\pm(k-1)}\left(2\pi-\frac{p_1}{T}\right)&=\pm i\pi
-\g^\pm_{\R 1}+\mathcal{O}(T^{-2})\,,\\
\gamma_{\R}^{\pm}\left(\frac{p_2}{T}\right)&=\qquad+\g^\pm_{\R 2}+\mathcal{O}(T^{-2})\,.
\end{aligned}
\end{equation}
In the expression above we used the quantities early defined in~\eqref{eq:grpm_expansion_pos_mom} and omitted terms irrelevant to one loop. The shift by $\pm i\pi$ yields a tree-level contribution. Schematically
\begin{equation}
\begin{aligned}
    &\frac{1}{i}\log\frac{R^{2}(\gamma^{+(k-1) \, +}_{\R\R})R^{2}(\gamma^{-(k-1) \, -}_{\R\R})}{R^{2}(\gamma^{+(k-1) \, -}_{\R\R})R^{2}(\gamma^{-(k-1) \, +}_{\R\R})}\\
    &= \frac{1}{i}\log\frac{\cosh^2(\tfrac{\g^+_{\R 1}+\g^+_{\R 2}}{2})}{\cosh^2(\tfrac{\g^+_{\R 1}+\g^-_{\R 2}}{2})}+\frac{1}{i}\log
    \frac{R^{2}(-\g^+_{\R 1}-\g^+_{\R 2}- i \pi)R^{2}(-\g^-_{\R 1}-\g^-_{\R 2}- i \pi)}{R^{2}(-\g^+_{\R 1}-\g^-_{\R 2}- i \pi)R^{2}(-\g^-_{\R 1}-\g^+_{\R 2}- i \pi)}\,,
\end{aligned}
\end{equation}
Since $\g^\a_{\R 1}+\g^\b_{\R 2}$ ($\a=\pm$, $\b=\pm$) differ by $\mathcal{O}(T^{-1})$, the last term on the r.h.s.\ of the expression above starts  contributing at $\mathcal{O}(T^{-2})$, that is at one loop. Its expansion can be readily computed by recalling that
\begin{equation}
   \frac{1}{i} \frac{d^2}{dz^2}\log R^2(z)= \frac{z-\sinh z}{4\pi\,\sinh^2(z/2)}\,.
\end{equation}
As for the ratio of cosh, this term precisely cancels the contribution of the $\alpha_{\R}$ functions in~\eqref{eq:app_SZbarY_neg_mom}. An explicit computation then confirms that the result of the expansion at negative $p_1$ matches perfectly with that at positive $p_1$.

\paragraph{Scattering of $\bar{Y}$ with $\bar{Y}$, obtained from $m=k+1$ bound state scattering.}

The computation relative to the second expression in~\eqref{eq:negativeprelations_new} is very similar to the one which we have just discussed. For the BES phase and the bound-state normalisation, we just need to substitute $x^{\pm(k-1)}_{\R1} \to x^{\pm(k+1)}_{\R1}$ in~\eqref{eq:BESstringQboundstate_new}, and $A^{k-1, 1}_{\bar{Y} \bar{Y}} (u_1, u_2) \to A^{k+1, 1}_{\bar{Y} \bar{Y}} (u_1, u_2) $, respectively.

The computation of the HL phase is slightly different. Now we are crossing the integration contour between $-\infty$ and $-\xi$ on the $x$ plane, which corresponds to approaching the cut with imaginary part $-\kappa$ in the $u$-plane from below, rather than from above. With this in  mind, we find for the HL phase
\begin{equation}
\begin{aligned}
&2 \tilde{\theta}^{\hl}_{\R\R}(x^{\pm (k+1)}_{\R 1},\tx^{\pm}_{ \R2})=2\tilde{\varPhi}_{\R\R}^{\hl, \, \cup}(x^{\pm(k+1)}_{\R 1},\tx^\pm_{\R2})\\
&
+i \log
\frac{x_{\R 1}^{+(k+1)} - \tx_{\R 2}^+}{x_{ \R}^{+(k+1)} - \tx_{\R 2}^-} \
\frac{x_{\R 1}^{-(k+1)} - \tx_{\R 2}^-}{x_{\R 1}^{-(k+1)} - \tx_{\R 2}^+} \
\frac{\frac{1}{x_{\L 1}^{+(k+1)}} - \tx_{\R 2}^-}{\frac{1}{x_{\L 1}^{+(k+1)}} - \tx_{\R 2}^+} \
\frac{x_{\R 1}^{+(k-1)} - \tx_{\R 2}^+}{x_{\R 1}^{+(k-1)} - \tx_{\R 2}^-}\,.
\end{aligned}
\end{equation}
Using this result and with minimal modification of the BES, normalisation and odd pieces, we get
\begin{equation}
\begin{aligned}
&\frac{1}{i} \log  S_{\bar{Y}\bar{Y}}(-p_1/T, p_2/T ) =\\
&-2\tilde{\varPhi}_{\R\R}^{\bes, \, \cup}(x^{\pm (k+1)}_{\R1},\tx^{\pm}_{\R2})+2\tilde{\varPhi}_{\R\R}^{\hl,\cup}(x^{\pm(k+1)}_{\R1},\tx^\pm_{\R2})-2\tilde{\varPhi}_{\R\R}^\barnes(x^{+(k+1)}_{\R1},\tx^{+}_{\R2})\\
&-i\log
\frac{x_{\R1}^{+(k+1)}}{x_{\R1}^{-(k+1)}}\frac{\tx_{\R2}^-}{\tx_{\R2}^+}
\frac{(x^{-(k+1)}_{\R1}-\tx_{\R2}^+)^2}{(x_{\R1}^{+(k+1)} - \tx_{\R 2}^-)^2}
\,.
\end{aligned}
\end{equation}
The perturbative expansion of this formula is fairly straightforward. One major difference with the previous case is that the expansion of the $\gamma$-rapidities of the bound state is
\begin{equation}
    \gamma_{\R}^{\pm(k+1)}\left(2\pi-\frac{p_1}{T}\right)=
\g^\pm_{\L 1}+\mathcal{O}(T^{-2})\,,
\end{equation}
without any shift of $\pm i\pi$, cf.~\eqref{eq:expansiongammakminus1}. The variables $\g^\pm_{\L i}$ are defined in~\eqref{eq:glpm_expansion_pos_mom}. As a result, the odd phase contributes only from one loop. An explicit computation then confirms that the expansion agrees with the one at positive momentum.

\section{Relativistic limit}\label{app:rel_limit}

\subsection{Result from boostrap in the relativistic model}

In~\cite{Frolov:2023lwd} we studied a particular relativistic limit of this model. The limit was obtained by keeping $k\in\mathbb{Z}$ fixed and $h \ll 1$, and expanding the momenta of the particles around the minimum of the dispersion relation, as written in~\eqref{eq:relativisticpexpansion}.
A universal formula for the dressing factors of left particles was obtained in this limit, with S~matrix elements for the scattering of $Y$ particles of arbitrary quantum numbers $m_1, \, m_2=1, \dots, k-1$ given by
\bal
S^{m_1 \, m_2}_{YY}(\theta_{12})=\Phi(m_1,m_2;\theta_{12})\sigma(m_1,m_2;\theta_{12})^{-2} \quad, \qquad m_1, m_2=1, 2, \dots, k-1 \,.
\eal
The expression above is composed of  a minimal solution to the crossing equations
 \bal
 \sigma(m_1 , m_2;\theta)^{-2} =  \frac{R\left(\theta-\frac{i\pi(m_1+m_2)}{k}\right)^2\,R\left(\theta+\frac{i\pi(m_1+m_2)}{k}\right)^2}{R\left(\theta-\frac{i\pi(m_1-m_2)}{k}\right)^2 \,R\left(\theta+\frac{i\pi(m_1-m_2)}{k}\right)^2}
 \eal
and a CDD-like factor of Toda theories of $A^{(1)}_{k-1}$ type
\begin{equation}
    \label{general_formula_for_CDD_factor}
     \Phi(m_1,m_2;\theta)= [m_1+m_2]_\theta [m_1+m_2-2]^2_\theta 
     \dots [|m_1-m_2|+2]^2_\theta [|m_1-m_2|]_\theta \,,
\end{equation}
where
\begin{equation}
\label{definition_building_block}
    [m]_\theta\equiv \frac{\sinh{\Bigl( \frac{\theta}{2} + \frac{i \pi m}{2k} \Bigl)}}{\sinh{\Bigl( \frac{\theta}{2} - \frac{i \pi m}{2k} \Bigl)}} \,.
\end{equation}
In~\cite{Frolov:2023lwd} it was shown that the scattering of left particles of quantum numbers $m_1, \, m_2=1, \dots, k-1$ is closed under the bootstrap fusion relations~\cite{Zamolodchikov:1989fp} and that the scattering of right particles is equivalent to the one of left particles under the periodicity $m \to k-m$. In particular, it holds that 
\bal
S^{k-m_1 \, k-m_2}_{YY}(\theta)=S^{m_1, m_2}_{\bar{Z} \bar{Z}}(\theta)\,, \qquad m_1, m_2 = 1, \dots, k-1\,.
\eal
From the relation above we can then write the scattering of right fundamental particles in the relativistic limit as
\bal
\label{eq:app_rel_limit_right}
S^{1 1}_{\bar{Z} \bar{Z}}(\theta)=\frac{\sinh{\Bigl( \frac{\theta}{2} - \frac{i \pi}{k} \Bigl)}}{\sinh{\Bigl( \frac{\theta}{2} + \frac{i \pi}{k} \Bigl)}} \ \frac{R\left(\theta+ \frac{2i \pi}{k} \right)^2\,R\left(\theta- \frac{2i \pi}{k}\right)^2}{R\left(\theta\right)^4}\,.
\eal

In this appendix we consider the relativistic limit of the S~matrix proposed in this paper, which for the scattering of right fundamental particles is given by
\bal
\label{eq:S_mat_right_particles_before_rel_lim}
 S^{11}_{\bar{Z} \bar{Z}} (u_1, u_2) &=\frac{u_{\R}(x^+_{\R1})-u_{\R}(x^-_{\R2})}{u_{\R}(x^-_{\R1})-u_{\R}(x^+_{\R2})}\big(\Sigma^{11}_{\R\R}\big)^{-2} (u_1, u_2)\\
 &=\frac{u_{\R}(x^+_{\R1})-u_{\R}(x^-_{\R2})}{u_{\R}(x^-_{\R1})-u_{\R}(x^+_{\R2})}\big(\Sigma^{\barnes}_{\R\R}\big)^{-2} (u_1, u_2) \big(\Sigma^{\besratio}_{\R\R}\big)^{-2} (u_1, u_2)\,,
 \eal
and show that its relativistic limit matches expression~\eqref{eq:app_rel_limit_right} expected from~\cite{Frolov:2023lwd}.
We perform the comparison for right fundamental particles since their momenta in the relativistic limit are in the interval $(0 , 2 \pi)$; they are therefore in the fundamental region of our $u$ plane and the limit is particularly simple for these particles. It is however possible to move away from this fundamental region by using the periodicity relations in~(\eqref{eq:2pishift-nomonodromy}, \eqref{eq:2pishift-nomonodromy2}, \eqref{eq:2pishift-yesmonodromy}), which hold at the level of the exact S~matrix (and therefore also in the relativistic limit).

\subsection{Relativistic limit of the normalisation}
In the relativistic limit (at the leading order in $h$) it holds that
\begin{equation}
\label{eq:xmpm_in_rel_lim_app}
x^{\pm m}_{\R} = e^{-\theta \pm \frac{i \pi}{k} m} \ \qquad m=1, 2, \dots \,.
\end{equation}
Since in this limit we consider $h \ll1$ then we have 
\begin{equation}
u_{\R}(x^{\pm m}_{\R}) \simeq \frac{k}{\pi h} \log \bigl( e^{-\theta \pm \frac{i \pi}{k} m} \bigl) = -\frac{k}{\pi h} \theta \pm i \frac{m}{h} \,.
\end{equation}
The last equality has been obtained by considering the infinite cover of the $\log$. On the principal branch of the $\log$ the equality above is satisfied if $m=1, \dots, k-1$ and $\theta \in \mathbb{R}$.
Then in this limit (as already mentioned in section~\ref{sec:rel_limit}) the S~matrix normalisation becomes
\bal
\label{eq:normalisation_final_rel_limit}
\frac{u_{\R}(x^+_{\R1})-u_{\R}(x^-_{\R2})}{u_{\R}(x^-_{\R1})-u_{\R}(x^+_{\R2})} \to \frac{\theta_{12} - {2i \pi \ov k}}{\theta_{12} + {2i \pi \ov k}}\,,
\eal
where $\theta_{12} \equiv \theta_1 - \theta_2$ is the difference of the relativistic rapidities of the scattered particles.

\subsection{Relativistic limit  of the odd phase}

We recall that in the string model $x_{\R}$ variables are connected to $\g_{\R}$ variables through
 \bal
 x^{\pm}_{\R} = \frac{1+ \xbr e^{\g^\pm_{\R}}}{ \xbr - e^{\g^\pm_{\R}}} \,.
 \eal
 In the relativistic limit $\xbr \to + \infty$ and therefore
  \bal
 x^{\pm}_{\R} =  e^{\g^\pm_{\R}}\,.
 \eal
 From~\eqref{eq:xmpm_in_rel_lim_app} we then identify
 \bal
 \g^\pm_{\R}=-\theta \pm \frac{i \pi}{k}
 \eal
 and the odd part of the dressing factor becomes
 \bal
 \label{eq:Barnes_final_rel_limit}
 \big(\Sigma^{\barnes}_{\R\R}\big)^{-2} (u_1, u_2)=\frac{R^2(\g^{--}_{\R\R}) R^2(\g^{++}_{\R\R})}{R^2(\g^{-+}_{\R\R}) R^2(\g^{+-}_{\R\R})} &\to  \frac{R^4(\theta_{21})}{R^2(\theta_{21} - \frac{2i \pi}{k}) R^2(\theta_{21} + \frac{2i \pi}{k})}\\
 &=\frac{R^2(\theta_{12} + \frac{2i \pi}{k}) R^2(\theta_{12} - \frac{2i \pi}{k})}{R^4(\theta_{12})} \,.
 \eal

\subsection{Relativistic limit of the HL phase.}
The remaining (even) phase is composed of a BES and HL piece; we compute these two pieces separately, starting from the HL piece.
The continuation of the HL phase to the string region is given by~\eqref{eq:newHL_string_aa} and for right fundamental particles it becomes
\bal
\tilde{\theta}^{\hl}_{\R\R}(x^{\pm}_{ \R1},x^{\pm}_{ \R2})=&+\tPhi_{\R\R}^{\hl}(x^+_{\R1},x^+_{\R2})+\tPhi_{\R\R}^{\hl}(x^-_{\R1},x^-_{\R2})-\tPhi_{\R\R}^{\hl}(x^+_{\R1},x^-_{\R2})-\tPhi_{\R\R}^{\hl}(x^-_{\R1},x^+_{\R2})\\
&+ \frac{i}{2} \log  \frac{1 - \frac{1}{x^+_{\L1} x^-_{\R2}}}{1 - \frac{1}{x^-_{\R1}x^+_{\L 2}}}  \ \frac{1 - \frac{1}{x^+_{\R1} x^+_{\L 2}}}{1 - \frac{1}{x^+_{\L 1}x^+_{\R 2}}}  \ \frac{x^-_{\R1} - x^+_{\R2}}{ x^-_{\R2} -x^+_{\R1}}  \,.
\eal
In the limit $h \ll 1$ then $\xbr \to + \infty$ and the string region of left particles is pushed to infinity. Indeed the boundary between the string and anti-string region of left particle (i.e. the contour on the r.h.s. of figure~\ref{fig:leftplanes} in the main text) blows up to infinity in the limit. 
Due to this fact in the relativistic limit $x^\pm_{\L1} \to \infty$ and $x^\pm_{\L2} \to \infty$, and the term arising from the continuation to the string region becomes
\bal
& \frac{1 - \frac{1}{x^+_{\L1} x^-_{\R2}}}{1 - \frac{1}{x^-_{\R1}x^+_{\L 2}}}  \ \frac{1 - \frac{1}{x^+_{\R1} x^+_{\L 2}}}{1 - \frac{1}{x^+_{\L 1}x^+_{\R 2}}}  \ \frac{x^-_{\R1} - x^+_{\R2}}{x^-_{\R2} - x^+_{\R1}} \to  - \frac{\sinh \left(\frac{\theta_{12}}{2} + i\frac{\pi}{k} \right) }{\sinh \left(\frac{\theta_{12}}{2} - i\frac{\pi}{k} \right)}\,,
\eal
where the only contribution comes from $\left(\frac{x^-_{\R1} - x^+_{\R2}}{x^-_{\R2} -x^+_{\R1}} \right)$.
Then in the relativistic limit the HL phase reduces to
\bal
\label{eq:HL_for_rel_limit_int_step}
\tilde{\theta}^{\hl}_{\R\R}(x^{\pm}_{ \R1},x^{\pm}_{ \R2})=&+\tPhi_{\R\R}^{\hl}(x^+_{\R1},x^+_{\R2})+\tPhi_{\R\R}^{\hl}(x^-_{\R1},x^-_{\R2})-\tPhi_{\R\R}^{\hl}(x^+_{\R1},x^-_{\R2})-\tPhi_{\R\R}^{\hl}(x^-_{\R1},x^+_{\R2})\\
&- \frac{1}{2i}\log  \left(-\frac{\sinh \left(\frac{\theta_{12}}{2} + i\frac{\pi}{k} \right) }{\sinh \left(\frac{\theta_{12}}{2} - i\frac{\pi}{k} \right)} \right) \,.
\eal
Using the antisymmetry of $\tPhi^\hl_{\R\R}$ we can write the building blocks of the HL phase~\eqref{eq:PhiHL_as_functions_of_IHL1} as\footnote{An anti-symmetrised boundary contribution 
$
-\frac{i}{8} \sgn (\Im(x_1)) \log(-x_2)+\frac{i}{8} \sgn (\Im(x_2)) \log(-x_1)
$
is also present. However, in the relativistic limit, these boundary terms cancel in the full HL phase.
}
\bal
\tPhi_{\R\R}^{\hl}(x_{1},x_{2})
=& - \frac{1}{8 \pi}   \lint_{\widetilde{ \rm cuts}} {\rm d} v \frac{\tx'_\R(v)}{\tx_\R(v) - x_{1}} \left( \log\left(\tx_\R(v - i \eps)-x_{2} \right) - \log\left(\tx_\R(v + i \eps)-x_{2} \right) \right)\\
& - (x_1 \leftrightarrow x_2)
\eal
We work with the mixed derivative of the HL phase and integrate the result in the end. We have
\bal
{\stackrel{\prime\prime}{\tPhi}}{}^{\hl}_{\R\R}(x_{1},x_{2}) =& + \frac{1}{8 \pi}   \lint_{\widetilde{ \rm cuts}} {\rm d} v \frac{\tx'_\R(v)}{\left( \tx_\R(v) - x_{1} \right)^2} \left( \frac{1}{\tx_\R(v - i \eps)-x_{2} } - \frac{1}{\tx_\R(v + i \eps)-x_{2} } \right)\\
& - (x_1 \leftrightarrow x_2)\,.
\eal
In the limit $h \ll1$, in the right $u$-plane the upper side of the main cut is mapped to the interval $0<x<\frac{1}{\xbr}$ and shrinks to zero. The lower side of the $-\ka$ cut is instead mapped to the interval $-\infty<x<-\xbr$ and goes to $-\infty$ in the limit.  Then in the limit $h \to 0$ we can approximate ${\stackrel{\prime\prime}{\tPhi}}{}^{\hl}_{\R\R}(x_{1},x_{2})$ with
\bal
{\stackrel{\prime\prime}{\tPhi}}{}^{\hl}_{\R\R}(x_{1},x_{2}) =& + \frac{1}{8 \pi}   \lint_{{ \rm lower \ side \ main \ cut}} {\rm d} v \frac{\tx'_\R(v)}{\left( \tx_\R(v) - x_{1} \right)^2} \left(\frac{1}{\tx_\R(v - i \eps)-x_{2} } \right)\\
&- \frac{1}{8 \pi}   \lint_{{ \rm upper \ side \ -\ka \ cut}} {\rm d} v \frac{\tx'_\R(v)}{\left( \tx_\R(v) - x_{1} \right)^2} \left(\frac{1}{\tx_\R(v + i \eps)-x_{2} } \right)\\
& - (x_1 \leftrightarrow x_2)\\
=& + \frac{1}{8 \pi}   \int^{+\infty}_0 {\rm d} \tx \frac{1}{\left( \tx - x_{1} \right)^2} \ \frac{1}{\tx-x_{2} } - \frac{1}{8 \pi}   \int^0_{-\infty} {\rm d} \tx \frac{1}{\left( \tx - x_{1} \right)^2} \  \frac{1}{\tx-x_{2} } \\
& - (x_1 \leftrightarrow x_2)\,.
\eal
The expression above can be easily integrated and we obtain
\bal
{\stackrel{\prime\prime}{\tPhi}}{}^{\hl}_{\R\R}&(x_{1},x_{2})
\\
&= -\frac{1}{4 \pi} \frac{1}{x_1 x_2} \left( \frac{x_1+x_2}{x_1 - x_2} + \frac{x_1 x_2}{(x_1 - x_2)^2} \left( \log(x_2)+ \log(-x_2)- \log(x_1)- \log(-x_1) \right)\right)\,.
\eal
In the relativistic limit
\bal
\frac{\partial^2 {\tPhi}{}^{\hl}_{\R\R}}{\partial \theta_1 \partial \theta_2} =\frac{\partial^2 {\tPhi}{}^{\hl}_{\R\R}}{\partial x_1 \partial x_2}  \frac{\partial x_1}{\partial \theta_1}  \frac{\partial x_2}{\partial \theta_2} = {\stackrel{\prime\prime}{\tPhi}}{}^{\hl}_{\R\R}(x_{1},x_{2})  x_1 x_2
\eal
where in the last equality we used~\eqref{eq:xmpm_in_rel_lim_app}, with $x_i=x^+_{\R i}$ or $x_i=x^-_{\R i}$.
Therefore it holds that
\bal
\frac{\partial^2 {\tPhi}{}^{\hl}_{\R\R}}{\partial \theta_1 \partial \theta_2}=-\frac{1}{4 \pi}  \left( \frac{x_1+x_2}{x_1 - x_2} + \frac{x_1 x_2}{(x_1 - x_2)^2} \left( \log(x_2)+ \log(-x_2)- \log(x_1)- \log(-x_1) \right)\right)\,.
\eal
We evaluate the expression above for  ${\tPhi}{}^{\hl}_{\R\R}$ as a function of $x_1^\pm, x_2^\pm$. We obtain
\bal
\frac{\partial^2 {\tPhi}{}^{\hl}_{\R\R}(x^+_1, x^+_2)}{\partial \theta_1 \partial \theta_2}=\frac{\partial^2 {\tPhi}{}^{\hl}_{\R\R}(x^-_1, x^-_2)}{\partial \theta_1 \partial \theta_2}= -\frac{1}{8\pi} \frac{\theta_{12}-\sinh{\theta_{12}}}{ \sinh^2 \frac{\theta_{12}}{2}} \,,
\eal
\bal
\frac{\partial^2 {\tPhi}{}^{\hl}_{\R\R}(x^+_1, x^-_2)}{\partial \theta_1 \partial \theta_2} = -\frac{1}{8\pi} \frac{ \left(\theta_{12} - \frac{2i \pi}{k}\right) -\sinh{ \left(\theta_{12}  - \frac{2i \pi}{k} \right)} }{ \sinh^2 \left(\frac{\theta_{12}}{2} - \frac{i \pi}{k} \right)} - \frac{i}{8} \frac{1}{\sinh^2 \left(\frac{\theta_{12}}{2} - \frac{i \pi}{k} \right)}\,,
\eal
\bal
\frac{\partial^2 {\tPhi}{}^{\hl}_{\R\R}(x^-_1, x^+_2)}{\partial \theta_1 \partial \theta_2} = -\frac{1}{8\pi} \frac{ \left(\theta_{12} + \frac{2i \pi}{k}\right) -\sinh{ \left(\theta_{12}  + \frac{2i \pi}{k} \right)} }{ \sinh^2 \left(\frac{\theta_{12}}{2} + \frac{i \pi}{k} \right)} + \frac{i}{8} \frac{1}{\sinh^2 \left(\frac{\theta_{12}}{2} + \frac{i \pi}{k} \right)}\,.
\eal
Integrating with respect to both $\theta_1$ and $\theta_2$, with the further requirement that the full phase must be zero for $\theta_2=\theta_1$ (due to the fact that it is antisymmetric), we obtain
\bal
\label{eq:diff_phiHL_rel_limit}
&\tPhi_{\R\R}^{\hl}(x^+_1, x^+_2)+\tPhi_{\R\R}^{\hl}(x^-_1, x^-_2) -\tPhi_{\R\R}^{\hl}(x^+_1, x^-_2) -\tPhi_{\R\R}^{\hl}(x^-_1, x^+_2) \\
&=-\frac{1}{2i}  \log  \frac{R^2 (\theta_{12} - \frac{2i \pi}{k}) R^2 (\theta_{12} + \frac{2i \pi}{k}) }{R^4 (\theta_{12})} +\frac{1}{2i} \log \left( -\frac{\sinh \left( \frac{\theta_{12}}{2} +\frac{i \pi}{k} \right) }{\sinh \left( \frac{\theta_{12}}{2} -\frac{i \pi}{k} \right)} \right)\,.
\eal
Pllugging~\eqref{eq:diff_phiHL_rel_limit} into~\eqref{eq:HL_for_rel_limit_int_step} we obtain the following relativistic limit for the HL phase
\bal
\label{eq:rel_lim_HL}
\exp \left[2i \tilde{\theta}^{\hl}_{\R\R}(x^{\pm}_{ \R1},x^{\pm}_{ \R2}) \right] \to  \frac{R^4 (\theta_{12})}{R^2 (\theta_{12} - \frac{2i \pi}{k}) R^2 (\theta_{12} + \frac{2i \pi}{k}) } \,.
\eal

\subsection{Relativistic limit of the BES phase}

Recall that the improved BES phase of right fundamental particles continued to the string region is given by 
\bal
\label{eq:theta_starting_for_rel_limit}
\tilde{\theta}_{\R \R}^{\bes}(x^\pm_{\R1}, x^\pm_{\R2})&=\tPhi_{\R\R}(x^+_{\R1},x^+_{\R2})+\tPhi_{\R\R}(x^-_{\R1},x^-_{\R2})-\tPhi_{\R\R}(x^+_{\R1},x^-_{\R2})-\tPhi_{\R\R}(x^-_{\R1},x^+_{\R2})\\
&+\tPsi_{\R}(x^+_{\R2}, x^+_{\R1})-\tPsi_{\R}(x^+_{\R1}, x^+_{\R2})+\tPsi_{\R}(x^+_{\R1}, x^-_{\R2})-\tPsi_{\R}(x^+_{\R2}, x^-_{\R1})\\
&-  K^\bes(u_\R(x^+_{\R1}) - u_\R(x^+_{\R2}))\,.
\eal
In the relativistic limit $\ka \to +\infty$ and therefore we are in the regime $\ka> \ka_{\rm cr}$. As discussed in section~\ref{mirrorPhi}, to avoid branch points on the integration contours we work with the principal value prescription, which corresponds to modify the kernel by a regulator that we send to zero after integrating (see~\eqref{eq:Kbes_pv_prescription_kgkc}). In the case $k$ even this regulator plays an important role in the computation of $\tPhi$ functions, while for $k$ odd it plays an important role in the computations of $\tPsi$ functions. However, we find that the final result for BES, obtained by combining the different terms into~\eqref{eq:theta_starting_for_rel_limit} is independent on whether $k$ is even or odd. 
In the following, we work out the relativistic limit of the BES phase for $k=2l$ even. The case $k$ odd can be performed similarly and leads to the same result.

To evaluate the relativistic limit of the improved BES phase we parameterise the integration variables of the $w$ plane as
\bal
w_1 = e^{t_1} \,, \qquad w_2 = e^{t_2}\,.
\eal
For $\Im(t_1) \in (-\pi , + \pi)$ and $h \ll1$, using the parameterisation above it holds that
\bal
u_\R(w_1)\simeq \frac{k}{\pi h} \log(e^{t_1})= \frac{k}{h \pi} t_1
\eal
In this case the strip $\Im(t_1) \in (-\pi , 0)$ describes the portion of the $u$ plane
\bal
- \ka<\Im(u_\R(w_1))<0
\eal
in the relativistic limit. Note that for $h \to 0$ we have that  $\ka \to + \infty$. Then this strip occupies the entire lower half of the $u$ plane in the limit.
The strip $\Im(t_1) \in (0, + \pi)$ describes instead the portion of the $u$ plane defined by 
\bal
0<\Im(u_\R(w_1))<+\ka
\eal
and in the relativistic limit occupies the upper half of the $u$.

\paragraph{Limit of $\tPsi$ functions.}
We begin by deriving the relativistic limit of the $\tPsi$ functions appearing in~\eqref{eq:theta_starting_for_rel_limit}.
Defining $n_1= \pm$ and $n_2= \pm$ it holds that
\bal
\tPsi_{\R}^{-}(x^{n_1}_{\R1},x^{n_2}_{\R2})=&-\int^{+\infty}_{-\infty}\frac{{\rm
d}t_2}{2\pi i}\,\frac{e^{t_2}}{e^{t_2}-e^{-\theta_2+\frac{i \pi}{k} n_2}} i \log \left[ \frac{\Gamma \left(1-\frac{n_1}{2} - \frac{ik}{2 \pi} (\theta_1+t_2) \right)}{\Gamma \left(1+\frac{n_1}{2} + \frac{ik}{2 \pi} (\theta_1+ t_2) \right)} \right]\\
&-\int^{-\infty- i \pi}_{+\infty - i\pi}\frac{{\rm
d}t_2}{2\pi i}\,\frac{e^{\theta_4}}{e^{t_2}-e^{-\theta_2+\frac{i \pi}{k} n_2}} i \log \left[ \frac{\Gamma \left(1-\frac{n_1}{2} - \frac{ik}{2 \pi} (\theta_1+ t_2) \right)}{\Gamma \left(1+\frac{n_1}{2} + \frac{ik}{2 \pi} (\theta_1+ t_2) \right)} \right]\,,
\eal
\bal
\tPsi_{\R}^{+}(x^{n_1}_{\R1},x^{n_2}_{\R2})=&-\int^{+\infty}_{-\infty}\frac{{\rm
d}\theta_4}{2\pi i}\,\frac{e^{t_2}}{e^{t_2}-e^{-\theta_2+\frac{i \pi}{k} n_2}} i \log \left[ \frac{\Gamma \left(1-\frac{n_1}{2} - \frac{ik}{2 \pi} (\theta_1+ t_2) \right)}{\Gamma \left(1+\frac{n_1}{2} + \frac{ik}{2 \pi} (\theta_1+ t_2) \right)} \right]\\
&-\int^{-\infty+ i \pi}_{+\infty + i\pi}\frac{{\rm
d}\theta_4}{2\pi i}\,\frac{e^{t_2}}{e^{\theta_4}-e^{-\theta_2+\frac{i \pi}{k} n_2}} i \log \left[ \frac{\Gamma \left(1-\frac{n_1}{2} - \frac{ik}{2 \pi} (\theta_1+ t_2) \right)}{\Gamma \left(1+\frac{n_1}{2} + \frac{ik}{2 \pi} (\theta_1+ t_2) \right)} \right]\,.
\eal
If $t_2 \to - \infty$ the integrands are exponentially suppressed. If $t_2 \to + \infty$ the integrands are not suppressed; however, they are exponentially suppressed in the difference
\bal
\tPsi^{\pm}_{\R} (x^+_{\R1},x^-_{\R2})-\tPsi^{ \pm}_{\R} (x^+_{\R1},x^+_{\R2})
\eal
and therefore they are exponentially suppressed in the full phase. Due to this fact we can replace the integration lines with the following closed rectangular contours:
\begin{enumerate}
\item $\Gamma^{-}$: it is a clockwise rectangle having the upper side on the real axis and the lower side at constant imaginary part $-i \pi$.
\item $\Gamma^{+}$:   it is a counterclockwise rectangle having the lower side on the real axis and the upper side at constant imaginary part $+i \pi$.
\end{enumerate}
Working with the derivative of the phase we can then write the $\tPsi$ functions in the following way
\bal
&\frac{d}{d \theta_1}\tPsi_{\R}^{-}(x^{n_1}_{\R1},x^{n_2}_{\R2})=\\
&-\frac{k}{2 \pi}\oint\limits_{\Gamma^{-}}\frac{{\rm d}t_2}{2\pi i}\,\frac{e^{t_2}}{e^{t_2}-e^{-\theta_2+\frac{i \pi}{k} n_2}}  \left[ \psi \left(1-\frac{n_1}{2} - \frac{ik}{2 \pi} (\theta_1+ t_2) \right) + \psi \left(1+\frac{n_1}{2} + \frac{ik}{2 \pi} (\theta_1+ t_2) \right) \right]\,,
\eal
\bal
&\frac{d}{d \theta_1}\tPsi_{\R}^{+}(x^{n_1}_{\R1},x^{n_2}_{\R2})=\\
&-\frac{k}{2 \pi}\oint\limits_{\Gamma^{+}}\frac{{\rm d}\theta_4}{2\pi i}\,\frac{e^{t_2}}{e^{t_2}-e^{-\theta_2+\frac{i \pi}{k} n_2}}  \left[ \psi \left(1-\frac{n_1}{2} - \frac{ik}{2 \pi} (\theta_1+ t_2) \right) + \psi \left(1+\frac{n_1}{2} + \frac{ik}{2 \pi} (\theta_1+ t_2) \right) \right] \,.
\eal
The $\psi$ functions have the following poles.
\begin{enumerate}
\item The poles of $\psi \left(1-\frac{n_1}{2} - \frac{ik}{2 \pi} (\theta_1+ t_2) \right)$ are located at
\bal
t_2=-\theta_1+ \frac{2 \pi i}{k} (\frac{n_1}{2} - j) \,, \quad j=1, 2, \dots
\eal
and contribute to $\tPsi_{\R}^{-}$. In particular if $n_1=+1$ then the poles associated with $j=1, \dots, l$ are in the contour $\Gamma^-$; instead for $n_1=-1$ the poles inside $\Gamma^-$ are the ones associated with $j=1, \dots, l-1$.
\item The poles of $\psi \left(1+\frac{n_1}{2} + \frac{ik}{2 \pi} (\theta_1+ t_2) \right)$ are located at
\bal
t_2=-\theta_1+ \frac{2 \pi i}{k} (\frac{n_1}{2} + j) \,, \quad j=1, 2, \dots
\eal
and contribute to $\tPsi_{\R}^{+}$. For $n_1=+1$ the poles associated with $j=1, \dots, l-1$ are in the contour $\Gamma^+$; for $n_1=-1$ the poles in the contour are the ones associated with $j=1, \dots, l$.
\end{enumerate}
Additionally there is a pole at 
\bal
t_2=-\theta_2+\frac{i \pi}{k} n_2
\eal
that for $n_2=-1$ is in the integration contour $\Gamma^{-}$ and for $n_2=+1$ is in the integration contour of $\Gamma^{+}$.
Summing over the poles we obtain
\bal
&\frac{d}{d \theta_1}\tPsi_{\R}^{-}(x^{n_1}_{\R1},x^{n_2}_{\R2})=\\
&+\frac{k}{2 \pi} \theta(-n_2) \left[ \psi \left(1-\frac{ik}{2 \pi} \theta_{12} - \frac{1}{2}(n_1 - n_2)  \right) + \psi \left(1+\frac{ik}{2 \pi} \theta_{12} + \frac{1}{2}(n_1 - n_2)  \right) \right]\\
&- i \theta(n_1) \frac{   e^{- \theta_1 + \frac{i \pi}{k}}   }{e^{- \theta_1 + \frac{i \pi}{k}} + e^{-\theta_2 + \frac{i \pi}{k}n_2}     }  - i \sum^{l-1}_{j=1} \frac{   e^{- \theta_1 + \frac{i \pi}{k} n_1 - \frac{2\pi i}{k}j}   }{e^{- \theta_1 + \frac{i \pi}{k}n_1-\frac{2\pi i}{k}j}  -  e^{-\theta_2 + \frac{i \pi}{k} n_2}     } \,,
\eal
\bal
&\frac{d}{d \theta_1}\tPsi_{\R}^{+}(x^{n_1}_{\R1},x^{n_2}_{\R2})=\\
&-\frac{k}{2 \pi} \theta(+n_2) \left[ \psi \left(1-\frac{ik}{2 \pi} \theta_{12} - \frac{1}{2}(n_1 - n_2)  \right) + \psi \left(1+\frac{ik}{2 \pi} \theta_{12} + \frac{1}{2}(n_1 - n_2)  \right) \right]\\
&- i \theta(-n_1) \frac{   e^{- \theta_1 - \frac{i \pi}{k}}   }{e^{- \theta_1 - \frac{i \pi}{k}} + e^{-\theta_2 + \frac{i \pi}{k}n_2}     }  - i \sum^{l-1}_{j=1} \frac{   e^{- \theta_1 + \frac{i \pi}{k} n_1 + \frac{2\pi i}{k}j}   }{e^{- \theta_1 + \frac{i \pi}{k}n_1+\frac{2\pi i}{k}j}  - e^{-\theta_2 + \frac{i \pi}{k} n_2}     } \,.
\eal
Combining the four $\tPsi$ functions (we recall that each one of them is given by the average of  $\tPsi_{\R}^{-}$ and $\tPsi_{\R}^{+}$) most of the terms in the sums cancel and using that $l=\frac{k}{2}$ we end up with 
\bal
\label{eq:rel_lim_psi_K}
\tPsi_{\R}(x^+_{\R2},x^{+}_{\R1}) &-\tPsi_{\R}(x^+_{\R1},x^{+}_{\R2}) +\tPsi_{\R}(x^+_{\R1},x^{-}_{\R2}) - \tPsi_{\R}(x^+_{\R2},x^{-}_{\R1})-K^\bes(x^+_{\R1},x^{+}_{\R2}) \\
&= \frac{1}{2i} \log \left[ \frac{\theta_{12} - \frac{2 \pi i}{k}}{\theta_{12} + \frac{2 \pi i}{k}} \, \frac{\sinh \left( \frac{\theta_{12}}{2} - \frac{i \pi}{k} \right) } {\sinh \left( \frac{\theta_{12}}{2} + \frac{i \pi}{k} \right)} 
\left(\frac{\Gamma[1+\frac{ik}{2 \pi} \theta_{12}]}{\Gamma[1-\frac{ik}{2 \pi} \theta_{12}]}\right)^2 \right]  \,.
\eal

\paragraph{Limit of $\tPhi$ functions.}

As before, we use the parameterisation $w_1 = e^{t_1}$, $w_2 = e^{t_2}$. Due to the suppression of the integrand for $t_1, t_2 \to \pm \infty$, even in this case we can substitute the integration lines of $\tPhi_{\R\R}$ with the closed rectangular contours $\Gamma^-$ and $\Gamma^+$ defined above. After differentiating with respect to $\theta_1$ and integrating by parts we obtain the following relativistic limit for the $\tPhi$ functions
 \bal
&\frac{d}{d \theta_1}\tPhi^{--}_{\R\R}(x^{n_1}_{\R1}, x^{n_2}_{\R2}) = -  \frac{k}{2 \pi}\oint_{\Gamma^-} \frac{d t_1}{2 \pi i}  \oint_{\Gamma^-} \frac{d t_2}{2 \pi i} \\
&\frac{e^{-\theta_1+ \frac{i\pi}{k}n_1}e^{t_2}}{\left(e^{-\theta_1+\frac{i \pi}{k} n_1} - e^{t_1} \right)  \left(e^{-\theta_2+\frac{i \pi}{k} n_2} - e^{t_2} \right)}\left(\psi \left( 1+ \frac{i k}{2 \pi} t_{12}  \right) +\psi \left( 1- \frac{i k}{2 \pi} t_{12}  \right) \right)\,,
\eal
 \bal
&\frac{d}{d \theta_1}\tPhi^{++}_{\R\R}(x^{n_1}_{\R1}, x^{n_2}_{\R2}) = -  \frac{k}{2 \pi}\oint_{\Gamma^+} \frac{d t_1}{2 \pi i}  \oint_{\Gamma^+} \frac{d t_2}{2 \pi i} \\
&\frac{e^{-\theta_1+ \frac{i\pi}{k}n_1}e^{t_2}}{\left(e^{-\theta_1+\frac{i \pi}{k} n_1} - e^{t_1} \right)  \left(e^{-\theta_2+\frac{i \pi}{k} n_2} - e^{t_2} \right)}\left(\psi \left( 1+ \frac{i k}{2 \pi} t_{12}  \right) +\psi \left( 1- \frac{i k}{2 \pi} t_{12}  \right) \right)\,.
\eal
The $\psi$ functions have the following poles:
\begin{enumerate}
\item $\psi(1- \frac{ik}{2 \pi} t_{12})$ has poles located at
\bal
t_1=t_2-\frac{2 i \pi}{k} j\,, \qquad j=1, 2, \dots \,.
\eal
If $t_2 \in \mathbb{R}$ then the poles associated with $j=1, \dots, l-1$ are inside the contour $\Gamma^-$. The pole associated with $j=l$ is on the lower side of the contour $\Gamma^-$. If $t_2 \in \mathbb{R}+ i \pi$ then the poles associated with $j=1, \dots, l-1$ are inside the contour $\Gamma^+$. In this case the pole associated with $j=l$ is on the lower side of the contour $\Gamma^+$.

\item $\psi(1+ \frac{ik}{2 \pi} t_{12})$ has poles located at
\bal
t_1=t_2 +\frac{2 i \pi}{k} j\,, \qquad j=1, 2, \dots \,.
\eal
If $t_2 \in \mathbb{R}$ the poles associated with $j=1, \dots, l-1$ are inside the contour $\Gamma^+$ and the pole associated with $j=l$ is on the upper side of the contour $\Gamma^+$. If $t_2 \in \mathbb{R}- i \pi$ the poles associated with $j=1, \dots, l-1$ are inside the contour $\Gamma^-$ and the pole associated with $j=l$ is on the upper side of the contour $\Gamma^-$.
\end{enumerate}
As already mentioned, we deal with the poles on the contour by using the principal value prescription.
We split the result for 
$$
\tPhi_{\R\R}=\frac{1}{2} \left(\tPhi^{--}_{\R\R}+\tPhi^{++}_{\R\R}\right)
$$
into a contribution arising from the poles of $\psi$ functions (which we label by $f_0^{(n_1, n_2)}$) and a contribution arising from the poles of the trigonometric prefactors (which we label by $\tPhi^{\psi}$). In the principal value prescription the poles on the contour contribute with half of their residues and we obtain
\bal
\frac{d}{d \theta_1}\tPhi_{\R\R}(x^{n_1}_{\R1}, x^{n_2}_{\R2}) &=\tPhi^{\psi}(x^{n_1}_{\R1}, x^{n_2}_{\R2})+ \int^{+\infty}_{-\infty} \frac{d t_2}{2 \pi i} f^{(n_1, n_2)}(t_2)
\eal
where
\bal
\label{eq:Phi_psi}
\tPhi^{\psi}(x^{n_1}_{\R1}, x^{n_2}_{\R2}) \equiv& -  \frac{k}{4 \pi} \theta(-n_1)  \oint_{\Gamma^-} \frac{d \theta_4}{2 \pi i} \frac{e^{t_2}}{e^{-\theta_2+\frac{i \pi}{k} n_2} - e^{t_2} }\left(\psi \left( 1- \frac{i k}{2 \pi} (\theta_1+t_2) - \frac{n_1}{2} \right)\right.\\
&\hspace{6.5cm}\left.+\psi \left( 1+ \frac{i k}{2 \pi} (\theta_1+t_2) + \frac{n_1}{2} \right) \right.
\\
&+  \frac{k}{4 \pi} \theta(n_1)  \oint_{\Gamma^+}\frac{d \theta_4}{2 \pi i} \frac{e^{t_2}}{e^{-\theta_2+\frac{i \pi}{k} n_2} - e^{t_2} }\left(\psi \left( 1- \frac{i k}{2 \pi} (\theta_1+t_2) - \frac{n_1}{2} \right)\right.\\
&\hspace{6.5cm}\left.+\psi \left( 1+ \frac{i k}{2 \pi} (\theta_1+t_2) + \frac{n_1}{2} \right) \right.\,,
\eal
and
\bal
\label{eq:app_fn1n2_defin}
&f^{(n_1, n_2)}(t_2) \equiv  \frac{i\, e^{-\theta_1+ \frac{i\pi}{k} n_1} e^{t_2}}{\left(e^{-\theta_1+ \frac{i\pi}{k} n_1} -e^{t_2} \right) \left(e^{-\theta_2+ \frac{i\pi}{k} n_2} +e^{t_2} \right)}- \frac{i\, e^{-\theta_1+ \frac{i\pi}{k} n_1} e^{t_2}}{\left(e^{-\theta_1+ \frac{i\pi}{k} n_1} +e^{t_2} \right) \left(e^{-\theta_2+ \frac{i\pi}{k} n_2} -e^{t_2} \right)}\\
&+\frac{i}{2} \sum^{l-1}_{j=1} \left[ \frac{e^{-\theta_1+ \frac{i\pi}{k} n_1} e^{t_2}}{\left(e^{-\theta_1+ \frac{i\pi}{k} n_1} +e^{t_2+ \frac{2i\pi}{k} j} \right) \left(e^{-\theta_2+ \frac{i\pi}{k} n_2} +e^{t_2} \right)} \right.
\\
&- \frac{e^{-\theta_1+ \frac{i\pi}{k} n_1} e^{t_2}}{\left(e^{-\theta_1+ \frac{i\pi}{k} n_1} -e^{t_2- \frac{2i\pi}{k} j} \right) \left(e^{-\theta_2+ \frac{i\pi}{k} n_2} -e^{t_2} \right)} +\frac{e^{-\theta_1+ \frac{i\pi}{k} n_1} e^{t_2}}{\left(e^{-\theta_1+ \frac{i\pi}{k} n_1} +e^{t_2- \frac{2i\pi}{k} j} \right) \left(e^{-\theta_2+ \frac{i\pi}{k} n_2} +e^{t_2} \right)}
\\
&\hspace{5cm}\left. - \frac{e^{-\theta_1+ \frac{i\pi}{k} n_1} e^{t_2}}{\left(e^{-\theta_1+ \frac{i\pi}{k} n_1} -e^{t_2+ \frac{2i\pi}{k} j} \right) \left(e^{-\theta_2+ \frac{i\pi}{k} n_2} -e^{t_2} \right)}\right]\,.
\eal
The contributions to $f^{(n_1, n_2)}$ arise from poles contained in the integration contours (which are collected in the sum over $l-1$ terms in the expression above) and poles on the contours (which are collected on the first line of the expression above and have been computed with the principal value prescription).

The expression in~\eqref{eq:Phi_psi} is immediately computed using the results obtained previously for the $\tPsi$ functions. We find that
\bal
&\tPhi^{\psi}(x^{n_1}_{\R1}, x^{n_2}_{\R2}) =  \frac{1}{2}\frac{d}{d \theta_1} \left( \theta (n_1)\tPsi^{++}_{\R\R}(x^{n_1}_{\R1}, x^{n_2}_{\R2}) -  \theta (-n_1)\tPsi^{--}_{\R\R}(x^{n_1}_{\R1}, x^{n_2}_{\R2}) \right)=\\
& +\frac{1}{2i} \left( \theta(-n_1) \theta(-n_2) + \theta(n_1) \theta(n_2) \right) \frac{d}{d \theta_{12}} \log \left( \frac{\Gamma \left(1- \frac{ik}{2 \pi} \theta_{12} - \frac{1}{2} (n_1-n_2) \right)}{\Gamma \left(1+ \frac{ik}{2 \pi} \theta_{12} + \frac{1}{2} (n_1-n_2) \right)} \right) \\
&  - \frac{i}{2} \theta(n_1) \sum^{l-1}_{j=1} \frac{   e^{- \theta_1 + \frac{i \pi}{k} n_1 + \frac{2\pi i}{k}j}   }{e^{- \theta_1 + \frac{i \pi}{k}n_1+\frac{2\pi i}{k}j} - e^{-\theta_2 + \frac{i \pi}{k} n_2}     }   + \frac{i}{2} \theta(-n_1) \sum^{l-1}_{j=1} \frac{   e^{- \theta_1 + \frac{i \pi}{k} n_1 - \frac{2\pi i}{k}j}   }{e^{- \theta_1 + \frac{i \pi}{k}n_1-\frac{2\pi i}{k}j} - e^{-\theta_2 + \frac{i \pi}{k} n_2}     } \,.
\eal
The sums over $l-1$ terms simplify in the full phase and we obtain
\bal
\label{eq:PhiPsicombined}
&\tPhi^{\psi}(x^{+}_{\R1}, x^{+}_{\R2})+\tPhi^{\psi}(x^{-}_{\R1}, x^{-}_{\R2})-\tPhi^{\psi}(x^{+}_{\R1}, x^{-}_{\R2})-\tPhi^{\psi}(x^{-}_{\R1}, x^{+}_{\R2})=\\
& \frac{1}{2i} \frac{d}{d \theta_{12}} \log \left( \frac{\Gamma^2 \left( 1 - \frac{i k}{2 \pi} \theta_{12} \right)}{\Gamma^2 \left( 1 + \frac{i k}{2 \pi} \theta_{12} \right)} \right)  - \frac{1}{2} \frac{\sin \left( \frac{2 \pi}{k} \right)}{\cosh \theta_{12}  - \cos \left( \frac{2 \pi}{k} \right)}\,.
\eal
To compute the remaining part of the improved BES phase we introduce the function
\bal
f^{(n_2)}(t_2) &\equiv f^{(+, n_2)}(t_2) - f^{(-, n_2)}(t_2)\\
&=\frac{i}{2} \frac{e^{ (t_2 - \theta_1)}}{e^{-\theta_2+ \frac{i \pi}{k} n_2} - e^{t_2}} \left[ \frac{1}{e^{-\theta_1} - e^{t_2 - \frac{i \pi}{k}}} - \frac{1}{e^{-\theta_1} - e^{t_2 + \frac{i \pi}{k}}}  \right]\\
&+\frac{i}{2} \frac{e^{ (t_2 - \theta_1)}}{e^{-\theta_2+ \frac{i \pi}{k} n_2} + e^{t_2}} \left[ \frac{1}{e^{-\theta_1} + e^{t_2 + \frac{i \pi}{k}}} - \frac{1}{e^{-\theta_1} + e^{t_2 - \frac{i \pi}{k}}}  \right] \,,
\eal
where the last equality in the expression above has been obtained by substituting~\eqref{eq:app_fn1n2_defin} and simplifying the sums over $l-1$ terms.
Integrating then we obtain
\bal
\label{eq:app_integral_fpminusfm}
&\int^{+\infty}_{-\infty} \frac{d t_2}{2 \pi i} \left(f^{(+)}(t_2) - f^{(-)}(t_2) \right)=-\frac{1}{2} \frac{\sin\frac{2\pi}{k}}{\cosh(\theta_{12}) -\cos\frac{2\pi}{k}}\\
&+\frac{1}{4\pi} \left(
(\theta_{12} + \tfrac{2 \pi i}{k})\coth\frac{\theta_{12} + \frac{2 \pi i}{k}}{2}+(\theta_{12} - \tfrac{2 \pi i}{k})\coth\frac{\theta_{12} - \frac{2 \pi i}{k}}{2}
-2\,\theta_{12} \coth\frac{\theta_{12}}{2}
\right) \,.
\eal
Combining~\eqref{eq:PhiPsicombined} and~\eqref{eq:app_integral_fpminusfm} we obtain
\bal
\label{eq:tPhi_final_rel_limit}
&\frac{d}{d \theta_{12}} \left(\tPhi_{\R\R}(x^{+}_{\R1}, x^{+}_{\R2})+\tPhi_{\R\R}(x^{-}_{\R1}, x^{-}_{\R2}) - \tPhi_{\R\R}(x^{+}_{\R1}, x^{-}_{\R2}) - \tPhi_{\R\R}(x^{-}_{\R1}, x^{+}_{\R2}) \right)=\\
&  +\frac{1}{2i} \frac{d}{d \theta_{12}} \log \left(
\left(\frac{\sinh(\frac{\theta_{12}}{2}+\frac{i\pi }{k})}{\sinh(\frac{\theta_{12}}{2}-\frac{i\pi}{k})}\right)^2
\frac{\Gamma^2 \left( 1 - \frac{i k}{2 \pi} \theta_{12} \right)}{\Gamma^2 \left( 1 + \frac{i k}{2 \pi} \theta_{12} \right)}
\frac{R^4(\theta)}{R^2(\theta+\frac{2\pi i}{k})R^2(\theta-\frac{2\pi i}{k})}\right).
\eal
Finally using~\eqref{eq:rel_lim_psi_K} and~\eqref{eq:tPhi_final_rel_limit}, and integrating wrt $\theta_{12}$, we have
\bal
\label{eq:rel_lim_BES}
&\exp \left[ -2i \tilde{\theta}^{\bes}_{\R\R}(x^{\pm}_{ \R1},x^{\pm}_{ \R2}) \right] \to   \frac{\theta_{12} + \frac{2 \pi i}{k}}{\theta_{12} - \frac{2 \pi i}{k}} \ 
\frac{\sinh(\frac{\theta_{12}}{2}-\frac{i\pi }{k})}{\sinh(\frac{\theta_{12}}{2}+ \frac{i\pi}{k})} \ \frac{R^2(\theta_{12}+\frac{2\pi i}{k})R^2(\theta_{12}-\frac{2\pi i}{k})}{R^4(\theta_{12})} \,.
\eal

From~\eqref{eq:rel_lim_BES} and~\eqref{eq:rel_lim_HL} then we obtain the following relativistic limit for the even part of the Right-Right dressing factor
\bal
\label{eq:rel_limit_even}
 \big(\Sigma^{\besratio}_{\R\R}\big)^{-2} (u_1, u_2)&= \exp \left[ -2i \tilde{\theta}^{\bes}_{\R\R}(x^{\pm}_{ \R1},x^{\pm}_{ \R2}) +2i \tilde{\theta}^{\hl}_{\R\R}(x^{\pm}_{ \R1},x^{\pm}_{ \R2})\right]\\
 & \to \frac{\theta_{12} - \frac{2 \pi i}{k}}{\theta_{12} + \frac{2 \pi i}{k}} \ 
\frac{\sinh(\frac{\theta_{12}}{2}+\frac{i\pi }{k})}{\sinh(\frac{\theta_{12}}{2}-\frac{i\pi}{k})} \,.
\eal
If we now substitute~\eqref{eq:normalisation_final_rel_limit}, \eqref{eq:Barnes_final_rel_limit} and~\eqref{eq:rel_limit_even} into~\eqref{eq:S_mat_right_particles_before_rel_lim} we obtain the S-matrix element~\eqref{eq:app_rel_limit_right}, in agreement with the result of~\cite{Frolov:2023lwd}.

\bibliographystyle{JHEP}
\bibliography{refs}
\end{document}